\documentclass[]{aa}  
\usepackage[varg]{txfonts}
\usepackage{comment}
\usepackage{graphicx}
\usepackage{txfonts}
\usepackage{xcolor}
\usepackage{natbib,twoopt}
\usepackage{dcolumn}
\usepackage{booktabs,textcomp}
\usepackage[normalem]{ulem}
\usepackage{multirow}
\usepackage{pdflscape}
\usepackage{lscape}
\usepackage{longtable}
\usepackage{soul}
\usepackage[breaklinks=true]{hyperref}
    \hypersetup{
        colorlinks=true,
        linkcolor=blue,
        filecolor=magenta,      
        urlcolor=blue,
        citecolor=blue
    }
\hypersetup{draft}
\definecolor{olivegreen}{rgb}{0.02, 0.8, 0.24}
\newcommand\stoutAna{\bgroup\markoverwith{\textcolor{olivegreen}{\rule[0.5ex]{2pt}{0.4pt}}}\ULon}

\newcommand\stoutNick{\bgroup\markoverwith{\textcolor{blue}{\rule[0.5ex]{2pt}{0.4pt}}}\ULon}

\begin{document}
\newcommand{\csixty}{C$_{60}$\xspace}
\newcommand{\cseventy}{C$_{70}$\xspace}
\newcommand{\csixtyplus}{C$_{60}^{+}$\xspace}
\newcommand{\andres}[1]{\textcolor{purple}{{Andres: #1}}}
\newcommand{\anna}[1]{\textcolor{orange}{{Anna: #1}}}
\newcommand{\meriem}[1]{\textcolor{green}{{Meriem: #1}}}
\newcommand{\keith}[1]{\textcolor{red}{{Keith: #1}}}
\newcommand{\henri}[1]{\textcolor{brown}{{HB: #1}}}
\title{A high resolution study of near-IR diffuse interstellar bands, search for small scale structure, time variability and stellar features}
\titlerunning{Near Infrared DIBs}       
\author{J. V. Smoker\inst{1,2}
\and
A. M{\"u}ller\inst{3}
\and
A. Monreal Ibero\inst{4,5,6}
\and
M. Elyajouri\inst{7}
\and
C. J. Evans\inst{2,8}
\and
F.~Najarro\inst{9}
\and
A. Farhang\inst{10}
\and
N. L. J. Cox\inst{11}
\and
J.~Minniti \inst{12}
\and
K.~T.~Smith\inst{13} 
\and
J. Pritchard\inst{14} 
\and
R. Lallement\inst{7}
\and
A. Smette\inst{1}
\and
H. M. J. Boffin\inst{14} 
\and
M. Cordiner\inst{15}
\and
J.~Cami\inst{16, 17, 18}
}
\institute{
European Southern Observatory, Alonso de Cordova 3107, Vitacura, Santiago, Chile
\and  
UK Astronomy Technology Centre, Royal Observatory, Blackford Hill, Edinburgh EH9 3HJ, UK
\and
Max-Planck-Institut fur Astronomie,  Königstuhl 17. 69117 Heidelberg, Germany
\and 
Leiden Observatory, Leiden University, PO Box 9513, 2300 RA Leiden, The Netherlands 
\and
Instituto de Astrof\'isica de Canarias, C/ V\'{\i}a L\'actea s/n, E-38205 La Laguna, Tenerife, Spain
\and
Departamento de Astrof\'isica, Universidad de La Laguna, E-38200 La Laguna, Tenerife, Spain
\and
GEPI, Observatoire de Paris, PSL Research University, CNRS, Place Jules Janssen, 92190 Meudon, France
\and
European Space Agency, ESA Office, Space Telescope Science Institute, 3700 San Martin Drive, Baltimore, MD 21218, USA 
\and
Departamento de Astrof\'isica, 
Centro de Astrobiolog\'ia, 
CSIC-INTA, 
Ctra. Torregjon a Ajalvir km. 4,
28850-Madrid,
Spain
\and
School of Astronomy, Institute for Research in Fundamental Sciences, 19395-5531 Tehran, Iran
\and
ACRI-ST, Centre d’Etudes et de Recherche de Grasse (CERGA), 10 Av. Nicolas Copernic, 06130 Grasse, France
\and
Nicolaus Copernicus Astronomical Center, Polish Academy of Sciences, Bartycka 18, 00-716 Warsaw, Poland
\and
AAAS Science International, Clarendon House, Clarendon Road, Cambridge CB2 8FH, UK
\and
European Southern Observatory, Karl-Schwarzschild-Str. 2 85748 Garching bei München, Germany
\and
NASA Goddard Space Flight Center, 8800 Greenbelt Road, Greenbelt, MD 20771, USA
\and
Department of Physics and Astronomy, The University of Western Ontario, London, ON N6A 3K7, Canada
\and
Institute for Earth and Space Exploration, The University of Western Ontario, London, ON N6A 3K7, Canada
\and
SETI Institute, 189 Bernardo Ave, Suite 100, Mountain View, CA 94043, USA
}

\abstract{The diffuse interstellar bands (DIBs) are a set of hundreds of unidentified absorption features that appear near ubiquitously throughout the interstellar medium. Most DIBs appear at optical wavelengths, but some are in the near-infrared.}
{We aim to characterise near-infrared DIBs at high spectral resolving power towards multiple targets.}
{We observed 76 early-type stars at a resolving power of 50,000 (velocity resolution $\sim$6 kms$^{-1}$) and signal to noise ratios of several hundreds using the CRyogenic high-resolution InfraRed Echelle Spectrograph (CRIRES). These data allow us to investigate the DIBs around 1318.1, 1527.4, 1561.1, 1565.1, 1567.0, 1574.4 and/or 1624.2~nm. We detect a total of six DIB features and 17 likely stellar features assisted by comparisons with a model spectrum computed with {\sc cmfgen}. 
Additionally, we measured equivalent widths of the DIBs at 1318.1 and 1527.4~nm using observations with X-shooter towards ten very highly reddened (3.2 $< E(B-V) < $ 6.5) Cepheid variable stars, towards four stars observed at low values of precipitable water vapour and using other archive data.}
{We measure correlations (correlation coefficient $r\sim$0.73--0.96, depending on the sub-sample used) between DIB equivalent width and reddening for the DIBs at 
1318.1, 1561.1, 1565.1 and 1567.0~nm. Comparing the near-infrared DIBs with 50 of the strongest optical DIBs, we find correlations $r$ > 0.8 between the 1318, 1527, 1561, 1565 or 1567~nm 
and optical DIBs 5705, 5780, 6203, 6283 and 6269~\AA. The 5797~\AA\, DIB is less well correlated with the near-infrared DIBs. The DIB at 9632.1~\AA, (likely C$_{60}^+$), is not well correlated with the 1318.1~nm DIB. Partial correlation coefficients using $E(B-V)$ as the covariate were also determined.
For stars earlier than B2, the 1318.1~nm DIB is affected by an emission line on its blue wing that is likely stellar in nature, although we cannot rule out an interstellar or circumstellar origin for this line caused by for example a DIB in emission. The 1318.1~nm DIB also has an extended red wing. The line is reasonably well fitted by two gaussian components although neither the component ratios nor separation between them are obviously correlated with indicators of such as $\lambda\lambda$5780/5797 and reddening. EW(1318~nm) correlates with H\,{\sc i} with EW(1318~nm)/$E(B-V)$ decreasing with f(H$_{2}$). 

Five pairs of stars within one arcmin of each other show very similar 1318.1~nm DIB profiles. Possible variation in the 1318.1~nm feature is seen between HD\,145501 and HD\,145502 (separated by 41~arcsec, equivalent to 7200~au) and HD\,168607 and HD\,168625 (separated by 67~arcsec, equivalent to 0.52~pc on the plane of the sky).  
Seventeen sightlines have repeat CRIRES observations separated by 6--14 months and two sightlines have repeat X-shooter observations separated by 9.9 years. No time-variability is detected in the 1318.1~nm DIB in the CRIRES data or the 5780.5, 5797.1 1318.1 and 1527.4~nm DIBs. Tentative time variation is observed in the "C$_{60}^+$" DIBs at 9577 and 9632~\AA \, towards HD\,183143 although very close to the noise level and with confirmation required.}
{The NIRDIBs observed occur more in more UV-irradiated regions than the 5797~\AA \, DIB allowing the study of heavily reddened sightlines. Future searches for time variability in DIBs require higher quality data and/or larger intervals between epochs.}
\maketitle
\section{Introduction}
\label{intro}
The diffuse interstellar bands (DIBs) were discovered around a century ago as unidentified absorption lines in the spectra of reddened stars \citep[e.g.][]{Heger22}.
The lines are generally several angstroms wide, so cannot be atomic, and were initially hypothesised to arise from dust grains or large molecules (see \citealt{Cox11} for a review). 
There is a correlation between reddening and DIB strength (although with large scatter at high reddening and with $E(B-V) \sim$0.2--0.3; \citealt{Wu1981}, \citealt{Herbig1993}, \citealt{Vos2011}), which initially favoured the dust hypothesis. However, 
a dust carrier was expected to produce emission in the wings of the DIB profiles \citep{van1946} which has not been detected, wavelength shifts and profile variations due to changes in the composition, size and structure of the dust grains that are not observed, and polarisation in the absorption lines which has also been ruled out to a level of less than 0.01 percent for 45 DIBs \citep{Cox2011a}.

The most supported hypothesis is that the DIB carriers are large gas-phase molecules.
Evidence for this includes the complex component structure of several DIBs which are reminiscent of molecular rotational band profiles \citep{Kerr1996}.
Potential carriers include the polycyclic aromatic hydrocarbon (PAH) molecules, which are known to be abundant in space from their mid-infrared bands \citep{Allamandola1989, Salama1996, Meeus2001}.
Several hundred DIBs have been identified in the interstellar medium (ISM), including in the Milky Way \citep[e.g.][]{Hobbs09} and nearby galaxies such as the Andromeda Galaxy \citep{Cordiner11}, Messier~33 \citep{Cordiner08b} the Large Magellanic Cloud \citep{Cox2006} and the Antennae Galaxies \citep{Monreal2018}.

Identifying the carriers via laboratory experiments is impractical as there are more than 1.2 million different species of PAHs with 100 or less carbon atoms which would need to be tested \citep{Cami2018}.
The only DIBs to have been convincingly assigned to a carrier are two (possibly four) DIBs due to ionised buckminsterfullerene C$_{60}^{+}$ (\citealt{Campbell15,Campbell2016,Kuhn16,Cordiner17,Cordiner18}; although see discussion by \citealt{Gala17,Lallement2018a}). 

Most studies of DIBs have concentrated on optical wavelengths \citep[e.g.][]{Merrill1938,Raimond12,Cox2017}\footnote{In this paper we use Angstroms for DIBs bluewards of 1000~nm and nm for redder wavelengths.}. However, laboratory measurements show that larger PAHs (more carbon atoms) have their strongest absorption lines in the near-infrared (NIR) \citep{Mattioda2005,2005A&A...432..515R}. Indeed, 
several DIBs have been observed in this region  (\citealt{Joblin90,Geballe11,Rawlings2011,Cox2014,Elyajouri2017a}).
These include twenty NIR DIBs that have been identified in observations of nine stars at a spectral resolving power ($R$) of $\sim$10,000 between 950 and 2500~nm \citep{Cox2014}. No DIBs were found redwards of $\sim$1800~nm.
The strongest correlations were found between the equivalent widths of some NIR DIBs (e.g. that at 1318.1~nm) and the optical DIB at 5780~\AA\ (correlation coefficient $r$=0.98). Additionally, many NIRDIBs were also found to correlate with reddening with $r\sim$0.9. 

Other studies include those investigating the 1527~nm DIB on a statistical basis towards $\sim$60,000 mainly late-type stars at $R$\,$\sim$\,22,000 using data from the APOGEE survey \citep{Zasowski15}. They found that this DIB persists in the ISM for long timescales 
and correlates with dust, but has 
a scale-height above the Galactic plane of $\sim$100~pc, twice that of dust.
\citet{2019A&A...628A..67E} also investigated this DIB using a larger sample from the APOGEE survey, providing measurements for 124\,064 sightlines and demonstrating that the carrier is concentrated in the outskirts of interstellar clouds. 
\cite{Hamano15,Hamano16} presented DIB observations from 0.91--1.32~$\mu$m at $R$=28,000.
They concluded that the 1078.0, 1079.2 and 1179.7~nm features likely belong to related carriers and correlate with the 5780~\AA\ DIB. 
\citet{Gala2017b} studied fourteen DIBs in the 1.45 to 2.45~$\mu$m range and measured basic correlations between the observed bands. \citet{Elyajouri2017a} confirmed several weak DIBs and found that the 1527~nm DIB is most abundant in diffuse interstellar clouds with low density and high ultraviolet radiation fields (known as $\sigma$-type clouds, with $\zeta$-type clouds having opposite properties). \cite{Hamano2022} looked at DIBs in the Y and J bands to postulate that DIBs at longer wavelength are caused by larger molecules. Finally, \cite{Ebenbichler2022_CRIRES} observed four stars at $R$=100,000 with CRIRES to confirm 17 new DIBs and identify a possible 12 additional DIBs.

We seek to extend previous work by observing multiple NIR DIBs at high spectral resolving power ($R\sim50,000$) and to better constrain the known correlations using data of higher signal to noise ratios (S/N).
Some optical DIBs are known to vary on small scales in the ISM (see Section \ref{smallscale} for references), so we also searched for such small-scale structure in the NIR DIBs. No optical DIBs are known to vary (convincingly) with time, but even so we also searched for time variability in the NIR DIBs using twin epoch observations of a number of sightlines.

This paper is organised as follows.
Section~\ref{sample} describes our sample of targets, observations, data reduction and telluric line removal.
Section~\ref{analysis} outlines the analysis of the data, including profile fitting.
The discussion is presented in Section~\ref{discussion}, including 
discrimination between stellar and interstellar or circumstellar features, the shape of the DIB profiles, a comparison of NIR DIB line strengths with optical DIBs, atomic lines and reddening, and finally a search for small-scale and time-variability in the NIR DIBs studied.
In Section~\ref{summary} we summarise our results and make suggestions for future work. 

\section{Sample, observations and data reduction}
\label{sample}

\subsection{The Sample -- CRIRES observations and data reduction}

Most of the sample was observed using CRIRES \citep{Kaeufl2004}, mounted on the 8.2-m Unit Telescope One (Antu) at the Very Large Telescope (VLT) under programme IDs 091.C-0655 (epoch 1) and 093.C-0480 (epoch 2). In epoch 1 the central wavelengths on detector 3 (of the four CRIRES detectors) were 1318.1~nm and/or 1574.4~nm and 1624.2~nm. In epoch 2 we observed many fewer stars but also including 1527.0~nm and 1568.8~nm settings for a few sightlines.
The wavelength coverage in each setting was small, ranging from 31~nm for the 1318.1~nm setting to 37~nm for the 1624.2~nm setting, with gaps of around 2~nm between the four detectors on the mosaic.
The vacuum wavelengths observed for each star are listed in Table~\ref{t_Sample}.
Observations were scheduled during poor weather conditions, so we used the widest slit possible with CRIRES of 0.4\arcsec\ 
and no adaptive optics for the majority of the stars, yielding a resolving power of $R\equiv\lambda/\Delta\lambda\sim50,000$, as measured from the telluric lines.

The CRIRES sample includes 67 O- and B-type stars (median magnitude in the $H$-band of 5.9) for observation between April and September 2013, and follow up of 17 stars with additional observations between March and June 2014. The latter group are either repeat observations of stars observed in the first epoch (to search for time-varying DIB strength), or companion stars in double or triple systems intended to search for small-scale structure in the ISM. The total sample comprised 76 stars. We favoured early-type stars as they typically have few stellar lines at the wavelengths of the NIR DIBs and are often fast rotators \citep[e.g.][]{Stutz2006}, making it easier to distinguish between stellar and interstellar features.
The CRIRES-observed sightlines have interstellar reddening ranging from 0 $\simeq E(B-V) \simeq$ 1.5~mag (median 0.6~mag), estimated from their observed $(B-V)$ colours and the intrinsic colour of the relevant spectral type taken from \citet{Fitzgerald1970} and \citet{Wegner1994}.

Total integration times ranged from 60~s to 1200~s depending on target brightness, with a median of 600~s.
An ABBA nodding pattern was used to facilitate sky subtraction.
For a handful of targets which had close companions, adaptive optics was used to separate the components. 
Data were reduced using the original CRIRES data reduction pipeline \footnote{https://www.eso.org/sci/software/pipelines/crires/crires-pipe-recipes.html} 
using instrument calibration data taken the morning after the observations and including a correction for detector non-linear response. The initial wavelength calibration was performed using exposures of a ThAr lamp, then refined during telluric correction (see below). The median S/N ratios of the reduced spectra in the 1318.1~nm setup were $\sim$300 per pixel ($\sim$550 per resolution element) with similar values for the other wavelength settings.

\subsection{X-shooter observations and archival data}

To increase the range of $E(B-V)$ sampled in this study, X-shooter  \citep{Vernet2011} spectra of ten stars with $m_{J}< 16.0$~mag. were taken from \citet{Minniti2020}, who studied Cepheid variables towards the Galactic Centre.
The S/N ratios were between 10 and 50 per pixel at 1318~nm and higher at 1527~nm, with $R\sim8,000$. These targets have extinction in the Ks band ranging from 1.13 to 2.25 mag, corresponding to $E(B-V)$ values between 3.13 and 6.51 ($A_V$ from $\sim$9.7 to 20.2), assuming the extinction law of \cite{Cardelli1989}.
Their distances range from 8.0 to 31.5~kpc.

Additionally, four stars were observed with X-shooter under dry weather conditions (precipitable water vapour PWV of less than 0.5~mm), with the aim of searching for time variation in the C$_{60}^+$ and NIR DIBs. Finally, we downloaded reduced spectra for a further nineteen X-shooter targets from the ESO archive to investigate the dependence on NIRDIB strength with fractional H$_{2}$ content. 

\begin{figure*}[]
   \resizebox{\hsize}{!}{\includegraphics{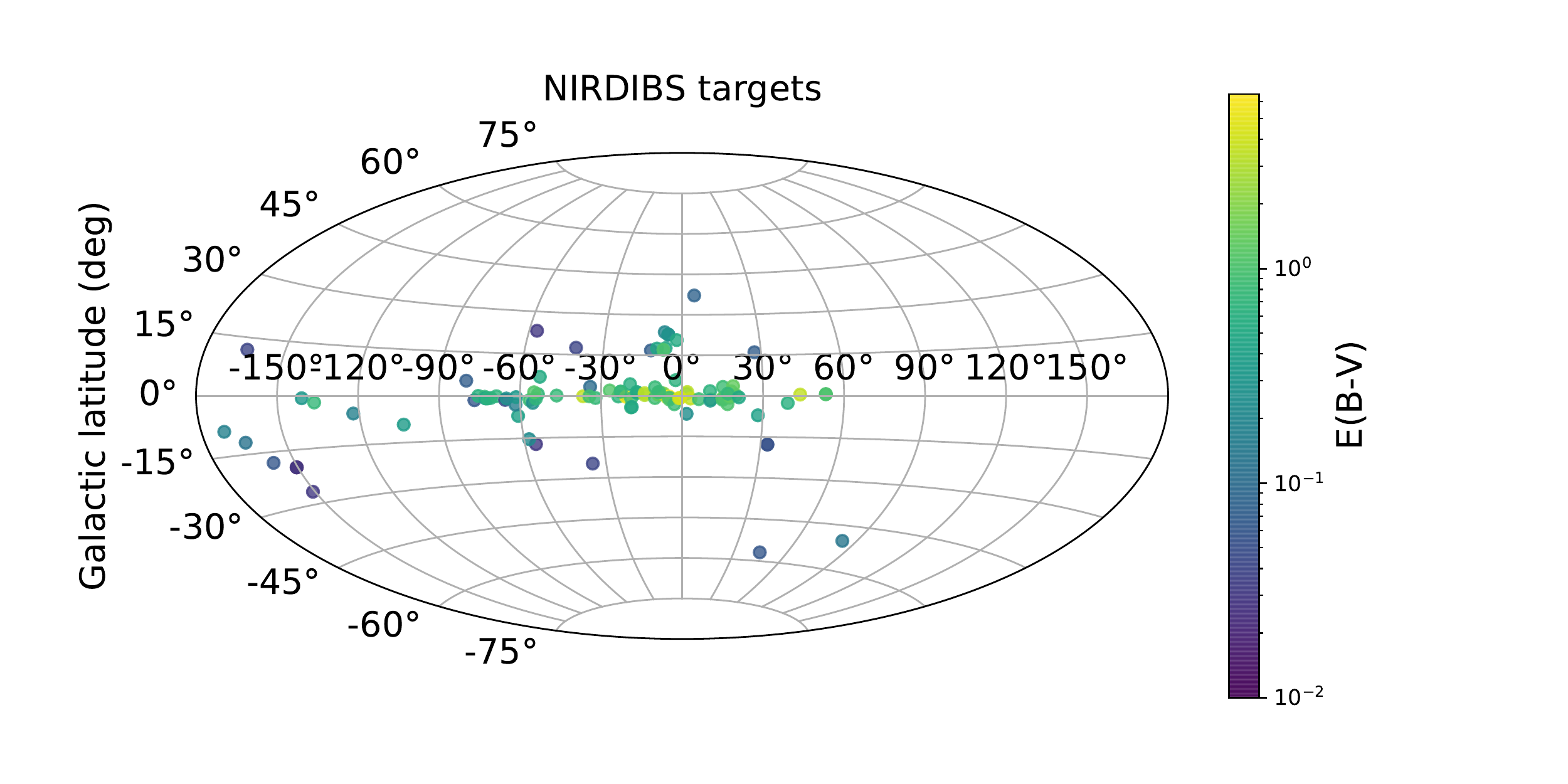}}
\caption{The distribution of our sample in Galactic coordinates, colour coded by reddening. Most targets are close to the Galactic plane, with the highest reddenings near the Galactic Centre.}
\label{f_sample}
\end{figure*}

Figure~\ref{f_sample} shows the distribution of the sample on the sky for the CRIRES and X-shooter spectra with  Table~\ref{t_Sample} listing basic information for all the observed stars.

\subsection{Telluric correction}

\begin{figure*}[tpb]
\resizebox{\hsize}{!}{\includegraphics{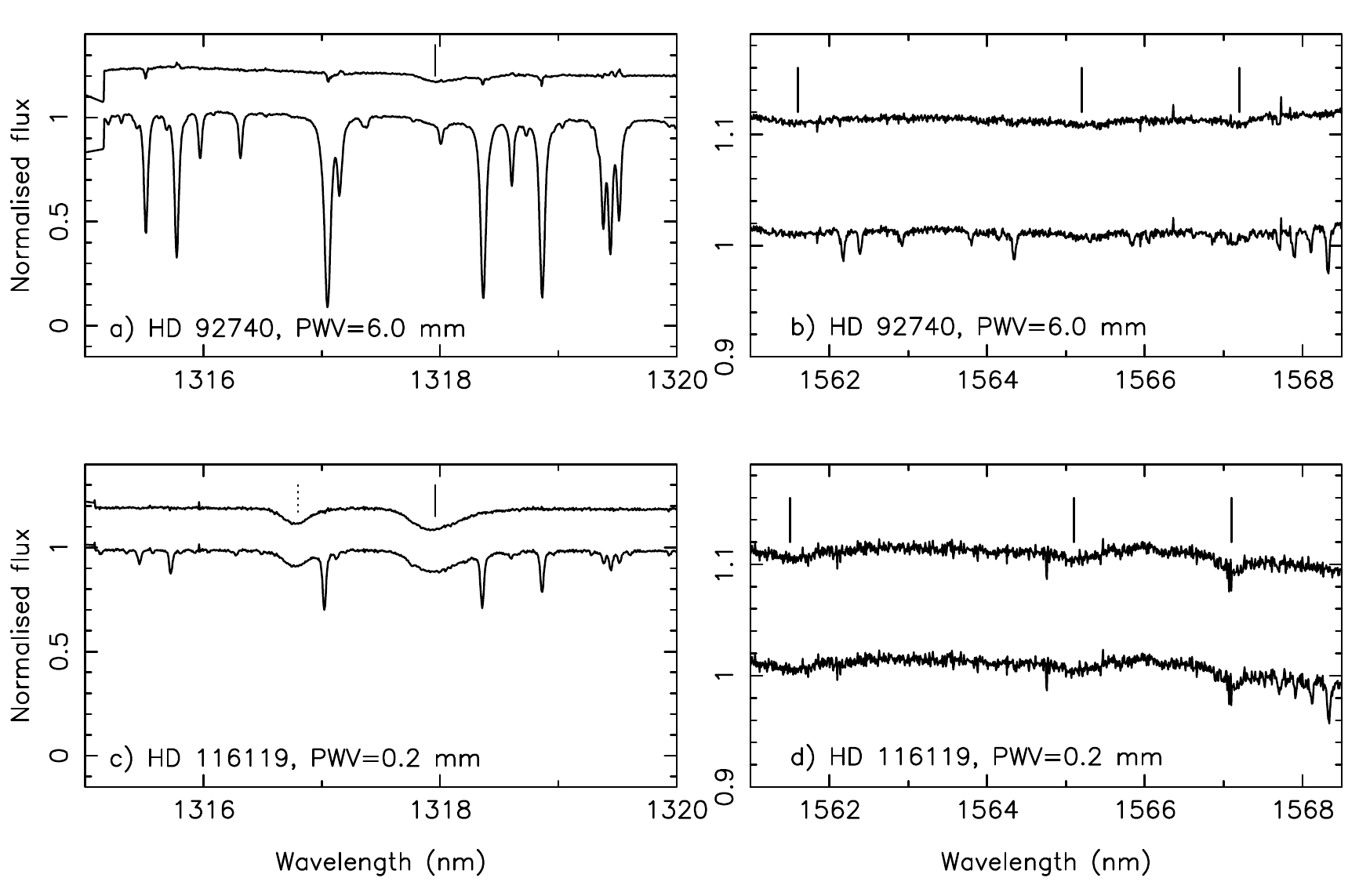}}
\caption{Examples for two wavelength regions of the use of {\sc molecfit} to remove telluric lines from spectra when the precipitable water vapour (PWV) was $\sim$6.0\,mm (plots (a) and (b)) and $\sim$0.2~mm (plots (c) and (d)), for HD\,92740 and HD\,116119. Spectra are normalised. 
The median night-time PWV for Paranal is 2.4~mm \citep{Kerber2014}. The lower spectra in each panel are
the reduced data, with the upper spectra the same objects after telluric correction (offset in the ordinate
for clarity). Vertical lines above the corrected spectra indicate the DIBs at 1318.1, 1561.1, 1565.1 and 1567.0~nm. Vacuum wavelengths are shown. The dashed vertical line displayed for HD\,116119 at a wavelength of 1316.8~nm indicates a likely stellar feature caused by an O\,{\sc i} triplet.}

\label{f_Molecfit_Example}
\end{figure*}

Correction for telluric lines was performed using the \textsc {molecfit} package\footnote{http://www.eso.org/sci/software/pipelines/skytools/molecfit} \citep{Smette2015}. The precipitable water vapour was fitted by the code. 
Example spectra before and after telluric line correction are shown in Fig.~\ref{f_Molecfit_Example} in two wavelength settings, each at low (0.2\,mm) and high (6.0\,mm) precipitable water vapour. 

\section{Data analysis}
\label{analysis}

The data analysis was straightforward and consisted in normalising the spectra with 
polynomials of order 3--4 either side of the DIB features and measurement of the 
equivalent widths (EWs) and full-width half-maximum (FWHM)  velocity-widths by fitting Gaussians to the detected components. Errors were calculated according to equation (10) of \cite{Ebbets1995}. Data were transformed to the Heliocentric restframe using {\sc rv}\footnote{http://www.starlink.ac.uk/docs/sun78.htx/sun78.html}.
Where the expected DIBs were not detected, five-sigma upper limits to the EWs were estimated 
according to:

\begin{math}
EW_{\rm upperlim} = 5.0 \times \frac{\sqrt{N_s} \times \Delta\lambda}{S/N},
\end{math}

where S/N is the signal to noise ratio per pixel, $\Delta\lambda$ is the width of each intensity sample in Angstroms and $N_s$ is the number of samples. 
The derived errors were compared with 
those estimated in twelve cases where the same star was observed on different 
nights. The standard deviation on the measurements is 8 percent ($N$=12) for EWs $\gtrsim$ 15~m\AA, with the largest difference in measurements at 
different epochs of 23 percent being towards HD\,145482 with a (small) EW of $\sim$4~m\AA\ for the 1318.1~nm line. 

Excluding the handful of sightlines contaminated by stellar lines, the detection rates for the DIBs approached 90 per cent for the 1318, 1561.1, 1565.1 and 1567.0~nm features. 
Finally, no clear DIBs were detected in the CRIRES setting around 1624.2~nm. Table \ref{t_NIRDIB_EWs} shows the equivalent width results for the current sample as well as the 5780 and 5797~\AA\,DIBs from optical spectra discussed in Sect.~\ref{NIRDIB_correlations}.


\section{Discussion}
\label{discussion}

\begin{figure*}[hpt!]
\resizebox{17cm}{!}{\includegraphics[angle=90]{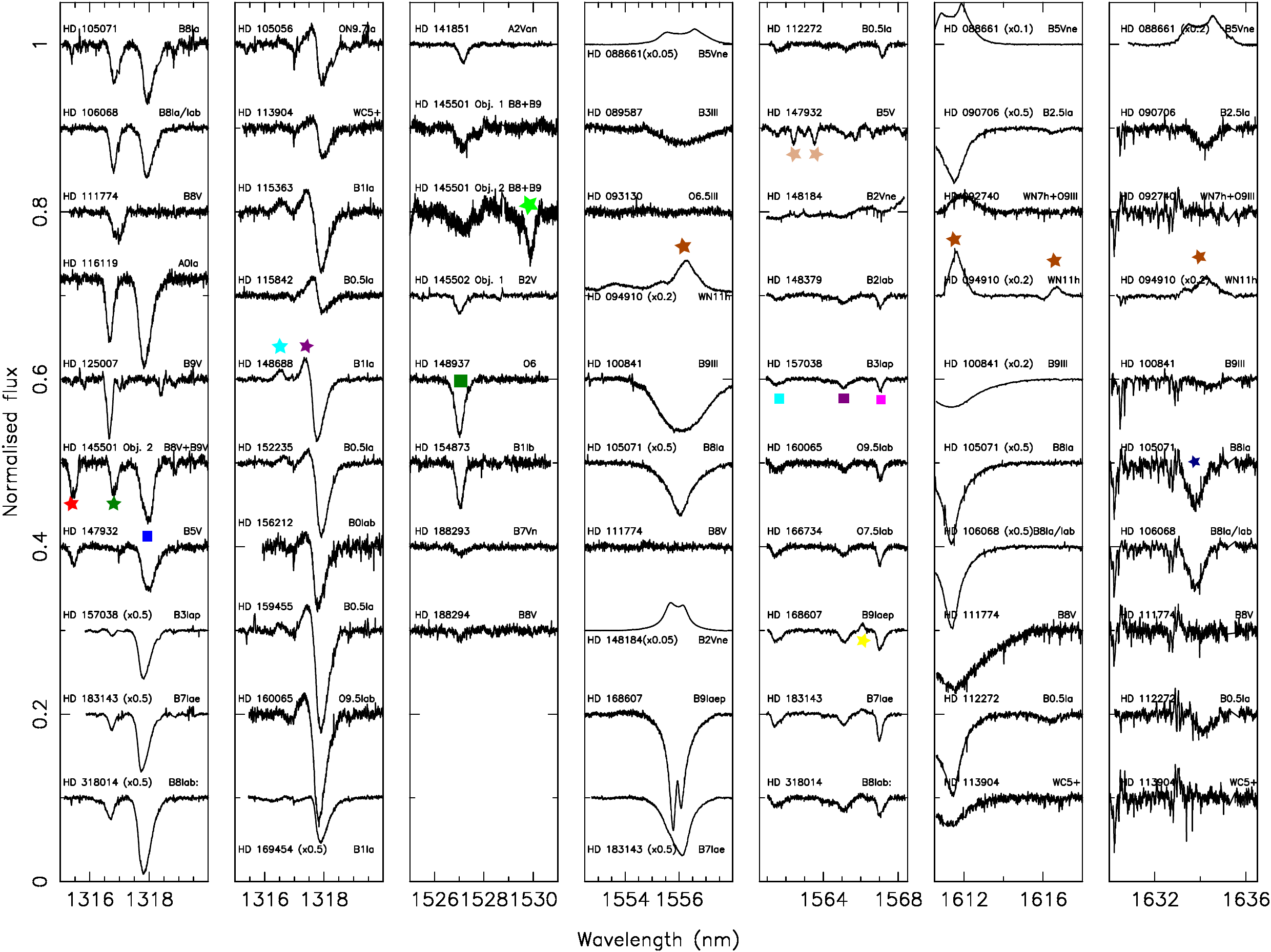}}
\caption{Normalised CRIRES spectra showing examples of all stellar and interstellar features observed. The ordinate is the normalised flux plus an offset with the abscissa being the wavelength in nm. Stellar lines: Red stars: 1315.4~nm, Green: 1316.8~nm, Cyan: 1316.5~nm (EmL), Purple: 1317.3~nm (EmL, uncertain), Light Green: 1529.9~nm, Yellow: 1566.1~nm (EmL), Gold: 1562.4, 1563.0, 1563.5~nm), Navy: 1633.8; Black: 1610.9~nm; Brown: 1555.7, 1616.7, 1625.3, 1626.1, 1634.2~nm (EmL). DIBs: Blue squares: 1318.1~nm, Green: 1527.1~nm, Cyan: 1561.1~nm, Purple: 1565.1~nm, Pink: 1567.0~nm. Not shown: broad and weak feature around 1564.4~nm.}
\label{f_CRIRES_exaample_spectra_1}
\end{figure*}

We detected a total of 23 non-telluric features, of which 17 are likely to be stellar 
and six interstellar (or possibly circumstellar) in nature. Figure \ref{f_CRIRES_exaample_spectra_1} 
shows examples of each, with all spectra being shown in Fig. \ref{f_CRIRES_spectra_1}. These are discussed in turn. Table~\ref{t_wavelengths} shows all of the detected lines. 

%
%

\begin{table*}[hpt!]
\begin{center}
\caption[]{Rest wavelengths in vacuum of lines detected in the CRIRES spectra. We also show lines taken from the NIST database \footnote{$https://physics.nist.gov/PhysRefData/ASD/lines_form.html$} within $\pm$0.2~nm in vacuum wavelength.}



\label{t_wavelengths}
\begin{tabular}{crrll}
\hline
\hline
$\lambda$       &   Stellar     & Abs. or        &  Notes                &   Lines from the NIST database within $\pm$0.2~nm                                          \\    
  (nm)          &   or IS       & EmL            &                       &                                                    \\
\hline
\hline
   1315.4       &       S      &    Abs.         &  (1)                 &   Zn\,{\sc i} (1315.42~nm), Al\,{\sc i} (1315.43~nm),  Th\,{\sc i} (1315.45~nm), Fe\,{\sc i} (1315.52~nm)    \\
                &              &                 &                      &   Fe\,{\sc i} (1315.52~nm), Cs\,{\sc ii} (1315.53~nm)     \\
   \hline
   1316.5       &       S       &    EmL          &  (2)                 &  C\,{\sc i} (1316.42~nm), Fe\,{\sc ii} (1316.60~nm),  Fe\,{\sc ii} (1316.65~nm), Cs\,{\sc ii} (1316.68~nm)  \\
   \hline
   1316.8       &       S?      &    Abs.         &                      &  O\,{\sc i} (1316.75, 1316.85, 1316.87~nm),  Fe\,{\sc ii} (1316.82~nm) \\
   \hline
   1317.3       &       S       &    EmL          &  (3)                 &  Th\,{\sc i} (1317.14, 1317.19~nm), Fe\,{\sc i} (1317.39, 1317.42~nm),  V\,{\sc i} (1317.43~nm),            \\
                &               &                 &                      &  Th\,{\sc ii} (1317.52~nm)        \\
   \hline
   1318.1       &      IS       &    Abs.         &  DIB                 &  Th\,{\sc i} (1317.97, 1318.20~nm),  Si\,{\sc i} (1318.05~nm), Fe\,{\sc i} (1318.06~nm)                  \\
                &               &                 &                      &  Kr\,{\sc i} (1318.10~nm), Ti\,{\sc i} 1318.20~nm),  Th\,{\sc ii} (1318.29~nm), V\,{\sc i} 1318.29~nm)   \\
   \hline
   1527.0       &      IS       &    Abs.         &  DIB                 & Cr\,{\sc ii} (1526.97~nm), Fe\,{\sc i} (1526.84, 1527.12~nm), Kr\,{\sc i} 1527.01~nm), Fe\,{\sc ii} (1527.10~nm) \\
   \hline
   1529.9       &       S       &    Abs.         &  (4)                 & Th\,{\sc ii} (1529.72~nm), Fe\,{\sc ii} (1529.73, 1529.87, 1529.90, 1529.98~nm), Cs\,{\sc ii} (1529.80~nm),      \\
                &               &                 &                      & V\,{\sc i} (1529.82~nm), Th\,{\sc i} (1529.93, 1529.98, 1530.04~nm), Fe\,{\sc ii} (1529.98~nm), \\
                &               &                 &                      & Hg\,{\sc i} (1530.00~nm), Fe\,{\sc i} (1530.00, 1530.03, 1530.06, 1530.10~nm)                   \\ 
   \hline
   1555.7       &       S       &    Abs. \& EmL  &  Br line             & Fe\,{\sc i} (1555.64, 1555.78, 1555.88~nm), Th\,{\sc i} (1555.59, 1555.74~nm), Ti\,{\sc ii} (1555.63~nm),   \\
                &               &                 &                      & Fe\,{\sc ii} (1555.81~nm)                                                                                   \\
   \hline
   1561.1       &      IS       &    Abs.         &  DIB                 & Ti\,{\sc i} (1560.91, 1560.94~nm), Fe\,{\sc ii} (1560.96~nm), Cr\,{\sc ii} (1561.14~nm), Al\,{\sc ii} (1561.14~nm) \\
                &               &                 &                      & Th\,{\sc i} (1561.24~nm) \\
   \hline
   1562.4       &       S       &    Abs.         &  (4)                 & Se\,{\sc i} (1562.27~nm), Ti\,{\sc ii} (1562.38~nm), Th\,{\sc i} (1562.44~nm), Fe\,{\sc ii} (1562.54, 1562.59~nm)   \\
   \hline
   1563.0       &       S       &    Abs.         &  (4)                  & Fe\,{\sc ii} (1562.84~nm), Th\,{\sc ii} (1562.89~nm), B\,{\sc i} (1562.90~nm)                                 \\
   \hline
   1563.5       &       S       &    Abs.         &  (4)                  & B\,{\sc i} (1563.33~nm), Fe\,{\sc i} (1563.36, 1563.39, 1563.53~nm, Th\,{\sc ii} (1563.58~nm)                              \\
   \hline
   1564.4       &     ???       &    Abs.         &  Weak                  & Fe\,{\sc i} (1564.32, 1564.38~nm), Th\,{\sc ii} (1564.47, 1564.56, 1564.61~nm), V\,{\sc i} (1564.47~nm),                          \\
                &               &                 &                      & Fe\,{\sc ii} (1564.52~nm)  \\
   \hline
   1565.1       &      IS       &    Abs.         &  DIB                 & Fe\,{\sc i} (1564.92, 1565.28~nm)  Fe\,{\sc ii} (1564.96, 1565.21~nm) Ar\,{\sc ii} (1565.21~nm)   \\
   \hline
   1566.1       &       S       &    EmL          &  (5)                 & V\,{\sc i} (1566.1~nm), Fe\,{\sc i} (1566.2~nm)        \\
   \hline
   1567.0       &      IS       &    Abs.         &  DIB                 & Ti\,{\sc i} (1566.05~nm), V\,{\sc i} (1566.07~nm)                                                        \\
   \hline
   1610.9       &       S       &    Abs. \& EmL  &  Br. line            & Ti\,{\sc i} (1619.73, 1610.99, 1611.03~nm), Th\,{\sc i} (1610.84, 1610.98~nm), Ar\,{\sc ii} (1610.94~nm)                    \\ 
   \hline
   1616.4       &       S       &   Abs.    &                            & Ti\,{\sc i} (1616.21, 1616.47, 1616.56), Be\,{\sc i} (1616.21)                                                                            \\
                &               &           &                            & Fe\,{\sc i} (1616.21, 1616.34~nm), Fe\,{\sc ii} (1616.31, 1616.39~nm)    \\
   \hline
   1616.7       &       S       &    Abs. \& EmL  &  (6)                 & Ti\,{\sc i} (1616.56, 1616.84~nm), Fe\,{\sc i} (1616.72~nm), Th\,{\sc ii} (1616.78~nm), Si\,{\sc i} (1616.81~nm) \\ 
   \hline
   1625.3       &       S       &    EmL          &                      & Al\,{\sc ii} (1625.11, 1625.23~nm), Fe\,{\sc i} (1625.14, 1625.33~nm)                                                     \\
   \hline
   1626.1       &       S       &    EmL          & --                   & Fe\,{\sc i} (1625.91~nm), P\,{\sc i} (1625.92~nm), Fe\,{\sc i} (1625.34~nm), Ti\,{\sc i} (1626.37~nm)       \\
   \hline
   1633.8       &       S?      &    Abs.         &                      & Fe\,{\sc ii} (1633.74~nm), Fe\,{\sc i} (1633.76~nm), C\,{\sc i} (1633.84~nm)    \\
   \hline
   1634.2       &       S       &    EmL          & --                   & Fe\,{\sc i} (1634.16~nm)                                                        \\
\hline
\hline
\end{tabular}
\end{center}
{\small Notes: (1) Only seen towards magnetic stars. (2) Only seen towards blue supergiants. (3) Seen in B2 and earlier spectral types. Contaminates 1318.1~nm DIB. (4) Only seen towards one magnetically-peculiar star. (5) Only seen towards blue supergiant and emission-line stars. (6) Only seen towards blue supergiant and Wolf--Rayet stars.}
\end{table*}

\subsection{Likely stellar features}

A number of likely stellar features are listed in the spectra  
in Table~\ref{t_wavelengths} with examples shown in Fig.~\ref{f_CRIRES_exaample_spectra_1}. 
We classified lines as being stellar based on their width compared with other
metallic stellar lines. For these lines we also searched the NIST database (www.nist.gov) for close matches in wavelength. 


\subsubsection{1315.4~nm (absorption)}

This line is seen towards only two stars in our sample,  HD\,145501 B and HD\,147932 (B5V). 


The estimated vsin($i$) for HD\,147932 is 153 km\,s$^{-1}$ \citep{Brown1997}, although this was perhaps affected by a secondary component that makes the spectrum difficult to analyse. HD\,145501 A is a magnetic chemically-peculiar star of type CP2 \citep[][and refs. therein]{Kochukhov2006,Wraight2012} with HD\,147932 also being a magnetic (and rotationally variable) star \citep{Alecian2014}.






\subsubsection{1316.5~nm (emission)}

A weak emission line is seen in a handful of the early-type stellar spectra, for example towards the blue supergiants HD\,115363 (B1Ia), HD\,148688 (B1Ia). HD\, 152235 (B0.5 Ia) and HD\,159455 (B0.5Ia). 
%
%
The FWHMs from single-component Gaussian fits to the observed lines are 59 km\,s$^{-1}$ for HD\,115363, 61 km\,s$^{-1}$ for HD\,148688,  92 km\,s$^{-1}$ for HD\,152235 and 66 km\,s$^{-1}$ for HD\,159455 although the feature does not appear Gaussian in this case. The lines observed (if stellar) are likely to be formed at the base of the wind, close to the photosphere, and the broadening they display is basically caused by rotation and macroturbulence and do not measure the wind terminal speed. The FWHM of the lines are broadly consistent with this picture and indicate that the lines do not form in the stellar wind as if they were they would be much broader \citep{Vink2021}. We also note that in the case of HD\,152235 the H$\alpha$ line at 6556~\AA \, is approximately twice as wide (in km\,s$^{-1}$) as this IR emission line. Finally, the difference in the wavelength of the peak emission of the 1316.5~nm line and 1317.3~nm line is not the same for the HD\,152235 and HD\,159455 sightlines, although the measurement is difficult.



\subsubsection{1316.8~nm (absorption)}

This line is seen towards several stars which are typically late B-types: HD\,105071 (B8Ia), HD\,106068 (B8Ia), HD\,115088 (B9.5), HD\,116119 (A0Ia), HD\,111774 (B8V),  HD\,125007 (B9V), HD\,145501 (B8), HD\,157038 (B3 Iap), HD\,164865 (B9Ia), HD\,183143 (B6Ia) and HD\,318014 (B8Ia). The measured FWHMs range from 44 km\,s$^{-1}$ for HD\,157038, with a FWHM of 50 km\,s$^{-1}$ from Si\,{\sc iii} (4567$\mbox{\AA}$), to 71 km\,s$^{-1}$ for HD\,105071, which has somewhat lower FWHMs in the optical of 55 km\,s$^{-1}$ (Si\,{\sc ii} 4128 and 4130$\mbox{\AA}$) and 48~km\,s$^{-1}$ (N\,{\sc ii} 3995$\mbox{\AA}$) from archival UVES spectra \citep{UVES1,UVES2}. \cite{Malkan2002} noted the presence of a blend of Al\,{\sc i} at 1312.6~nm and Si\,{\sc i} at 1318.0~nm in cooler stars, although only detectable in their data for stars of type A4 or later. 

Finally, HD\,111774 has a FWHM of 100~km\,s$^{-1}$ for visible lines at 4549 and 4555$\mbox{\AA}$, while the NIR line has a FWHM of 88~km\,s$^{-1}$. Overall, although there are a few discrepant values, the FWHM values indicate that this line is stellar and not interstellar. There is no correlation between the line strength and reddening for these sightlines, which have a narrow range of 130 to 150~m\AA\ for all but three sightlines, HD\,157038 (EW=20~m\AA), HD\,164875 (EW=25~m\AA) and HD\,183143 (EW=8~m\AA).

\subsubsection{1317.3~nm (possible stellar emission merged with DIB causing inverted P-Cygni profile)}

A number of sightlines show inverted P-Cygni like profiles at around 1317.3~nm, bluewards of 
the main DIB at 1318.1~nm. Such emission lines were predicted 
in solid state carriers by Van de Hulst (1948) associated with a red wing. The features are weakly visible in the X-shooter spectra of \cite{Cox2014} although not commented on and much more obvious with the current higher spectral resolution data. The Wolf--Rayet stars in the Galactic Centre observed by \cite{Najarro2017} do not show this emission-line at 1317.3~nm, although again the resolving power of those data is only $R$\,$=$\,1200--5000 so it would be difficult to detect (although they do clearly detect the 1318~nm DIBs).

\begin{figure}
\resizebox{\hsize}{!}{\includegraphics{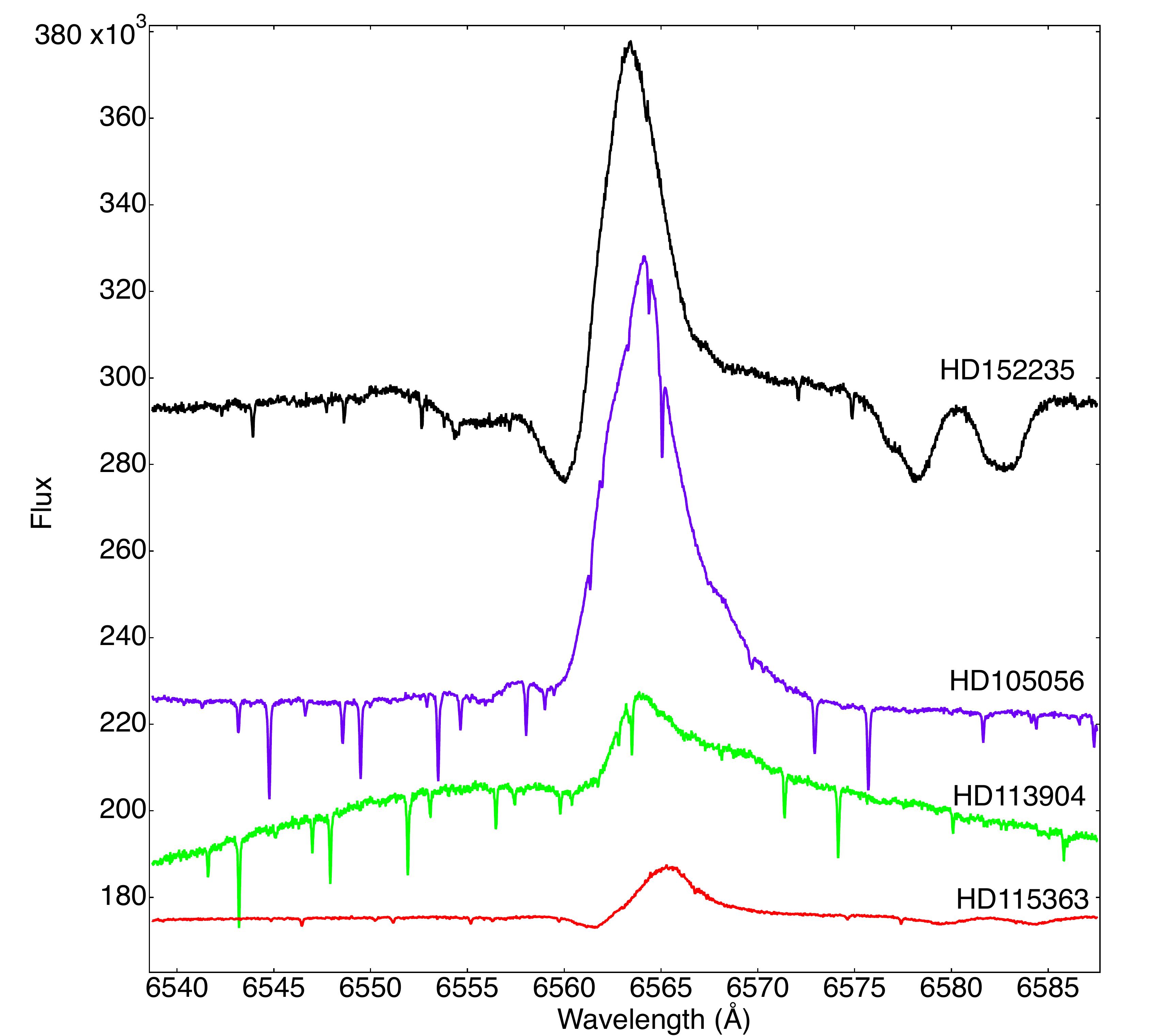}}
\caption{Archival UVES spectra showing H$\alpha$ emission for four stars towards which an emission feature is also seen near to the 1318~nm DIB.} 
\label{f_H_Alpha}
\end{figure}

As the features are present mainly in stars of spectra of type B2 and earlier, there is the strong likelihood that in fact we are seeing a mix of a stellar emission line in hot stars combined with the interstellar DIB at 1318.1~nm. We note that one third of our sample have spectral types later than B2, in which no inverted P-Cygni profiles are seen. Additionally, a search for emission lines in the optical in a handful of these objects showing the 1317.3~nm emission line feature revealed strong H$\alpha$ (Fig.~\ref{f_H_Alpha}).


To investigate the nature of the emission, we (a) compared the emission line velocities with stellar and DIB velocities and (b) computed a model spectrum of the region for an early B-type star. For the former, we determined the stellar velocities of the sightlines where the emission feature was seen, either using values from the literature or our own measurements from archive data. Figure~\ref{f_EmissionLine_1318nm} shows spectra towards eight stars with emission lines bluewards of 1318~nm, with their stellar velocities marked.  

\begin{figure}
\resizebox{\hsize}{!}{\includegraphics{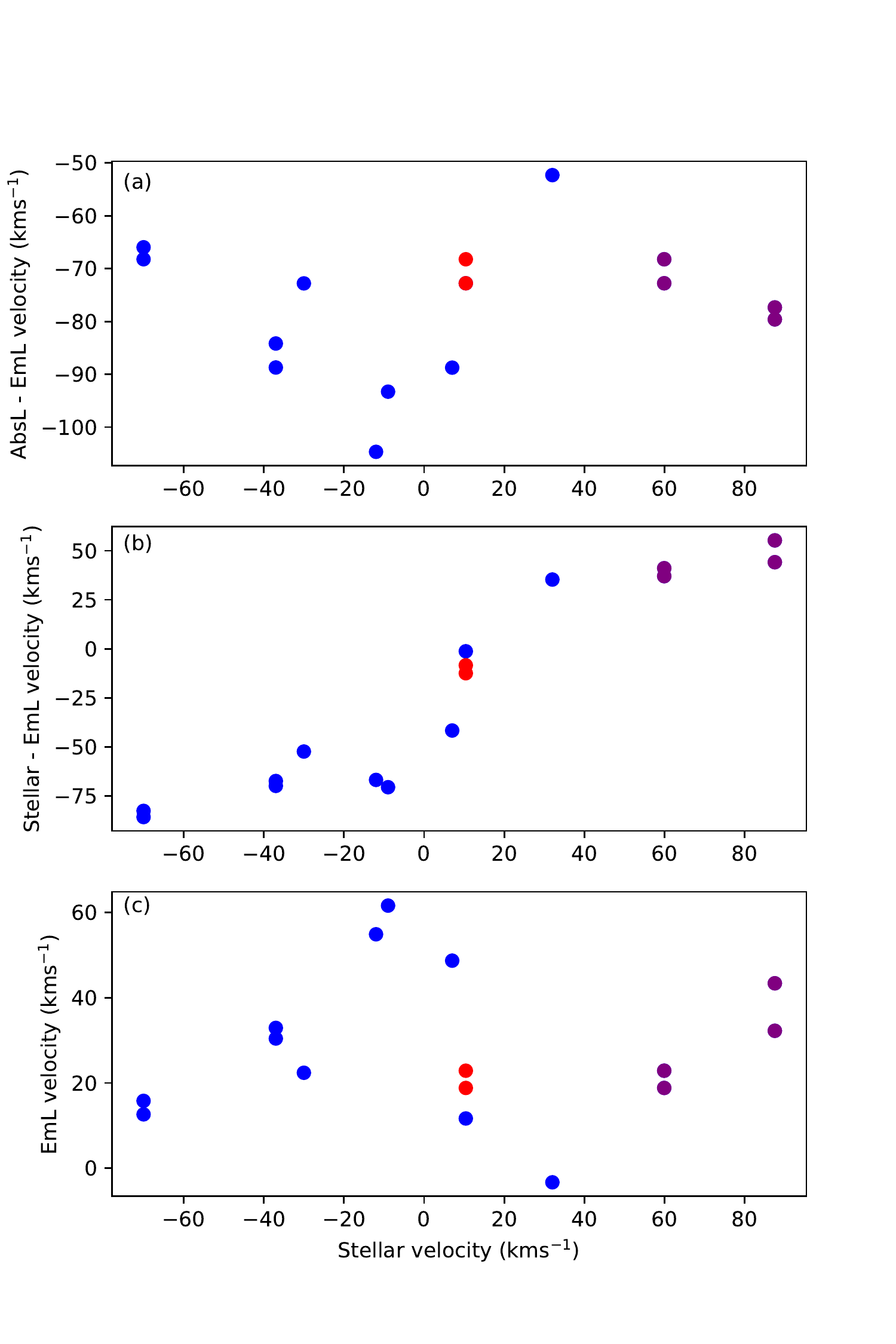}}
\caption{(a) Difference in absorption and emission line peak velocity around 1318~nm plotted against the stellar velocity. (b) Ditto for stellar velocity minus emission line velocity. (c) Emission line velocity vs stellar velocity, assuming the emission line has a rest wavelength of 1317.37~nm. Results with the same stellar velocities are from repeat observations of a given star. Purple points are from Gaia DR2 with Red points from Gaia DR3 for HD\,160065}.
\label{Emission_minus_Stellar_Velocity}
\end{figure}

%
%
\begin{figure}
\resizebox{\hsize}{!}{\includegraphics{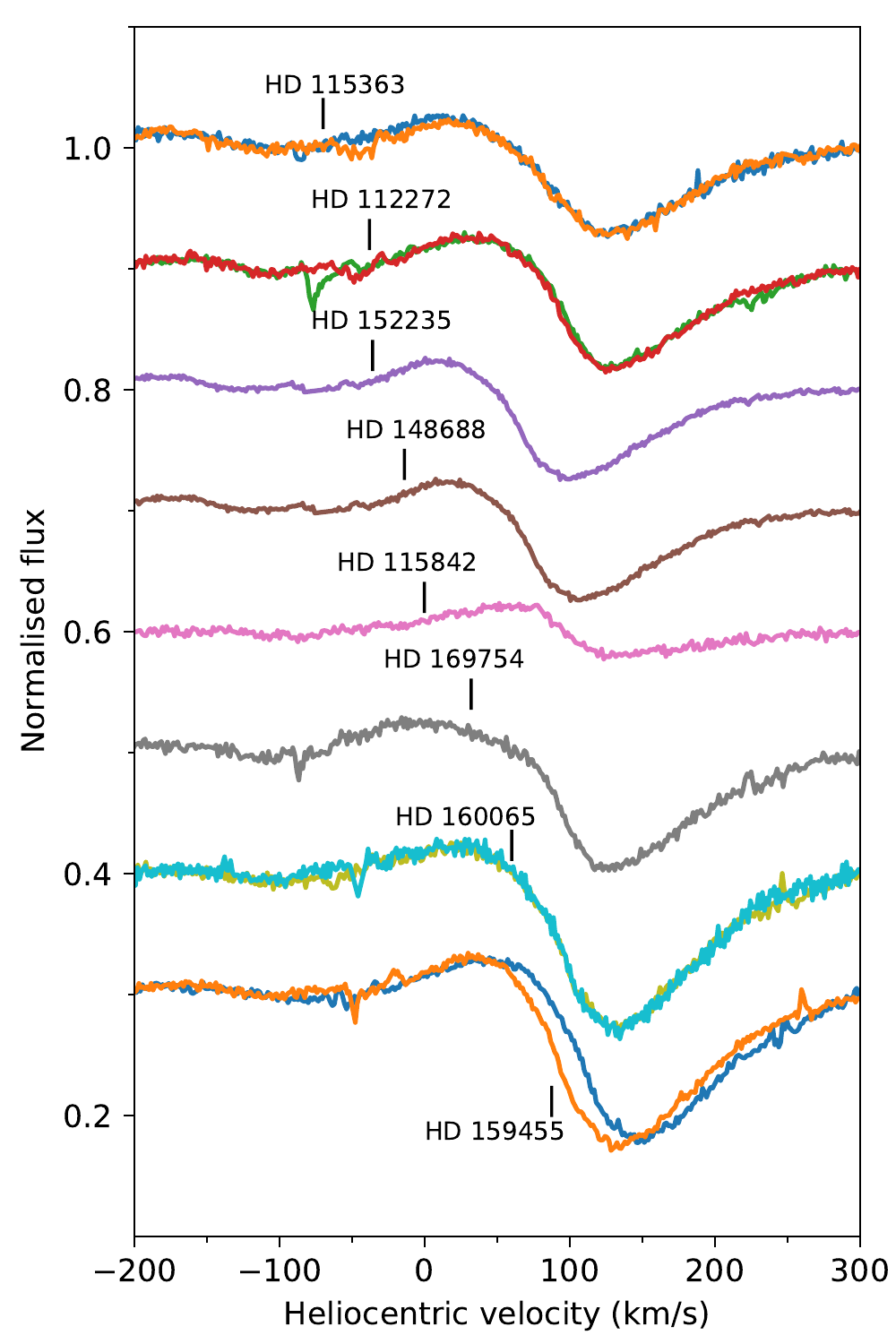}}
\caption{Example spectra showing the bluewards emission feature towards the 1318~nm DIB with Heliocentric stellar velocities marked by vertical black lines. For HD\,115363, HD\,112272, HD\,159455 and HD\,160065 two spectra are shown taken at different epochs.  HD\,159455 has a shift in radial velocity between the two epochs. \cite{Kervella2019} use Gaia data to determine that this object has a tangential velocity anomaly of 2.4 kms$^{-1}$ and is hence likely a binary although this should not affect the DIB absorption profile. The rest wavelength of the emission line is assumed to be 1317.30~nm.}
\label{f_EmissionLine_1318nm}
\end{figure}

Of particular interest are three stars, HD\,115363, HD\,159455 and HD\,160065 which have Heliocentric radial velocities of $-$70, +88 and +60 km\,s$^{-1}$, respectively. {\em If} the emission lines were at the stellar velocities then one would expect an offset of approximately 0.7~nm (158~km\,s$^{-1}$) between the position of the emission line peaks when comparing HD\,115363 and HD\,159455. Similarly, the offset between the emission line peak from the star and the absorption line trough from the DIBs would also be expected to change. The fact that the emission-absorption offset of 0.51~nm for HD\,115363 and 0.46~nm for HD\,159455 is similar for both stars could indicate that the emission line does not move at the stellar velocity. 
%
%
To investigate this further, Fig.~\ref{Emission_minus_Stellar_Velocity} shows the difference in emission and absorption line velocities (a), the difference in stellar and emission line velocities (b), and the emission line velocity (c), all plotted against the stellar velocity. The simple trough or peak of the absorption or emission lines were used for the velocity measurements and are shown in Table~\ref{t_1318nm_emission_data}. Additionally we performed a fit to the data comprising an emission-line component and a 1318~nm DIB profile taken from Fig. \ref{f_1318_1527_1568_nm_profiles}, shifted in wavelength, scaled in flux and stretched in wavelength by up to 5 percent. The weighted wavelength was derived for the two components by averaging the wavelength bins weighted by the absolute value of the absorption line depth, $f(\lambda)$ vis:

\begin{equation}
   \lambda_{\rm weighted} =  \frac{\int{((1.0-f(\lambda) \times \lambda)) \, d\lambda}}{\int{\lambda} \, d\lambda}
\end{equation}

\begin{figure}
\resizebox{\hsize}{!}{\includegraphics{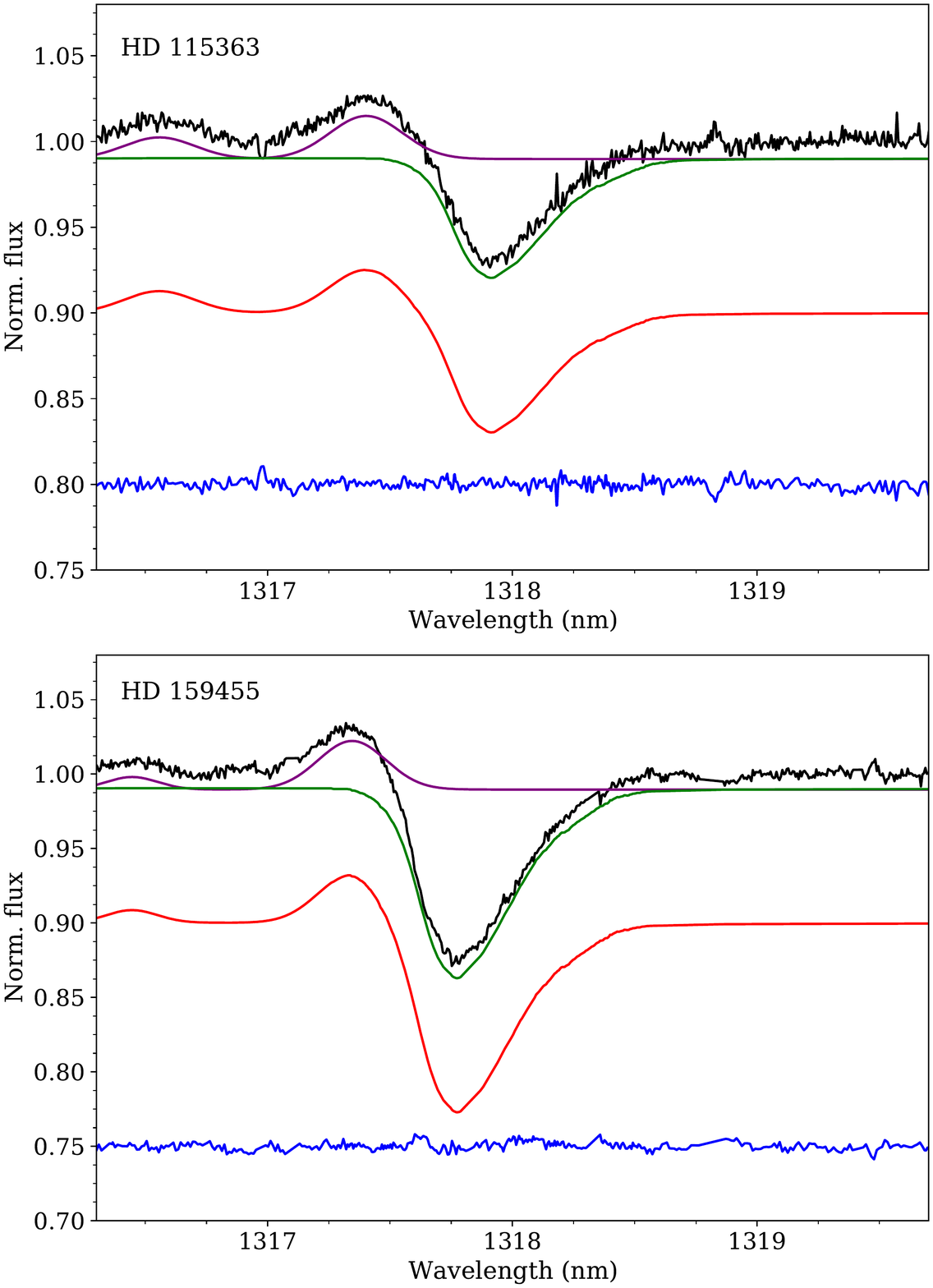}}
\caption{Normalised absorption and emission around 1318~nm towards HD\,115363 (B1 Ia) and HD\,159455 (B0.5 Ia). Black lines: data. Purple lines: Twin-component Gaussian fits to the unidentified emission lines. Green line: Scaled, shifted and wavelength stretched absorption line DIB profile. Red line: Combined Gaussian fit plus DIB model. Blue line: Data divided by model. Lines are offset in the ordinate for clarity.}
\label{Plot_CRIRES_HD115363_HD159455_Data_EmL_Plus_Model}
\end{figure}

The values of the Gaussian fit wavelengths are also included in Table~\ref{t_1318nm_emission_data}. Figure \ref{Plot_CRIRES_HD115363_HD159455_Data_EmL_Plus_Model} shows the model for two stars (HD\,115363 and HD\,159455) as well as the data/model residual. 

The emission line minus absorption line velocity difference shown in Fig.~\ref{Emission_minus_Stellar_Velocity}(a) shows no strong trend given the likely large uncertainties in the velocities indicating they could be formed in the same medium. 
$A priori$ it would be expected that if the emission line were stellar in nature it should be correlated with the stellar velocity. However, panel~\ref{Emission_minus_Stellar_Velocity}(c) implies that this is not the case. Indeed, panel~\ref{Emission_minus_Stellar_Velocity}(b) shows that the difference in the stellar and emission line velocity is correlated with the stellar velocity when a constant offset may have been expected. 
We caution that measuring the centre of the emission peak is not straightforward and these errors could dominate. 

\begin{figure}
\resizebox{\hsize}{!}{\includegraphics{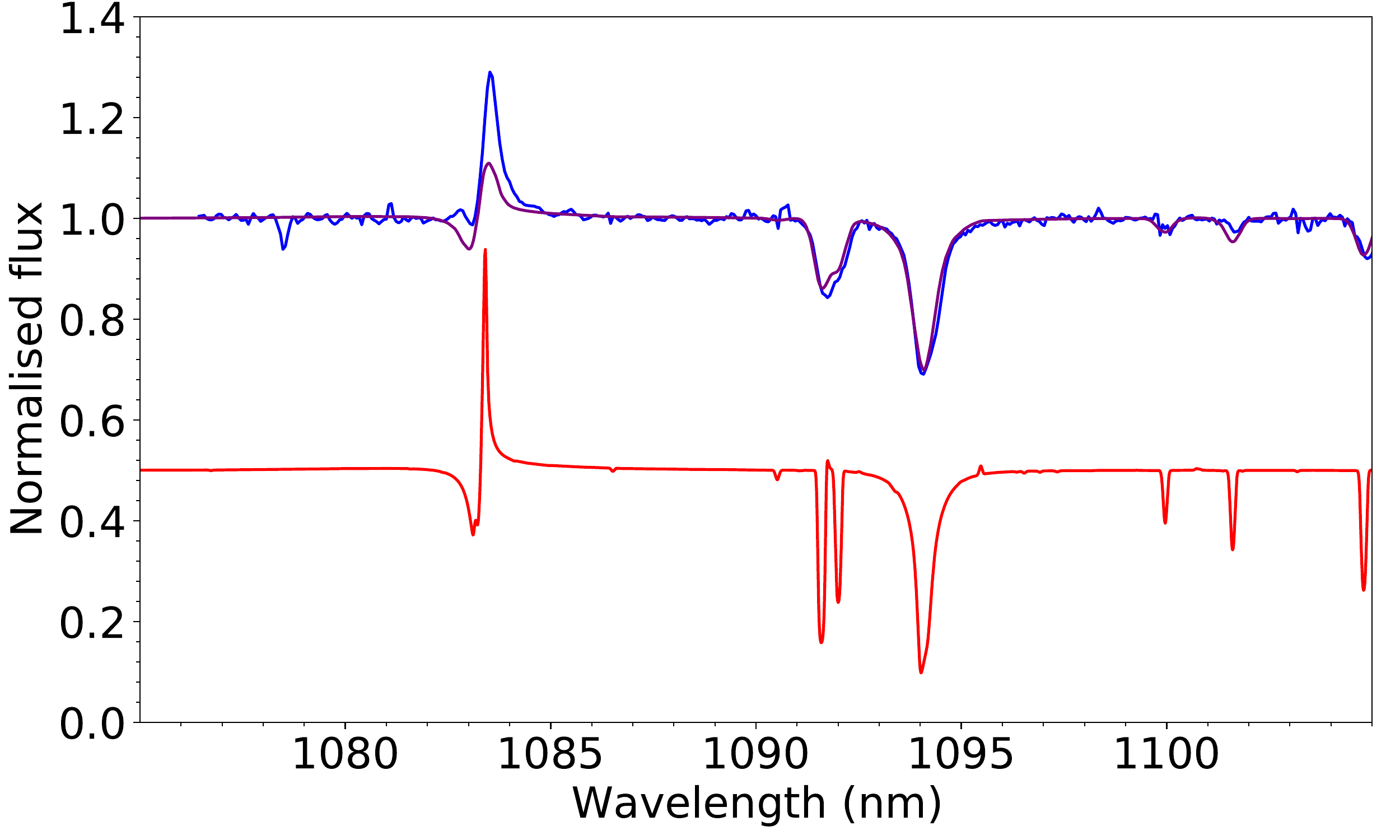}}
\caption{Blue line: X-shooter data of HD\,152235 from 1075 to 1105~nm showing an emission line from the star at 1083~nm (He\,{\sc i}) and absorption at 1094~nm (H\,{\sc i}). Red line: Stellar model in vacuum wavelength obtained using {\sc cmfgen} of a B0Ia star of solar metalicity, $T_{\rm eff}$=21,800 K, log($g$)=2.55 and slightly enhanced He/H=0.135 by number, offset in the ordinate for clarity. The purple line shows the model spectrum Gaussian smoothed to a FWHM of 0.91~nm ($\sim$250 km\,s$^{-1}$) to fit the absorption-line data. }
\label{B0Ia_Teff26500_logg3p0_CMFGEN_1}
\end{figure}

A final attempt to determine the the origin of the line was performed by computing a stellar model of a  B0.5Ia blue supergiant star HD\,152235 using {\sc cmfgen} \citep{Hillier2001} and comparing it with observations. A solar metalicity was assumed, effective temperature of 21,800~K, log($g$)=2.55 and slightly enhanced He/H=0.135 by number. In Fig.~\ref{B0Ia_Teff26500_logg3p0_CMFGEN_1} we show the emission and absorption lines from the X-shooter data and our model in the 1075 to 1105~nm region. In particular, the model predicts emission from the He\,{\sc i} P-Cygni profile line at 1083~nm with an EW of 820~m\AA\ (integrated from 1072 to 1090~nm), somewhat lower than the observed value of 1660$\pm$100 m\AA. The corresponding integrated line strength for the model and data for the absorption line doublet at 1091.6 and 1092.0~nm are 1000 and 1180$\pm$70 m\AA, and finally for the Hydrogen Paschen line at 1094.0~nm the model and data have EWs of 2700 and 2750$\pm$100~m\AA \, respectively.

\begin{figure*}
\resizebox{\hsize}{!}{\includegraphics{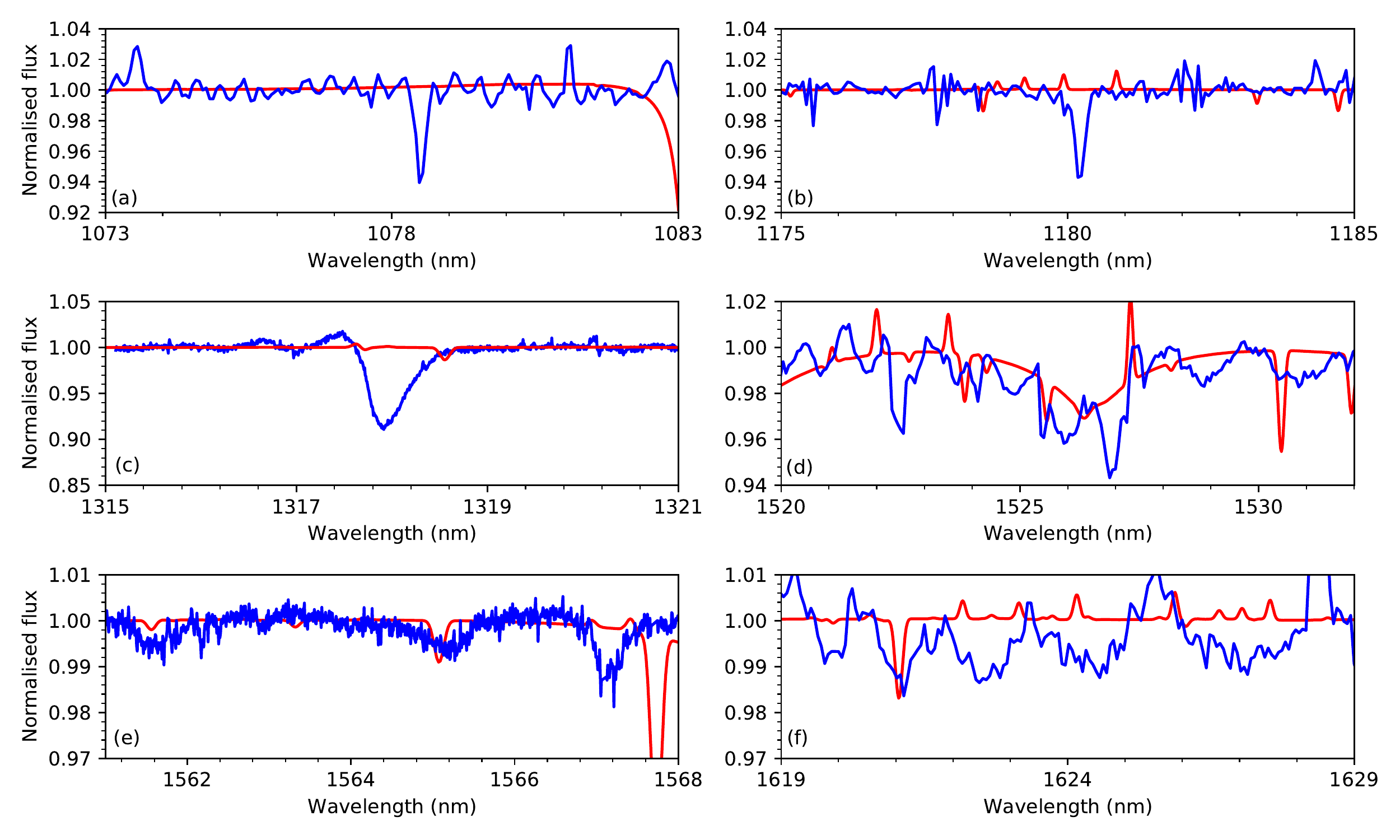}}
\caption{Red lines: The same stellar model as shown in Fig. \ref{B0Ia_Teff26500_logg3p0_CMFGEN_1}, plotted at the wavelength regions of the NIR DIBs shown in this paper plus the NIR DIBs around 1078 and 1180~nm. Blue lines: observations of HD\,152235 either from CRIRES (panels c and e) or X-shooter (panels a, b, d, f).}
\label{B0Ia_Teff26500_logg3p0_CMFGEN_2}
\end{figure*}

Figure \ref{B0Ia_Teff26500_logg3p0_CMFGEN_2} shows the stellar model and CRIRES or X-shooter data for HD\,152235 for the DIB regions studied in this paper plus the 1078 and 1080~nm DIBs studied by \citet{Cox2014}. Panel (c) of this figure shows that the model only predicts very weak emission (5 m\AA) in the bluewards emission P-Cygni feature of the 1318~nm DIB compared with the actual value of 40$\pm$5~m\AA, indicating a non-stellar nature. Although the model also predicts lower values for emission in the 1075 to 1090-nm regions the difference is less pronounced than for the 1317.3 nm feature. Given the weakness of the predicted emission compared to that observed, a perhaps more intriguing possibility is that we are seeing a PAH in emission near to the hot star, turning to absorption further out. Such behaviour has been claimed towards the Red Rectangle in the DIBs around 5797$\mbox{\AA}$\, (\citealt{Fossey1991}, \citealt{Sarre1991} although see \citealt{Glinski2009}). The lack of the emission line feature in later stars in this scenario would be explained by a higher UV flux from the earlier type stars. On the other hand, such emission would require a large amount of cooler and dustier circumstellar material. Additionally, Section~\ref{NIRDIB_correlations} and previous work such as that by \cite{Hamano15} and \cite{Hamano16} have found that the 1318~nm DIB acts more like a 5780~\AA\ DIB than the 5797~\AA\ DIB variety, and in the Red Rectangle no emission is seen around 5780~\AA. Furthermore, in the Red Rectangle the emission is redwards not bluewards of the DIB at 5797~\AA. Hence it seems unlikely that we are seeing PAH emission from the DIB although follow-up observations of the red rectangle using the upgraded CRIRES or X-shooter might be interesting. We note that circumstellar absorption has been claimed for the 1527~nm DIB towards the Red Square Nebula by \cite{ZasowskiRedSquare2015}. Figure 7 shows two emission line features towards HD\,115363 at around --200 and +25 km\,s$^{-1}$, that could be explained by circumstellar material rotating the central star and that could produce a double peak in velocity. However, the strengths of the two peaks are different and emission line velocities are not symmetric about the stellar velocity of 70 km\,s$^{-1}$.

In conclusion, the fact that the emission is only seen in stars of B2 or earlier, and that these stars also show H$\alpha$ in emission, points to a stellar or circumstellar origin for the line, but the strength of the emission line compared with a stellar model and lack of obvious correlation with the stellar velocity point to an interstellar explanation.  To conclusively eliminate an ISM or circumstellar nature would need an investigation into the wind speed of the targets.

\begin{table*}
\begin{center}
\caption{Stellar velocities, velocity of the peak of the emission line and velocity at peak of absorption line towards stars where the 1318~nm showed strong emission. The standard of rest is Heliocentric. Values in brackets are flux weighted wavelengths determined using fitting as shown in Fig. \ref{Plot_CRIRES_HD115363_HD159455_Data_EmL_Plus_Model}.} 
\label{t_1318nm_emission_data}
\begin{tabular}{crrccccr}
\hline
HD       & Stellar                    &  Ref.     &  MJD       &  EmL, Abs.           & $\delta\lambda$  & EmL, Abs.        & Hel.  RV        \\
Number   & velocity                   &           &            &  Wavelength (nm)     &     (nm)         & norm. flux peak  & corr            \\ 
         &  (km\,s$^{-1}$)            &           &            &  at $|$peak$|$       &                  & (abs. value)     & (km\,s$^{-1}$)  \\ 
         &                            &           &            &                      &                  &                  &                 \\ 
\hline 
105056   & +5$\pm$6,  --9.0$\pm$3.1,  &  (1), (4) & 56411.272  &  1317.56, 1317.95    &     0.39         &  1.020, 0.955    &  +2.4        \\   
   "     & +14.5 $\pm$1.8             &  (5)      &            & (1317.61, 1318.05)   &     0.44         &                  &              \\ 
112272   & --36$\pm$2, --38.0$\pm$4.4 &  (1)      & 56440.172  &  1317.46, 1317.89    &     0.43         &  1.025, 0.918    & --6.0        \\   
   "     & --37.1$\pm$1.8             &  (5)      &            & (1317.47, 1318.02)   &     0.55         &                  &              \\
   "     &                            &           & 56732.333  &  1317.38, 1317.79    &     0.41         &  1.025, 0.917    &  +14.7       \\  
   "     &                            &           &            & (1317.39, 1317.92)   &     0.53         &                  &              \\
113904   &  +7$\pm$5, +4.4$\pm$1.5    &  (1), (5) & 56459.143  &  1317.56, 1317.97    &     0.41         &  1.010, 0.965    & --10.5       \\    
   "     &                            &           &            & (1317.58, 1318.07)   &     0.49         &                  &              \\
115363   & --70$\pm$2, --71.0$\pm$4.5 &  (1), (4) & 56465.029  &  1317.41, 1317.92    &     0.51         &  1.023, 0.929    & --12.4      \\    
   "     & --70.4$\pm$1.4             &  (5)      &            & (1317.39, 1318.02)   &     0.63         &                  &             \\
   "     &                            &           & 56732.346  &  1317.30, 1317.80    &     0.50         &  1.021, 0.929    &  +15.8      \\  
   "     &                            &           &            & (1317.31, 1317.90)   &     0.59         &                  &             \\
115842   & --21$\pm$2 --3.0$\pm$4.3   &  (1), (4) & 56465.058  &  1317.61, 1317.95    &     0.34         &  1.020, 0.980    & --15.7      \\    
   "     & --21.1$\pm$1.7             &  (5)      &            & (1317.57, 1318.07)   &     0.50         &                  &             \\
152235   & --25$\pm$1, --35.2$\pm$2.3 &  (1), (4) & 56467.098  &  1317.43, 1317.91    &     0.48         &  1.015, 0.913    & --7.2       \\    
   "     & --35.9$\pm$1.8             &  (5)      &            & (1317.50, 1318.01)   &     0.51         &                  &             \\
159455   & 87.5$\pm$1.9               &  (2)      & 56435.263  &  1317.44, 1317.90    &     0.46         &  1.029, 0.880    &  +11.5      \\    
   "     &                            &           &            & (1317.46, 1318.00)   &     0.54         &                  &             \\
   "     &                            &           & 56764.186  &  1317.33, 1317.78    &     0.45         &  1.032, 0.875    &  +25.4      \\
   "     &                            &           &            & (1317.34, 1317.93)   &     0.59         &                  &             \\
160065   & 59.9$\pm$6.4, 10.4$\pm$2.7 &  (2), (6) & 56434.396  &  1317.33, 1317.83    &     0.50         &  1.022, 0.873    &  +12.0      \\    
   "     &                            &           &            & (1317.32, 1317.94)   &     0.62         &                  &             \\
   "     &                            &           & 56746.394  &  1317.27, 1317.75    &     0.48         &  1.022, 0.870    &  +29.7      \\
   "     &                            &           &            & (1317.22, 1317.68)   &     0.46         &                  &             \\
169754   & 32.0$\pm$7.4               &  (3)      & 56497.240  &  1317.34, 1317.91    &     0.57         &  1.025, 0.904    & --12.4      \\    
   "     &                            &           &            & (1317.37, 1318.04)   &     0.67         &                  &             \\
\hline
\end{tabular}
\end{center}
\tablebib{References: (1) POP, UVES pop survey \citep{Bagnulo2003} (2) Gaia DR2  (3) \cite{Kharchenko2007} (4) \cite{Gontcharov2006} (5) This paper with the stellar velocity measured using He\,{\sc i} at $\lambda$=4387~\AA \, (6) Gaia DR3 \cite{GAIADR3}. There is a large (50 km\, s$^{-1}$) difference in the Gaia DR2 and DR3 velocities for the one star (HD\,160065) for which data are available for both releases}.
\end{table*}






\subsubsection{1529.9~nm (absorption line)}

A single sightline, the magnetically chemical peculiar star HD\,145501 (B8V+B9VpSi) displays an absorption line at 1529.9~nm. Given the fact that it is the only time the line is seen in the sample, with the reddening towards the star a "normal" value of $E(B-V)$=0.22~mag., the line is very likely stellar. The line is not seen towards a close companion.  



\subsubsection{1555.7~nm (emission and absorption lines)}

The 16--4 Brackett line at 1555.65~nm is observed in many spectra\footnote{See  http://www.gemini.edu/sciops/instruments/nearir-resources/astronomical-lines/h-lines for a list of Hydrogen lines.}, both in absorption and emission. 
The lines have a range of shapes including: strong, symmetric absorption (HD\,105071, B8Ia), asymmetric absorption (HD\,183143, B7Ia), twin absorption components (possibly from super-imposed central emission in HD\,168607, B9 Iaep) and double-peaked emission in the Be-type objects (HD\,88661 and HD\,148184), while several stars do not show the line at all (e.g. HD\,111774, B8V; HD\,93130, O6.5 III).


 



\subsubsection{1562.4~nm, 1563.0~nm and 1563.5~nm (absorption)}

The B5V star HD\,147932 ($\rho$ Oph C), displays several absorption lines around 1563~nm, the strongest being at 1562.4, 1563.0 and 1563.5~nm with FWHM widths of 49~km\,s$^{-1}$. This compares with the FWHM of  the Mg\,{\sc ii} line at 4481~\AA\ of 180~km\,s$^{-1}$, indicating that if the lines are stellar they likely originate from a cool spectroscopic binary companion as indicated from Gaia observations \citep{Kervella2019}. The star is magnetic \citep{Alecian2014} and shows possible variability over timescales of 20.7 hours \citep{Koen2002} although with possible contamination of the spectra by the brighter $\rho$ Oph \citep{David2019}.

\subsubsection{1566.1~nm (emission)}

Weak emission at 1566.1~nm is seen towards the supergiant and emission-line stars HD\,148184 (B2Vne; broad and hard to fit the continuum) HD\,168607 (B9Iaep, FWHM=0.26~nm), HD\,183143 (B7Iae, and often used as a `template' for DIBs), and HD\,318014 (B8Iab). 

\subsubsection{1610.9~nm (absorption and emission)}

A number of spectra show absorption or emission near to 1610.9~nm, which is the 13--4
Hydrogen Brackett line.






\subsubsection{1616.4~nm (absorption line)}

An absorption feature at 1616.4~nm is visible towards the supergiant stars HD\,112272 (B0.5Ia), HD\,148379 (B2Iab), HD\,148688 (B1Iae), HD\,168987 (B1Ia), HD\,169454 (B1Ia), HD\,318014 (B8Ia), close to the emission lines noted in the next section. 

\subsubsection{1616.7, 1625.3, 1626.1 and 1634.2~nm (emission lines)}
\label{s_emlstars_1616_1625_1626_1634}

The Wolf--Rayet star HD\,94910 shows a rich set of emission lines at 1616.7, 1626.1 and 1634.2~nm, with the supergiant HD\,148688 (B1Ia) showing an emission line at 1625.3~nm. 

\subsubsection{1633.8~nm (absorption line)}
\label{s_1633p8}

Absorption line features are seen at 1633.8~nm towards HD\,90706 (B2.5Ia), HD\,94910 (WN11 type star, near emission-line feature, creating P-Cygni profile), HD\,105071 (B8Ia), HD 106068 (B8Ia/Iab), HD\,112272 (B0.5Ia), HD\,148379 (B9Iaep), HD\,148688 (B1Iaeqp), HD\,164865 (B9Ia), HD\,168607 (B9Iaep, twin absorption line profiles), HD\,168625  (B6Iap), HD\,169454 (B1Ia), HD\,170938  (B1Ia), HD\,183143 (B6Ia), HD\,318014 (B8Ia). At a similar wavelength several of the stars in the sample display prominent emission (Fig.~\ref{f_CRIRES_exaample_spectra_1} and Sect.~\ref{s_emlstars_1616_1625_1626_1634}). Neither \cite{Geballe11} or \cite{Cox2014} identified this feature. It is likely stellar although we could not identify the line. Both N\,{\sc ii} (1634~nm) and Ne\,{\sc i} were candidates but not visible in the B0.5Ia model (nor in a model for a cooler B8Ia star).



\subsection{Likely interstellar or circumstellar features}
\label{s_is_cs_features}

A number of insterstellar or cirucumstellar features were detected and are described below. 

\subsubsection{Wavelengths of detected features}
Interstellar or circumstellar features were detected at the following wavelengths; 
1318.1~nm, 1527.9~nm (partly contaminated by stellar absorption as seen in the {\sc cmfgen} model of HD\,152235 in  Fig. \ref{B0Ia_Teff26500_logg3p0_CMFGEN_2}), 1561.1~nm, 1564.4~nm (broad and weak, possibly stellar), 1565.1~nm (also partly contaminated by stellar absorption), 1567.0~nm and possibly at 1634.2~nm 
although it seems likely stellar (see Sect. \ref{s_1633p8}). 
We had hoped to detect the broad DIB at 1622.7~nm as previously observed towards the highly-reddened Galactic centre \citep{Geballe11}. This DIB was also observed at 1622.7~nm in 4U~1907+09, the most heavily reddened object in the sample of nine stars observed by \citet{Cox2014}. We do not detect the absorption in any of our higher resolution CRIRES spectra nor the more highly-reddenened X-shooter Cepheid data; there is nearby weak telluric absorption at 1623.3~nm but no obvious DIB absorption. 

\subsubsection{Shape of the DIB profiles and Gaussian fitting} 
\label{gaussfits}

\begin{figure*}
\begin{center}
   \resizebox{17cm}{!}{\includegraphics{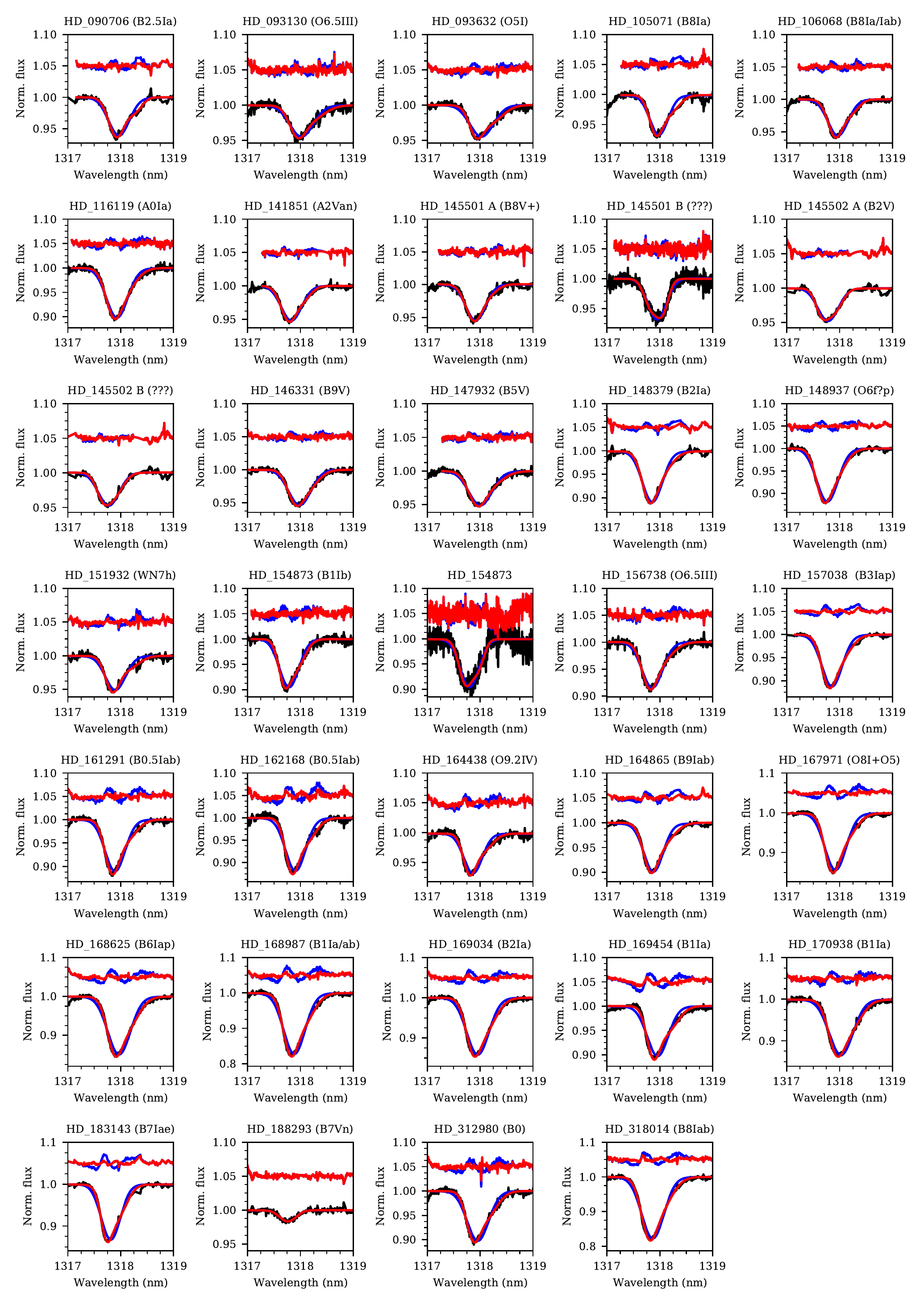}}
\end{center}
\caption{CRIRES DIB observations around 1318~nm corrected for telluric absorption. Blue lines: one component Gaussian fit plus the residual (1.0+data-model). Red lines: ditto for the two component Gaussian fits. The spectral resolution is $\sim$0.026~nm. 
%
%
}
\label{f_1318nm_Gauss_fits}
\end{figure*}

Example fits to the sample are shown in Fig.~\ref{f_1318nm_Gauss_fits}, using one and two Gaussian components. 
For the DIB at 1318~nm the majority of the lines are best fitted by twin Gaussians due to the extended red wing. A single-fit Gaussian leaves larger residuals. 

An attempt was made to search for correlations between the ratio of the integrated flux in the twin Gaussian components and the difference in their peak wavelengths with 5780, 5797~\AA \, equivalent widths,  $\lambda\lambda$5780/5797 DIB ratio and reddening $E(B-V)$. No obvious correlations were found. 

For the 1318~nm line there is a range of FWHM fitted with a single component Gaussian of 
0.36$\pm$0.01~nm (82$\pm$2.5 km\,s$^{-1}$) for HD\,188293 to 0.53$\pm$0.01~nm (120$\pm$2.5  km\,s$^{-1}$) for HD\,170938. This compares with FWHM values of 0.45$\pm$0.12~nm from \cite{Cox2014} and 0.40$\pm$0.05~nm from \cite{Joblin90} in lower-resolution spectra. For the current observations, the 
instrumental resolution of CRIRES was 6~km\,s$^{-1}$, so the DIB lines appear to be always 
resolved. Historically, the blue wing seen towards the 4428~\AA\, DIB was thought to provide evidence for a solid state carrier \citep{Wickramasinghe1968}, although this explanation has fallen out of favour for other reasons, principally being the lack of polarisation seen in DIBs. In contrast, the steep blue side (or red wing) seen in many DIBs is currently explained by the short-wavelength spectral limit of the R-branch being reached in molecular carriers \citep[e.g.,][]{Sarre2014}. 
In the current data even the single-component dominated sightlines show the wing in the 1318~nm DIB.
Finally, we note that no substructure is seen in the 1318~nm DIB at a spectral resolution of 50,000. In the optical DIB at 6614~\AA, such structure has been used to infer the rotational constant, temperature and possible excitation of the DIB via rotational contour modelling \citep{Kerr1996,Cami04,Marshall2015,MacIsaac2022}. Higher spectral resolution CRIRES observations currently being analysed by Cox et al. ($R$=100,000) may be useful in finding such substructure.

\begin{figure*}
   \resizebox{\hsize}{!}{\includegraphics{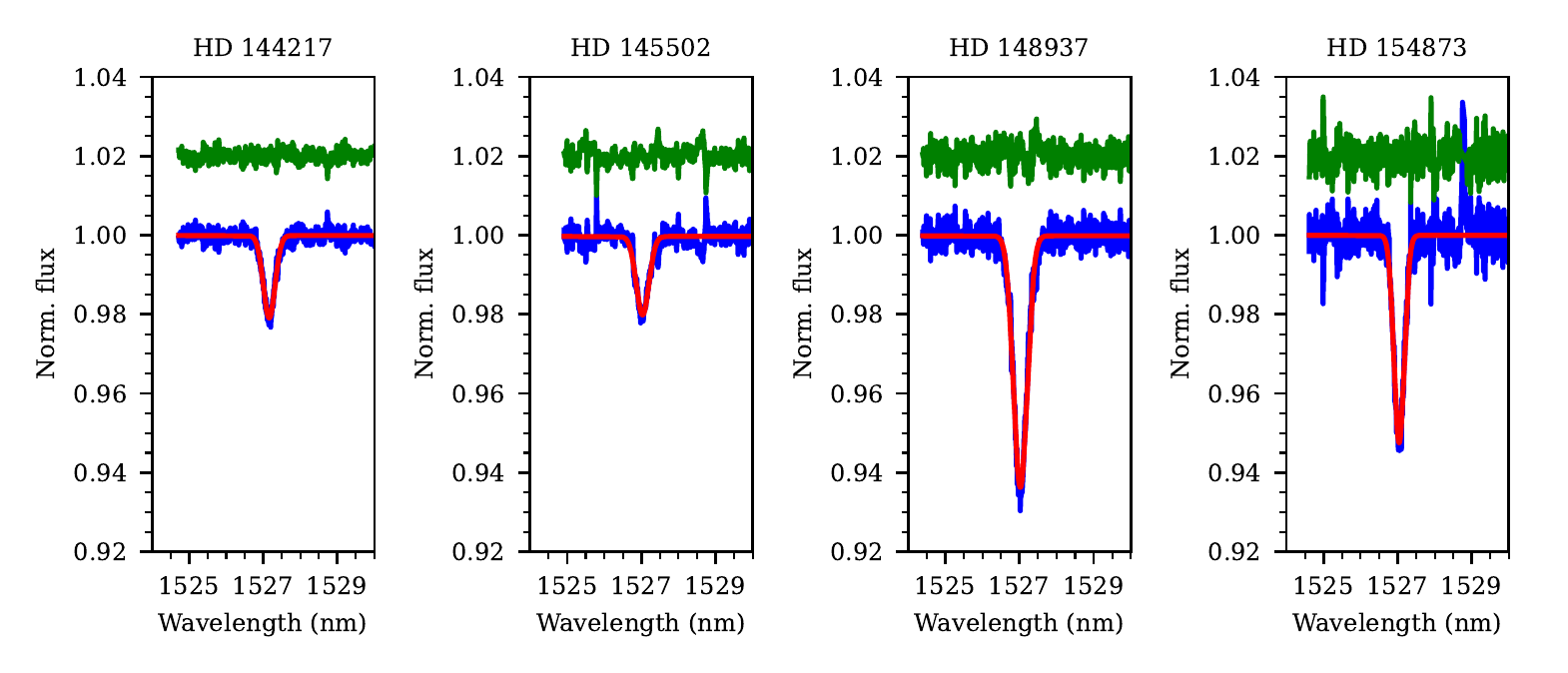}}
\caption{CRIRES DIB observations around 1527~nm corrected for telluric absorption. Blue lines: data. Red lines: one component Gaussian fit. Green lines: residual (1.0+data-model), offset in the ordinate for clarity. The spectral resolution is $\sim$0.030~nm. We recall that HD\,145502 is a binary.}
\label{f_1527nm_Gauss_fits}
\end{figure*}

Figure~\ref{f_1527nm_Gauss_fits} shows the 1527~nm DIB towards four sightlines. At the S/N ratio and resolution of the current observations, a Gaussian fit is sufficient to fit the profile, although in each case the trough of the profile is slightly deeper than the fit implies and in the highest S/N spectra towards HD\,148937 there are some residuals present. 

The three DIBs at 1561, 1565 and 1567~nm have been extensively studied by \citet{Elyajouri2017a} at a spectral resolution of 21,500. They measure mean FWHM values for the three DIBs of 0.44, 0.57 and 0.37~nm, compared with the current work of 0.53, 0.53 and 0.35~nm, respectively. As for the 1318~nm DIB, the current work finds that the 1561 and 1567~nm DIBs are better fitted with twin rather than single Gaussian components.

\begin{figure*}
\begin{center}
   \resizebox{16cm}{!}{\includegraphics{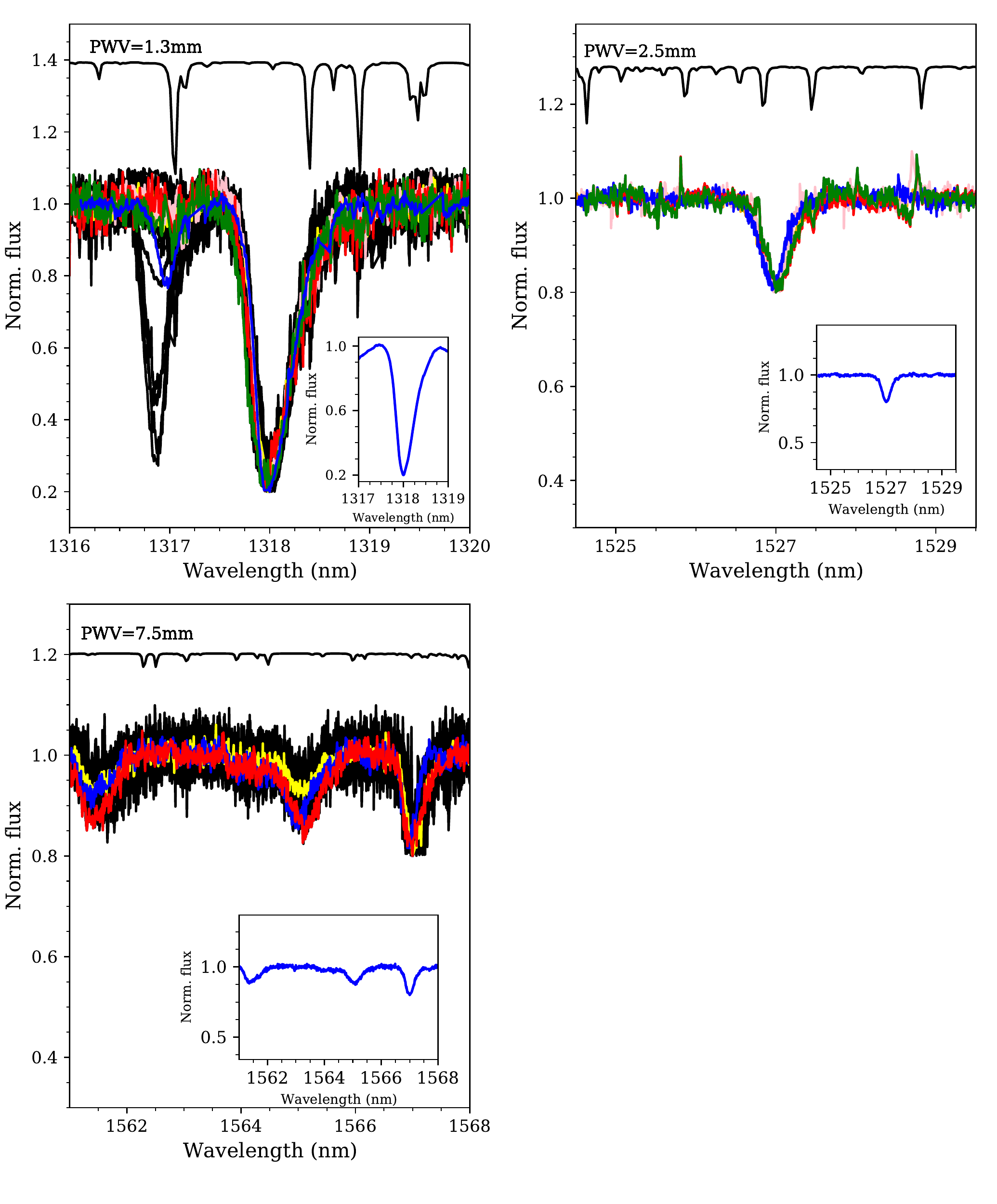}}
\end{center}
\caption{CRIRES DIB spectra around 1318, 1527 and 1561-1567~nm. Main panels: Observations towards all stars, shifted in wavelength to 1318, 1527 and 1567~nm, respectively and the absorption line trough normalised to 0.2. Top black line: Model of telluric absorption assuming PWV values of 1.3, 2.5 and 7.5~mm, respectively with $R$=50,000. Inset panels: averaged profiles for the DIBs with the strongest lines and for the 1318~nm DIBs excluding sightlines showing inverted P-Cygni type profiles.}
\label{f_1318_1527_1568_nm_profiles}
\end{figure*}

Figure~\ref{f_1318_1527_1568_nm_profiles} shows a number of individual profiles for the 1318, 1527, 1561, 1565 and 1567~nm DIBs plus the co-added profile for spectra with the highest S/N ratios. For the DIBs from 1561--1567~nm the spectrum appears to show an extra broad component at around 1564.4~\AA, but the S/N ratio is low. Continuum placement is hence difficult, and such a feature is not mentioned in the case of the higher S/N but lower spectral resolution data of \citep{Elyajouri2017a}. On the other hand, a close look at their Figure 4 shows that a similarly broad and shallow absorption is not precluded. The feature is not telluric in nature.

\subsection{Comparison of Near IR DIB strengths with other DIBs, atomic lines and reddening}
\label{NIRDIB_correlations}


Correlation studies between DIBs and other DIBs, atomic and molecular features serve two purposes. Firstly, to search for lines that near-perfectly correlate, providing evidence that the carrier producing the two lines is the same, and secondly to determine the physical and/or chemical conditions in which certain DIBs exist and hence can be used in other work as probes of the environmental conditions. We note that just as in atomic lines a 1:1 correlation does not mean that the carriers are the same (as in the case of the near perfect correlation between Na\,{\sc i} and K\,{\sc i} equivalent width), there is some evidence that the lack of a 1:1 correlation between different lines at different wavelength may not rule out a common carrier as in the case of the postulated C${60}^+$ DIBs at 9577 and 9632~\AA \, due to issues in telluric line fitting, possible overlapping bands and stellar contamination \citep{Walker17}.

We searched for optical data Advanced Data Products (ADPs) towards our NIRDIB sightlines using the ESO archive\footnote{archive.eso.org} for the FEROS, HARPS \citep{HARPS} and UVES instruments. Some ADPs did not exist for FEROS observations so the data were re-reduced. Additionally we performed a search of the NARVAL/ESPADONS online archives but no relevant data were found. One star (HD\,157778) was used from historical observations using the SOPHIE spectrometer at Observatoire the Haut Provence \citep{Perruchot2008SPIE}. A total of 57 out of 76 of the objects for which we have CRIRES spectra had optical data. 

In each sightline we measured the EW of the strongest 50 DIBs in the catalogue of DIBs in the spectrum of HD\,204827 \citep{Hobbs2008,Hobbs09} as well as the EWs derived from the current spectra. Many of the weaker DIBs were not detected in the archive spectra due to insufficient S/N ratios. Telluric correction was performed by using scaled versions of synthetic spectra produced by {\sc skycalc}. CSV files containing the results of all of the equivalent width measurements and derived correlations are available in CDS/Vizier. Aside from the DIBs, we also measured a number of atomic, ionised and molecular lines. The full list is shown in Table~\ref{t_opticallinesused} with Fig.~\ref{f_Optical_Data} displaying the Na\,{\sc i} (5895~\AA) and K\,{\sc i} (7698~\AA) data for sightlines where available, as well as the DIBs at 5780, 5797, 6196, 6203, 6269, 6283, 6379 and 6614~\AA. 

\begin{table*}[h!]
\begin{center}
\caption[]{Optical lines for which equivalent widths were measured for available spectra.}
\label{t_opticallinesused}
\begin{tabular}{rrr}
\hline
   Species, $\lambda$($\mbox{\AA}$) &  Species, $\lambda$($\mbox{\AA}$) &  Species, $\lambda$($\mbox{\AA}$) \\
\hline   
   NaI   3302.36         &       DIB 5236.18       &         DIB 6204.49    \\
   NaI   3302.97         &       DIB 5418.87       &         DIB 6269.85    \\
   TiII  3383.75         &       DIB 5450.60       &         DIB 6283.84    \\
   FeI   3719.93         &       DIB 5452.40       &         DIB 6376.08    \\
   FeI   3859.91         &       DIB 5494.10       &         DIB 6379.32    \\
   CNI   3873.99         &       DIB 5542.20       &         DIB 6397.01    \\
   CNI   3874.60         &       DIB 5545.40       &         DIB 6439.48    \\
   CNI   3875.76         &       DIB 5546.90       &         DIB 6445.28    \\
   CaII  3933.66         &       DIB 5705.08       &         DIB 6449.22    \\
   KI    4044.14         &       DIB 5711.60       &         DIB 6521.30    \\
   C3    4051.60         &       DIB 5763.20       &         DIB 6597.40    \\
   CaI   4226.72         &       DIB 5766.10       &         DIB 6614.00    \\
   CH+   4232.54         &       DIB 5780.48       &         DIB 6993.13    \\
   CH    4300.31         &       DIB 5797.06       &         DIB 7224.03    \\
   DIB   4501.79         &       DIB 5849.81       &         DIB 7562.25    \\
   DIB   4726.83         &       NaI 5889.95       &         KI  7698.96    \\
   DIB   4762.61         &       NaI 5895.92       &         C2  7714.20    \\
   DIB   4780.02         &       DIB 6089.85       &         C2  7724.20    \\
   DIB   4963.88         &       DIB 6113.18       &         Rb  7800.30    \\
   DIB   4984.79         &       DIB 6195.98       &         C2  8757.68    \\
   DIB   4987.42         &       DIB 6203.05       &         C2  8761.19    \\
   DIB   5074.47         &       DIB 6204.49       &         C2  8763.74    \\
   DIB   5176.04         &       DIB 6269.85       &  C$_{60}^{+}$ 9632.10  \\
\hline
\end{tabular}
\end{center}
\end{table*}                                                                                 
 In work on optical DIBs, \cite{Moutou1999} classified a Pearson correlation coefficient $r$ of less than 0.7 as weak, 0.7 to 0.95 as good, and anything exceeding 0.95 as strong \citep[also see][who suggest two quantities may have a physical relation if $r$ exceeds 0.86--0.88]{Fan2017,Friedman11}. This is due to the fact that within the Galaxy increasing distance along the line of sight will typically result in more absorbing material, so some correlation is natural.  Most DIBs correlate (if only weakly), although to date the only near perfect 1:1 correlation between two DIBs is for those at 6196.0 and 6613.6$\mbox{\AA}$ \, \citep{McCall2010}. 

\begin{figure}
\resizebox{\hsize}{!}{\includegraphics[trim=0cm 0cm 1.0cm 0cm]{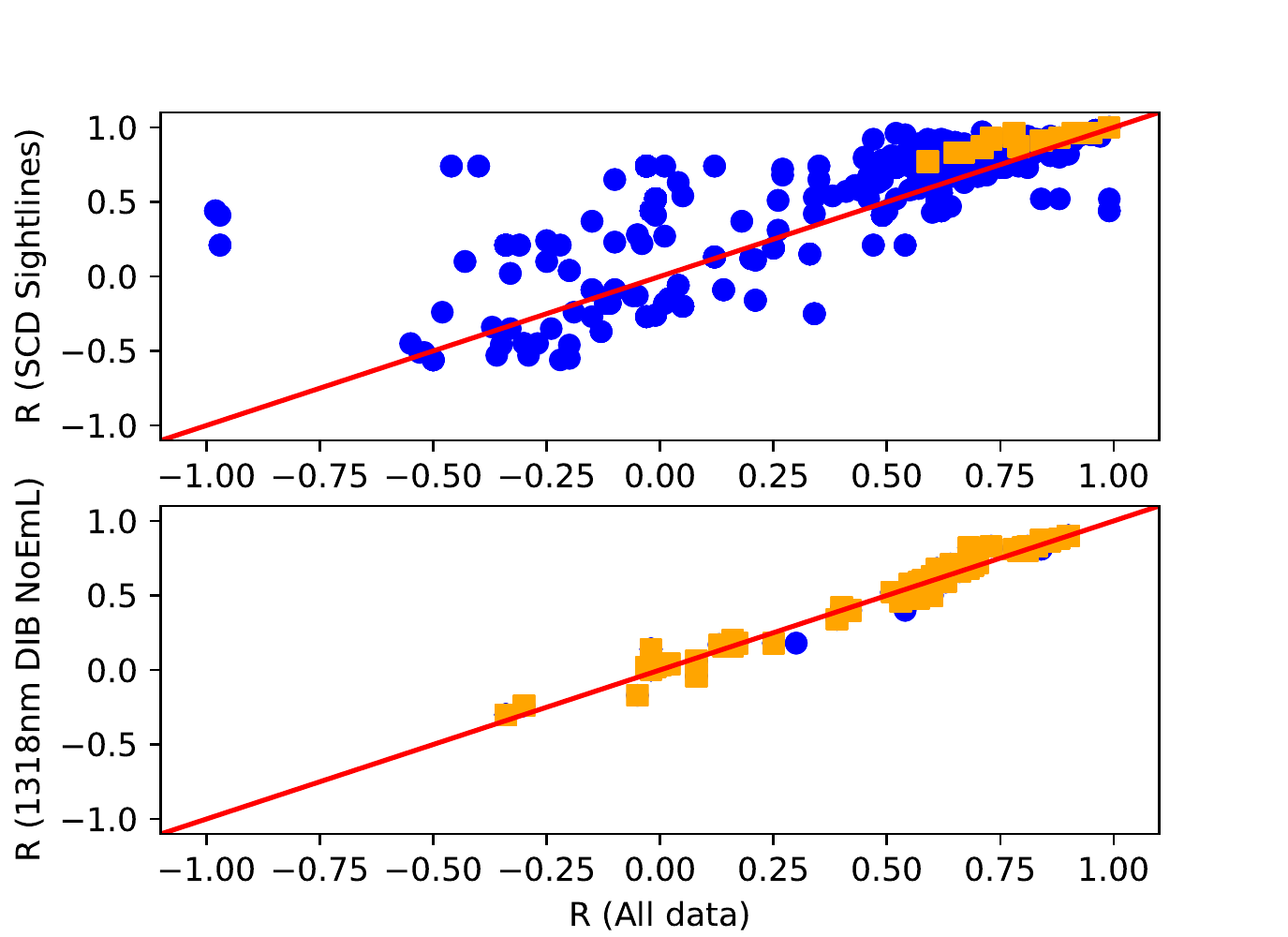}}
\caption{Top panel: Comparison of correlation coefficients derived using the whole sample (abscissa) with those determined for the Single Cloud Dominated (SCD) sightlines (ordinate). There is a tendency for the correlation coefficients for the latter to be higher. The stars that we consider to be SCD from their K\,{\sc i} profiles are HD numbers 36861, 36862, 88661, 89587, 93576, 137753, 144217, 145501, 145502, 147932, 148184, 156738, 164438, 164865, 167264, 169454, 188294 and 318014. Blue and orange symbols: greater than or equal to 10 or 15 datapoints, respectively. Bottom panel: Correlation coefficients comparing the 1318~nm sample excluding emission line stars with the whole sample. }
\label{fig_Corr_Coefficients_Single_Dominating_Sightlines_vs_Full}
\end{figure}

Table~\ref{DIB_corr_coeff} shows the Pearson correlation coefficients for those DIBs that showed at least one instance of a correlation $r$ exceeding 0.87 with another feature (the criteria of \cite{Friedman11} to indicate that "two quantities are correlated at a significant level") for the whole sample, plus some optical transitions. Note that the correlation coefficients should be considered lower limits as when sightlines with only one dominating component are considered (SCDs), typically the correlation coefficient improves as shown in Fig.~\ref{fig_Corr_Coefficients_Single_Dominating_Sightlines_vs_Full} and Table~\ref{DIB_corr_coeff_SDC}. The disadvantage of such sightlines is that they are few in number in the current sample. Figure \ref{f_Correlation_fits} shows the correlations themselves where $r\ge$0.87.

Following the referees suggestion, the method of partial correlation coefficients was used to investigate the confidence level in the derived correlation coefficients. Partial correlation coefficients take into account the fact that even when there is correlation between two variables, it may be partly caused by a third variable, the so-called covariate or controlling variable \citep{Vio2020}. For the current work the reddening $E(B-V)$ was used as the covariate. The implementation of \cite{Vallat2018} was used to derive the 95 percent confidence limits around the partial correlation coefficient. These are also shown in Tables~\ref{DIB_corr_coeff} and \ref{DIB_corr_coeff_SDC}. If the 95 percent ranges are large then the chance of the two main variables not being physically correlated is increased. We note that results for correlations are shown only where there are 5 data points or more and with so few samples the significance is low. This is particularly the case for the sightlines which are single cloud dominated where there are fewer data available.

%
%

%
%

\subsubsection{Correlations between NIRDIB strength and reddening}

Figure~\ref{fig_NIRdibs_vs_Reddening} shows the correlation between DIB EWs and reddening for the lines studied in this paper. In the optical wavelength range, at low values of $E(B-V)$ the correlation between EW and reddening is reasonable \citep[e.g.][]{Lan2015}, but reaches saturation at higher values, partly due to the skin effect by which DIB carriers appear to disappear in the centre of dense clouds \cite{Snow1974}. One DIB relatively well correlated with reddening is the "Gaia DIB" at 8620.4$\mbox{\AA}$ \, \citep{Munari2000} although even it shows some intrinsic scatter.

There are reasonable correlations for the CRIRES-observed targets with reddening for all of the the NIRDIBs studied. Correlations between the 1318, 1527, 1561, 1565 and 1567~nm DIBs and $E(B-V)$ are in the range 0.76 to 0.92 for the full sample and 0.74 to 0.96 for the single cloud dominated sample. These are similar to the values observed by \cite{Cox2014} for the 1318 and 1527~nm DIBs. We note that in the current dataset the scatter in the measurements above $E(B-V)$ of around 2-3 mags becomes larger. Of course, this could partly be due to stellar contamination of the lines in the high-reddening X-shooter Cepheid sample. Stellar contamination in the 1318~nm and 1527~nm lines for the Cepheid variables was estimated from X-shooter spectra of six Cepheid variable stars taken from the X-shooter spectral library of \cite{Chen2014}. Five of these are in the Galactic disc with reddening $E(B-V)$ values between 0.04 and 0.89 and one towards the low-metallicity cluster $\omega$~Cen with $E(B-V)$=0.12\ . 



\begin{figure*}
\resizebox{\hsize}{!}{\includegraphics{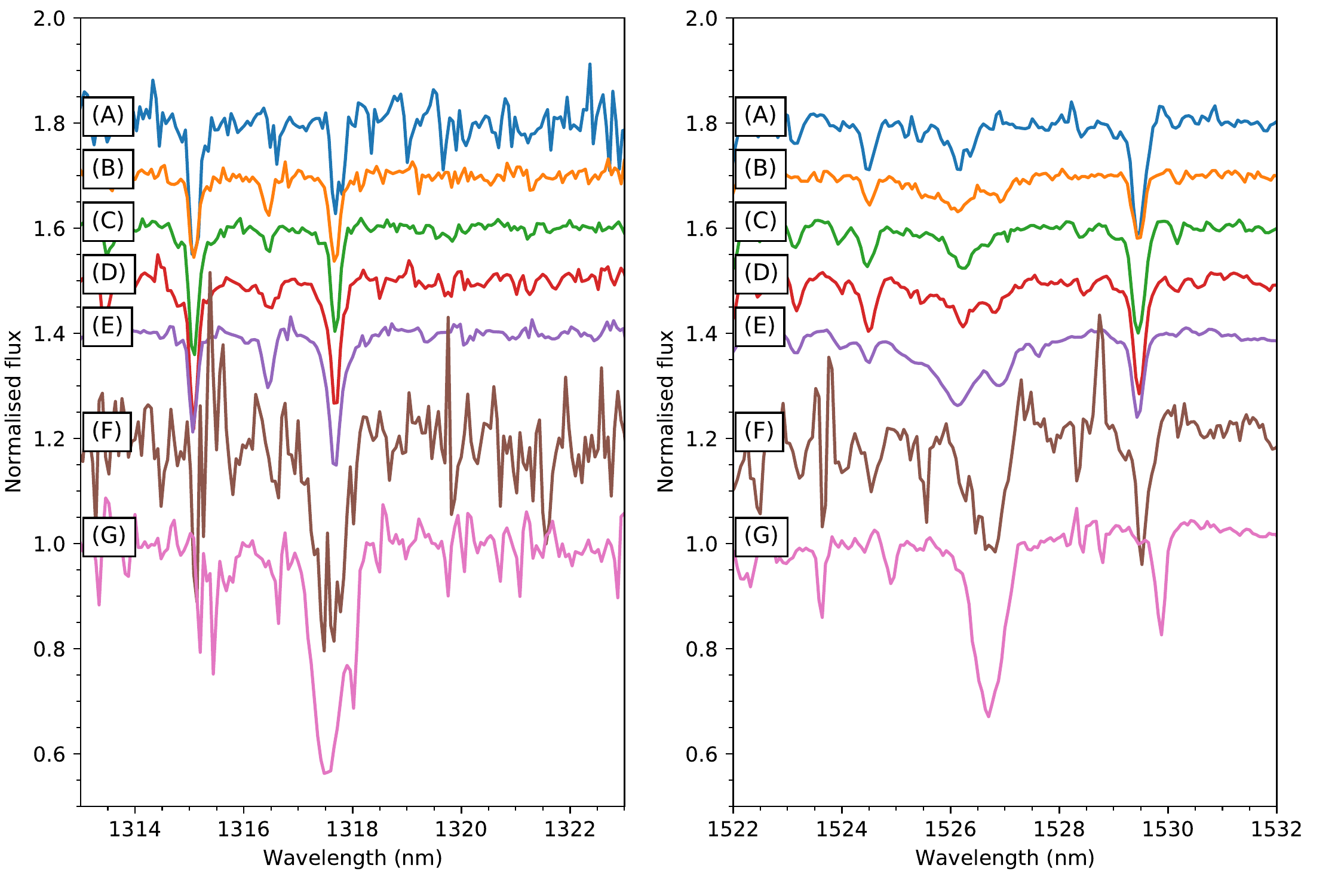}}
\caption{X-shooter spectra of Cepheid variables around 1318~nm (left panels) and 1527~nm (right panels) taken from \cite{Chen2014} and \cite{Minniti2020}. Reddening increases downwards. (A) HD\,052973, $E(B-V)$=0.04, EW(1318~nm)=368$\pm$30 m\AA, EW(1527~nm)=450$\pm$50 m\AA. (B) HD\,101602, $E(B-V)$=0,25, EW(1318~nm)=419$\pm$30, EW(1527~nm)=840$\pm$50 m\AA. (C) HD\, 178287, $E(B-V)$=0.40, EW(1318~nm)=470$\pm$20 m\AA, EW(1527~nm)=570$\pm$60 m\AA. (D) HD\,179315, $E(B-V)$=0.59, EW(1318~nm)=691$\pm$50 m\AA, EW(1527~nm)=890$\pm$70 m\AA. (E) HD\,139717, $E(B-V)$=0.89, EW(1318~nm)=913$\pm$30 m\AA, EW(1527~nm)=2300$\pm$600 m\AA. (F) B19, $E(B-V)$=3.27, EW(1318~nm)=2300$\pm$250 m\AA, EW(1527~nm)=1400$\pm$300 m\AA. (G) D20, $E(B-V)$=3.57, EW(1318~nm)=2120$\pm$110 m\AA, EW(1527~nm)=3300$\pm$100 m\AA.}
\label{fig_xshooter_cepheids}
\end{figure*}

\begin{figure*}
\resizebox{\hsize}{!}{\includegraphics{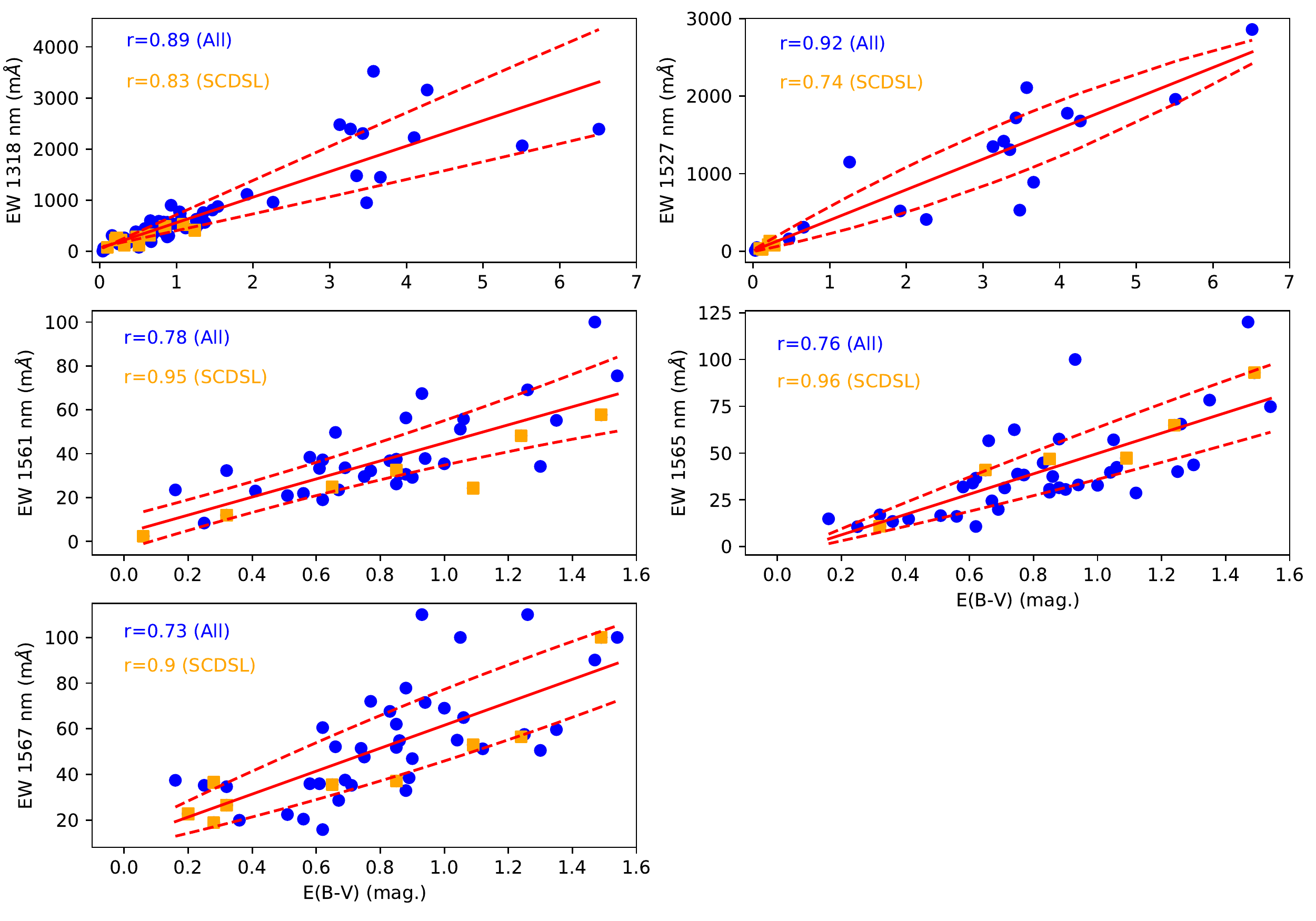}}
\caption{Reddening $E(B-V)$ (mag.) vs. equivalent width in m\AA \, for the DIBs in the current sample. Blue points show all of the data with orange points showing only single cloud dominated sightlines. For the 1318 and 1527~nm DIBs for reddening exceeding 3.13 mag. the data are from the X-shooter Cepheid sample which suffer from stellar contamination of around 500~m\AA\ for unreddened sightlines. The solid red line is the best fit y=mx+c with the dashed red curves showing the upper and lower 1 sigma error bounds.}
\label{fig_NIRdibs_vs_Reddening}
\end{figure*}

Figure~\ref{fig_xshooter_cepheids} shows the \citeauthor{Chen2014} spectra (at a range of reddenings) and two targets from \cite{Minniti2020}. The EW values of the stellar lines around 1318~nm and 1527~nm DIBs for the Galactic plane Cepheid with $E(B-V)$ of 0.12 are 368 and 651~m\AA, respectively. Hence, at $E(B-V)$ $\approx$ 3 magnitudes the stellar contribution in the Cepheid data could be up to 20 percent in both 1318~nm and 1527~nm DIBs. We note that the Cepheid towards the low-metallicity cluster Omega Cen (NGC 5139 LEID 32029) with $E(B-V)$=0.12 has a 1318~nm EW of 135~m\AA \, although the region around 1527~nm is very crowded with lines with the total EW from 1524 to 1530~nm of $\sim$1.5~\AA. 

These results compare to \cite{Cox2014} who found $r$=0.91 for the 1318.1~nm DIB (with fewer datapoints). No correlation coefficients are shown for $E(B-V)$ in the paper of \citep{Elyajouri2017a} for this DIB.

\subsubsection{Correlation between strength of NIR DIBs}

\begin{figure*}
\resizebox{\hsize}{!}{\includegraphics{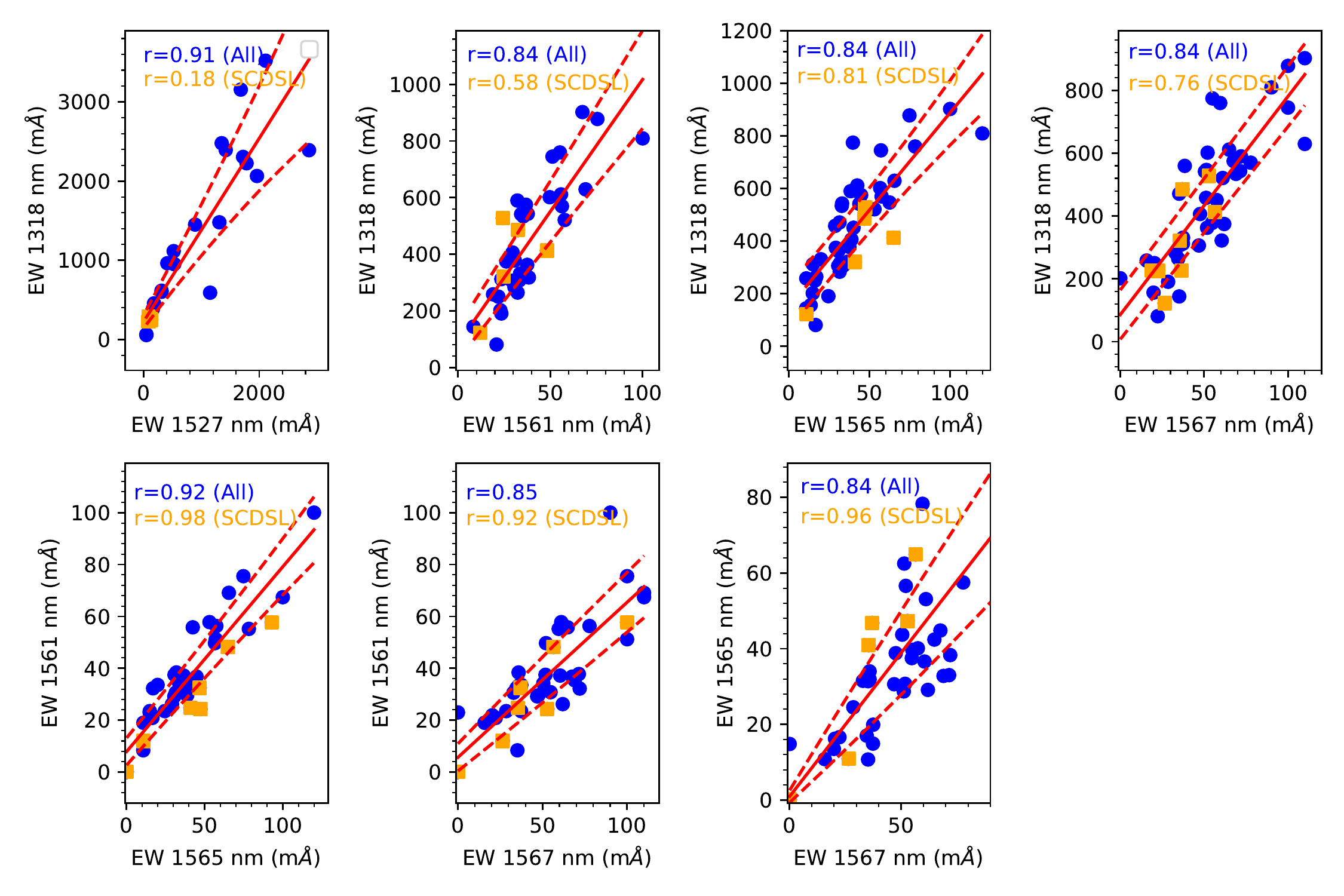}}
\caption{Correlations in equivalent width in m\AA \, for the NIR DIBS in the current sample. Both the whole sample (blue points) and the Single Cloud Dominant subsample (orange points) are shown. The dashed red curves show the upper and lower 1 sigma error bounds.}
\label{fig_NIRdibs_vs_NIRdibs_EW}
\end{figure*}

Figure \ref{fig_NIRdibs_vs_NIRdibs_EW} shows correlations between the line strength of 
DIBs studied in this paper. The correlations between the 1318, 1527, 1561, 1565 and 1567~nm DIBs range from 0.84 to 0.92 for the entire sample and from 0.18 (few stars) to 0.98 for the single cloud dominated sample. Recall that these are lower limits due to the S/N limitations and difficulty in the baseline fitting especially for the 1561~nm DIB. \cite{Cox2014} determined $r$=0.93 when comparing the 1318 and 1527~nm DIBs, similar to our value of 0.90. 
%
%

\subsubsection{Correlation between 1318~nm DIB strength, H\,{\sc i}, H$_{2}$ and f(H$_2$)}

\begin{figure*}
\resizebox{\hsize}{!}{\includegraphics[clip,trim=0cm 0.0cm 0.0cm 0cm]{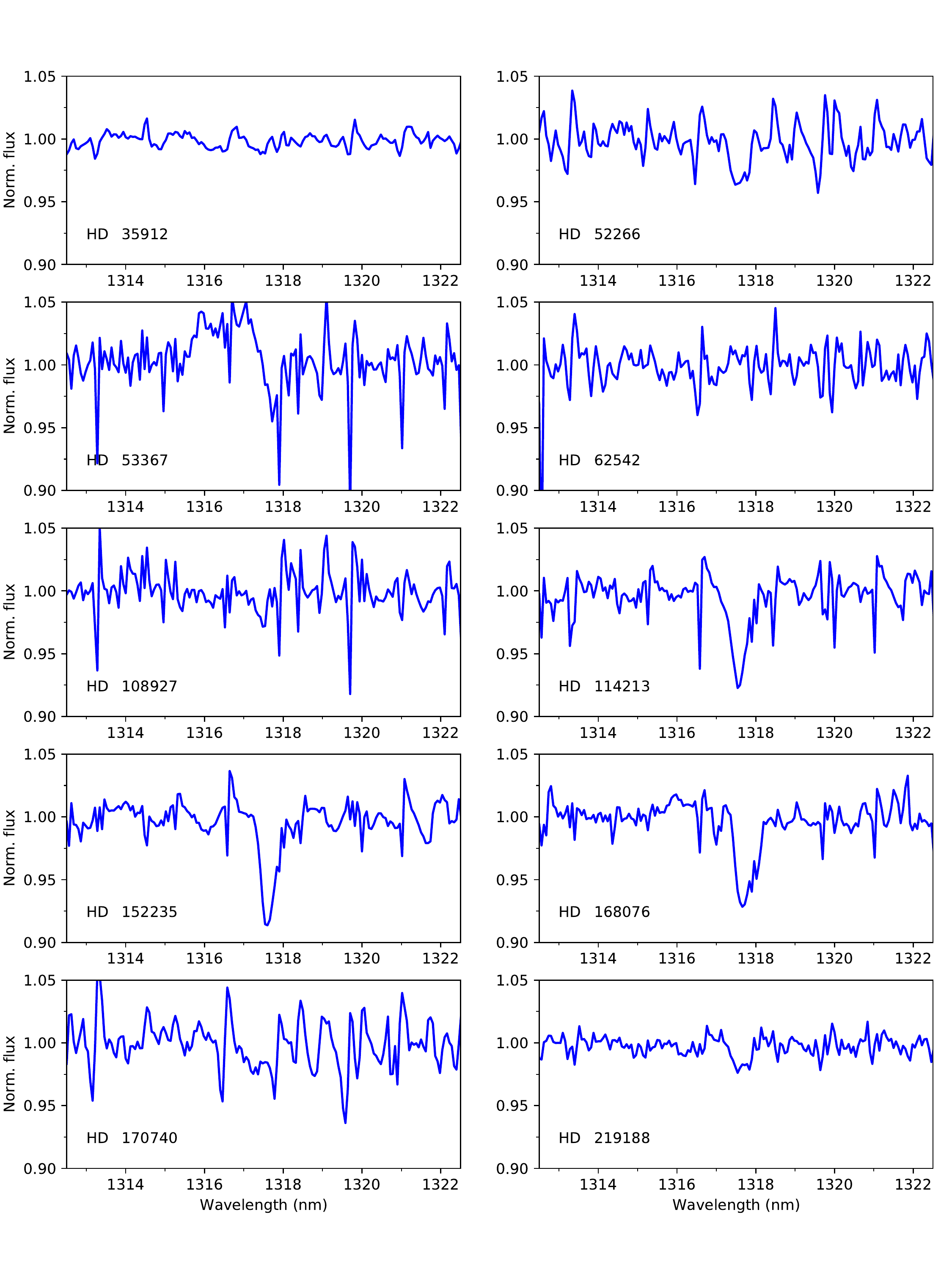}}
\caption{X-shooter data around 1318~nm corrected for telluric contamination using {\sc molecfit} for which f(H$_2$) is available in the literature.}
\label{fig_XSHOOTER_1318nm_DIBS}
\end{figure*}

Figure~\ref{fig_XSHOOTER_1318nm_DIBS} shows normalised X-shooter spectra, corrected for telluric contamination using {\sc molecfit} for early-type stars for which $N$(H\,{\sc i}), $N$(H$_{2}$) and f(H$_{2}$) data were available from the literature (\cite{Fan2017} and refs therein). To these data we added our CRIRES 1318.1~nm observations, using either literature values for H\,{\sc i}, H$_{2}$ and f(H$_{2}$), or, in the case where these did not exist, estimates of H$_{2}$ from the CH(4300~\AA) line derived from log($N$(H$_{2}$)) = 0.8875 x log($N$(CH)) +  8.88 using data from \cite{Welty06}. The DIB at 5780~\AA \, was used to estimate $N$(H\,{\sc i}) via log($N$(H\,{\sc i})( = 0.82 x log(EW(5780~\AA \, DIB)) + 19.34 \citep{Friedman11}. The results are shown in Table~\ref{f_H_2_vs_NIRDIB_strength}.

\begin{figure*}
\resizebox{\hsize}{!}{\includegraphics[clip,trim=0cm 0.4cm 0cm 0.4cm]{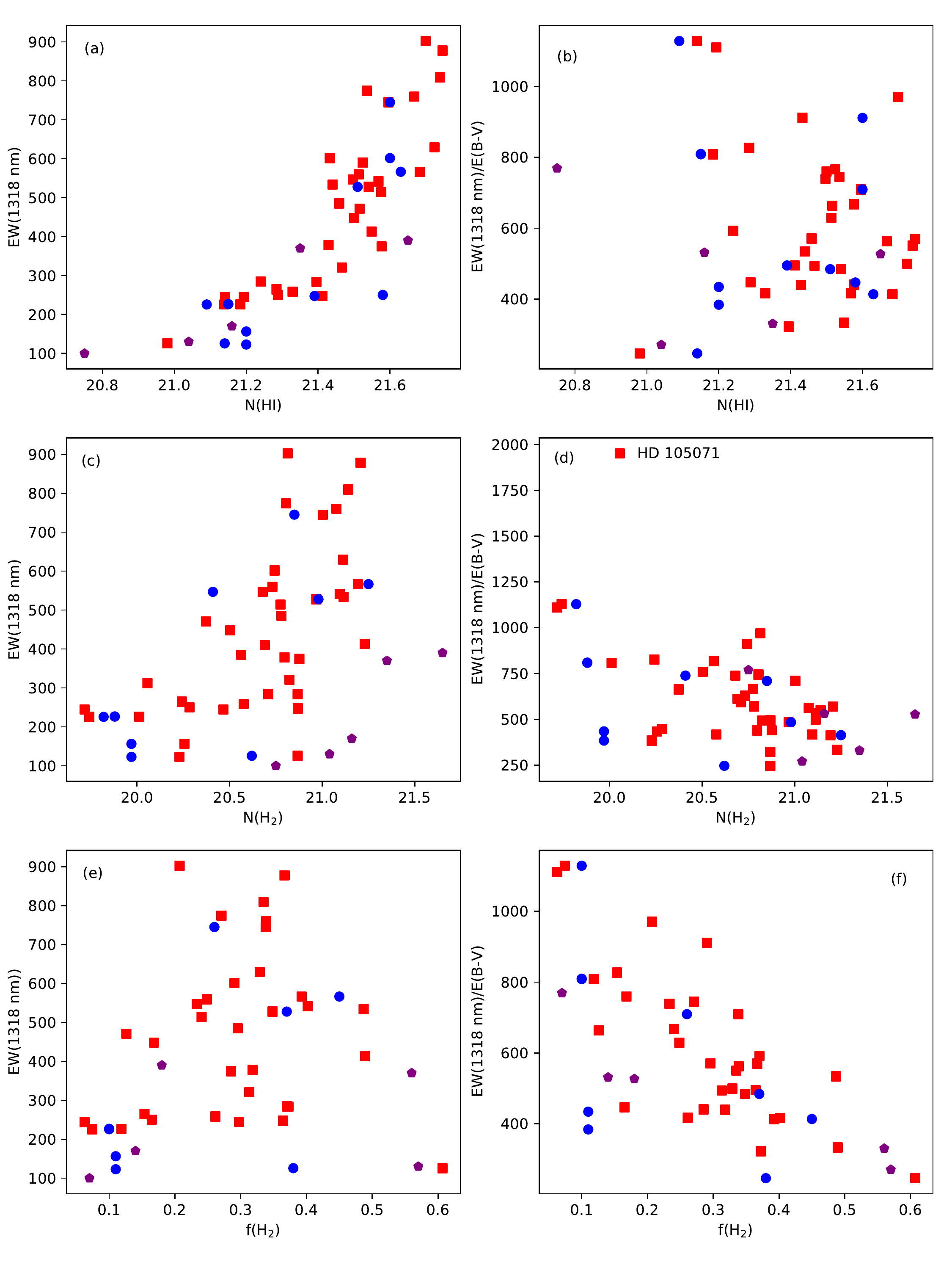}}
\caption{EW(1318~nm) (plots a,c,e) or EW(1318~nm)/$E(B-V)$ (plots b,d,f) plotted against $N$(\ion{H}{i}), $N$(H$_{2}$) and f(H$_{2}$). Stars with emission lines near 1318~nm were excluded. Blue circles: CRIRES data with $N$(\ion{H}{i}), $N$(H$_{2}$), f(H$_{2}$) from the literature, red squares: CRIRES data with \ion{H}{i} calculated from the 5780 DIB and N(H$_{2}$) from CH, purple pentagons: X-shooter with $N$(\ion{H}{i}), $N$(H$_{2}$), f(H$_{2}$) from the literature. Outlier HD~105071 is labelled.
}
\label{fig_EW1318_HI_H2_fH2}
\end{figure*}

Figure~\ref{fig_EW1318_HI_H2_fH2} shows either the EW of the 1318~nm line or the EW divided by the $E(B-V)$ plotted against $N(\ion{H}{i})$, $N(H_{2})$ and f(H$_{2}$). The aim of these plots was to search for the lambda-shaped trend which shows that atomic and diffuse molecular clouds show dramatically different balances in the creation and destruction of DIBs \citep{Fan2017}. Although there are hints of such a trend when plotting the EW of the 1318~nm DIB against molecular fraction, when this is normalised by the reddening, the shape disappears and what is left is a noisy but slowly decreasing EW(1318~nm)/$E(B-V)$ trend with increasing molecular fraction. In the straight correlation we find better correlation between the 1318~nm DIB and \ion{H}{i} than with H$_{2}$, although still with a lot of scatter. 

%
%
%

%
%
\begin{table*}
\begin{center}
\caption{Values of the equivalent widths of the 1318~nm DIB, CH(4300~\AA) and the 5780~\AA \, DIB for CRIRES (C) and X-shooter (X) sightlines for which f(H2) is available or can be estimated. }
\label{f_H_2_vs_NIRDIB_strength}
\begin{tabular}{rlrrrrrrrrr}
\hline
HD        & Inst.   & EW(1318)      & EW(CH4300) & EW(5780)   & N(H$_2$)    & N(H$_2$)c   & N(H{\sc i})  &    NH{\sc i}c &    fH$_2$     &   fH$_2$c     \\
          &         & (m\AA)        & (m\AA)     & (m\AA)     & (cm$^{-2}$) & (cm$^{-2}$) & (cm$^{-2}$)  &   (cm$^{-2}$) &   (cm$^{-2}$) &   (cm$^{-2}$) \\
\hline
    35912      & X       &  \ldots            &       \ldots       &      \ldots        &  20.35 &   \ldots       & 20.35 &     \ldots      &   \ldots       &   0.12 \\
    36408      & C       &  Det. (HWV)  &     \ldots         &     \ldots         &     \ldots       &   \ldots        &    \ldots       &    \ldots        &    \ldots       &     \ldots      \\
    36861      & C       &    $<$       &      1.76 &     48.29 &   19.65 &  19.11 &  20.77 &   20.72 &   0.14 &   0.04 \\
    36862      & C       &      \ldots           &      2.44 &     34.27 &   19.80 &  \ldots         &    \ldots       &   20.60 &   0.24 &  \ldots         \\
    36959      & C       &  $<$ (HWV)  &       \ldots       &     \ldots         &   \ldots         &    \ldots       &  20.40 &   \ldots         &   \ldots        &     \ldots      \\
    52266      & X       &       170.00 &      \ldots        &    \ldots          &  21.16 &   \ldots       & 21.16 &   \ldots        &    \ldots      &   0.14 \\
    53367      & X       &      \ldots &     \ldots         &    \ldots          &  21.32 &   \ldots       & 21.32 &    \ldots       &   \ldots       &   0.51 \\
    57060      & X       &      \ldots &     \ldots         &     \ldots         &  20.78 &    \ldots      & 20.78 &   \ldots        &   \ldots       &    \ldots       \\
    62542      & X       &      \ldots &     \ldots         &     \ldots         &  20.90 &   \ldots       & 20.90 &    \ldots       &    \ldots      &   0.62 \\
    88661      & C       &  $<$ (EmL) &     \ldots         &     13.21 &    \ldots        &    \ldots       &  \ldots         &   20.26 &   \ldots        &        \ldots   \\
    90706      & C       &       308.10 &     \ldots         &      \ldots        &    \ldots        &    \ldots       &   \ldots        &    \ldots        &     \ldots      &    \ldots       \\
    92740      & C       &       156.30 &      6.38 &     \ldots         &   20.26 &  19.97 &  21.20 &    \ldots        &   \ldots        &   0.11 \\
    93130      & C       &       250.20 &      6.76 &    237.90 &   20.28 &     \ldots      &  21.58 &   21.29 &   0.17 &   \ldots        \\
    93632      & C       &       258.40 &     12.55 &    266.51 &   20.58 &    \ldots       &   \ldots        &   21.33 &   0.26 &    \ldots       \\
    94910      & C       &       448.10 &     10.76 &    430.55 &   20.50 &    \ldots       &   \ldots        &   21.50 &   0.17 &    \ldots       \\
    97707      & C       &       190.80 &     \ldots         &      \ldots        &    \ldots        &    \ldots       &    \ldots       &        \ldots    &    \ldots       &    \ldots       \\
   105056      & C       &       202.00 &      3.36 &     \ldots         &   19.95 &   \ldots        &  21.15 &     \ldots       &    \ldots       &    \ldots       \\
   105071      & C       &       312.20 &      4.18 &     \ldots         &   20.06 &    \ldots       &   \ldots        &    \ldots        &    \ldots       &     \ldots      \\
   106068      & C       &       264.60 &      6.19 &    234.96 &   20.24 &   \ldots        &    \ldots       &   21.28 &   0.15 &    \ldots       \\
    108927      & X       &       \ldots          &     \ldots         &    \ldots          &  20.86 &   \ldots       & 20.86 &     \ldots      &   \ldots       &   0.46 \\
   111774      & X       &     \ldots            &      1.00 &     \ldots         &   19.38 &   \ldots        &     \ldots      &    \ldots        &   1.00 &     \ldots      \\
   112272      & C       &       362.70 &     39.21 &    441.12 &   21.12 &   \ldots        &   \ldots        &   21.51 &   0.45 &    \ldots       \\
   113904      & X       &       144.20 &      1.66 &    126.35 &   19.62 &  19.84 &  21.08 &   21.06 &   0.07 &   0.10 \\
   114213      & X       &       370.00 &    \ldots          &       \ldots       &  21.35 &    \ldots      & 21.35 &  \ldots         &    \ldots      &   0.56 \\
   115088     & C       &    $<$       &    \ldots          &     13.75 &  \ldots          &   \ldots        &    \ldots       &   20.27 &   \ldots        &     \ldots      \\
   115363      & C       &       331.10 &     27.87 &     \ldots         &   20.95 &    \ldots       &    \ldots       &    \ldots        &    \ldots       &     \ldots      \\
   115842      & C       &        81.20 &     21.83 &    234.59 &   20.84 &   \ldots        &  21.15 &   21.28 &   0.42 &      \ldots     \\
   116119      & C       &       471.00 &      8.17 &    449.75 &   20.37 &   \ldots        &    \ldots       &   21.52 &   0.13 &   \ldots        \\
   125007      & C       &      \ldots           &       \ldots       &     18.28 &     \ldots       &     \ldots      &   \ldots        &   20.37 &    \ldots       &    \ldots       \\
   125241      & C       &       514.10 &     19.10 &    532.96 &   20.78 &    \ldots       &    \ldots       &   21.58 &   0.24 &    \ldots       \\
   142468      & C       &       589.80 &    \ldots          &    461.15 &   \ldots         &    \ldots       &    \ldots       &   21.52 &        \ldots   &     \ldots      \\
   144217      & C       &       225.70 &      2.15 &    156.24 &   19.74 &  19.82 &  21.09 &   21.14 &   0.07 &   0.10 \\
   144218      & C       &       244.60 &      9.95 &    157.34 &   20.47 &    \ldots       &  \ldots        &   21.14 &   0.30 &     \ldots      \\
   145501      & C       &       244.30 &      2.04 &    182.11 &   19.72 &   \ldots        &   \ldots        &   21.19 &   0.06 &     \ldots      \\
   145502      & C       &       226.40 &      3.80 &    177.17 &   20.01 &  19.88 &  21.15 &   21.18 &   0.12 &   0.10 \\
   147889      & X       &      \ldots &   \ldots           &     \ldots         &  21.80 &   \ldots       & 21.80 &   \ldots        &   \ldots        &   0.48 \\
   147932      & C       &       284.30 &     16.59 &    207.86 &   20.71 &   \ldots        &     \ldots      &   21.24 &   0.37 &     \ldots      \\
   148184      & X       &       125.70 &     23.21 &    100.11 &   20.87 &  20.62 &  21.14 &   20.98 &   0.61 &   0.38 \\
   148379      & C       &       546.70 &     15.60 &    427.06 &   20.68 &  20.41 &    \ldots       &   21.50 &   0.23 &    \ldots       \\
   148688      & C       &       318.10 &     13.20 &     \ldots         &   20.60 &   \ldots        &   \ldots       &    \ldots        &    \ldots       &     \ldots      \\
   148937      & X       &       601.60 &     17.85 &    356.50 &   20.74 &    \ldots       &  21.60 &   21.43 &   0.29 &    \ldots       \\
   151932      & C       &       247.40 &     23.27 &    336.08 &   20.87 &    \ldots       &  21.39 &   21.41 &   0.36 &    \ldots       \\
   152235      & X       &       406.30 &     29.13 &    418.07 &   20.98 &  21.00 &  21.42 &   21.49 &   0.38 &   0.43 \\
   152386      & C       &       378.00 &     19.97 &    352.84 &   20.80 &   \ldots        &   \ldots        &   21.43 &   0.32 &    \ldots       \\
   154368      & X       &   $<$        &    \ldots          &    \ldots          &  21.00 &  \ldots       & 21.00 &   \ldots        &     \ldots     &   0.74 \\
   154811      & C       &       409.70 &     15.96 &    \ldots          &   20.69 &    \ldots       &    \ldots       &    \ldots        &   \ldots        &    \ldots       \\
   154873      & C       &       384.60 &     12.18 &   \ldots           &   20.56 &    \ldots       &   \ldots        &   \ldots         &    \ldots       &   \ldots        \\
 \hline
\end{tabular}
\end{center}
\end{table*}
  
\setcounter{table}{3}
\begin{table*}
\begin{center}
\caption{continued.}
\begin{tabular}{rlrrrrrrrrr}
\hline
HD        & Inst.   & EW(1318)      & EW(CH4300) & EW(5780)   & N(H$_2$)   & N(H$_2$)c & N(H{\sc i})  &    NH{\sc i}c &    fH$_2$ &   fH$_2$c \\
          &         & (m\AA)        & (m\AA)     & (m\AA)     & (cm$^{-2}$)  & (cm$^{-2}$) & (cm$^{-2}$)   &(cm$^{-2}$) &        &        \\
\hline
   156134      & X       &       283.70 &     23.21 &    320.89 &   20.87 & \ldots & \ldots &   21.40 &   0.37 & \ldots \\
   156201      & C       &       306.20 &    \ldots &    318.77 &  \ldots & \ldots & \ldots &   21.39 & \ldots & \ldots \\
   156212      & X       &       322.10 &     30.49 &    375.79 &   21.00 & \ldots & \ldots &   21.45 &   0.41 & \ldots \\
   156738      & C       &       413.00 &     49.81 &    494.35 &   21.23 & \ldots & \ldots &   21.55 &   0.49 & \ldots \\
   157038      & C       &       559.60 &     17.40 &    447.04 &   20.73 & \ldots & \ldots &   21.51 &   0.25 & \ldots \\
   157778      & C       &    $<$       &    \ldots &     34.26 &  \ldots & \ldots & \ldots &   20.60 & \ldots & \ldots \\
   164438      & C       &       320.90 &     21.13 &    391.36 &   20.82 & \ldots & \ldots &   21.47 &   0.31 & \ldots \\
   164865      & C       &       485.20 &     19.28 &    383.28 &   20.78 & \ldots & \ldots &   21.46 &   0.30 & \ldots \\
   165319      & C       &       374.60 &     23.65 &    533.85 &   20.88 & \ldots & \ldots &   21.58 &   0.29 & \ldots \\
   166734      & X       &       566.50 &     46.12 &    720.74 &   21.19 &  21.25 &  21.63 &   21.68 &   0.39 &   0.45 \\
   167264      & X       &       122.90 &      6.01 &    \ldots &   20.23 &  19.97 &  21.20 &  \ldots & \ldots &   0.11 \\
   167971      & C       &       745.00 &     30.89 &    563.33 &   21.00 &  20.85 &  21.60 &   21.60 &   0.34 &   0.26 \\
   168076      & X       &       390.00 &    \ldots &    \ldots &   21.65 & \ldots &  21.65 &  \ldots & \ldots &   0.18 \\
   168607      & C       &       877.80 &     47.60 &    859.65 &   21.21 & \ldots & \ldots &   21.75 &   0.37 & \ldots \\
   168625      & C       &       809.30 &     41.27 &    842.82 &   21.14 & \ldots & \ldots &   21.74 &   0.34 & \ldots \\
   168987      & C       &       902.30 &     20.74 &    752.62 &   20.81 & \ldots & \ldots &   21.70 &   0.21 & \ldots \\
   169034      & X       &       759.60 &     36.03 &    689.54 &   21.08 & \ldots & \ldots &   21.67 &   0.34 & \ldots \\
   169454      & X       &       527.80 &     28.68 &    482.95 &   20.97 &  20.98 &  21.51 &   21.54 &   0.35 &   0.37 \\
   170740      & X       &       130.00 &    \ldots &    \ldots &   21.04 & \ldots &  21.04 &  \ldots & \ldots &   0.57 \\
   170938      & C       &       774.30 &     20.30 &    476.37 &   20.80 & \ldots & \ldots &   21.54 &   0.27 & \ldots \\
   171722      & C       &   $<$        &      0.47 &     20.88 &   19.02 & \ldots & \ldots &   20.42 &   0.07 & \ldots \\
   183143      & C       &       629.40 &     38.99 &    807.49 &   21.11 & \ldots & \ldots &   21.72 &   0.33 & \ldots \\
   214080      & X       &       \ldots &    \ldots &    \ldots &   20.58 & \ldots & 20.58  &  \ldots & \ldots &   0.01 \\
   219188      & X       &       100.00 &    \ldots &    \ldots &   20.75 & \ldots & 20.75  &  \ldots & \ldots &   0.07 \\
   318014      & C       &       813.00 &     42.88 &    675.39 &   21.16 & \ldots & \ldots &   21.66 &   0.39 & \ldots \\
   319699      & C       &       534.00 &     39.22 &    363.85 &   21.12 & \ldots & \ldots &   21.44 &   0.49 & \ldots \\
   319702      & C       &       541.70 &     37.45 &    521.04 &   21.09 & \ldots & \ldots &   21.57 &   0.40 & \ldots \\
\hline
\end{tabular}
\end{center}
Notes: A "<" corresponds to no detection. The log of the column densities of H$_2$ and H{\sc i} 
taken from the literature or derived from CH (N(H$_2$)c) and the 5780~\AA \, DIB (N(H{\sc i})c) are also shown, along with the fractional H$_2$ content or its value derived 
from CH and the 5780~\AA \, DIB, (fH$_2$c). HWV refers to sightlines that had high water vapour affecting the 1318~nm DIB measurements and EmL observations affected by emission lines in the DIB. \end{table*}

\subsubsection{Correlation between optical DIB strengths}

There is an extensive literature concerning correlations between the strengths of optical DIBs (e.g. \citealt{Cami1997}, \citealt{Moutou1999}, \citealt{McCall2010}, \citealt{Kos13} amongst many others). 

\begin{figure*}
   \resizebox{\hsize}{!}{\includegraphics{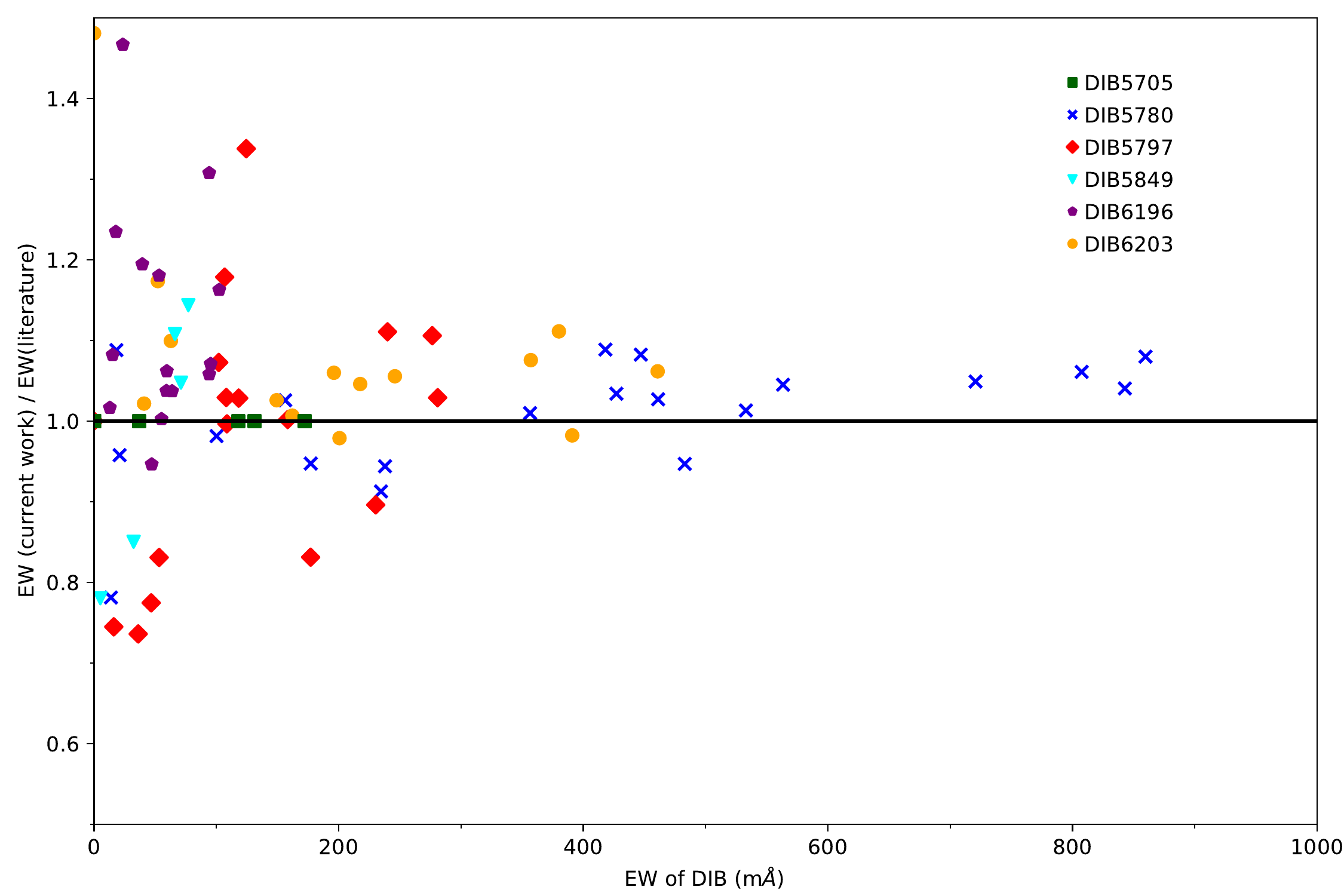}}
\caption{Ratio of the equivalent width from the current paper divided by values taken from the literature. Values are shown for several DIBs. The unity line is shown. }
\label{fig_CurrentPaper_vs_Reference_Optical_DIB_EWs}
\end{figure*}

We first check on the accuracy of our new measurements and estimation of correlations. Figure  \ref{fig_CurrentPaper_vs_Reference_Optical_DIB_EWs} shows the ratio of the EWs measured for several sightlines in the current paper with values taken from the literature. The current measurements have a median value some 4.6 percent larger than the literature values ($N$=86) with 76/86 of the measurements lying within 20 percent of each other. On the other hand, the correlation coefficient of the two lines of the Na\,{\sc i} UV doublet and the Na\,{\sc i} vs K\,{\sc i} are (as expected) close to 1.00, respectively, indicating no gross error in our analysis.

We next consider optical to optical correlations with values of $r\ge$0.95 and more than or equal to six data points. These include the 6614~\AA\ DIB correlated with 5780~\AA, 5797~\AA, 6269~\AA, 6376~\AA, 6439~\AA\ and 6693~\AA. Of these,  5780, 6196, 6203, 6269, and 6284~\AA\ were historically classified as a DIB ``family'' (\citealt{Krelowski1987} although see \citealt{Cami1997}). 
Additionally the 6439~\AA\ DIB correlates extremely well with 6376~\AA\ and the NIRDIB at 1527~nm although for the latter with only six data points. 

Finally, a number of DIBs show negative correlations when compared with other optical or IR DIBs. For correlations with greater than or equal to 10 datapoints these include the 4501.8 DIB compared with the infrared DIBs described in this paper, the 4762.6 nm, 5074.5 nm and 6993.1 nm DIBs compared with more than 10 other DIBs, and finally the 4780.0 DIB compared with 5418.9 DIB. The correlations between all DIBs are available online in VIZIER.

\subsubsection{Correlation between NIRDIB and Optical DIB strengths}

Of particular interest is the correlation between the NIRDIBs and "C$_{60}^+$"-DIBs whose wavelengths were measured in the lab by \cite{Campbell15,Campbell2016} of 9348.4, 9365.2, 9427.8, 9577.0, and 9632.1~\AA \, with an uncertainty of 0.2~\AA. The strongest lines are the reddest two. In this work we only measured the 9632.2~\AA\ DIB as it is less affected by telluric contamination. That said, there is a nearby Mg\,{\sc ii} stellar line that causes contamination of the DIB (\citealt{Gala17,Lallement2018a}). The correlation coefficient between the "C$_{60}^+$" 9632~\AA\ DIB and 1318~nm DIB is only $r$=0.50 (N=10), not correcting for possible stellar contamination. 

\begin{figure*}
\resizebox{\hsize}{!}{\includegraphics[clip,trim=0cm 0.0cm 0cm 1.0cm]{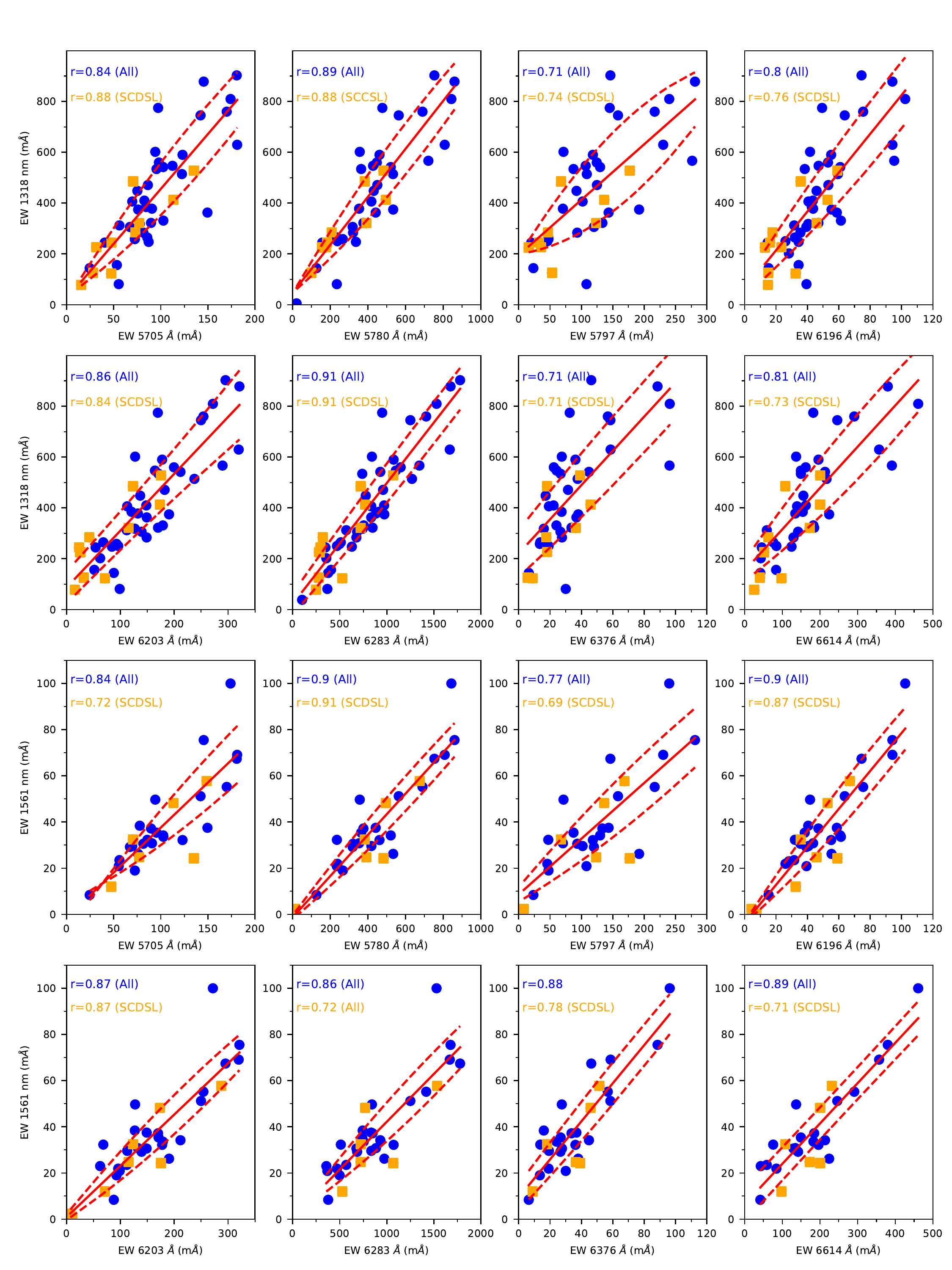}}
\caption{Correlations in equivalent width in m\AA \, for the 1318~nm and 1561~nm DIBs in the current sample towards the optical DIBs where the highest correlation coefficient was observed. The dashed lines show upper and lower error bounds obtained by fitting with a second order polynomial.}
\label{fig_NIRdibs_vs_Optical_EW}
\end{figure*}

The 1318, 1561 and 1567~nm DIBs show good correlation ($r=0.88, 0.86, 0.85$) when compared with the 5780~\AA\ DIB (\cite{Cox2014} determine $r$=0.98 for the 1318~nm DIB but with fewer data points) and similar values when comparing with the 6283~\AA\ DIB ($r$=0.91, 0.85, 0.89), 6203~\AA\ DIB ($r$=0.86, 0,85, 0.88), 5705~\AA\ DIB ($r$=0.84, 0.83, 0.80), 6196~\AA\, DIB ($r$=0.81, 0.88, 0.80), 6269~\AA\ DIB ($r$=0.79, 0.85, 0.81) and 6993~\AA\ DIB ($r$=0.78, 0.85, 0.83). For the 5797~\AA\ DIB the correlation is slightly lower with the NIRDIBs, being $r$=0.71, 0.77 and 0.73 for the 1318, 1561 and 1567~nm DIBs, respectively with typically around 30 data points in the samples. These results indicate that the NIRDIBs are more sigma type than eta type and supports the conclusion of \cite{Cox2014}  from the absence/weakness of NIRDIBs in the HD\,147889 sightline probing a translucent cloud. The correlation plots are shown in Fig. \ref{fig_NIRdibs_vs_Optical_EW}.

\subsubsection{Correlations between DIB ratios and $\lambda\lambda$5797/5780, Ca\,{\sc i}/Ca\,{\sc ii}, CH/CH+ and Na\,{\sc i}/Ca\,{\sc ii}}

The most commonly studied optical DIBs are at 5797~$\mbox{\AA}$ and 5780~$\mbox{\AA}$. The former, like the blue 4429~$\mbox{\AA}$ \, DIB, is known to trace the neutral ISM with the latter probing more ionised regions \citep[e.g.][]{Farhang15,Bailey16}. The 5797~\AA/5780~\AA\ DIB equivalent width ratio is therefore used probe the UV field, with values of less 0.35 probing regions with higher UV, and higher values being shielded from UV (e.g. \citealt{Cami1997, Vos2011} and refs therein). 

\begin{figure*}
    \begin{center}
    \resizebox{17cm}{!}{\includegraphics{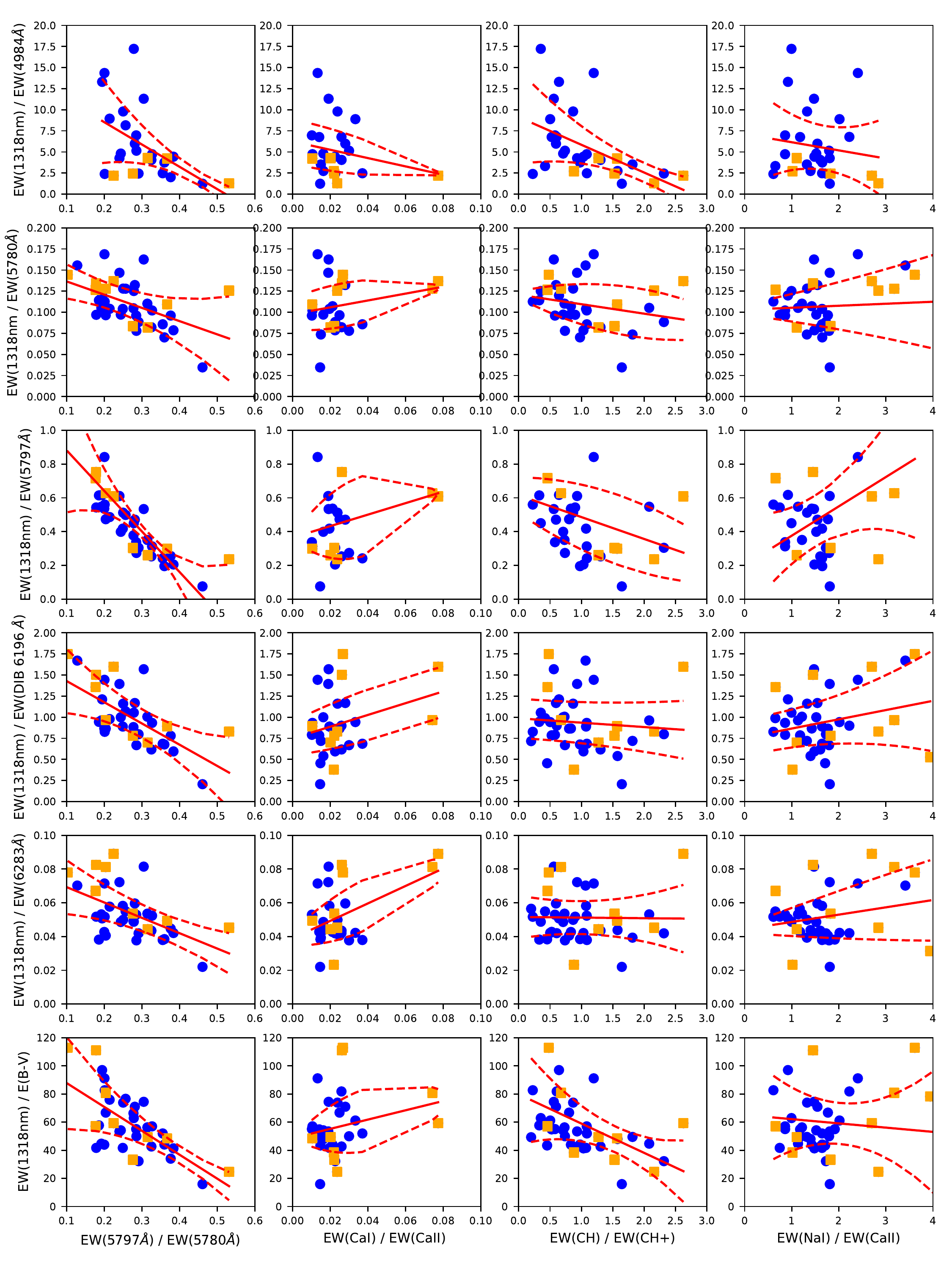}}
    \end{center}
    \caption{Ratio of the equivalent width of the 1318~nm NIRDIB divided by the EW of optical DIBs or $E(B-V)$ plotted against (1) $\lambda\lambda$5797/5780 ratio, (2) $\lambda\lambda$Ca\,{\sc i}/Ca\,{\sc ii}, (3) $\lambda\lambda$CH/CH+ and (4) Na\,{\sc i} (5895~\AA)/Ca\,{\sc ii} (3933~\AA). Blue dots are for the whole sample with orange points showing the single cloud dominated DIBs. The dashed lines show uppper and lower error bounds obtained by fitting with a second order polynomial.
    }
    \label{fig_1318_nm_Ratios}
\end{figure*}

Hence we also searched for correlations between ratios of the 1318~nm DIBs to a number of the stronger optical DIBs and reddening, the first part as presented by \citet{Hamano15,Hamano16} and \citet{Elyajouri2017a} in order to determine the relative dependence of the 1318~nm band on the UV radiation field. This was performed by plotting in Fig.~\ref{fig_1318_nm_Ratios} the ratio of equivalent width of the 1318~nm DIB divided by the 4984, 5780, 5797, 6196, or 6283~\AA \, EW against the $\lambda\lambda$5797/5780, ratio. As noted above, the latter ratio is an indicator of UV radiation field with the 5780~\AA \, DIB preferring regions where the UV radiation field is strong (the so-called $\sigma$ DIBs) and the 5797~\AA \,  DIB where the field is weaker (the $\zeta$ DIB).  Although our results are somewhat scattered, they are in line with previous work in that the 1318~nm DIB appears to favour regions of higher UV field and perhaps are cation molecules \citep{Hamano16}.

Finally, in the same figure we plot the 1318~nm / optical DIB ratio against a number of line ratios. The first is \ion{Ca}{i}/\ion{Ca}{ii} which has been used as an indicator of electron density (e.g. \citealt{Mcevoy2015}). The use of this line ratio has many issues, in particular the presence of perhaps dominating Ca\,{\sc iii} \citep{Sembach2000} and the assumption that Ca\,{\sc i} and Ca\,{\sc ii} trace the same parts of the ISM \citep{Crawford2002} (see also \citealt{Welty03}). Secondly, we show correlations with CH/CH$+$ which is an indicator of turbulent dissipation and warm H$_{2}$ \citep{Valdivia2017}. The last is \ion{Na}{i}/\ion{Ca}{ii} which is an indicator of shocks and dust content via the Routly-Spitzer effect \citep{RoutlySpitzer1952,Vallerga1993}. We note that saturation can be an issue in these lines and that there are no convincing trends in any of the relationships. 


\subsection{Search for correlations in normalised DIB spectra}

\cite{Cami1997} and more recently Fan et al. (2021) searched for correlations in DIB strengths by using the normalised DIB strength (EW/$E(B-V$)). One of the goals of the latter work was to search for anti-correlated DIBs and compare the changes in the correlation coefficients across DIB families. In the current dataset we find that no NIRDIBs show anti-correlations less than $\sim$--0.7 with more than ten datapoints. The best correlations are found between the 1318 and 1527~nm reddening-normalised DIB EWs where $r$ exceeds 0.9. The correlations between all DIBs are available online in VIZIER.

\subsection{Search for small scale structure and time-variability in NIRDIBs}\label{smallscale}

\subsubsection{Search for small-scale structure}

In neutral species such as Na\,{\sc i} and K\,{\sc i}, small-scale (pc-) structure 
is ubiquitous in the interstellar medium \citep{Meyer99}, with variations also seen (but smaller) in singly ionised elements such as \ion{Ca}{ii} and \ion{Ti}{ii} \citep{Smoker2011}. 
Variations have also been detected in five optical DIBs between 5780 and 6614~\AA\ by \citet{Cordiner13} towards the stellar binary $\rho$ Oph A and B, separated by 344~au and with variations in the DIB strengths of 5--9 per cent. Sometimes the variations 
seen are greater than in the corresponding difference between atomic and diatomic lines such as \ion{Na}{i}. Similarly, \cite{vanLoon13} observed DIBs in both Galactic and LMC gas, finding that small-scale structure increased from \ion{Na}{i} to the 5780~\AA\ DIB to the DIB at 5797~\AA. 
%
%
In detail, in Milky Way gas the 5797~\AA\, DIB displayed variation of order 10 times greater than that seen in \ion{Na}{i} on scales of 0.04~pc. \cite{Wendt2017} discuss small-scale variations over the face of NGC 6397 (a globular cluster), which are of the order of a few thousand au. 

Eleven of our CRIRES sightlines are in double, triple or quadruple systems. Table \ref{t_companions} shows the stars that have companions within 1$'$ that were observed to search for small-scale structure in DIBs. 
The subsample was derived in part from observations by \cite{Watson1996} and \cite{Lauroesch1999} who studied many of the sightlines in either \ion{Na}{i} or \ion{K}{i} and partly by chance when another object was visible on the slit viewer image. Of our sample of eleven fields, five show no DIB at 1318.1~nm and four show no difference in profile between the two closely separated objects. The spectra are shown in Figs.~\ref{f_sss_HD_144501_HD_144502} to ~\ref{f_sss}.

\begin{figure}
\includegraphics{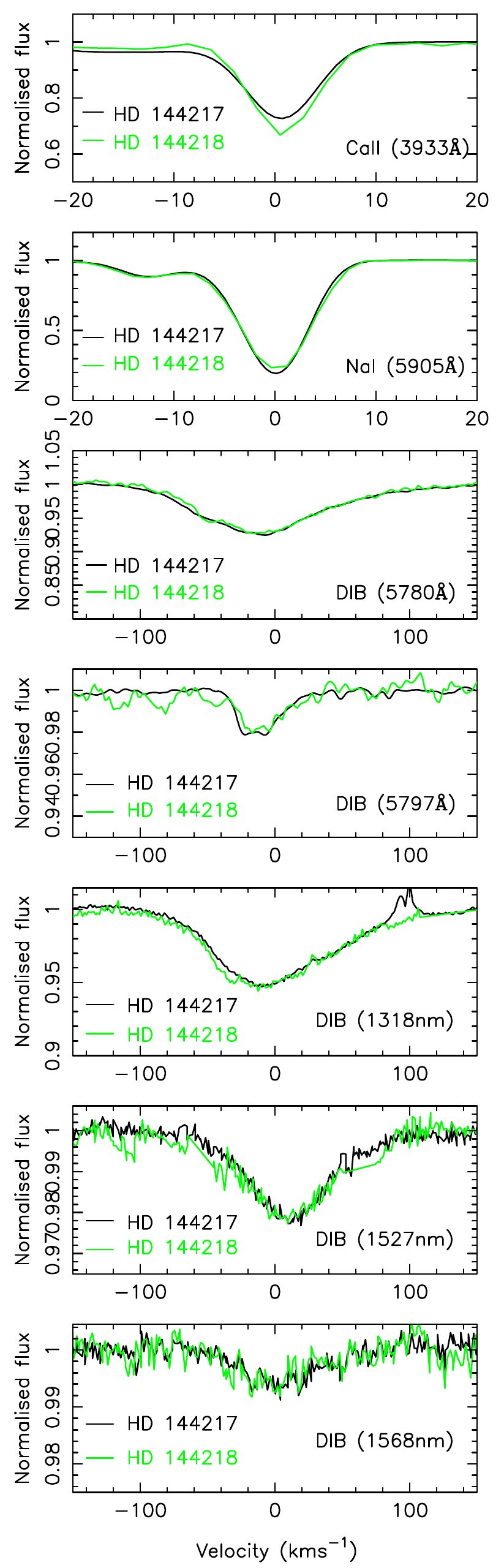}
\caption{Spectra of the binary stars HD\,144217 and HD\,144218 field in the Ca\,{\sc ii}, Na\,{\sc i} line and 5780~\AA, 5797~\AA \, 1318~nm, 1527~nm and 1567~nm DIBs.  }
\label{f_sss_HD144217_HD_144218}
\end{figure}

\begin{figure}
   \resizebox{\hsize}{!}{\includegraphics{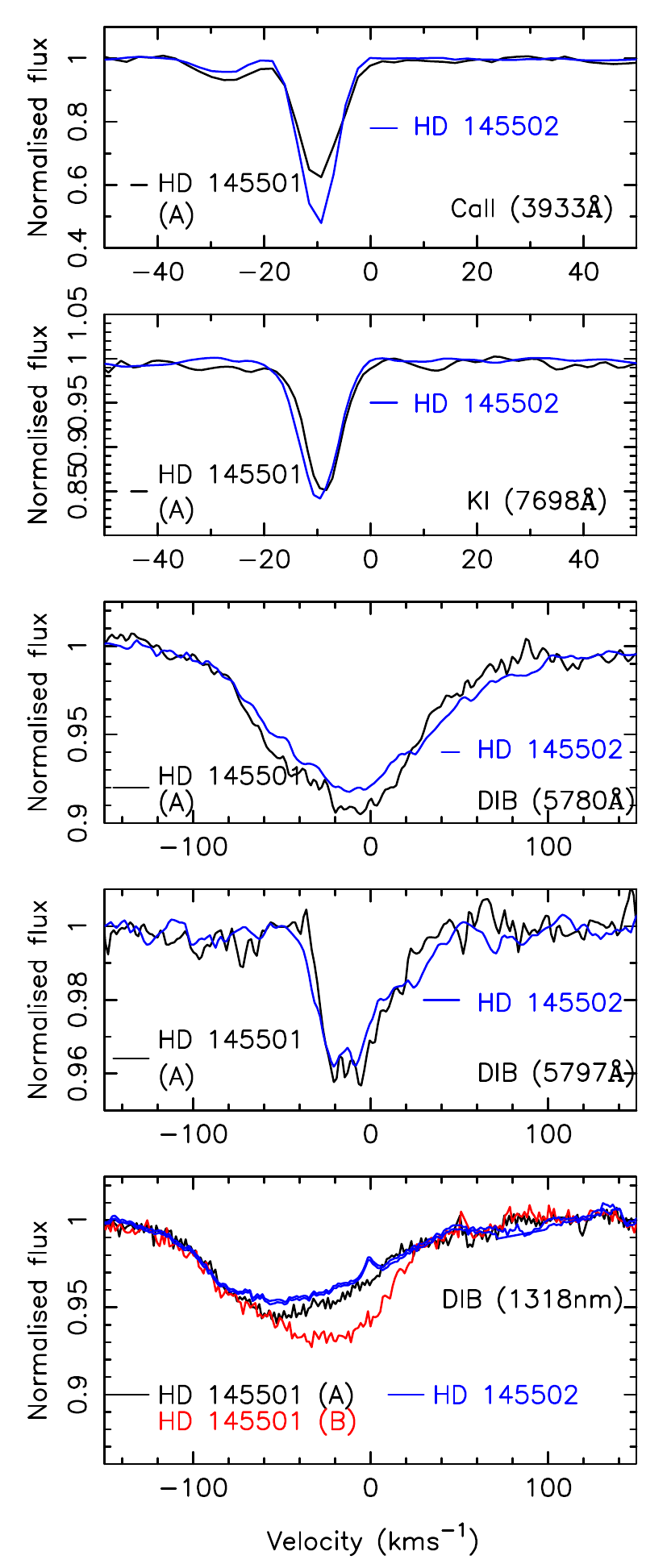}}
\caption{Spectra of the HD\,145501 / HD\,144502 field in the Ca\,{\sc ii} and K\,{\sc i} lines and 5780, 5797~\AA \, and 1318.1~nm DIBs. We recall HD\,145502 is a close binary.}
\label{f_sss_HD_144501_HD_144502}
\end{figure}

\begin{figure}
   \resizebox{\hsize}{!}{\includegraphics{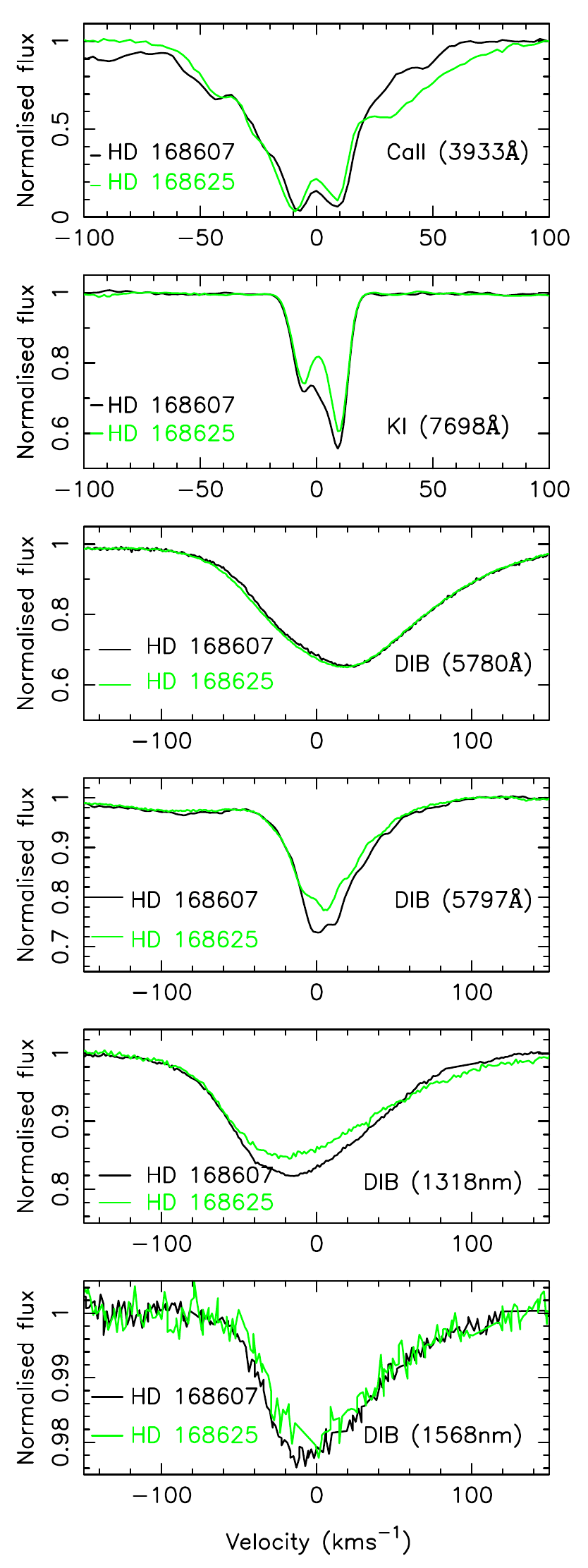}}
\caption{Spectra of the HD\,168607 /  HD\,168625 field in the Ca\,{\sc ii}, K\,{\sc i} lines and 5780, 5797~\AA \, and 1318 and 1568~nm DIBs.}
\label{f_sss_HD_168607_HD_168625}
\end{figure}

\begin{figure*}
   \resizebox{\hsize}{!}{\includegraphics{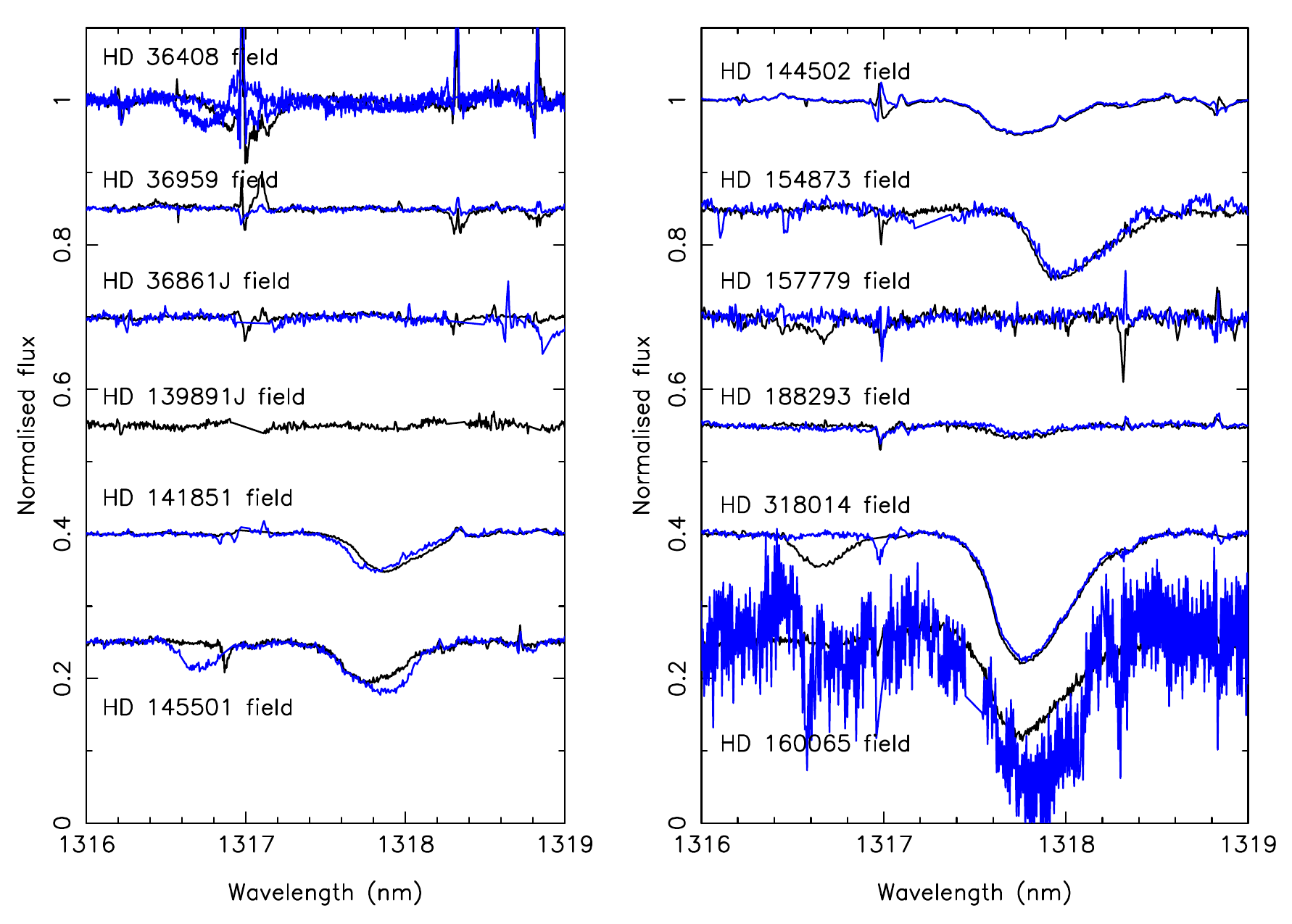}}
\caption{Spectra of binary and multiple systems observed around 1318~nm.}
\label{f_sss}
\end{figure*}

In particular, three stars have bright companions within 20 arcseconds; HD\,144217 and HD\,144218 that are separated by 13\farcs6 and have similar values of parallax (8.07$\pm$0.78 and 7.75$\pm$0.40 mas), the HD\,154873 A/B pair with separation of 8\farcs2 and similar parallaxes of 0.97$\pm$0.06 and 1.00$\pm$0.08 mas respectively and HD\,318014 which is a line of sight companion to TYC 7380-1056-1 at 20\farcs8. The parallaxes of these stars are somewhat different, being 0.42$\pm$0.05 and 0.18$\pm$0.06 mas, respectively. None of the stars above show convincing differences in the 1318~nm EW compared with their line of sight companions. In the case of the HD\,144217 and HD\,144218 pair some optical data exist in the ESO archives (see Fig.~\ref{f_sss_HD144217_HD_144218}). Similarly, there are no convincing small-scale variations observed in the optical lines.  



This leaves three fields in which possible variations in the 1318~nm line are seen. The first is the HD\,145501 / HD\,145502 field. This comprises the two main objects, each of which has a companion. The two main objects were observed in \ion{K}{i} by \cite{Lauroesch1999} as many of the lines were saturated in \ion{Na}{i}. They found a density contrast of between 1.1$\pm$0.1 and 1.8$\pm$0.8 in three different components observed, with the HD\,145501 sightline having stronger \ion{K}{i} lines. This contrast in the \ion{K}{i} components is not seen in data we downloaded from the ESO archive and shown in Fig.~\ref{f_sss_HD_144501_HD_144502}, likely due to the lower spectral resolving power in the archive data. 

In the 1318~nm spectra, the main component, HD\,145501~A ($\pi$=7.09$\pm$0.07 mas), separated by 41 arcs or $\sim$7200~au from HD\,145502 A, has a marginally deeper line profile than HD\,145502 A/B with EW of 25$\pm$1 m\AA\ compared with 22$\pm$1 m\AA. The secondary component HD\,145501 B, separated by 2\farcs5, has a different profile shape and stronger EW of 31$\pm$1 m\AA. However, its spectral type and parallax are unknown and the proper motion of $-$8.0$\pm$0.4, $-$24.0$\pm$0.3 mas yr$^{-1}$ is different to that of HD\,145501 A which has proper motion of $-$11.9$\pm$0.1 $-$24.4$\pm$0.1 mas yr$^{-1}$. Hence HD\,145501 B may well be a line-of-sight "companion". The blue absorption line components at 1315.5 and 1316.8~nm are only seen in the HD\,145501 B spectrum and likely stellar. Finally, there are some indications that the DIB at 1527~nm is also stronger in HD\,145501 A/B than HD\,145502 A/B, although the data are rather noisy. 

The two components of spectroscopic binary HD\,145502 (nu Sco A and nu Sco B) are separated by 1\farcs35 and no obvious difference is visible in  their 1318.1~nm DIB profiles which have equivalent widths of 22$\pm$1 m$\mbox{\AA}$. Their proper motions are $-$7.65$\pm$0.71 and $-$23.71$\pm$0.47 mas yr$^{-1}$ with parallax unknown for HD\,145501 and 6.88$\pm$0.76 mas for HD\,145502, both from Hipparcos data. The stars have no parallax or proper motion data in Gaia DR2. Figure \ref{f_sss_HD_144501_HD_144502} shows the available K\,{\sc i}, 5780~\AA \, 5797~\AA\, and 1318~nm data towards the field with Table \ref{t_sss_EWs} 
showing the NIR DIB EW measurements.

Comparing with optical data, the variation in between the main components of HD 145501 and HD 145502 in the 1318~nm DIB are also reflected in the 5780~\AA \, DIB, but not in the K\,{\sc i} line and only marginally in the 5797\AA \, DIB. We recall that in the data from \cite{Cordiner13}, 
the 5797~\AA \, DIB showed more variation than the 5780~\AA \, feature. The observed small-scale structure seen in HD\,145501 A/B may just be caused by the objects being at different distances. A spectroscopic parallax for this object would be useful. 

The second field where possible variation is seen in the 1318~nm line is the HD\,160065 field. HD\,160065 has $E(B-V)$ of 1.06 mag. and another star with separation of $\sim$18\farcs6 was put on the slit. The "companion" has no reference in SIMBAD.
The spectrum of the second star is far noisier than towards HD\,160065 but a clear difference in the DIB strength is seen. Follow-up work on this star would be needed to determine its distance. 

The final field where there is some evidence for infrared DIB structure is the region that contains the stars HD 168607 ($\pi$=0.6438$\pm$0.0603 mas) and HD 168625 ($\pi$=0.6212$\pm$0.0640 mas). These stars have similar reddenings ($E(B-V)$=1.54 and 1.47, respectively) and are separated on the sky by 67$''$ or around 0.52~pc. Possible variation in the 1318~nm line is seen. Clear variation is also seen in the K\,{\sc i} line and possible variation in the 5797~\AA \,DIB. Although the variation of this line presented in Fig. \ref{f_sss_HD_168607_HD_168625} looks convincing, the baseline fitting is not straightforward. No clear variation is seen in the 5780~\AA\ DIB.

Tables \ref{t_sss_EWs} and \ref{t_sss_variation} show the equivalent widths and percentage variations detected in the stars discussed above. 

In conclusion, possible variation in the 1318~nm DIB is only seen towards two double systems; HD\,145501 and HD\,145502 with separation of $\sim$7200~au, and towards HD\,168607 and HD\,168625 with separation of around 0.52~pc or 107,000~au.

%
%
\begin{table*}
\begin{center}
\caption{Objects with companions within one arcminute that were observed. CRIRES reference wavelengths are as follows: (a) 1318.1~nm (b) 1527.9~nm (c) 1568.8~nm (d) 1574.4~nm (e) 1624.2~nm. }
\label{t_companions}
\begin{tabular}{lrrrrrl}
\hline
  HD      &     Alternative        &    $l,b$            &  Companion  &     Separation      &  NIRDIB     &  Reference and comments                                           \\
  number  &        names           &    (deg.)           &             & (arcs, au)    &  setting(s) &                                                                   \\
          &                        &                     &             &               &  observed   &                                                                   \\   
\hline          
   36408  &    HR 1847             &  188.4942 -08.8854  & HD 36408B   &   9.5,  2100   &  a,c        &  WM96. No NIRDIB detection.                                       \\
   36959  &    YSO                 &  209.5670 -19.7206  & HD 36960    &  36.0, 21000   &  a,b,c      &  WM96. No NIRDIB detection.                                       \\ 
  36861 J &    lam Ori A           &  195.0519 -11.9951  & HD 36862    &   3.5,  2500   &  a,b        &  WM96. No NIRDIB detection.                                       \\
  139891  &    HR 5834             &  058.6978 +53.4129  & HD 139892   &   6.3,  --     &  a,b        &  LWV03. No NIRDIB detection.                                      \\
  144217  &    HR 5984, bet Sco    &  353.1929 +23.5996  & HD 144218   &  13.6,  2900   &  a,b,c      &  WM96.  NIRDIB detection.                                         \\
  145501  &    HR 6026             &  354.6140 +22.7107  & HD 145501 B &   2.5,  --     &  a,b,c      &  NIRDIB detection.                                                \\
  145501  &    HR 6026             &  354.6140 +22.7107  & HD 145502 A &  41.0,  7200   &  a,b,c      &  LM99.  NIRDIB detection.                                         \\
  145502  &    HR 6027, nu Sco A   &  354.6087 +22.7002  & HD 145502 B &   1.4,  --     &  a,b,c      &  LM99.  NIRDIB detection.                                         \\
  154873 A&    HIP 84010           &  341.3454 -04.1096  & HD 154873 B &   8.0, --      &  a,b        &  No optical data. NIRDIB det.                                     \\
  157779  &    HR 6485, rho Her A  &  061.4422 +32.7103  & HD 157778   &   4.3,   480   &  a,b,c      &  WM96. No NIRDIB detection.                                       \\
  160065  &    LS 4314             &  355.3166 -01.3562  & --          &  18.6,  --     &  a          &  Companion has no SIMBAD                                          \\
          &                        &                     &             &               &             &  reference. NIRDIB detection.                                     \\
  168607  &                        &  014.9679 -00.9397  & HD 168625   &  67.4, 108289 &             &  NIRDIB detection.                                                \\
  188293  &    57 Aql A            &  032.6612 -17.7609  & HD 188294   &  35.6, 5700   &  a,b,c      &  WM96. NIRDIB detection.                                          \\
  318014  &    LS 4255             &  355.0738 -00.7062  & LS 4256     &  20.5, --     &  a          &  NIRDIB detection.                                                \\
\hline
\end{tabular}
\end{center}
\tablebib{WM96 \citep{Watson1996}, LM99 \citep{Lauroesch1999}, LWV03 \citep{Lallement2003}.}
\end{table*}

%
%
\begin{table*}
\begin{center}
\caption{Equivalent widths of sightlines with companions within one arcminute that had a NIRDIB detection.}
\label{t_sss_EWs}
\begin{tabular}{lrrrrrrr}
\hline
   Star          &               EW          &      EW         &    EW      &          EW        &       EW        &    EW         &        EW       \\
                 &         Ca\,{\sc ii}      &  Na\,{\sc i}    & K\,{\sc i} &    DIB(5780~\AA)    &  DIB(5797~\AA)   &  DIB(1318~nm)  &    DIB(1527~nm)  \\   
                 &            (m\AA)         &    (m\AA)       &   (m\AA)   &         (m\AA)     &     (m\AA)      &    (m\AA)     &      (m\AA)     \\
\hline
 HD 144217       &              40$\pm$1     &  148$\pm$1      &      --    &       157$\pm$1    &    16$\pm$1     &  22$\pm$1     &    8.3$\pm$ 0.4 \\
 HD 144218       &              38$\pm$2     &  152$\pm$2      &      --    &       148$\pm$2    &    16$\pm$1     &  24$\pm$1     &    9.9$\pm$ 0.4 \\
                 &                           &                 &            &                    &                 &               &                 \\
 HD 145501 A     &              58$\pm$2     &      --         &  35$\pm$1  &       175$\pm$3    &    31$\pm$1     &  25$\pm$1     &      --         \\
 HD 145501 B     &                 --        &      --         &    --      &         --         &      --         &  31$\pm$1     &      --         \\
 HD 145502 A     &              62$\pm$1     &      --         &  37$\pm$1  &       172$\pm$2    &    33$\pm$1     &  22$\pm$1     &      --         \\
 HD 145502 B     &                 --        &      --         &    --      &          --        &      --         &  22$\pm$1     &      --         \\
                 &                           &                 &            &                    &                 &               &      --         \\
 HD 160065       &                 --        &      --         &    --      &          --        &      --         &  60$\pm$2     &      --         \\
 No name         &                 --        &      --         &    --      &          --        &      --         & 110$\pm$12    &      --         \\
                 &                           &                 &            &                    &                 &               &                 \\
 HD 168607       &             768$\pm$3     &  695$\pm$2      & 220$\pm$2  &       880$\pm$5    &    359$\pm$5    & 87$\pm$1      &      --         \\
 HD 168625       &             793$\pm$3     &  683$\pm$2      & 182$\pm$2  &       887$\pm$5    &    305$\pm$5    & 80$\pm$1      &      --         \\
                 &                           &                 &            &                    &                 &               &                 \\
 HD 188293       &               19$\pm$1    &      --         &    --      &          --        &      --         & 6.4$\pm$0.4   &     4.2$\pm$0.6 \\
 HD 188294       &               17$\pm$1    &   80$\pm$1      &  30$\pm$1  &          --        &      --         & 4.1$\pm$0.6   &     3.5$\pm$0.9 \\
                 &                           &                 &            &                    &                 &               &                 \\
 HD 318014       &             808$\pm$15    &      --         & 194$\pm$3  &      643$\pm$5     &      --         & 91$\pm$2      &     --          \\
 LS 4256         &                 --        &      --         &    --      &         --         &      --         & 88$\pm$1      &     --          \\
\hline
\end{tabular}
\end{center}
\end{table*}

\begin{table*}%
\begin{center}
\caption{Variation in equivalent width in percent for sightlines where NIRDIBs were detected. Where available, optical data are shown.}
\label{t_sss_variation}
\begin{tabular}{lrrrrrrrr}
\hline
Main         &  Companion   &      Var.     &    Var.     &     Var.    &   Var.        &  Var.           &  Var.         & Var.          \\
object       &  star        &  Ca\,{\sc ii} & Na\,{\sc i} & K\,{\sc i}  & DIB (5880~\AA) &  DIB (5897~\AA)  &  DIB (1318~nm) & DIB (1527~nm)  \\
             &              &   (\%)        &  (\%)       & (\%)        & (\%)          &  (\%)           & (\%)          & (\%)          \\
             \hline
             &              &               &             &             &               &                 &               &                \\
HD\,144217   & HD\,144218   &     5$\pm$7   & --3$\pm$2   &     --      &    6$\pm$2    &    0$\pm$12     &  --9$\pm$9    &  --16$\pm$9    \\
HD\,145501 A & HD\,145501 B &        --     &   --        &     --      &      --       &        --       & --20$\pm$7    &     --         \\
HD\,145501 A & HD\,145502 A &   --6$\pm$5   &   --        &  --6$\pm$5  &    2$\pm$3    &  --6$\pm$6      &   14$\pm$8    &     --         \\
HD\,145502 A & HD\,145502 B &        --     &   --        &     --      &      --       &        --       &    0$\pm$9    &     --         \\
HD\,160065   & No name      &        --     &   --        &     --      &      --       &        --       & --45$\pm$14   &     --         \\
HD\,168607   & HD\,168625   &   --3$\pm$1   &   2$\pm$1   &   20$\pm$2  &  --1$\pm$1    &   18$\pm$3      &    9$\pm$2    &     --         \\
HD\,188293   & HD\,188294   &    11$\pm$10  &   --        &     --      &      --       &        --       &   56$\pm$20   &   20$\pm$40    \\
HD\,318014   & LS\,4256     &        --     &   --        &     --      &      --       &        --       &    4$\pm$3    &     --         \\
\hline
\end{tabular}
\end{center}
\end{table*}

\subsubsection{Search for time-variability}

Seventeen of our sightlines were observed twice by CRIRES, with the epochs separated by six to fourteen months, with the aim of searching for time-variable structure in NIRDIBs. Additionally, two of our X-shooter sightlines observed in conditions of low water vapour in August 2019 were previously observed in 2010 and published by \citet{Cox2014}. These are additionally used to search for time variation in the 5780, 5797 and  C$_{60}^{+}$ 9577 and 9632~\AA \, DIBs.

Table~\ref{t_multiple_epochs} lists the stars that were observed at more than one epoch in order to check for time-variability in the ISM. The transverse distances derived from the Hipparcos proper motions in these time range from 1.7$\pm$0.3 to 24$\pm$3~au. Time variability, although uncommon, has been found in over 20 cases in atomic lines such as \ion{Na}{i} and singly ionised lines such as \ion{Ca}{ii} \citep[e.g.][]{Hobbs1991,Crawford2000,Smoker2011}; see \citet{Lauroesch2007} and \citet{Stanimirovic2018} for reviews). In the optical, a twin-epoch study of more than 100 early-type stars by \cite{Mcevoy2015} found that in spectra with $R$ = 80\,000 to 160\,000 and S/N ratios 
of several hundreds, typically a few percent of sightlines (corresponding to less than 1 percent of all components) in the diffuse interstellar medium show such time variation in optical lines on timescales from 6--20 years. In their survey, time-variable structure was only visible in sightlines with upper limits of 50~au or more of transverse separation. Previous claims of time-variability in DIBs have been claimed by \cite{Law2017} and, more recently, by Farhang (in preparation) towards the DIBs at 4735, 5547, 6196 and 6602~\AA, in high resolution UVES, FEROS and HARPS data towards early-type stars in the EDIBLES sample. The changes are typically very close to the noise level and will have to be followed-up by long-term monitoring. Non-detection of time variability in the 5780, 5797, 6613~\AA\, DIBs towards $\zeta$~Oph was recently reported by \cite{Cox2020}.

\begin{figure}
   \resizebox{\hsize}{!}{\includegraphics{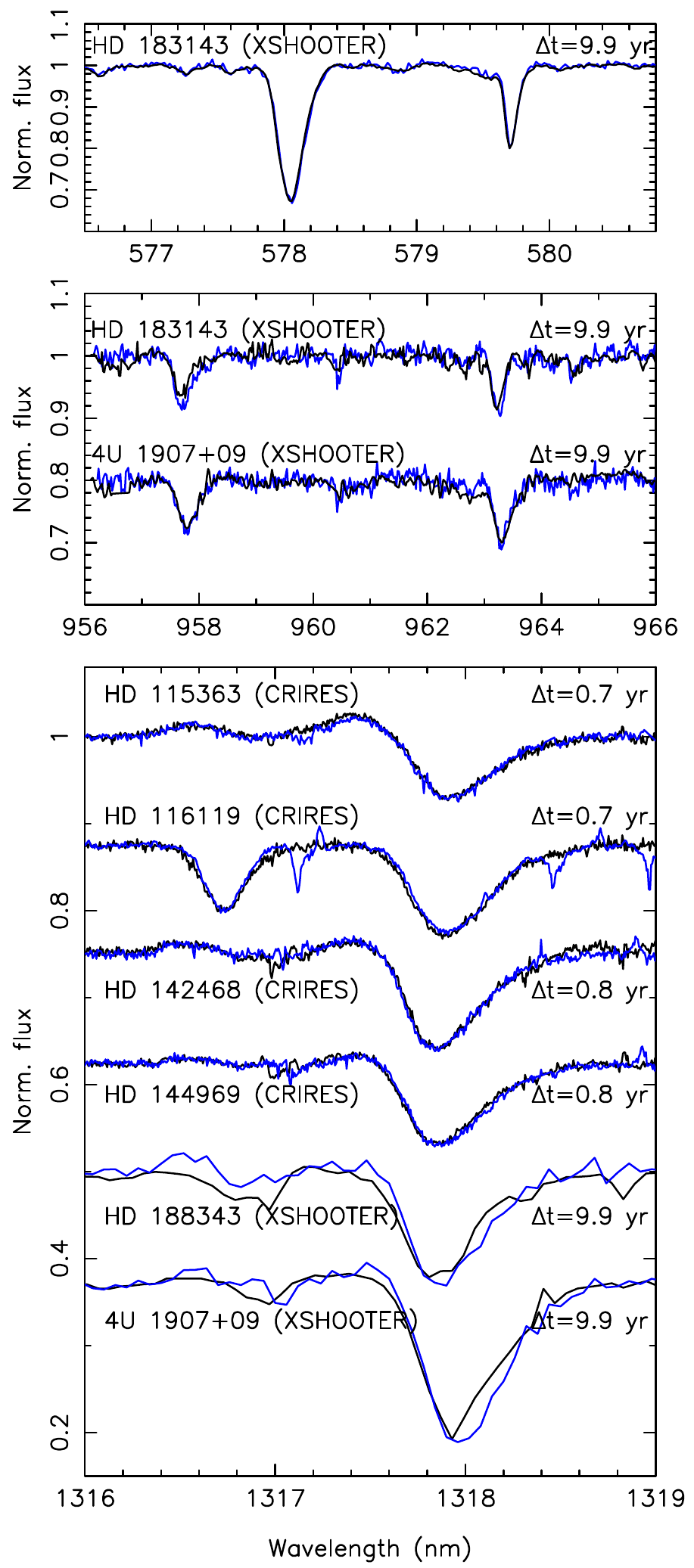}}
\caption{Examples of normalised spectra for which twin epochs were taken separated by more than six months. Top panel: 5797 and 5780~\AA \, DIBs observed with X-shooter at low PWV (4U 1907+09 not shown due to low S/N ratio). Middle panel: Ditto C$_{60}^+$ DIBs observed with X-shooter. Bottom panel: 1318.1~nm NIRDIBs.
The black line is the earlier epoch and the blue line the later epoch. For HD\,115363, HD\,116119, HD\,142468, HD\,144969 and HD\,183143 the transverse distances moved by the star between the 
two epochs are 24.0$\pm$3.4, 19.5$\pm$2.9, 7.2$\pm$0.7, 6.4$\pm$3.1 and 138$\pm$20~au,  respectively. The parallax and proper motions for 4U 1907+09 are not available in SIMBAD.}
\label{f_timevariable}
\end{figure}

Four examples of our CRIRES spectra and two observed with X-shooter at twin epochs are shown in Fig. \ref{f_timevariable}. 
No time-variability is detected in the 1318.1~nm DIB in the CRIRES data or the 5780.5, 5797.1~\AA, 1318.1 and 1527.4~nm DIBs. Tentative time variation is observed in the "C$_{60}^+$" DIBs at 9577 and 9632~\AA \, towards HD\,183143 although very close to the noise level and with confirmation required at a higher spectral resolution as continuum fitting and telluric line corrections are not straightforward. Although the transverse distance travelled in 9.9 years of 124 AU is the largest in our sample, variation is not seen in the $\lambda\lambda$5780, 5797 DIBS.


%
%
\begin{table*}
\begin{center}
\caption[]{Objects with repeat observations at 1318~nm (for CRIRES data 'C'), and with X-shooter low water vapour data ('XL') at 1318~nm and 1527~nm (for 4U 1907+09 and HD 183143 only) to search for time-variability in the ISM. Parallaxes and proper motions are from Gaia. The transverse distance travelled by the star between the first and last epoch is shown in the last column.}
\label{t_multiple_epochs}
\begin{tabular}{crrrrr}
\hline
  HD     &             Observation     & Inst.             &        PM (mas/year)          &  $\pi$         &  Transverse.    \\
  number &                 dates       &                   &        RA DEC                 &  (mas)         &  dist. (au)     \\
\hline
4U 1907+09&  2010-Jul-02 2019-Aug-10   & XL                & --                            &     --         &     --      \\
HD 106068 &  2013-May-27 2014-Mar-16   & C                 & -5.18$\pm$0.06 -0.24$\pm$0.06 & 0.58$\pm$0.04  &  7.1$\pm$1.9\\
HD 112272 &  2013-May-28 2014-Mar-16   & C                 & -8.49$\pm$0.05 -7.93$\pm$0.05 & 0.63$\pm$0.04  & 14.5$\pm$0.9\\
HD 115363 &  2013-Jun-22 2014-Mar-16   & C                 & -8.81$\pm$0.05 -1.92$\pm$0.05 & 0.27$\pm$0.04  & 24.1$\pm$3.4\\
HD 116119 &  2013-Jun-22 2014-Mar-17   & C                 & -5.71$\pm$0.04 -1.71$\pm$0.05 & 0.22$\pm$0.03  & 19.6$\pm$2.9\\
HD 125241 &  2013-May-16  2014-Mar-20  & C                 & -6.92$\pm$0.04 -1.99$\pm$0.06 & 0.41$\pm$0.03  & 14.6$\pm$1.3\\
HD 142468 &  2013-May-16 2014-Mar-20   & C                 & -2.52$\pm$0.07 -3.05$\pm$0.06 & 0.46$\pm$0.05  &  7.2$\pm$0.7\\
HD 144969 &  2013-May-16 2014-Mar-20   & C                 & 3.31$\pm$0.63  -1.85$\pm$0.32 & 0.50$\pm$0.21  &  6.3$\pm$3.1\\
HD 147932 &  2013-Jun-24 2014-Jun-17   & C                 & -9.34$\pm$0.15 -21.94$\pm$0.09 & 7.46$\pm$0.07 &  3.1$\pm$0.1 \\
HD 148379 &  2013-May-16 2014-Mar-22   & C                 & -2.97$\pm$0.19 -4.04$\pm$0.11 & 0.33$\pm$0.13  & 12.7$\pm$5.2\\
HD 148937 &  2013-May-21 2014-Mar-23   & C                 & 0.88$\pm$0.08  -3.38$\pm$0.04 & 0.88$\pm$0.05  &  3.3$\pm$0.3\\
HD 154873 &  2013-Sep-04 2014-Mar-17   & C                 & 2.57$\pm$0.09  -3.17$\pm$0.07 & 0.97$\pm$0.06  &  2.2$\pm$0.2\\
     "    &  2013-Sep-04 2014-Mar-30   & C                 &  "                            &     "          &  2.4$\pm$0.2 \\
HD 157038 &  2013-May-22 2014-May-15   & C                 & -0.20$\pm$0.11 -0.60$\pm$0.08 & 0.36$\pm$0.05  &  1.7$\pm$1.0\\
HD 159455 &  2013-May-23 2013-Jun-25   & C                 & 0.06$\pm$0.11 -0.90$\pm$0.08  & 0.34$\pm$0.06  &  0.2$\pm$0.0\\
     "    &  2013-May-23 2014-Apr-17   & C                 &  "                            &                &  2.4$\pm$0.3\\
HD 160065 &  2013-May-22 2014-Mar-30   & C                 & 1.18$\pm$0.11 -1.46$\pm$0.08 & 0.47$\pm$0.06   &  3.4$\pm$0.6\\
HD 162168 &  2013-Jun-02 2014-Apr-27   & C                 & -0.70$\pm$0.11 -1.13$\pm$0.09 & 0.70$\pm$0.06 &  1.7$\pm$0.3\\
HD 183143 &  2013-Apr-11 2014-Jun-23   & C                 & -1.09$\pm$0.06 -5.57$\pm$0.08 & 0.41$\pm$0.05 & 16.6$\pm$2.1\\
     "    &  2010-Jul-02 2019-Aug-09   & XL           &  "                           &  "            & 124$\pm$16  \\
HD 318014 &  2013-May-13 2014-Apr-17   & C                 & 0.16$\pm$0.10 -1.40$\pm$0.08  & 0.43$\pm$0.05  &  3.0$\pm$2.0  \\
\hline
\end{tabular}
\end{center}
\end{table*}
\normalsize

\section{Summary and future work}
\label{summary}

We have undertaken a mini-survey of several Near-Infrared Diffuse Interstellar Bands (NIRDIBs) using CRIRES with limited wavelength coverage. These data were supplemented by lower S/N and resolution X-shooter observations towards ten highly reddened Cepheid variables, nineteen archival X-shooter targets and towards four stars observed at very low values of water vapour in order to investigate time variability of the ``C$_{60}^+$''-DIBs and correlations with the NIRDIBs. 
Our findings can be stated as follows:

1) The NIRDIBs studied are reasonably well correlated with the strong optical bands, in particular those at 5705, 5780, 6203, 6283 and 6269~\AA. The 5797~\AA\, DIB is less well correlated with NIRDIB strengths. Using DIB ratios, and as found by previous authors, this implies that the NIRDIBs are $\sigma$ type (unshielded) DIBs as opposed to the $\zeta$ DIB that are affected by UV radiation. The NIRDIBs are not well correlated with the ``C$_{60}^+$''-DIB at 9632.1~\AA, neither is the 5797~\AA\, DIB although numbers are small. 

2) Correlations with reddening are fair to good with $r$ values for the whole dataset of $\sim$0.89 for the 1318~nm DIB.


3) The 1318~nm DIB does not have a Gaussian profile shape but has a red wing that can be fitted adequately using two Gaussian functions. The integrated flux ratio of the two components is not correlated with reddening or the $\lambda\lambda$5780/5797 flux ratio. 

4) Towards stars of type earlier than B2, the 1318~nm DIB often displays an emission feature to the blue. It is unclear if this is stellar, circumstellar or interstellar in nature. 

5) In spectra with S/N ratios of several hundreds and spectral resolving power of $\sim$50,000, variation in the 1318~nm DIB was found towards two near equidistant stellar systems (HD\,145501/HD 145502 and HD\,168607/HD\,168625). Variations of $\sim$14 and $\sim$9 percent in the 1318~nm line were observed with transverse separations of 7200~au and 0.52~pc. Two further sightlines (HD 145501 and line of sight companion separated by 2.5 arcsec, and HD 165501 with line of sight companion separated by 18.6 arcsec) also show small scale variation in the 1318~nm line, but the distance to the companion object is uncertain in both cases. The remaining stellar pairs show no difference in their 1318~nm profiles. 

6) No time variation was detected towards the 1318~nm DIB on timescales of 6--14 months towards 17 sightlines and towards two other stars observed with X-shooter over epochs separated by 9.9 years. One of these two objects (HD\,183143) showed tentative time variation for the ``C$_{60}^+$''-DIBs at 9577 and 9632~\AA\ although very close to the noise level. 

Future work will use the upgraded CRIRES spectrometer \citep{Dorn2014} with its expanded wavelength coverage to look for new DIBs in the infrared, check for emission in the 1318~nm DIBS towards the red rectangle, search for polarisation in the 1318~nm absorption and emission line DIBs and observe more sightlines in the 1527~nm band at high spectral resolution. 



\begin{acknowledgements}
We thank the referee Ed Jenkins for many useful corrections and suggestions to the paper. JVS thanks the UK Astronomy Technology Centre for hospitality.
AMI, KTS and ME thank ESO for funding under the auspices of the Director General's Discretionary Fund. AMI acknowledges support from the Spanish MINECO through project AYA2015-68217-P. Observations were performed as ESO programmes 091.C-0655, 093.C-0480 and 0103.C-0766(A) and archival data taken from ESO programmes  072.D-0286, 081.C-0475, 266.D-5655, 194.C-0833, 082.C-0566, 083.C-0503, 194.C-0833, 074.D-0240, 194.C-0833, 081.D-2002, 194.C-0833, 284.D-5048, 60.A-9036, 079.C-0597 and 095.A-9029.
This research has made use of the SIMBAD database operated by CDS in Strasbourg. 
\end{acknowledgements}


\bibliographystyle{aa}
\bibliography{mybib2.bib}

\appendix

\section{Appendix material to go on CDS}
\makeatletter

%
%
%
%
%
%
%
\begin{table*}[p!]
\begin{center}
\caption[]{
Stellar sample, sorted alphabetically and split into CRIRES and X-shooter subsections. }
\label{t_Sample}
\begin{tabular}{lllcccccccc}
\hline
      Target           &  Inst. &    Spectral    &  $l$,$b$       &   $m_B$  & $m_V$  & $m_H$  & \multicolumn{2}{c}{$(B-V)_{0}$}  & $E(B-V)$ &  $N_\text{Obs}$               \\
                       &        &      type      &  (deg.)        &  \multicolumn{3}{c}{(mag.)}    &  Fitz.  &  Weg. &          &    a       b      c     d       e    \\
\hline
    \object{HD 36408 A} &   C    &      B7-8IV-V  &  188.49 $-$08.89 &   6.10   & 6.09   & 6.02  &  $-0.13$   &   \ldots       &   \ldots      &      1 1 1 -- --      \\
    \object{HD 36408 B} &   C    &         B7II   &  188.49 $-$08.89 &   6.53   & 6.49   & 6.28  &  $-$0.13 &   \ldots   &   0.17  &      1      1      1     --     --    \\
    \object{HD 36861}   &   CAO  &         O8III  &  195.05 $-$12.00 &   3.48   & 3.66   & 3.77  &  $-$0.30 & $-$0.30  &   0.12  &      1      1     --     --     --    \\
    \object{HD 36862}   &   CAO  &         B0.5V  &  195.05 $-$11.99 &  \ldots  & 6.32   & \ldots &  \ldots      &  \ldots        &    \ldots     &  1  1  -- -- --  \\
    \object{HD 36959}   &   C    &       B1VvYSO  &  209.57 $-$19.72 &   5.44   & 5.67   & 6.24  &  $-$0.26 & $-$0.23  &   0.01  &      1      1      1     --     --    \\
    \object{HD 36960}   &   C    &     B1/2Ib/II  &  209.56 $-$19.71 &   4.50   & 4.72   & 5.38  &  $-$0.17 & $-$0.16  &  $\phantom{^1}$0.03$^{1}$ & 1      1      1     --     --    \\
    \object{HD 89587}   &   C    &         B3III  &  279.84 +05.19 &   6.74   & 6.87   & 7.20  &  $-$0.20  &  \ldots   &   0.07  &      1     --     --      1     --    \\
    \object{HD 88661}   &   C    &         B5Vne  &  283.08 $-$01.48 &   5.66   & 5.75   & 5.57  &  $-$0.16  & $-$0.15  &   0.06  &      1     --     --      1      1    \\
    \object{HD 90706}   &   C    &        B2.5Ia  &  284.52 +00.00 &   7.56  &  7.10   & 5.84  &  $-$0.15  & $-$0.14  &   0.61  &      1     --     --      1      1    \\
    \object{HD 92740}   &   C    &         WN7h+  &  287.17 $-$00.85 &   6.50   & 6.42   & 5.58  &  \ldots  &  \ldots   &   0.36  &      1     --     --      1      1    \\
                       &   C    &       O9III-V  &                &        &&               &            &    &         &                                       \\             
    \object{HD 93130}   &   C    &       O6.5III  &  287.56 $-$00.86 &   8.31   & 8.04  &  7.26  &  \ldots  &   \ldots   &  $\phantom{^2}$0.56$^{2}$ &      1     --     --      1     --    \\
    \object{HD 93632}   &   C    &           O5I  &  288.03 $-$00.87 &   8.57   & 9.10  & 7.06  &   \ldots    & \ldots   &  $\phantom{^2}$0.62$^{3}$ &  1     --     --      1     --    \\
    \object{HD 94910}   &   C    &         WN11h  &  289.18 $-$00.70 &   7.57   & 6.96  & 5.08  &   \ldots    &  \ldots   &   0.59  &      1     --     --      1      1    \\
    \object{HD 97707}   &   C    &          B11b  &  291.26 $-$00.11 &   8.57   & 8.07  & 6.77  &   \ldots    &  \ldots   &   0.67  &      1     --     --      1      1    \\
    \object{HD 100841}  &   C    &         B9III  &  294.47 $-$01.40 &   3.10   & 3.14  &  3.10  &  $-$0.08  &   \ldots   &   0.04  &      1     --     --      1      1    \\
    \object{HD 105056}  &   C    &       ON9.7Ia  &  298.96 $-$07.06 &   7.48   & 7.34  &  7.14  &  $-$0.27  & $-$0.27  &   0.41  &      1     --     --      1     --    \\
    \object{HD 105071}  &   C    &          B8Ia  &  298.24 $-$03.09 &   6.45   & 6.31  &  5.67  &  $-$0.01 &  $-$0.03  &   0.16  &      1     --     --      1      1    \\
    \object{HD 106068}  &   C    &      B8Ia/Iab  &  298.51 $-$00.42 &   6.28   & 5.98  & 5.04  &  $-$0.01 &  $-$0.03  &   0.32  &      2     --     --      1      1    \\
    \object{HD 111774}  &   C    &           B8V  &  303.04 +23.19 &   5.89   & 5.97  & 6.21  &  $-$0.11 &  $-$0.11  &   0.03  &      1     --     --      1      1    \\
    \object{HD 112272}  &   C    &        B0.5Ia  &  303.49 $-$01.49 &   8.03   & 7.39   & 5.36  &  $-$0.22 & $-$0.20  &   0.85  &      2     --     --      1      1    \\
    \object{HD 113904}  &   C    &     WC5+B0III  &  304.67 $-$02.49 &   5.50  & 5.53 &  \ldots  &  \ldots   &  \ldots   & $\phantom{^4}$0.25${^4}$  &      1     --     --      1      1    \\             
                       &   C    &          O9IV  &                &                       &             &   &         &                                       \\  
    \object{HD 115088}  &   C    &      B9.5/A0V  &  304.18 $-$17.17 &   6.31  & 6.33 &  6.47  &  $-$0.04 & $-$0.06  &   0.03  &      2     --     --      2      1    \\
    \object{HD 115363}  &   C    &          B1Ia  &  305.88 $-$00.96 &   8.32  &  7.82 & 6.13  &  $-$0.19 & $-$0.19  &   0.69  &      2     --     --      1      1    \\
    \object{HD 115842}  &   C    &        B0.5Ia  &  307.08 +06.83 &   6.39  & 6.09 &  5.30  &  $-$0.22 & $-$0.20  &   0.51  &      1     --     --      1      1    \\
    \object{HD 116119}  &   C    &          A0Ia  &  306.62 +00.63 &   8.58  & 7.87 &  5.88  & $\phantom{-}$0.00 &  $\phantom{-}$0.00  &   0.71  &      2     --     --      1      1    \\
    \object{HD 125007}  &   C    &           B9V  &  319.45 +17.49 &   7.00  & 7.03 &  7.03  &  $-$0.07 & $-$0.07  &   0.04  &      1     --     --      1     --    \\
    \object{HD 125241}  &   C    &        O8.5Ib  &  313.54 +00.14 &   9.08  & 8.60 &  6.99  &  $-$0.29 & $-$0.30  &   0.77  &      2     --     --      1      1    \\
    \object{HD 137753}  &   C    &          B7IV  &  325.93 +03.34 &   6.66  & 6.69 &  6.68  &  $-$0.13 &  \ldots   &   0.10  &      1     --     --      1      1    \\
    \object{HD 139891}  &   C    &           B9V  &  058.69 +53.41 &   5.82  & 5.93 & \ldots  &  $-$0.07 & $-$0.07  &   0.07  &      1      1     --     --     --    \\
    \object{HD 139892} &   C    &           B7V  &  058.69 +53.41 &   4.87   & 4.98 &   5.30  &   \ldots     &  \ldots      &   \ldots         &      1      1     --     --     --    \\  
    \object{HD 142468}  &   C    &      B1Ia/Iab  &  327.95 $-$00.76 &   8.48 &  7.90 &  6.41  &  $-$0.19 & $-$0.19  &   0.77  &      2     --     --      1      1    \\
    \object{HD 144217}  &   CAO   &           B1V  &  353.19 +23.60 &   2.55  & 2.62 &   2.80  & \ldots      &   \ldots    &   0.20  &      1      1      1     --     --    \\
    \object{HD 144218}  &   CAO   &           B2V  &  353.20 +23.60 &   4.87  & 4.89 &   4.47  &  \ldots     &   \ldots    &  \ldots       &      1      1      1     --     --    \\
    \object{HD 144969}  &   C    &          B1Ia  &  333.18 +02.01 &   9.32   & 8.39 &   5.68  &  $-$0.19 & $-$0.19  &   1.12  &      2     --     --      1      1    \\
    \object{HD 145482}  &   C    &           B2V  &  348.12 +16.84 &   4.42   & 4.57 &   5.01  &  $-$0.24 & $-$0.21  &   0.07  &      2     --     --      2      2    \\
    \object{HD 145501A} &   CAO  &          B8V+  &  354.61 +22.71 &   6.43   & 6.30 &   5.69  &  $-$0.09 & $-$0.09  &   0.22  &      1      1      1     --     --    \\
                       &   CAO  &        B9VpSi  &                &                       &            &    &         &                                       \\    
    \object{HD 145501B} &   CAO  &      \ldots  &  354.61 +22.71 &   7.54  & 7.23 & \ldots &   \ldots   &   \ldots  &     \ldots    &     1 1 1 --  --\\         
    \object{HD 145502A} &   C    &           B2V  &  354.61 +22.71 &   4.35  & 4.35 & \ldots &  $-$0.24  & $-$0.21  &   0.28  &      1      1      1     --     --    \\
    \object{HD 145502B} &   C    &            \ldots  &  354.61 +22.70 &   6.72   & 6.60  & \ldots  &     \ldots &  \ldots        &   \ldots &  1 1 1 -- -- \\
    \object{HD 146331}  &   C    &           B9V  &  350.41 +17.53 &   8.69   & 8.30 &  7.19  &  $-$0.07 & $-$0.07  &   0.46  &      1     --     --      1     --    \\
    \object{HD 147932}  &   CAO  &           B5V  &  353.72 +17.71 &   7.59   & 7.27 &  5.92  &  $-$0.16 & $-$0.15  &   0.48  &      1     --     --      1      1    \\
    \object{HD 148184}  &   C    &         B2Vne  &  357.93 +20.67 &   4.71   & 4.43 &  3.26  &  $-$0.24 & $-$0.21  &   0.51  &      1     --     --      1      1    \\
    \object{HD 148379}  &   C    &         B2Iab  &  337.25 +01.57 &   5.95   & 5.37 &  3.83  &  $-$0.17 & $-$0.16  &   0.74  &      2     --     --      1      1    \\
    \object{HD 148688}  &   C    &          B1Ia  &  340.72 +04.34 &   5.78   & 5.39 &  4.24  &  $-$0.19 & $-$0.19  &   0.58  &      1     --     --      1      1    \\
    \object{HD 148937}  &   C    &         O6f?p  &  336.37 $-$00.22 &   7.12  & 6.71 &  5.74  &  \ldots      &   \ldots     &   0.66  &                                       \\
    \object{HD 151932}  &   C    &          WN7h  &  343.22 +01.43 &   6.72   & 6.49 &  5.27  & \ldots   & \ldots  &   $\phantom{^4}$0.50${^5}$  &      1     --     --      1      1    \\
    \object{HD 152235}  &   C    &        B0.5Ia  &  343.31 +01.10 &   6.92   & 6.38 &  4.94  &  $-$0.22  & $-$0.20  &   0.75  &      1     --     --      1      1    \\
\hline
\end{tabular}
\end{center}
\end{table*}

\clearpage
\newpage
\setcounter{table}{0}
\begin{table*}[hp!]
\begin{center}
\caption[]{ctd}
\begin{tabular}{lllcccccccc}
\hline
   Target             &  Inst. & Spectral       &  $l$,$b$       &   $m_B$  & $m_V$  & $m_H$  &  \multicolumn{2}{c}{$ (B-V)_{0}$}  &   $E$ &      $N_\text{Obs}$  \\
                        &        & type           &  (deg.)        &  \multicolumn{3}{c}{(mag.)}  &  Fitz.  & Weg.  & $(B-V)$  &    a b c d e                         \\
\hline 
    \object{HD 152386}  &   C    &     O5/6I:fek  &  341.11 $-$00.94 &   8.64 &  8.13 &  6.62  &  $-$0.32 & $-$0.31  &   0.83  &      1     --     --      1      1    \\
    \object{HD 154811}  &   C    &      OC9.7Iab  &  341.06 $-$04.21 &   7.33 &  6.93 &  5.85  &  $-$0.27 & $-$0.27  &   0.67  &      1     --     --      1      1    \\
    \object{HD 154873A} &   C    &          B1Ib  &  341.35 $-$04.10 &   6.98 &  6.70 & \ldots  &  $-$0.19 & $-$0.19  &   0.47  &      3      1      2      2      1    \\
    \object{HD 154873B} &   C    &      \ldots    &  341.34 $-$04.11 &   8.49 &  8.23 & \ldots &  \ldots & \ldots  &            &      3      1      2      2      1    \\
    \object{HD 156134}  &   C    &      B0Iab...  &  351.20 +01.36   &   8.70 &  8.05 &  6.41  &  $-$0.24 & $-$0.22  &   0.88  &      1     --     --      1      1    \\      
    \object{HD 156201}  &   C    &        B0.5Ia  &  351.51 +01.48 &   8.70  & 8.01 &  6.10  &  $-$0.22 & $-$0.20  &   0.90  &      1     --     --      1      1    \\
    \object{HD 156212}  &   C    &         B0Iab  &  357.59 +05.82 &   8.34  & 7.95 &  6.57  &  $-$0.24 & $-$0.22  &   0.62  &      1     --     --      1      1    \\
    \object{HD 156738}  &   C    &       O6.5III  &  351.18 +00.48 &  10.27  & 9.35 &  6.92  &  $-$0.32 & \ldots  &   1.24  &      1     --     --      1     --    \\      
    \object{HD 157038}  &   C    &         B3Iap  &  349.95 $-$00.79 &   7.20 &  6.44 &  4.29  & $-$0.13 & $-$0.13  &   0.89  &      2     --     --      1      1    \\
    \object{HD 157778}  &   C    &          A0Vn  &  061.22 +32.71   &   5.40 &  5.40 & \ldots & \ldots  & \ldots & \ldots    &       1 1 1 -- --    \\
    \object{HD 157779}  &   C    &       B9.5III  &  061.44 +32.71   &   4.49 &  4.56 & \ldots  & $-$0.05 & \ldots  &   $\phantom{^6}$0.00$^{5}$ & 1      1      1     --     --    \\
    \object{HD 159455}  &   C    &        B0.5Ia  &  355.23 $-$00.61 &   9.11 &  8.44 &  6.25  &  $-$0.22 & $-$0.20  &   0.88  &      3     --     --      2      2    \\
    \object{HD 160065A} &   C    &       O9.5Iab  &  355.31 $-$01.35 &   9.41 &  8.60  & 6.17  &  $-$0.27 & $-$0.24  &   1.06  &      2     --     --      1      1    \\
    \object{HD 160065B} &   C    &         \ldots &  \ldots    & \ldots  & \ldots &  \ldots & \ldots   & \ldots  &  \ldots &      1     --     --     --     --    \\
    \object{HD 161289}  &   C    &           A0V  &  027.50 +16.03 &   6.60 &  6.54  & 6.38  &  $\phantom{-}$0.00 &  $\phantom{-}$0.00  &   0.07  &      1     --     --     --     --    \\      
    \object{HD 161291}  &   C    &       B0.5Iab  &  001.49 +00.85 &   9.61  & 8.88 &  7.04  &  $-$0.22 &  $-$0.20  &   0.94  &      1     --     --      1     --    \\
    \object{HD 162168}  &   C    &       B0.5Iab  &  357.13 $-$03.08 &   9.04 &  8.42 &  6.84  &  $-$0.22 & $-$0.20  &   0.83  &      2     --     --      1      1    \\
    \object{HD 164438}  &   C    &        O9.2IV  &  010.35 +01.78 &   8.89  & 8.76  & 6.65  & \ldots & \ldots  &   0.65  &   1 -- -- 1 1                       \\
    \object{HD 164865}  &   C    &         B9Iab  &  006.21 $-$01.19 &   8.47 &  7.63 &  4.93  &  $\phantom{-}$0.00 &  $-$0.01  &   0.85  &      1     --     --      1      1    \\
    \object{HD 165319}  &   C    &        O9.7Ib  &  015.12 +03.33 &   8.63 &  8.04 &  6.36  &  $-$0.27 & $-$0.24  &   0.85  &      1     --     --      1      1    \\
    \object{HD 166734}  &   C    &       O7.5Iab  &  018.92 +03.62 &   9.49 &  8.42 &  5.53  &  $-$0.31 & $-$0.30  &   1.37  &      1     --     --      1      1    \\
    \object{HD 167264}  &   C    &       O9.7Iab  &  010.45 -01.74 &   5.42 &  5.37 &  5.21  &  $-$0.27 & $-$0.27  &   0.32  &      1     --     --      1      1    \\
    \object{HD 167971}  &   C    &        O8I+O5  &  018.25 +01.68 &   8.27 &  7.50 &  5.32  &  \ldots & \ldots  &   $\phantom{^7}$1.05${^6}$  &      1     --     --      1      1    \\
    \object{HD 168607}  &   C    &        B9Iaep  &  014.97 $-$00.93 &   9.82 &  8.28 &  3.88  & $\phantom{-}$0.00 & $-$0.01  &   1.54  &      1     --     --      1      1    \\
    \object{HD 168625}  &   C    &         B6Iap  &  014.97 $-$00.95 &   9.78 &  8.37 &  4.54  &  $-$0.07 & $-$0.06  &   1.47  &      1     --     --      1      1    \\
    \object{HD 168987}  &   C    &       B1Ia/ab  &  014.99 $-$01.44 &   8.76 &  8.02 &  5.53  &  $-$0.19 & $-$0.19  &   0.93  &      1     --     --      1      1    \\
    \object{HD 169034}  &   C    &          B2Ia  &  017.66 $-$00.06 &   9.42 &  8.24 &  4.77  &  $-$0.17 & $-$0.16  &   1.35  &      1     --     --      1      1    \\
    \object{HD 169454}  &   C    &          B1Ia  &  017.53 $-$00.66 &   7.61 &  6.71 &  4.09  &  $-$0.19 & $-$0.19  &   1.09  &      1     --     --      1      1    \\
    \object{HD 169754}  &   C    &        B0.5Ia  &  020.02 +00.23 &   9.64 &  8.60 &  5.81  &  $-$0.22 & $-$0.20  &   1.25  &      1     --     --      1      1    \\
    \object{HD 170938}  &   C    &          B1Ia  &  016.84 $-$03.04 &   8.84 &  7.99 &  5.51  &  $-$0.19 & $-$0.19  &   1.04  &      1     --     --      1      1    \\
    \object{HD 171722}  &   C    &           B9V  &  324.68 $-$24.68 &   7.21 &  7.24 &  7.29  &  $-$0.07 & $-$0.07  &   0.04  &      1     --     --      2     --    \\
    \object{HD 183143}  &   C    &         B7Iae  &  053.24 +00.62 &   8.08 &  6.86  & \ldots  &  $-$0.04 & $-$0.04  &   1.26  &      2     --     --      2      2    \\
    \object{HD 188293}  &   C    &          B7Vn  &  032.66 $-$17.76 &   5.63 &  5.71 &  5.88  &  $-$0.13 & $-$0.13  &   0.05  &      1      1      --     --     --    \\    
    \object{HD 188294}  &   C    &           B8V  &  032.65 $-$17.76 &   6.42 &  6.44 &  6.51  &  $-$0.11 & $-$0.11  &   0.09  &      1      1      --     --     --    \\
    \object{HD 303188}  &   C    &            B3  &  286.91 $-$00.30 &   9.90 &  9.19 &  6.64  &  \ldots  & \ldots  &   0.75  &      1     --     --      1      1    \\
    \object{HD 312980}  &   C    &            B0  &  010.70 $-$01.29 &  10.31 &  9.63 &  7.19  &  \ldots  & \ldots  &  $\phantom{^8}$1.15$^{7}$  &      1     --     --      1     --    \\
    \object{HD 318014}  &   C    &        B8Iab:  &  355.07 $-$00.70 &  10.24 &  8.78 &  4.21  &  $-$0.02 & $-$0.03  &   1.49  &      2     --     --      1      1    \\
    \object{HD 319699}  &   C    &           O5V  &  351.32 +00.91 &  10.32 &  9.63 &  7.45  &  $-$0.32 & $-$0.30  &   1.00  &      1     --     --      1     --    \\
    \object{HD 319702}  &   C    &         O8III  &  351.34 +00.60 &  10.94 & 10.12 &  7.56  &  \ldots  & \ldots  &   $\phantom{^9}$1.30$^{8}$  &      2     --     --      2     --    \\
    \object{LS 4825}    &   C    &          B1Ib  &  001.67 $-$06.62 &  12.04 & 11.99  & \ldots  &  $-$0.19 & $-$0.19  &   0.24  &      1     --     --      1     --    \\  
                       &        &                &                &                       &                &                                                 \\
   \hline
\end{tabular}
\end{center}
\end{table*}
                    
\clearpage     
\newpage
\setcounter{table}{0}
\begin{table*}[hp!]
\begin{center}
\caption[]{ctd.}
\begin{tabular}{lllcccccccc}
\hline
   Target             &  Inst. & Spectral       &  $l$,$b$       &   $m_B$  & $m_V$  & $m_H$  &  \multicolumn{2}{c}{$ (B-V)_{0}$}  &   $E$ &      $N_\text{Obs}$  \\
                        &        & type           &  (deg.)        &  \multicolumn{3}{c}{(mag.)}  &  Fitz.  & Weg.  & $(B-V)$  &    a b c d e                         \\
\hline

    \object{4U 1907+09} &   XL    &       O8.5Iab  &  043.74 +00.48 &  19.41 & 16.35  & \ldots    & \ldots & \ldots &   3.48  &     --     --     --      --    --    \\
    \object{ALS 18752}  &   XL    &    O3.5III(f*) &  353.17 +00.89 &  13.33 & 11.84  & 7.28  &   \ldots & \ldots & $\phantom{^10}$1.92$^{9}$  & --     --     --      --    --    \\
    \object{ALS 19692}  &   XL   &     O5.5IV(f)  &  353.11 +00.65 &  13.53 & 11.82   & 6.93  &  \ldots & \ldots& $\phantom{^10}$2.26$^{9}$  & --     --     --      --    --    \\
    \object{B02}        &   X    &           T2C  &  001.93 +01.44 &     \ldots & \ldots & 14.00 &  \ldots & \ldots &   3.35  &     --     --     --      --    --    \\
    \object{B09}        &   X    &            CC  &  355.22 $-$00.19 &   \ldots & \ldots & 14.20 &  \ldots & \ldots &   6.51  &     --     --     --      --    --    \\
    \object{B19}        &   X    &            CC  &  354.37 $-$00.44 &   \ldots & \ldots & 13.76 &  \ldots & \ldots &   3.27  &     --     --     --      --    --    \\
    \object{B21}        &   X    &           T2C  &  358.86 $-$00.85 &   \ldots & \ldots & 13.45 &  \ldots & \ldots &   4.10  &     --     --     --      --    --    \\
    \object{B22}        &   X    &           T2C  &  359.29 $-$00.71 &   \ldots & \ldots & 12.75 & \ldots & \ldots &   4.27  &     --     --     --      --    --    \\
    \object{B24}        &   X    &           T2C  &  003.13 $-$00.97 &   \ldots & \ldots & 13.42 & \ldots & \ldots&   3.66  &     --     --     --      --    --    \\
    \object{D07}        &   X    &            CC  &  339.26 $-$00.32 &   \ldots & \ldots & 14.67 & \ldots & \ldots &   5.51  &     --     --     --      --    --    \\
    \object{D20}        &   X    &            CC  &  323.33 $-$00.13 &   \ldots & \ldots & 13.07 & \ldots & \ldots &   3.57  &     --     --     --      --    --    \\
    \object{D17}        &   X    &            CC  &  346.13 +01.00   &   \ldots & \ldots & 13.79 & \ldots & \ldots &   3.43  &     --     --     --      --    --    \\
    \object{D18}        &   X    &           T2C  &  346.16 +00.31   &   \ldots & \ldots & 13.84 & \ldots & \ldots &   3.13  &     --     --     --      --    --    \\
    \object{HD  35912}  &  XA    & B2/3V          &  201.88 $-$17.84   &    6.21 & 6.38  & 6.83 & \ldots & \ldots &   $\phantom{^11}$0.06$^{10}$  &     -- -- -- -- -- \\
    \object{HD  52266}  &  XA    & O9.5IIIn       &  219.13 $-$00.67   &    7.22  & 7.23   & 7.24 & \ldots & \ldots & $\phantom{^11}$0.32$^{10}$  &     -- -- -- -- -- \\   
    \object{HD  52973}  &  XA    & CC             &  195.75 +11.90     &    4.58  & 3.79   & 2.03 & \ldots & \ldots & $\phantom{^12}$0.04$^{11}$  &     -- -- -- -- -- \\
    \object{HD  53367}  &  XA    & B0IV/Ve        &  223.70 $-$01.90   &    7.40  & 6.96   & 6.22  & \ldots & \ldots & $\phantom{^11}$0.74$^{10}$  &     -- -- -- -- -- \\   
    \object{HD  57060}  &  XA    & O7Iafpvar      &  237.82 $-$05.36   &    4.80  & 4.95   & 5.19 & \ldots & \ldots & $\phantom{^11}$0.17$^{10}$  &     -- -- -- -- -- \\  
    \object{HD  62542}  &  XA    & B3V            &  255.91 $-$09.23   &    8.21  & 8.03   & 7.58 & \ldots & \ldots & $\phantom{^11}$0.35$^{10}$  &     -- -- -- -- -- \\  
    \object{HD 108927}  &  XA    & B5V            &  301.91 $-$15.35   &    7.83  & 7.76   & 7.58 & \ldots & \ldots & $\phantom{^11}$0.22$^{10}$  &     -- -- -- -- -- \\  
    \object{HD 101602}  &  XA    & CC             &  294.95 $-$00.92   &    9.60  & 8.30   & 6.96 & \ldots & \ldots & $\phantom{^12}$0.25$^{11}$  &     -- -- -- -- -- \\
    \object{HD 114213}  &  XA    & B1Ib           &  305.18 +01.31     &    9.88  & 8.96   & 6.46 & \ldots & \ldots & $\phantom{^11}$1.12$^{10}$  &     -- -- -- -- -- \\  
    \object{HD 139717}  &  XA    & CC             &  325.65 $-$00.16   &   10.68  & 9.22   & 5.18 & \ldots & \ldots & $\phantom{^12}$0.89$^{11}$  &     -- -- -- -- -- \\
    \object{HD 147889}  &  XA    & B2III/IV       &  352.85 +17.04   &    8.73  & 7.90   & 4.94 & \ldots & \ldots & $\phantom{^11}$1.07$^{10}$  &     -- -- -- -- -- \\  
    \object{HD 152235}  &  XA    & B0.5Ia         &  343.31 +01.10   &    6.92  & 6.38   & 4.94 & \ldots & \ldots & $\phantom{^11}$0.69$^{10}$  &     -- -- -- -- -- \\    
    \object{HD 154368}  &  XA    & O9.5Iab        &  349.97 +03.21   &    6.63  & 6.13   & 4.85 & \ldots & \ldots & $\phantom{^11}$0.78$^{10}$  &     -- -- -- -- -- \\    
    \object{HD 168076}  &  XA    & O4III(f)       &  016.93 +00.83   &    8.72  & 8.25   & 6.66  & \ldots & \ldots & $\phantom{^11}$0.74$^{10}$  &     -- -- -- -- -- \\    
    \object{HD 170740}  &  XA    & B2/3II         &  021.05 $-$00.52   &    5.96  & 5.72   & 5.25 & \ldots & \ldots & $\phantom{^11}$0.48$^{10}$  &     -- -- -- -- -- \\    
    \object{HD 178287}  &  XA    & CC             &  028.20 $-$07.12   &    8.88  & 7.79   & 5.14 & \ldots & \ldots & $\phantom{^12}$0.40$^{11}$  &     -- -- -- -- -- \\
    \object{HD 179315}  &  XA    & CC             &  039.15 $-$02.57   &    9.01  & 7.78   & 4.80 & \ldots & \ldots & $\phantom{^12}$0.59$^{11}$  &     -- -- -- -- -- \\
    \object{HD 183143}  &  XL    & B6Ia           &  053.24 +00.63   &    8.08  & 6.86   & \ldots& \ldots & \ldots & $\phantom{^11}$1.26$^{10}$  &     -- -- -- -- -- \\
    \object{HD 214080}  &  XA    & B1/2Ib         &  044.80 $-$56.91   &    6.81  & 6.93   & 7.25 & \ldots & \ldots & $\phantom{^11}$0.06$^{10}$  &     -- -- -- -- -- \\    
    \object{HD 219188}  &  XA    & K0III          &  083.02 $-$50.17   &    6.90  & 7.06   & 7.43 & \ldots & \ldots & $\phantom{^11}$0.13$^{10}$  &     -- -- -- -- -- \\     
\hline
\end{tabular}  
\end{center}
Notes: Targets are identified by HD number or alternative catalogue designations for those not in the HD catalogue. 
The instrument used is denoted by C for CRIRES (no adaptive optics), CAO for CRIRES using adaptive optics, X for data from \cite{Minniti2020}, XA for archival X-shooter data, and XL for the X-shooter low-water-vapour sample. Spectral types, Galactic coordinates and apparent magnitudes in the $B$, $V$ and $H$ bands are taken from SIMBAD. CC indicates a classical Cepheid and 
T2C a type 2 Cepheid; the intrinsic colours of these stars vary over the pulsation period. Cepheid names are from \citet{Minniti2020}, where "B" refers to a Bulge Cepheid and "D" to a Disc Cepheid.
Intrinsic ($B-V$)$_{0}$ colours for the spectral type are taken from \cite{Fitzgerald1970} and \cite{Wegner1994}; where the values differed we took an average. Reddening $E(B-V)$ was estimated 
by comparing the observed $(B-V)$ colours to the intrinsic colours, except these entries taken from the literature shown in the footnote and towards the Cepheids as explained in the text.
$N_\text{Obs}$ indicates the number of observations in each of the 
five wavelength regions observed using CRIRES, with central wavelengths (a) 1318.1~nm (b) 1527.9~nm (c) 1568.8~nm (d) 1574.4~nm and (e) 1624.2~nm. All wavelengths were observed in a single setting for X-shooter. There are no SIMBAD data for 145501 "B" nor 160065 "B" that were detected using CRIRES.
\tablebib{Published reddening values taken from: (1) \cite{Sonnentrucker2018}; 
(2) \cite{Weselak2014};
(3) \cite{Crowther2006}; 
(4) \cite{Zhekov2014};
(5) \cite{Swihart2017}; 
(6) \cite{Xiang2017};
(7) \cite{Wegner2003};
(8) \cite{Russeil2012}
(9) \cite{MaizApellaniz15}
(10) \cite{Fan2017}
(11) \cite{Chen2014}.}
\end{table*}

\clearpage
\newpage

\renewcommand{\thefigure}{A\@arabic\c@figure}
\makeatother
\setcounter{figure}{0}
\begin{figure*}
\resizebox{\hsize}{22cm}{\includegraphics[]{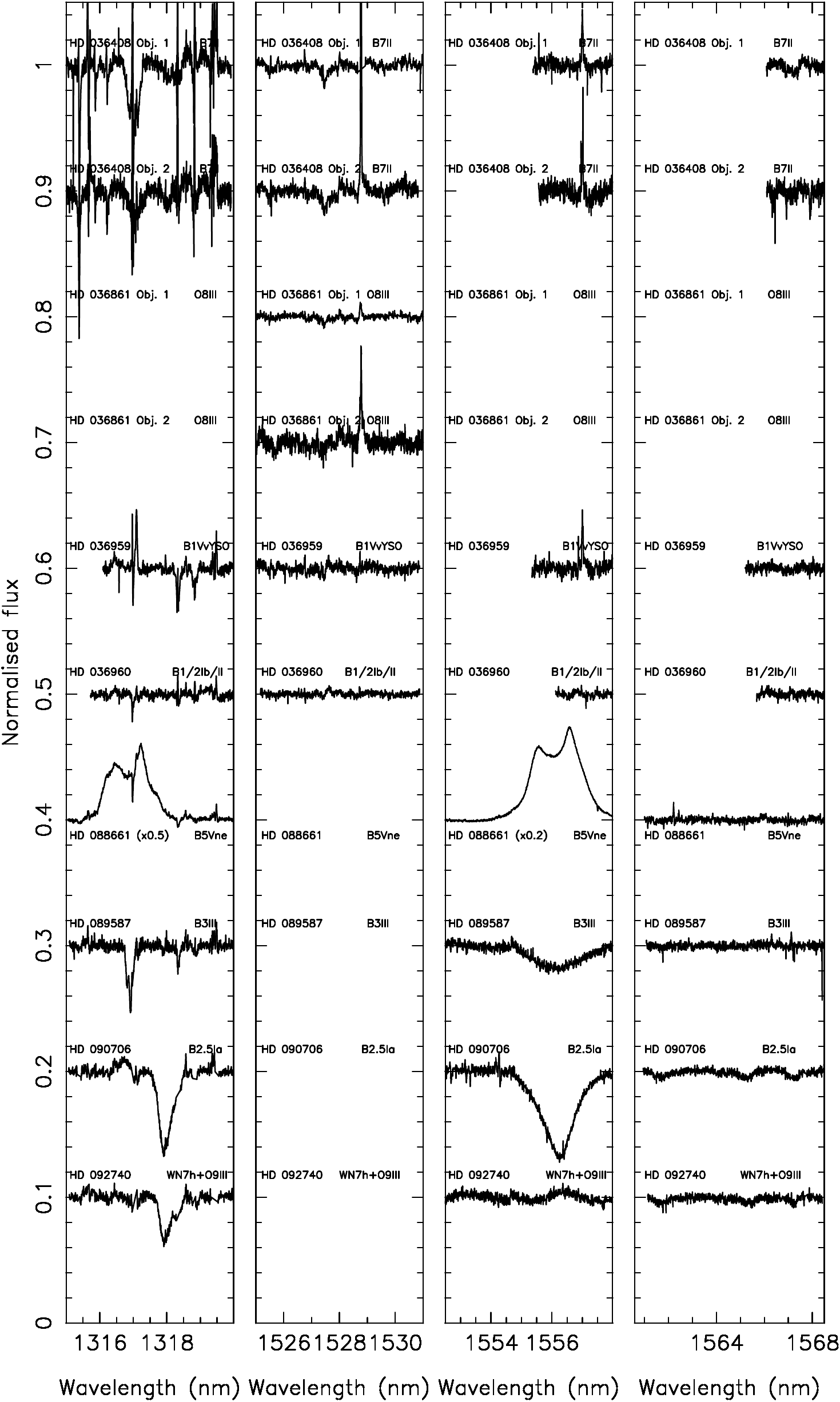}}
\caption{Normalised CRIRES spectra around 1318, 1528, 1555 or 1565~nm offset in the ordinate for clarity.}
\label{f_CRIRES_spectra_1}
\end{figure*}

\setcounter{figure}{0}
\begin{figure*}
   \resizebox{\hsize}{22cm}{\includegraphics[]{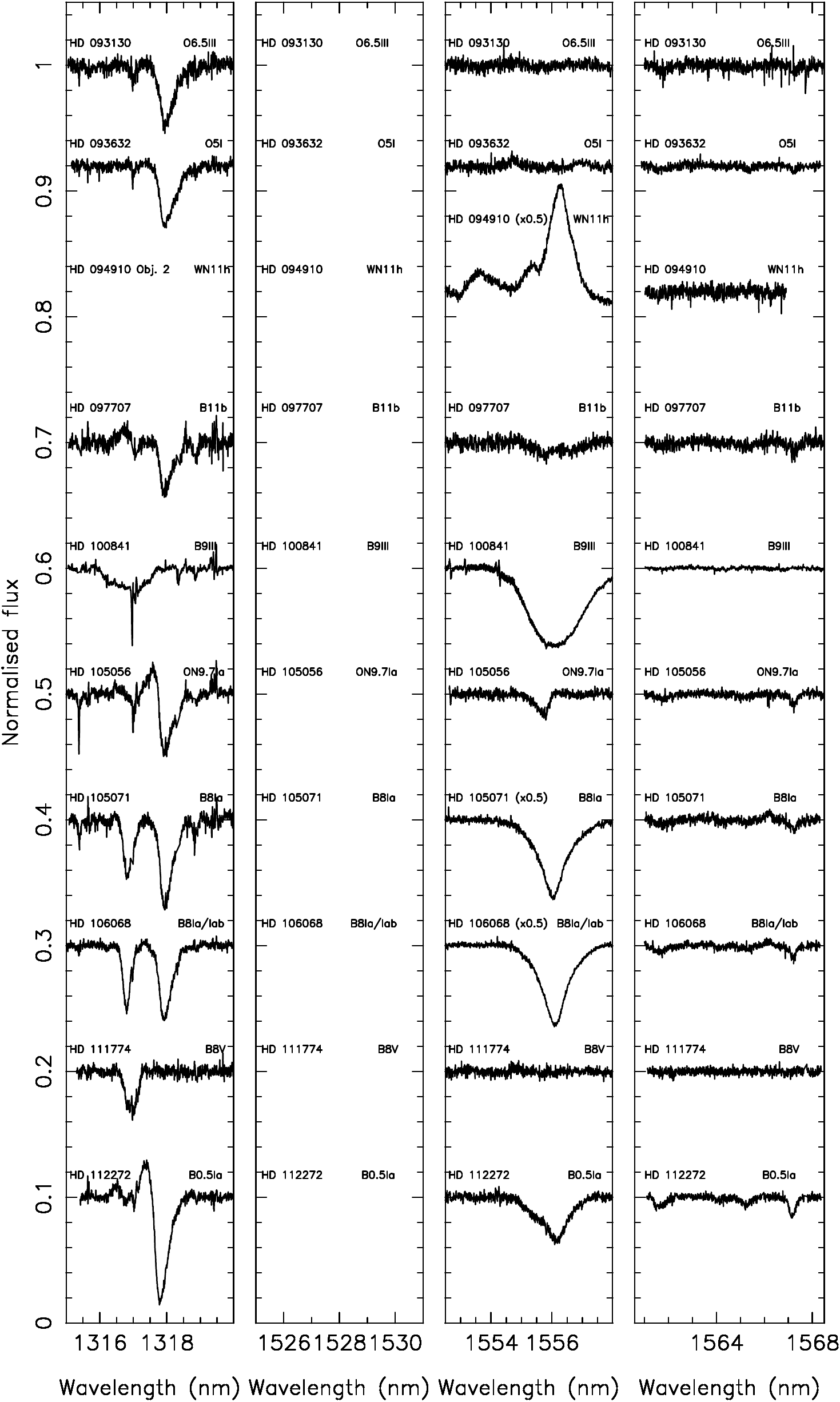}}
\caption{CRIRES spectra ctd.}
\end{figure*}

\setcounter{figure}{0}
\begin{figure*}
   \resizebox{\hsize}{22cm}{\includegraphics[]{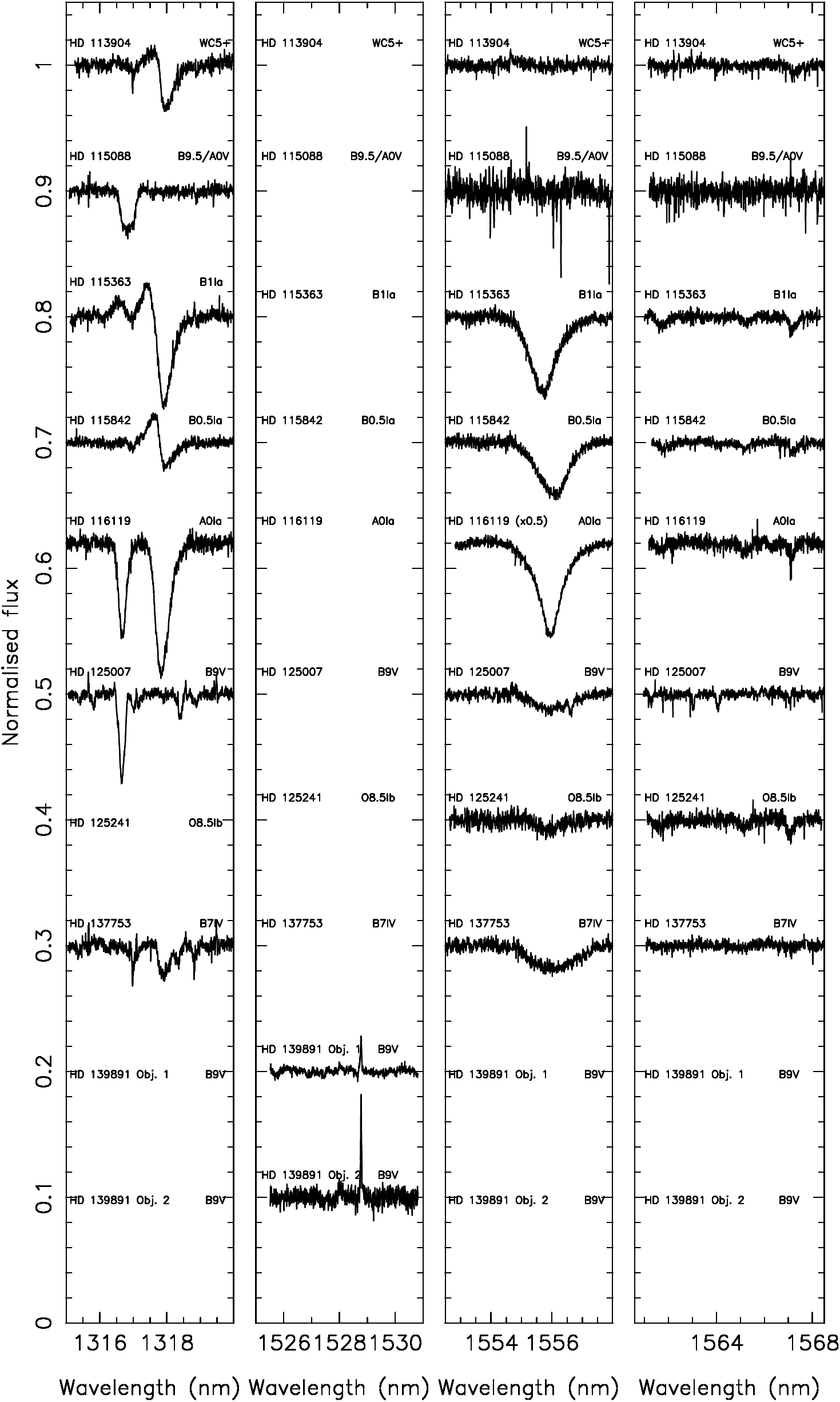}}
\caption{CRIRES spectra ctd.}
\end{figure*}

\setcounter{figure}{0}
\begin{figure*}
   \resizebox{\hsize}{22cm}{\includegraphics[]{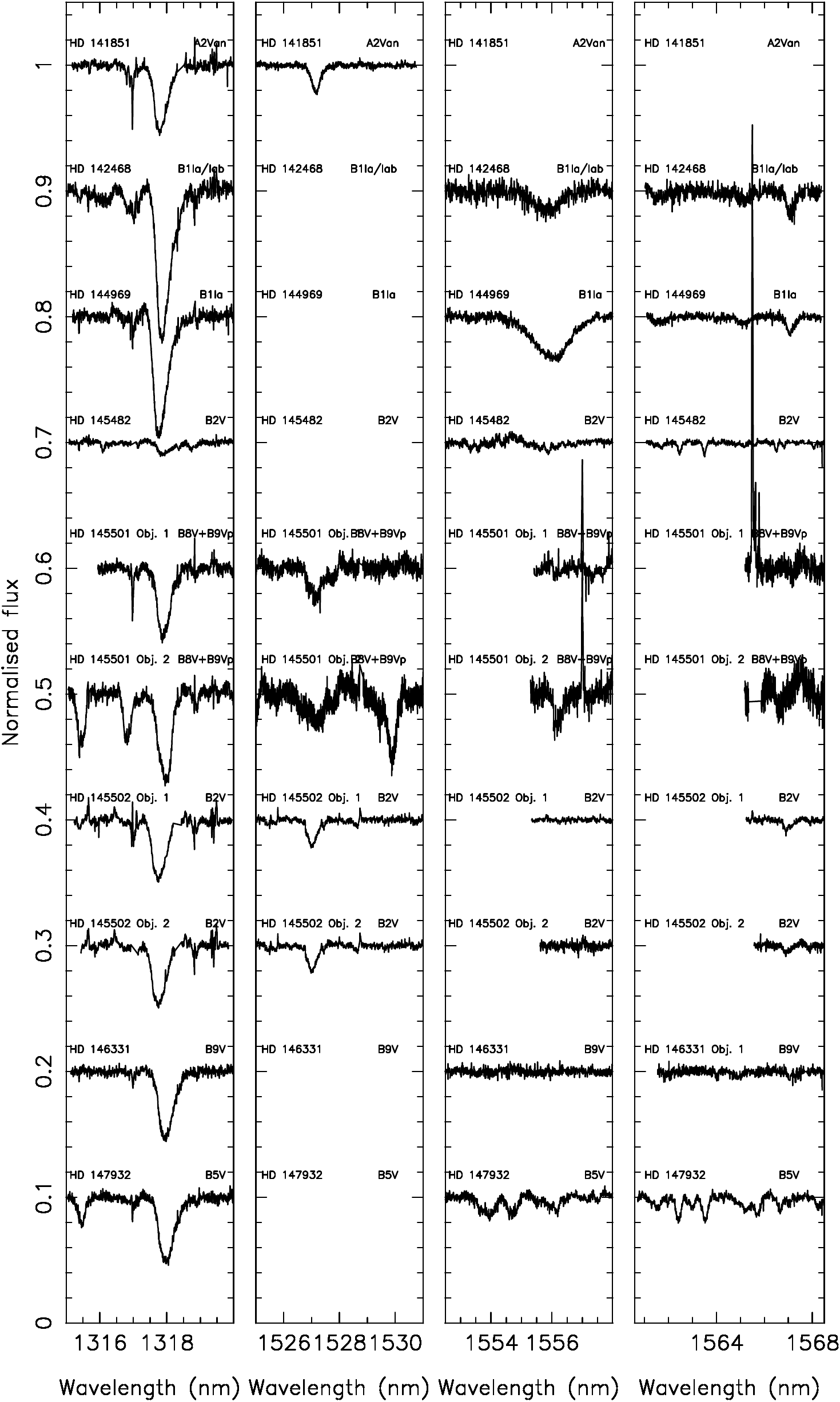}}
\caption{CRIRES spectra ctd.}
\end{figure*}

\setcounter{figure}{0}
\begin{figure*}
   \resizebox{\hsize}{22cm}{\includegraphics[]{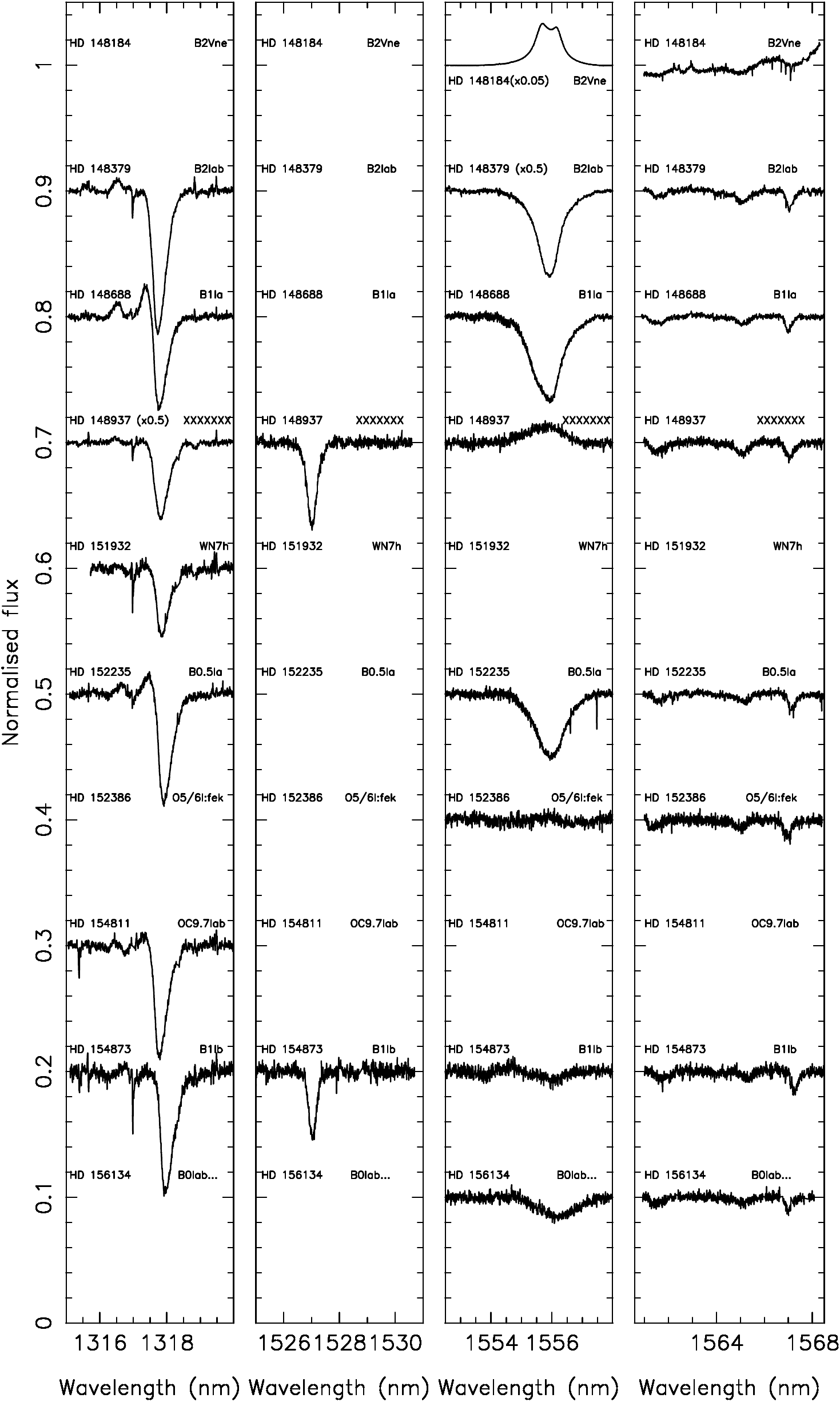}}
\caption{CRIRES spectra ctd.}
\end{figure*}

\setcounter{figure}{0}
\begin{figure*}
   \resizebox{\hsize}{22cm}{\includegraphics[]{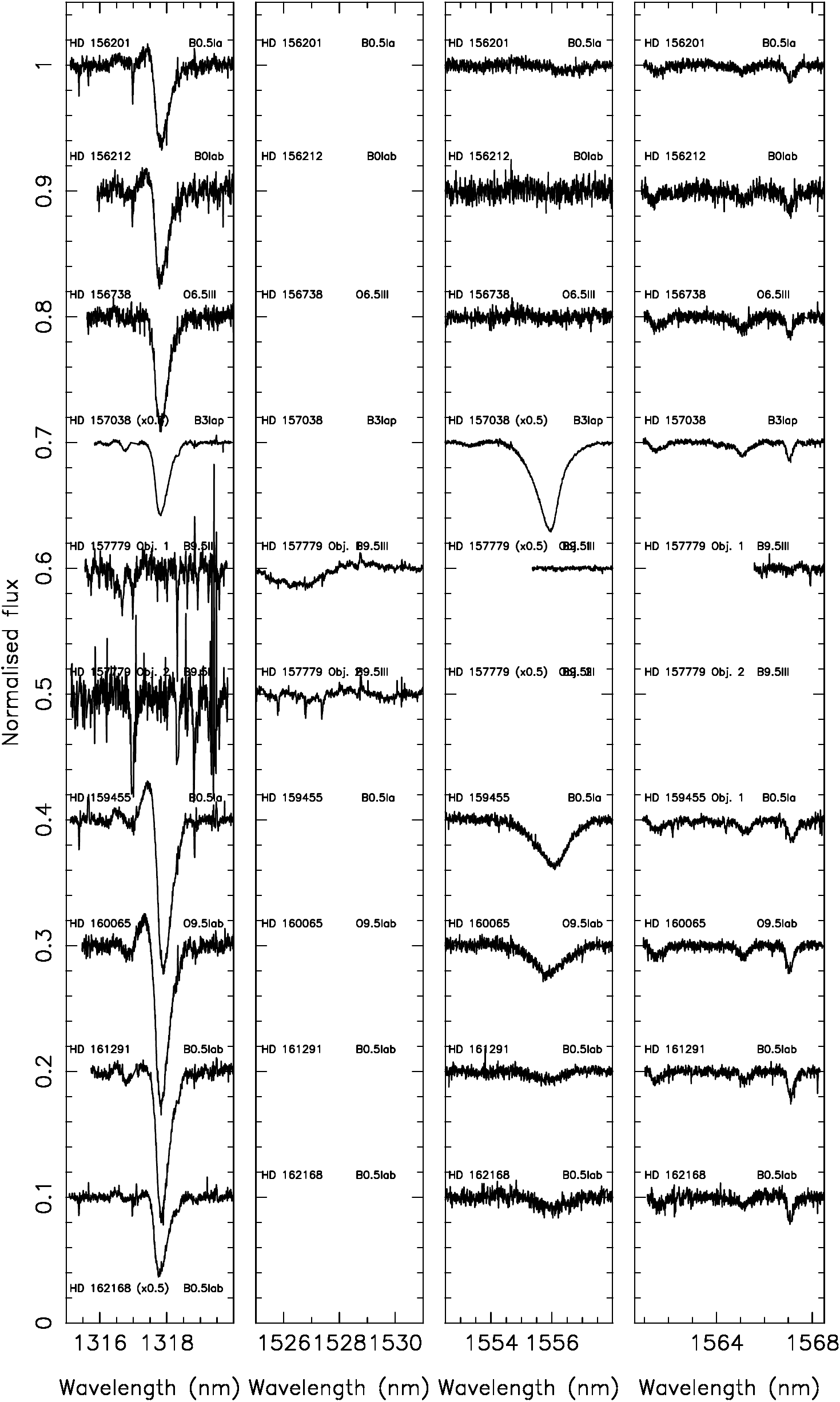}}
\caption{CRIRES spectra ctd.}
\end{figure*}

\setcounter{figure}{0}
\begin{figure*}
   \resizebox{\hsize}{22cm}{\includegraphics[]{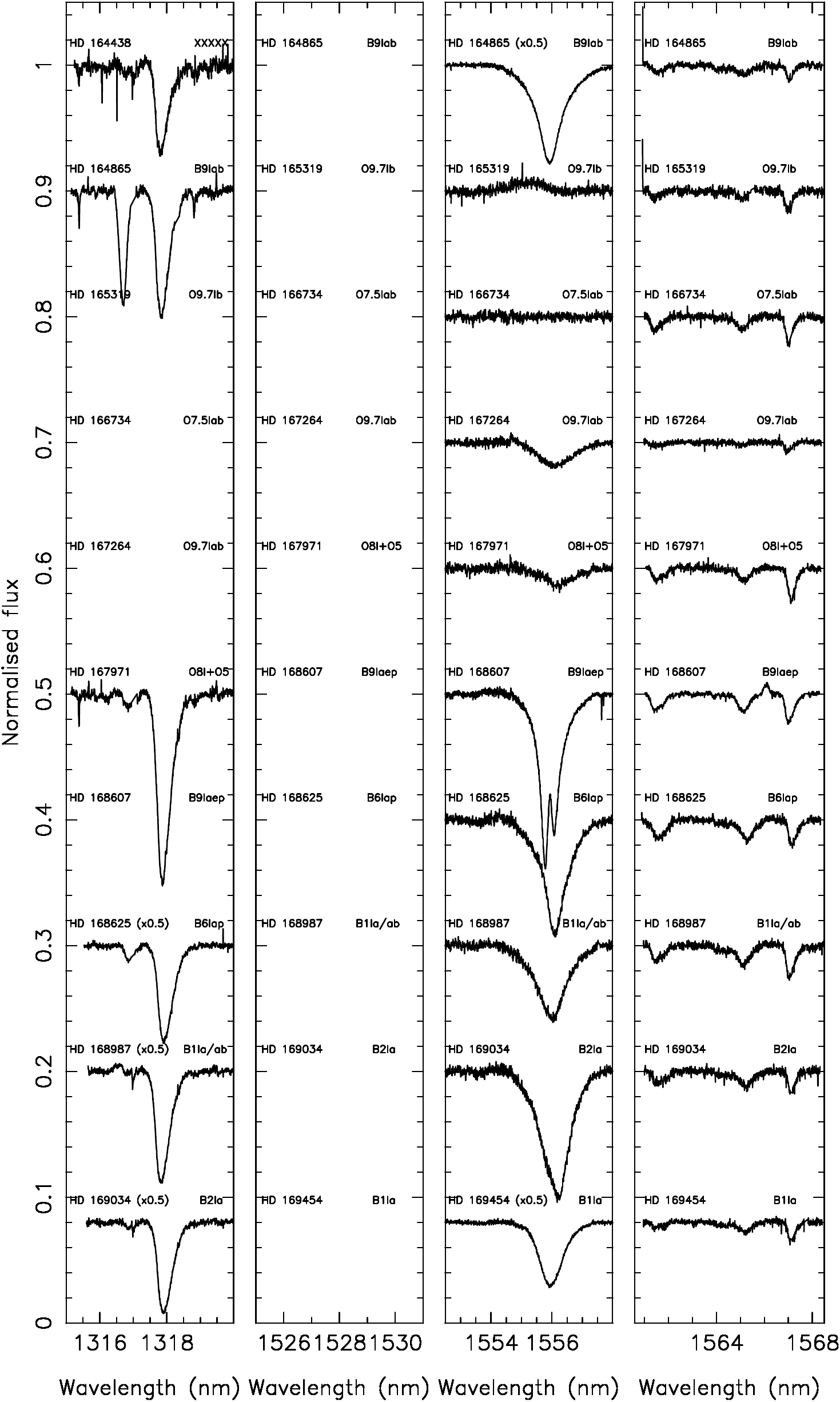}}
\caption{CRIRES spectra ctd.}
\end{figure*}

\setcounter{figure}{0}
\begin{figure*}
   \resizebox{\hsize}{22cm}{\includegraphics[]{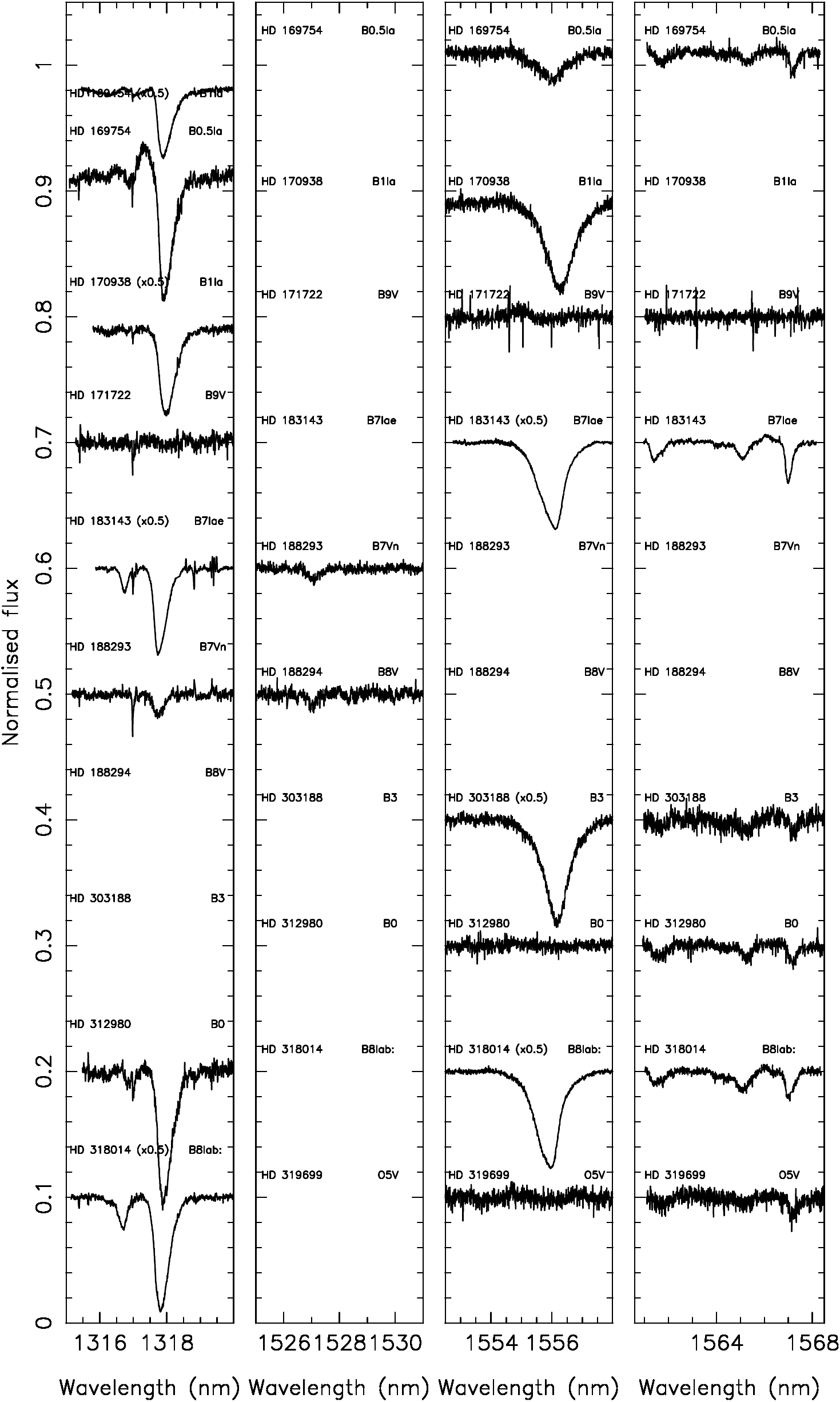}}
\caption{CRIRES spectra ctd.}
\end{figure*}

\clearpage
\newpage

%
%
%
%
%
%
%
%
%
%

\begin{table*}
\begin{center}
\caption{Equivalent width values for the NIR DIBs studied in this paper as well as the optical DIBs at 5780 and 5797~\AA. Columns with no entries either had no data taken or were affected by stellar lines or severe telluric absorption.}
\label{t_NIRDIB_EWs}
\begin{tabular}{rrrrrrrr}
\hline
   HD      &     EW     &    EW       &      EW     &      EW     &      EW     &      EW        &      EW       \\
  number   &   (1318~nm)&   (1527~nm) &    (1561~nm)&    (1565~nm)&    (1567~nm)&    (578.0~nm)  &   (579.7~nm)  \\
           &   (m\AA)   &   (m\AA)    &    (m\AA)   &    (m\AA)   &    (m\AA)   &     (m\AA)     &    (m\AA)     \\
\hline
     13989& $< $ &    $<$      &\ldots&\ldots&\ldots&\ldots&\ldots \\
     35912&\ldots& \ldots      &  55  &  $<$ & $<$  &\ldots&\ldots \\
     36408&   170&    55       &  $<$ &  $<$ & $<$  &\ldots&\ldots \\
     36861&\ldots&   $<$       &\ldots&\ldots&\ldots&   48 &     16\\
     36862&\ldots& \ldots      &\ldots&\ldots&\ldots&   34 &     17\\
     36959&\ldots&   $<$       &  $<$ &  $<$ & $<$  & $<$  &    8.2\\
     36960&   $<$&   $<$       &  $<$ &  $<$ & $<$  & 15.0 &    5.6\\
     52266&   170& \ldots      &\ldots&\ldots&\ldots&\ldots&\ldots \\
     88661&\ldots&   $<$       &  $<$ &  $<$ & $<$  & 13.2 &    8.2\\
     89587&\ldots&   $<$       &\ldots&  $<$ & $<$  &\ldots&\ldots \\
     90706&   308&  \ldots     &    33&    34&    36&\ldots&\ldots \\
     90772&\ldots&  \ldots     &\ldots&\ldots&\ldots&  222 &\ldots \\
     92740&   156&  \ldots     &\ldots&\ldots&\ldots&\ldots&\ldots \\
     93130&   250&  \ldots     &    22&\ldots&    20&  237 &     46\\
     93576&\ldots&  \ldots     &\ldots&\ldots&\ldots&  199 &     78\\
     93632&   258&  \ldots     &    19&\ldots&    16&  266 &     48\\
     94910&   448&  \ldots     &      &\ldots&\ldots&  430 &     92\\
     97707&   191&  \ldots     &    24&    25&    29&\ldots&\ldots \\
    100841&    $<$& \ldots     &   $<$&   $<$&  $<$ &\ldots&\ldots\\
    101065&      &  \ldots     &      &\ldots&\ldots&\ldots&\ldots \\
    105056&   202&  \ldots     &    23&    15&    28&\ldots&\ldots \\
    105071&   312&  \ldots     &    24&    15&    37&\ldots&\ldots \\
    106068&   265&  \ldots     &    32&    17&    35&  235 &     47\\
    111774&   $<$&  \ldots     &\ldots&\ldots&\ldots&\ldots&    $<$\\
    112272&   363&  \ldots     &    38&\ldots&\ldots&  441 &    143\\
    113904&   144&  \ldots     &     8&    11&    35&  126 &     24\\
    114213&   370&  \ldots     &\ldots&\ldots&\ldots&\ldots& \ldots\\
    115088&   $<$&  \ldots     &\ldots&\ldots&\ldots& 13.8 &    0.0\\
    115363&   331&  \ldots     &    34&    20&    38&\ldots& \ldots\\
    115842&    81&  \ldots     &    21&    17&    22&  234 &    108\\
    116119&   471&  \ldots     &\ldots&\ldots&\ldots&  449 &    125\\
    125007&   $<$&  \ldots     &\ldots&\ldots&\ldots&   18 &    <$<$\\
    125241&   514&  \ldots     &\ldots&\ldots&\ldots&  533 &    109\\
    137753&    78&  \ldots     &\ldots&\ldots&\ldots&\ldots& \ldots\\
    138014&   869&  \ldots     &\ldots&\ldots&\ldots&\ldots& \ldots\\
    139892&   $<$&  \ldots     &\ldots&\ldots&\ldots&\ldots& \ldots\\
    141850&\ldots&  \ldots     &\ldots&\ldots&\ldots&\ldots& \ldots\\
    142468&   590&  \ldots     &    32&    38&    72&  461 &    118\\
    144217&   226&    86       &\ldots&\ldots&    23&  156 &     16\\
    144218&   245&   100       &\ldots&\ldots&\ldots&  157 &     20\\
    144969&   458&  \ldots     &\ldots&    29&    51&\ldots& \ldots\\
    145482&    39&  \ldots     &\ldots&\ldots&\ldots&\ldots& \ldots\\
    145501&   244&   130       &\ldots&\ldots&\ldots&  182 &     32\\
\hline
\end{tabular}
\end{center}
\end{table*}

\setcounter{table}{0}
\begin{table*}
\begin{center}
\caption{continued.}
\begin{tabular}{rrrrrrrr}
\hline
   HD      &     EW       &    EW        &      EW     &      EW     &      EW     &      EW        &      EW       \\
  number   &   (1318~nm)  &   (1527~nm)  &    (1561~nm)&   (1565~nm) &    (1567~nm)&    (578.0~nm)  &   (579.7~nm)  \\
           &   (m\AA)     &   (m\AA)     &    (m\AA)   &    (m\AA)   &    (m\AA)   &     (m\AA)     &    (m\AA)     \\
\hline
    145501&   244 &   130 &\ldots& \ldots&\ldots&  182 &   32 \\
    145502&   226 & \ldots&\ldots& \ldots&  36  &  177 &   36 \\
    146331&   281 & \ldots&\ldots& \ldots&\ldots&\ldots&\ldots\\
    147932&   284 & \ldots&\ldots& \ldots&\ldots&  207 &   46 \\
    148184&   126 & \ldots&\ldots& \ldots&\ldots&  100 &   53 \\
    148379&   547 & \ldots&\ldots& \ldots&  51  &  427 &  106 \\
    148688&   318 & \ldots&    38& \ldots&  36  &\ldots&\ldots\\
    148739&   539 & \ldots&\ldots& \ldots&\ldots&\ldots&\ldots\\
    148937&   602 &   310 &    50& \ldots&  52  &  356 &   71 \\
    151932&   247 & \ldots&\ldots& \ldots&\ldots&  336 &\ldots\\
    152235&   406 & \ldots&    30& \ldots&  48  &  418 &  101 \\
    152270& \ldots& \ldots&\ldots& \ldots&\ldots&  353 &   66 \\
    152386&   378 & \ldots&    31& \ldots&  55  &  352 &   70 \\
    154811&   410 & \ldots&\ldots& \ldots&\ldots&\ldots&\ldots\\
    154873&   385 &   160 &\ldots& \ldots&\ldots&\ldots&\ldots\\
    156134&   284 & \ldots&    31&     31&  33  &  320 &   93 \\
    156201&   306 & \ldots&    29&     31&  47  &  318 &  119 \\
    156212&   322 & \ldots&    37&     37&  60  &  375 &  133 \\
    156738&   413 & \ldots&    48&     65&  57  &  494 &  136 \\
    157038&   560 & \ldots&\ldots& \ldots&  39  &  447 &  124 \\
    157778& \ldots& \ldots&\ldots& \ldots&\ldots&   34 &  $<$ \\
    157779&   $<$ &   $<$ &  $<$ &    $<$&\ldots&\ldots&\ldots\\
    159455&   570 & \ldots&    56&     58&  78  &\ldots&\ldots\\
    160065&   611 & \ldots&    56&     42&  65  &\ldots&\ldots\\
    161289& \ldots& \ldots&\ldots& \ldots&\ldots&\ldots&\ldots\\
    161291&   543 & \ldots&    38&     33&  72  &\ldots&\ldots\\
    162168&   575 & \ldots&    37&     45&  68  &\ldots&\ldots\\
    164438&   321 & \ldots&    25&     41&  35  &  391 &  123 \\
    164865&   485 & \ldots&    32&     47&  37  &  383 &   67 \\
    165319&   375 & \ldots&    26&     29&  62  &  533 &  191 \\
    166734&   566 & \ldots&\ldots& \ldots&\ldots&  720 &  276 \\
    167264&   123 & \ldots&    12&     11&  26  &\ldots&\ldots\\
    167971&   745 & \ldots&    51&     57& 100  &  563 &  158 \\
    168607&   878 & \ldots&    75&     75& 100  &  859 &  280 \\
    168625&   809 & \ldots&   100&    120&  90  &  842 &  239 \\
    168987&   902 & \ldots&    67&    100& 110  &  752 &  146 \\
    169034&   760 & \ldots&    55&     78&  60  &  689 &  216 \\
    169454&   528 & \ldots&    24&     47&  53  &  482 &  177 \\
    169754&   451 & \ldots&\ldots&     40&  58  &\ldots&\ldots\\
    170938&   774 & \ldots&\ldots&     40&  55  &  476 &  145 \\
    171722&   $<$ & \ldots&\ldots& \ldots&\ldots&   20 &  $<$ \\
    183143&   629 & \ldots&    69&     65& 110  &  807 &  230 \\
    188293&    64 &    48 &\ldots& \ldots&\ldots&\ldots&\ldots\\
    188294&    37 & \ldots&\ldots& \ldots&\ldots&\ldots&\ldots\\
    303188&   690 & \ldots&   $<$&    $<$& $<$  &\dots &\ldots\\
    305529& \ldots& \ldots&\ldots& \ldots&\ldots&  211 &   49 \\
    312980&   521 & \ldots&    58&     53&  61  &\ldots&\ldots\\
    318014&   998 & \ldots&    58&     93& 100  &  675 &  168 \\
    319699&   534 & \ldots&    35&     33&  69  &  363 &   87 \\
    319702&   542 & \ldots&    34&     44&  50  &  521 &  129 \\
\hline
\end{tabular}
\end{center}
\end{table*}

\clearpage
\newpage

%
%

\setcounter{figure}{1}
\begin{figure*}
\begin{center}
   \resizebox{17cm}{!}{\includegraphics{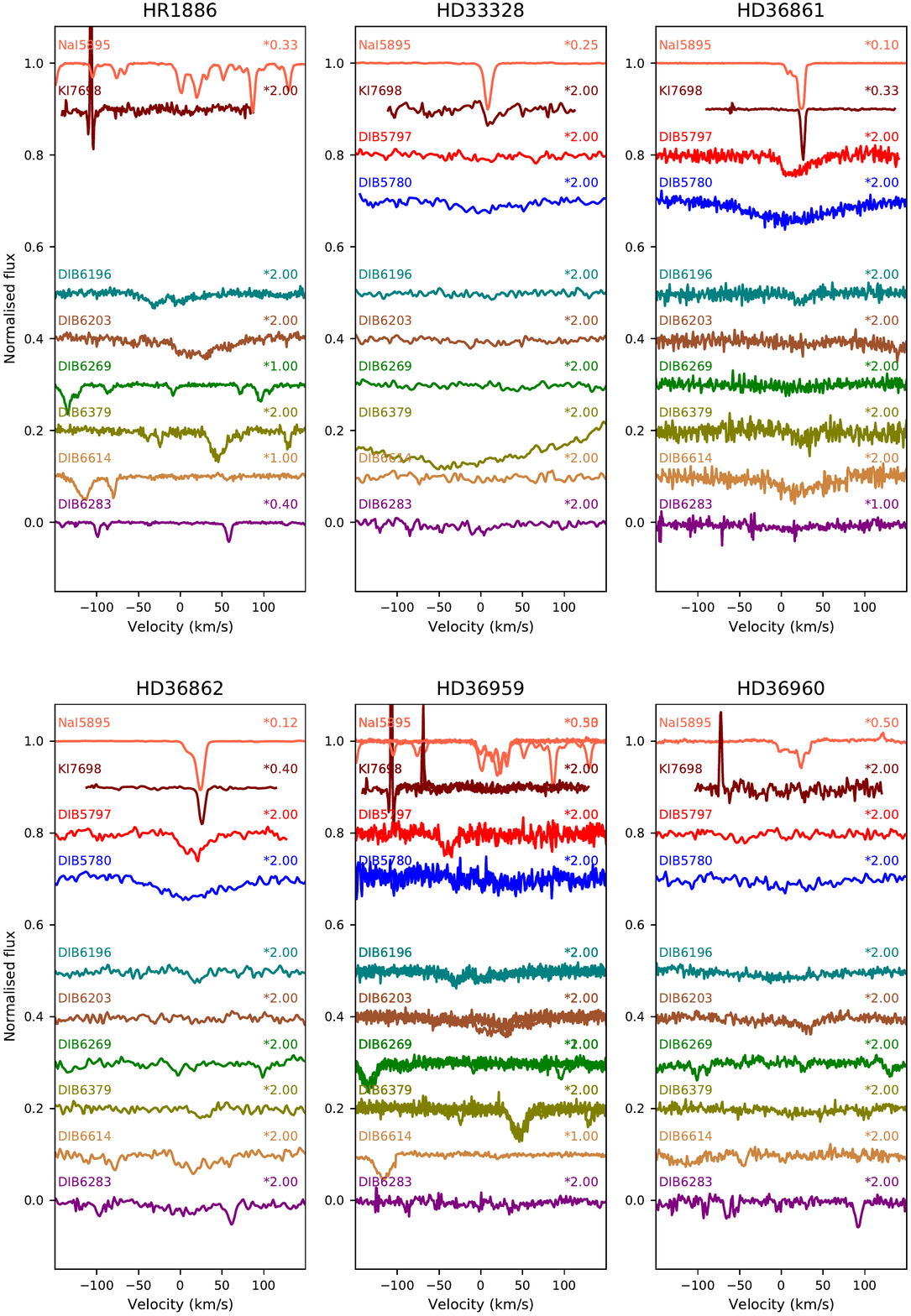}}
\end{center}
\caption{Normalised flux (scaled by the factor shown and offset in the ordinate for clarity) vs heliocentric velocity for Na\,{\sc i}, K\,{\sc i} and a handful of DIBs. }
\label{f_Optical_Data}
\end{figure*}

\setcounter{figure}{1}
\begin{figure*}
\begin{center}
   \resizebox{17cm}{!}{\includegraphics{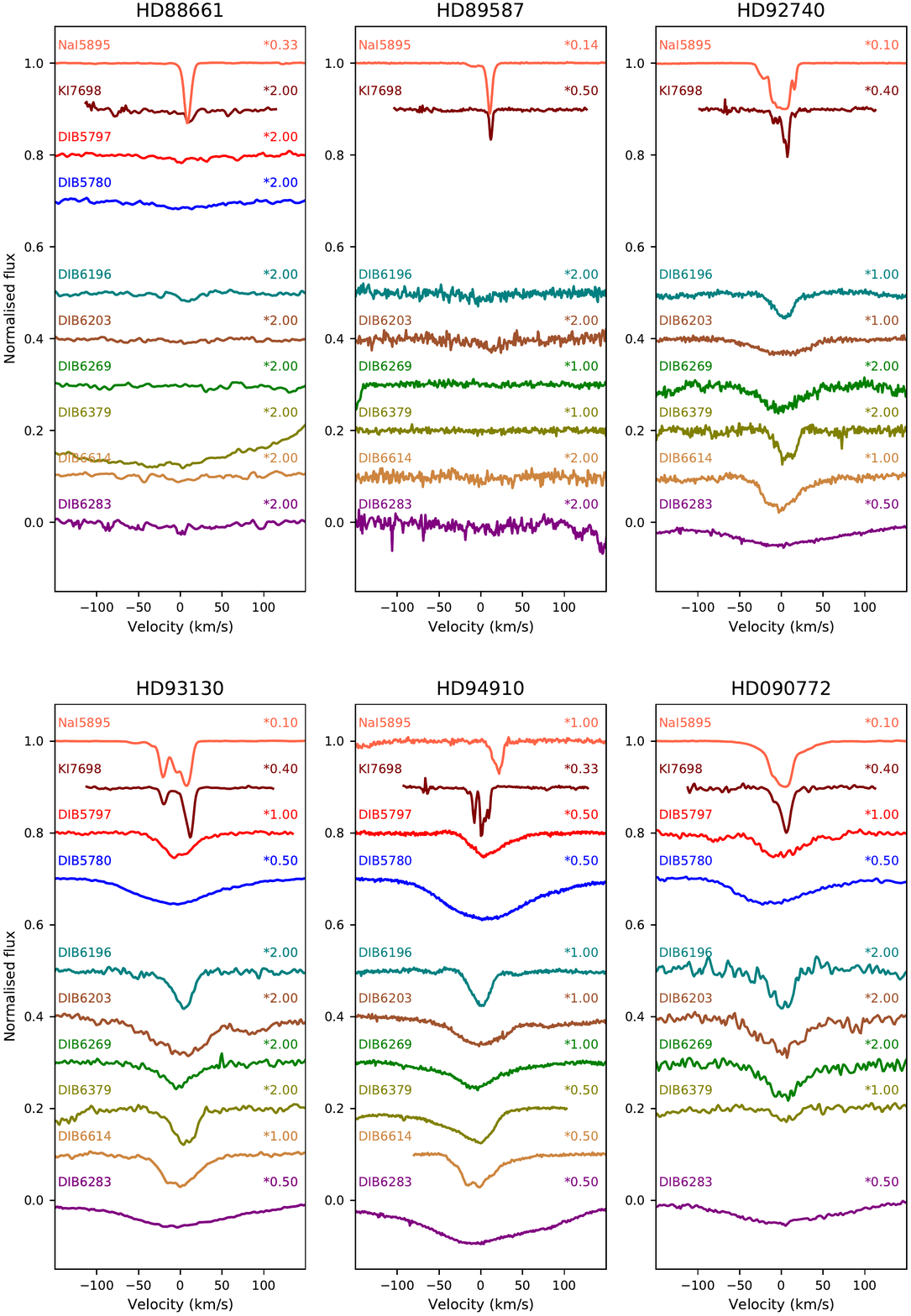}}
\end{center}
\caption{Optical data ctd.}
\end{figure*}

\setcounter{figure}{1}
\begin{figure*}
\begin{center}
   \resizebox{17cm}{!}{\includegraphics{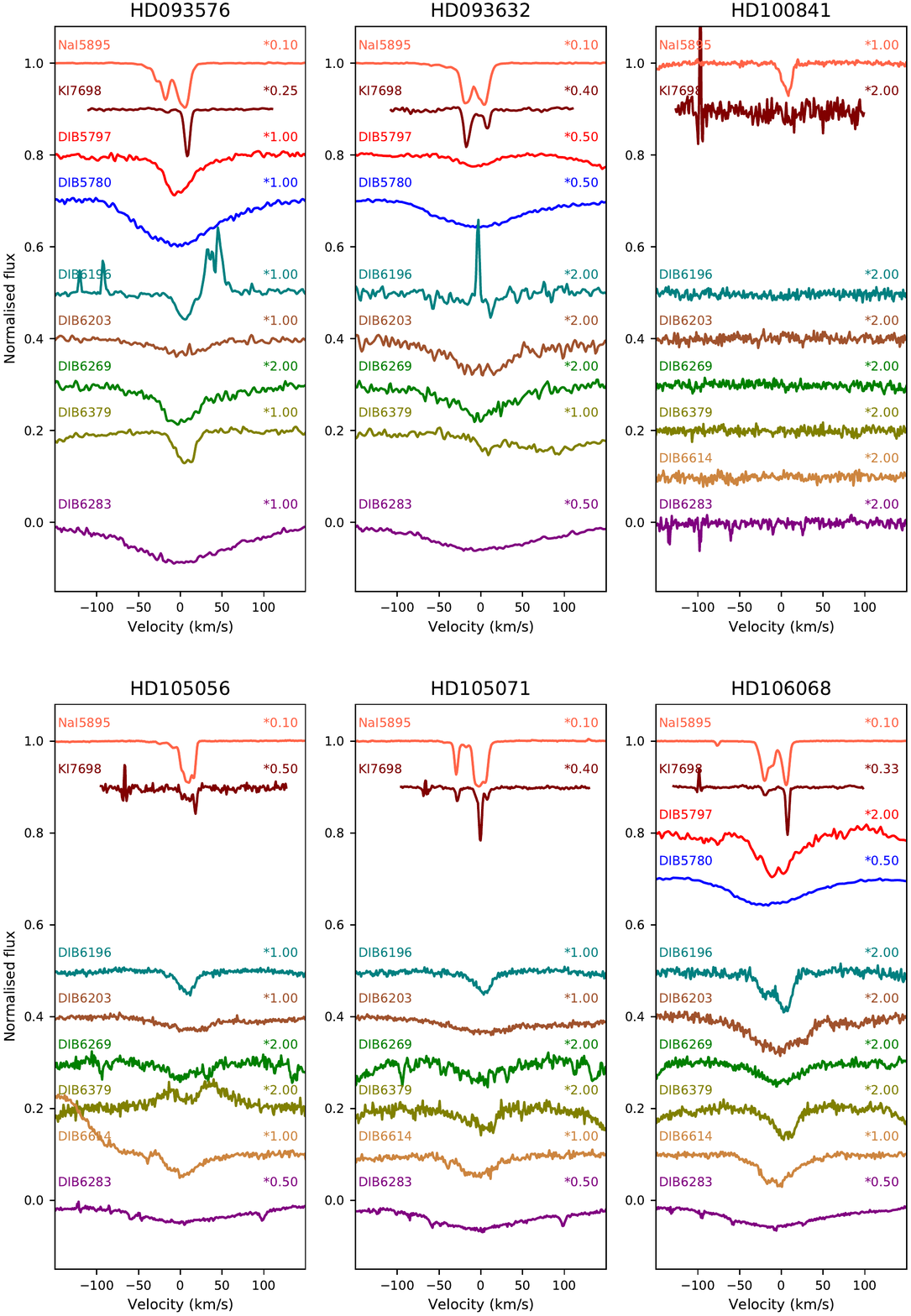}}
   \end{center}
\caption{Optical data ctd.}
\end{figure*}

\setcounter{figure}{1}
\begin{figure*}
\begin{center}
   \resizebox{\hsize}{17cm}{\includegraphics{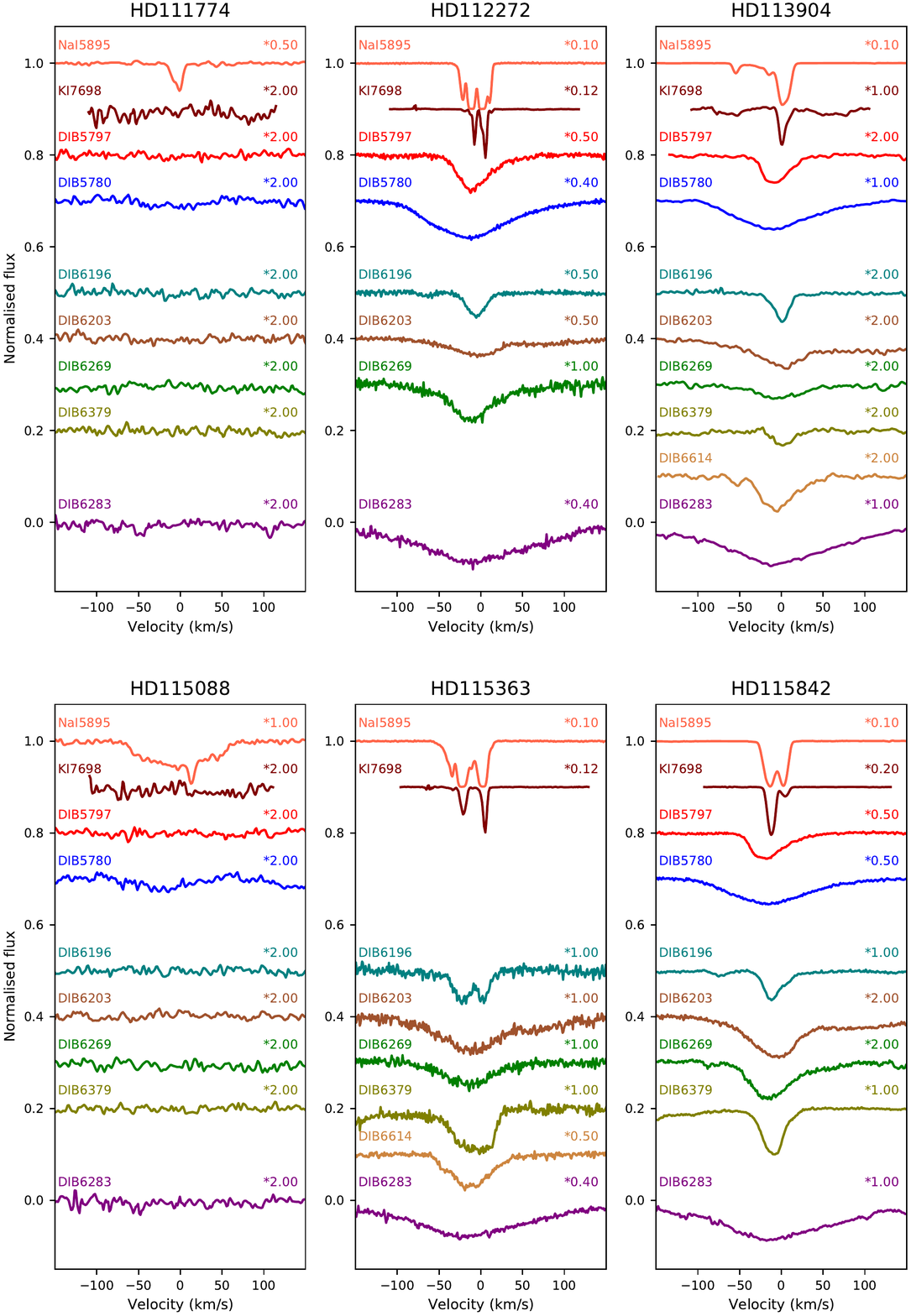}}
   \end{center}
\caption{Optical data ctd.}
\end{figure*}

\setcounter{figure}{1}
\begin{figure*}
\begin{center}
   \resizebox{17cm}{!}{\includegraphics{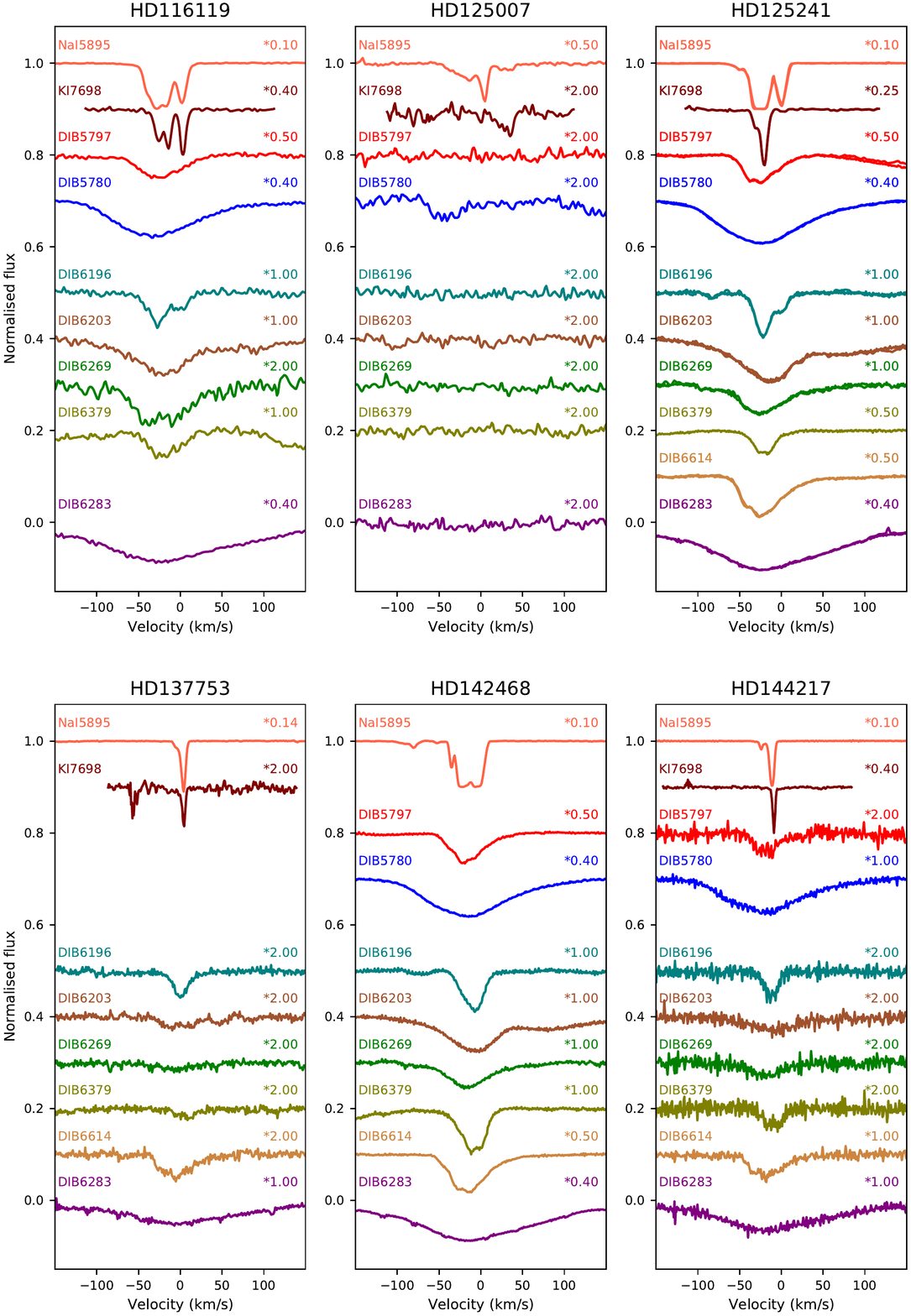}}
   \end{center}
\caption{Optical data ctd.}
\end{figure*}

\setcounter{figure}{1}
\begin{figure*}
\begin{center}
   \resizebox{17cm}{!}{\includegraphics{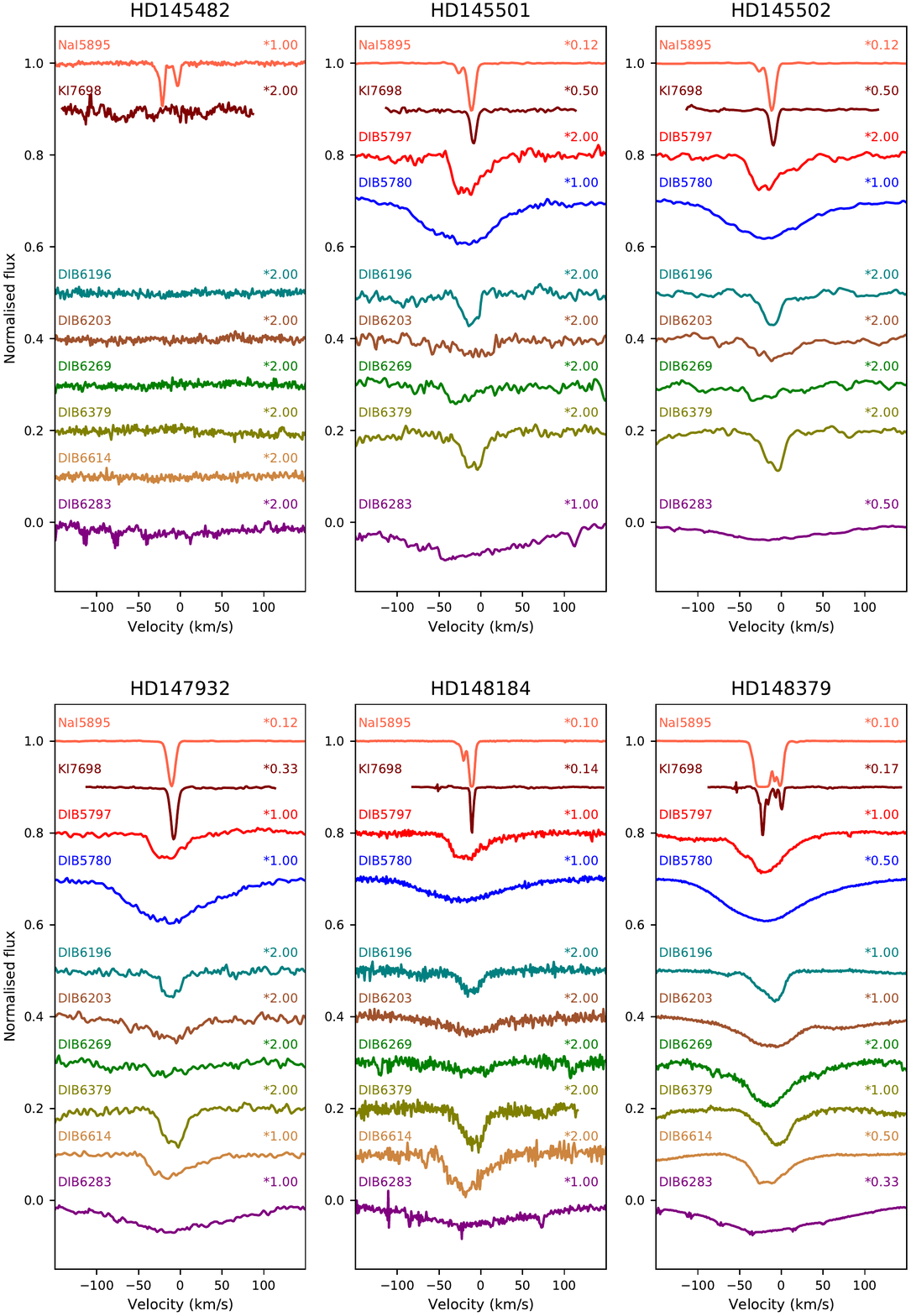}}
   \end{center}
\caption{Optical data ctd.}
\end{figure*}

\setcounter{figure}{1}
\begin{figure*}
\begin{center}
   \resizebox{\hsize}{17cm}{\includegraphics{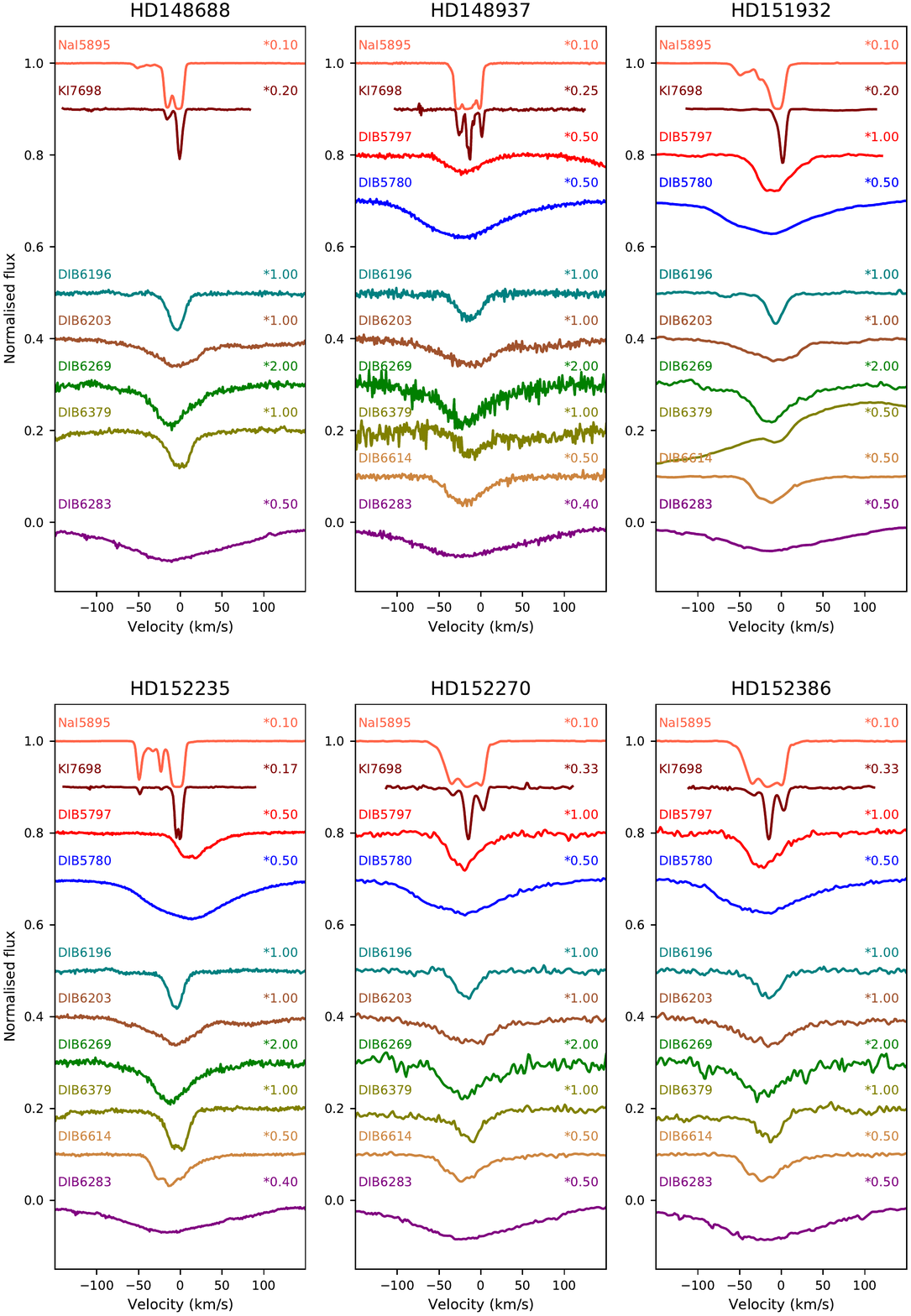}}
   \end{center}
\caption{Optical data ctd.}
\end{figure*}

\setcounter{figure}{1}
\begin{figure*}
\begin{center}
   \resizebox{17cm}{!}{\includegraphics{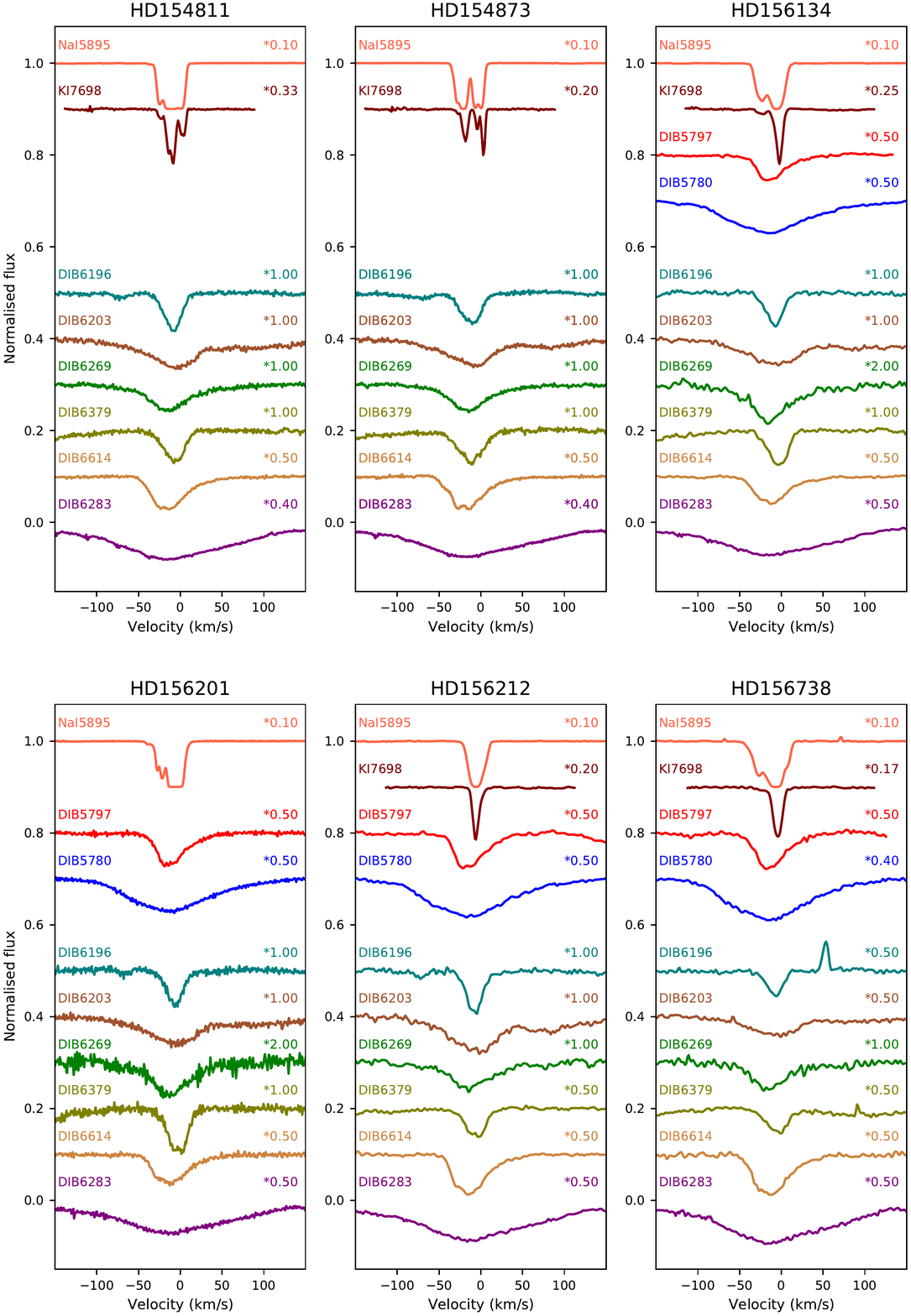}}
   \end{center}
\caption{Optical data ctd.}
\end{figure*}

\setcounter{figure}{1}
\begin{figure*}
\begin{center}
   \resizebox{17cm}{!}{\includegraphics{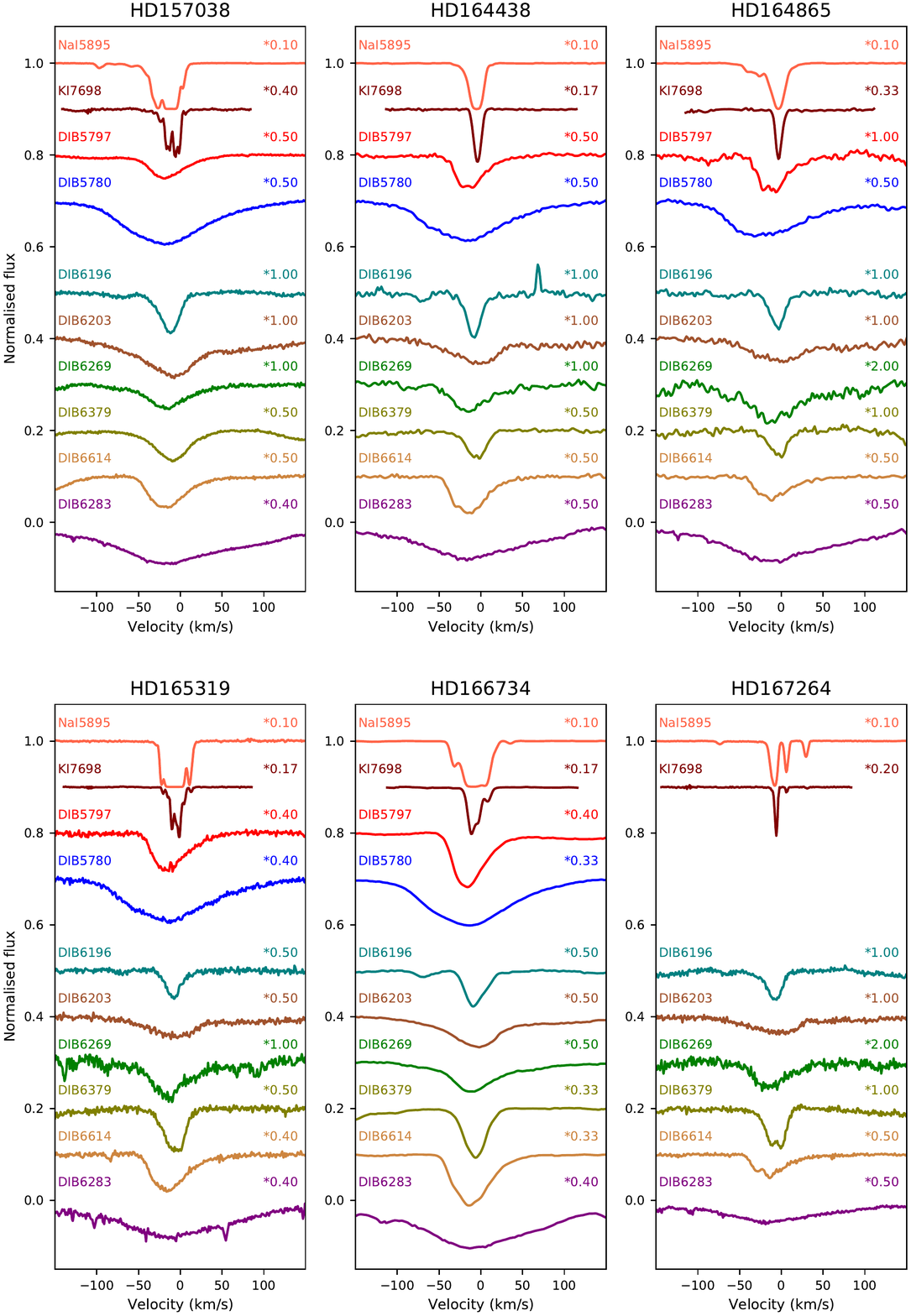}}
   \end{center}
\caption{Optical data ctd.}
\end{figure*}

\setcounter{figure}{1}
\begin{figure*}
\begin{center}
   \resizebox{17cm}{!}{\includegraphics{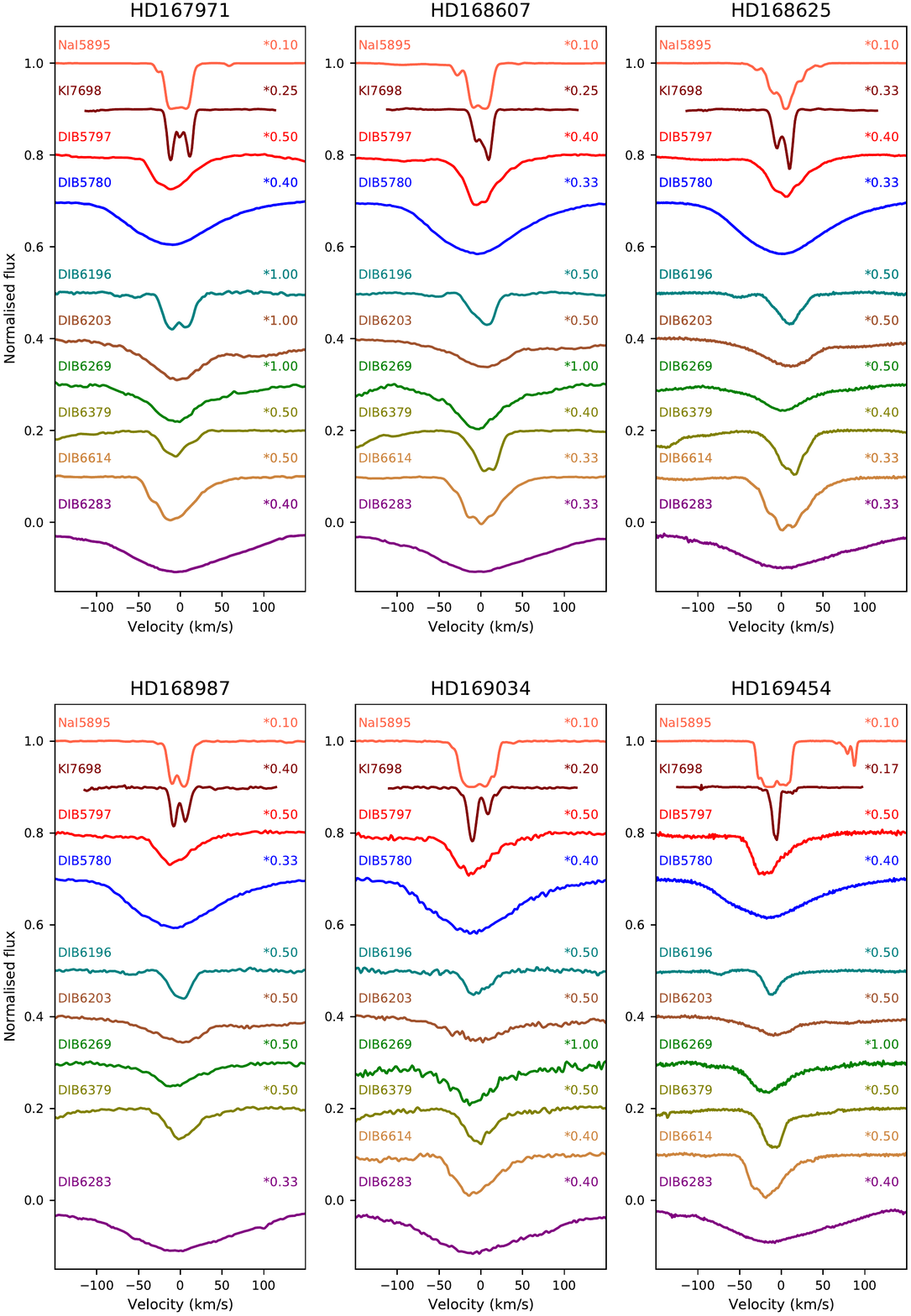}}
   \end{center}
\caption{Optical data ctd.}
\end{figure*}

\setcounter{figure}{1}
\begin{figure*}
\begin{center}
   \resizebox{17cm}{!}{\includegraphics{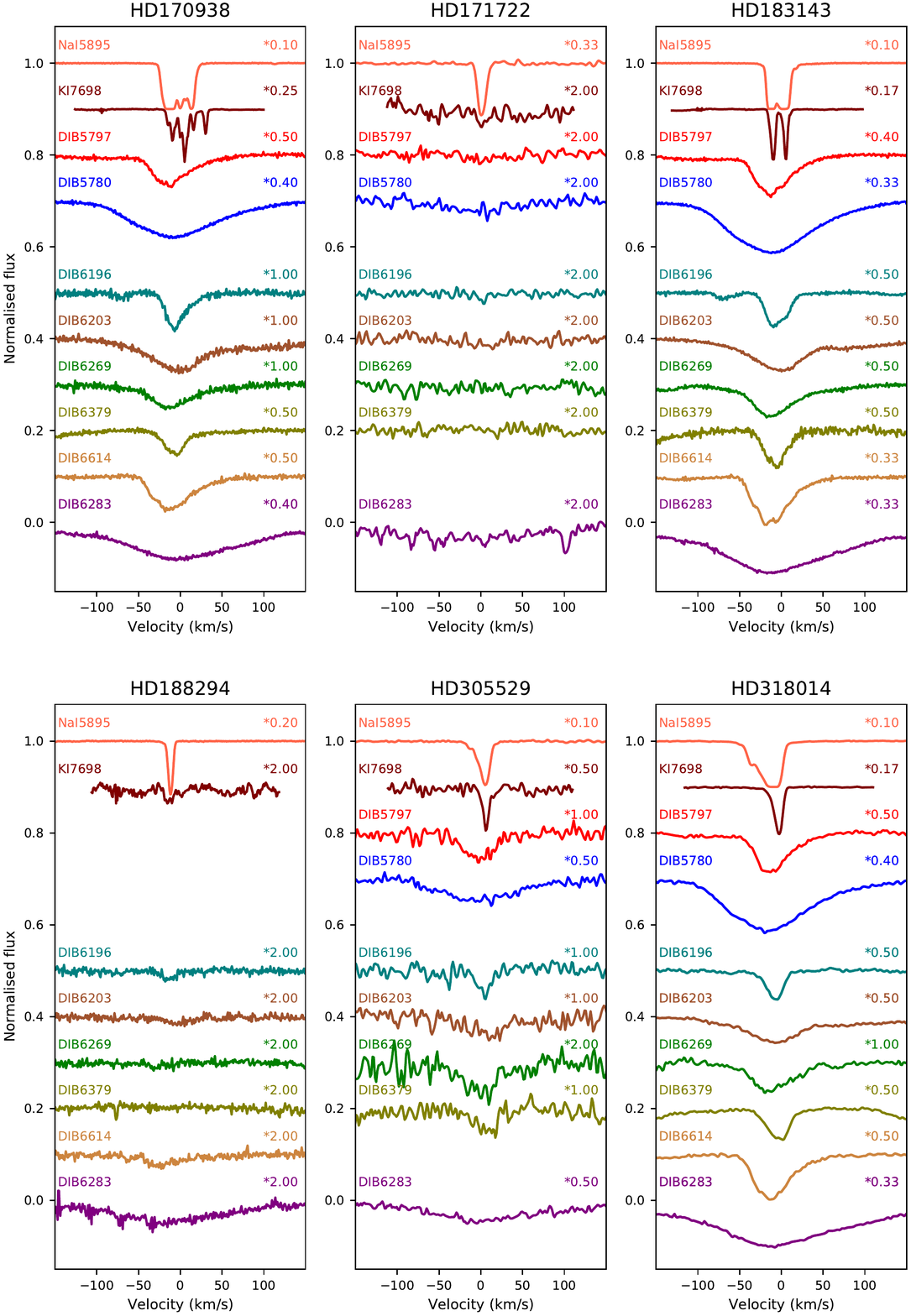}}
   \end{center}
\caption{Optical data ctd.}
\end{figure*}

\setcounter{figure}{1}
\begin{figure*}
\begin{center}
\resizebox{11.5cm}{!}{\includegraphics[clip,trim=0.0cm 15.0cm 7.0cm 0.0cm]{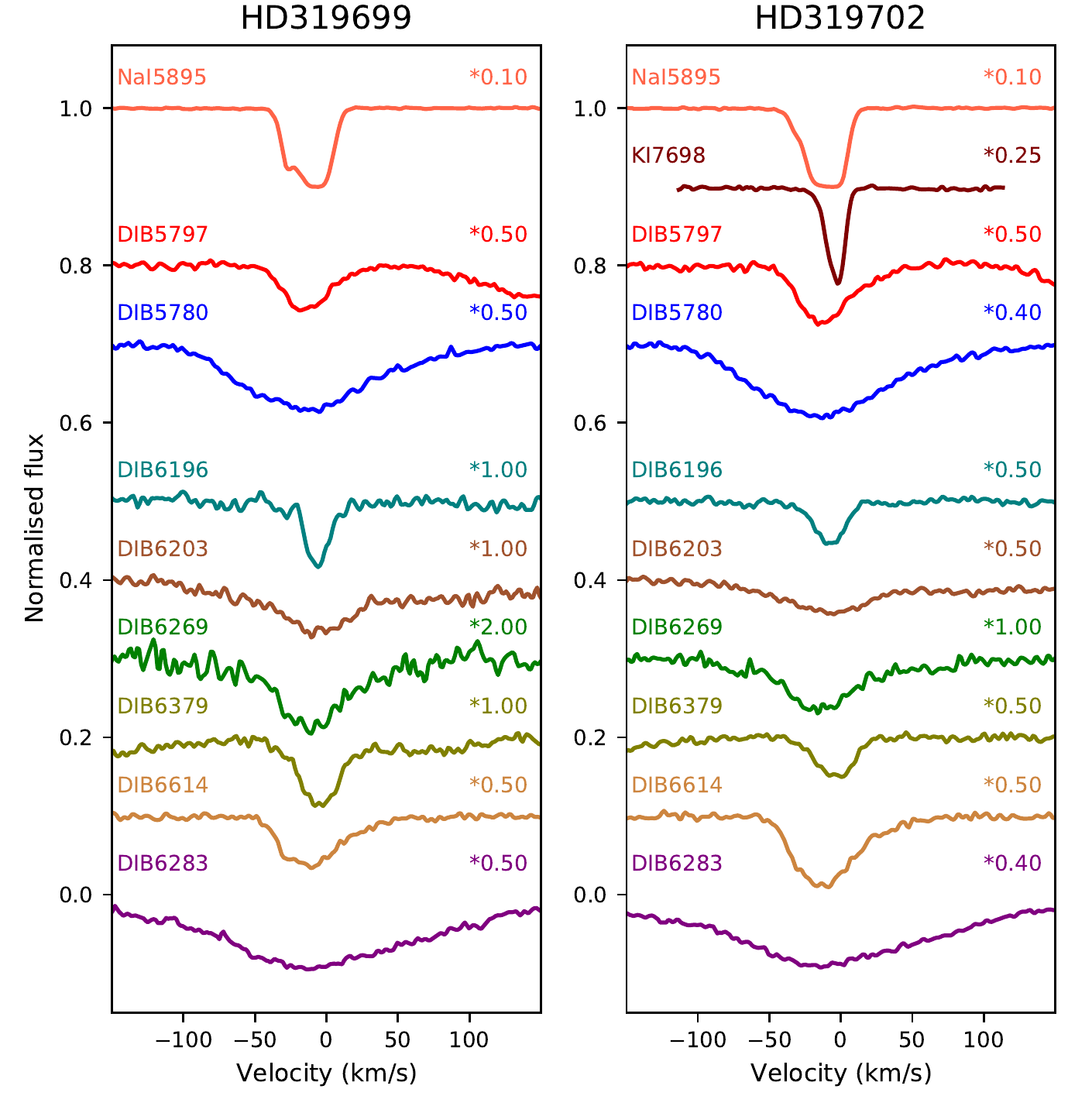}}
\end{center}
\caption{Optical data ctd.}
\end{figure*}

\setcounter{figure}{2}
\begin{figure*}
\begin{center}
   \resizebox{17cm}{!}{\includegraphics{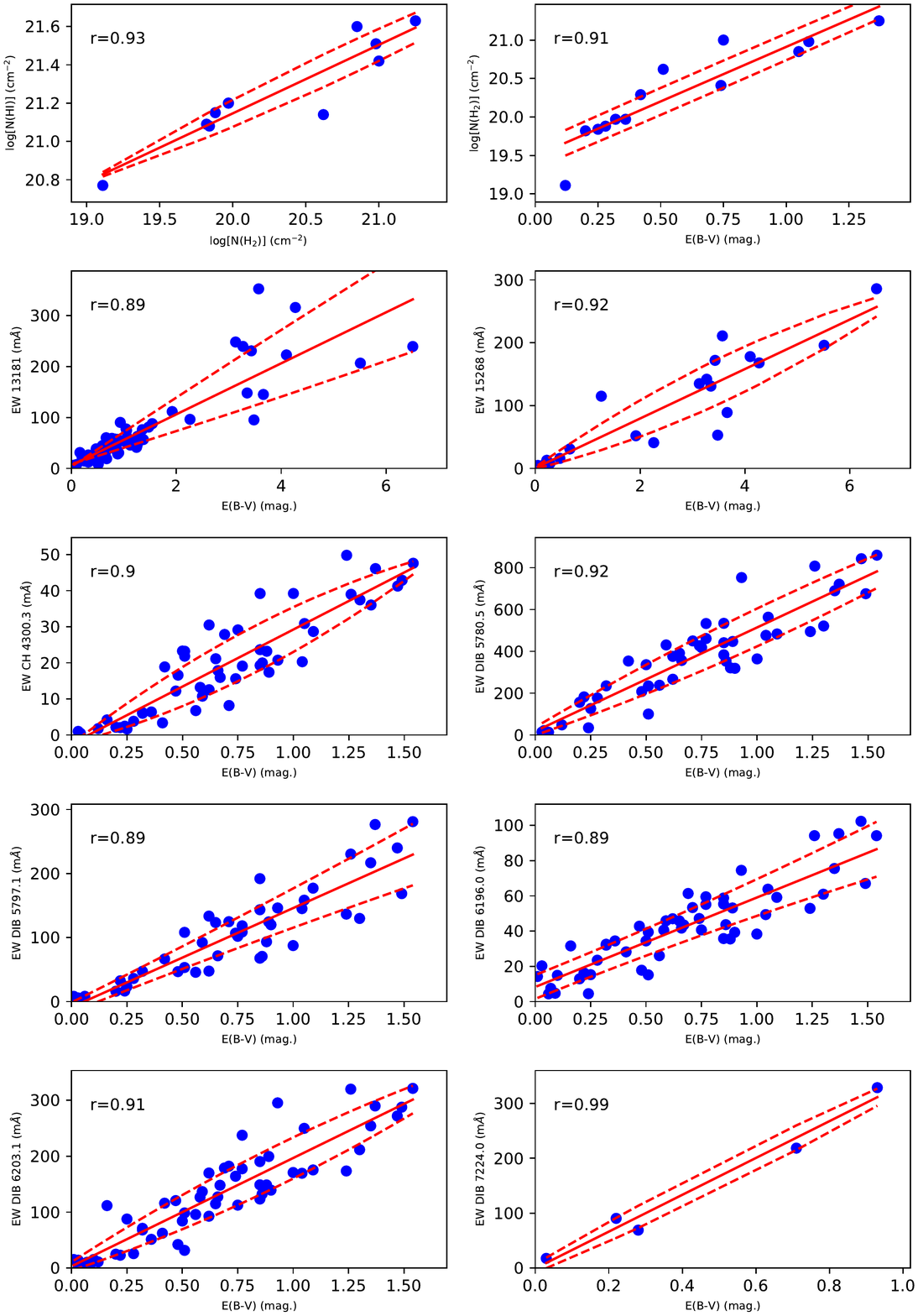}}
\end{center}
\caption{Correlations between equivalent width of DIBs or reddening $E(B-V)$ for which the Pearson correlation coefficient exceeded 0.87, plus some optical lines. EW is measured in m$\mbox{\AA}$\, with reddening in magnitudes. Red points have been excluded from the correlations. The dashed lines show upper and lower error bounds obtained byfitting with a second order polynomial.}
\label{f_Correlation_fits}
\end{figure*}

\setcounter{figure}{2}
\begin{figure*}
\begin{center}
   \resizebox{17cm}{!}{\includegraphics{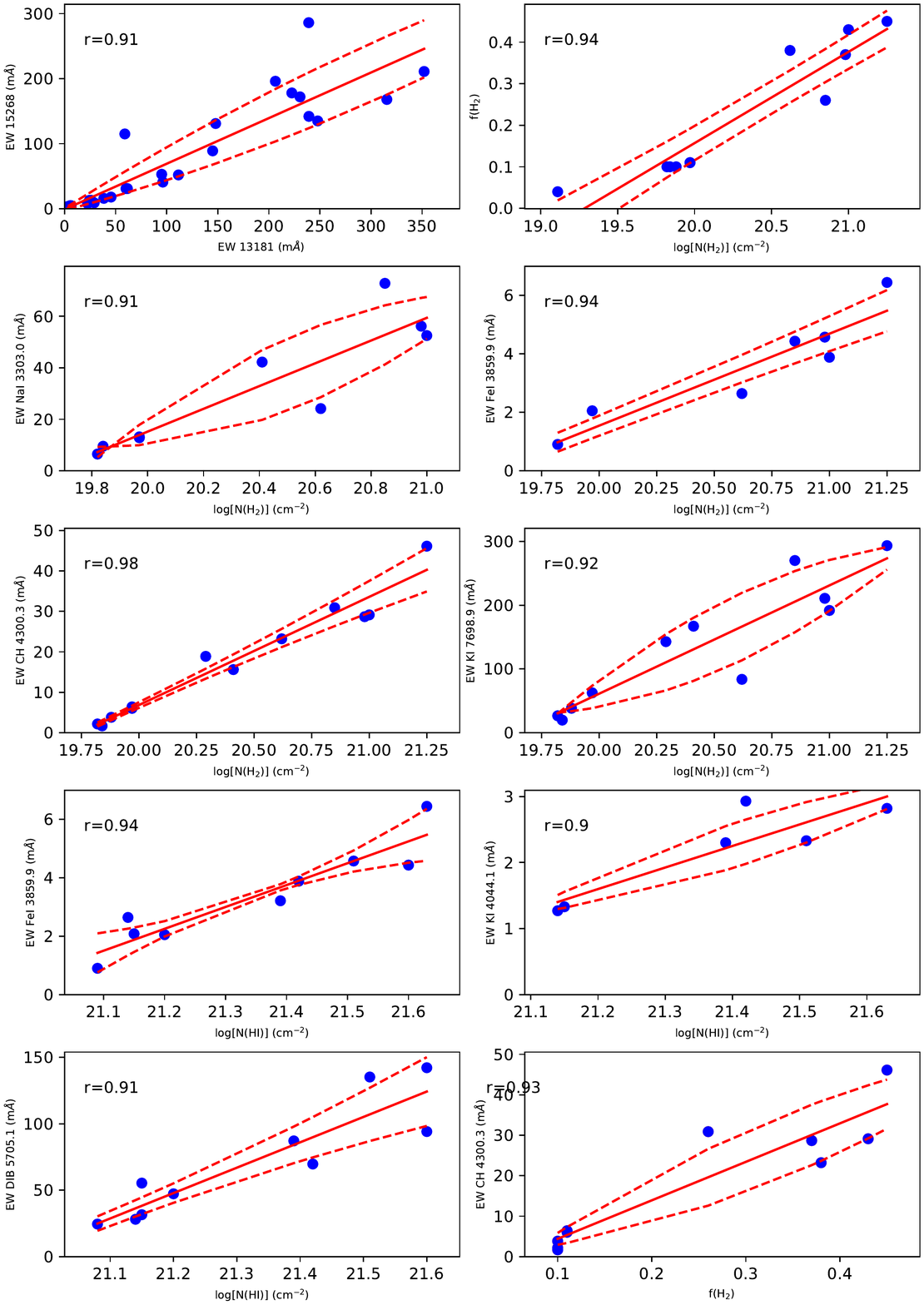}}
   \end{center}
\caption{Correlations ctd.}
\end{figure*}

\setcounter{figure}{2}
\begin{figure*}
\begin{center}
   \resizebox{17cm}{!}{\includegraphics{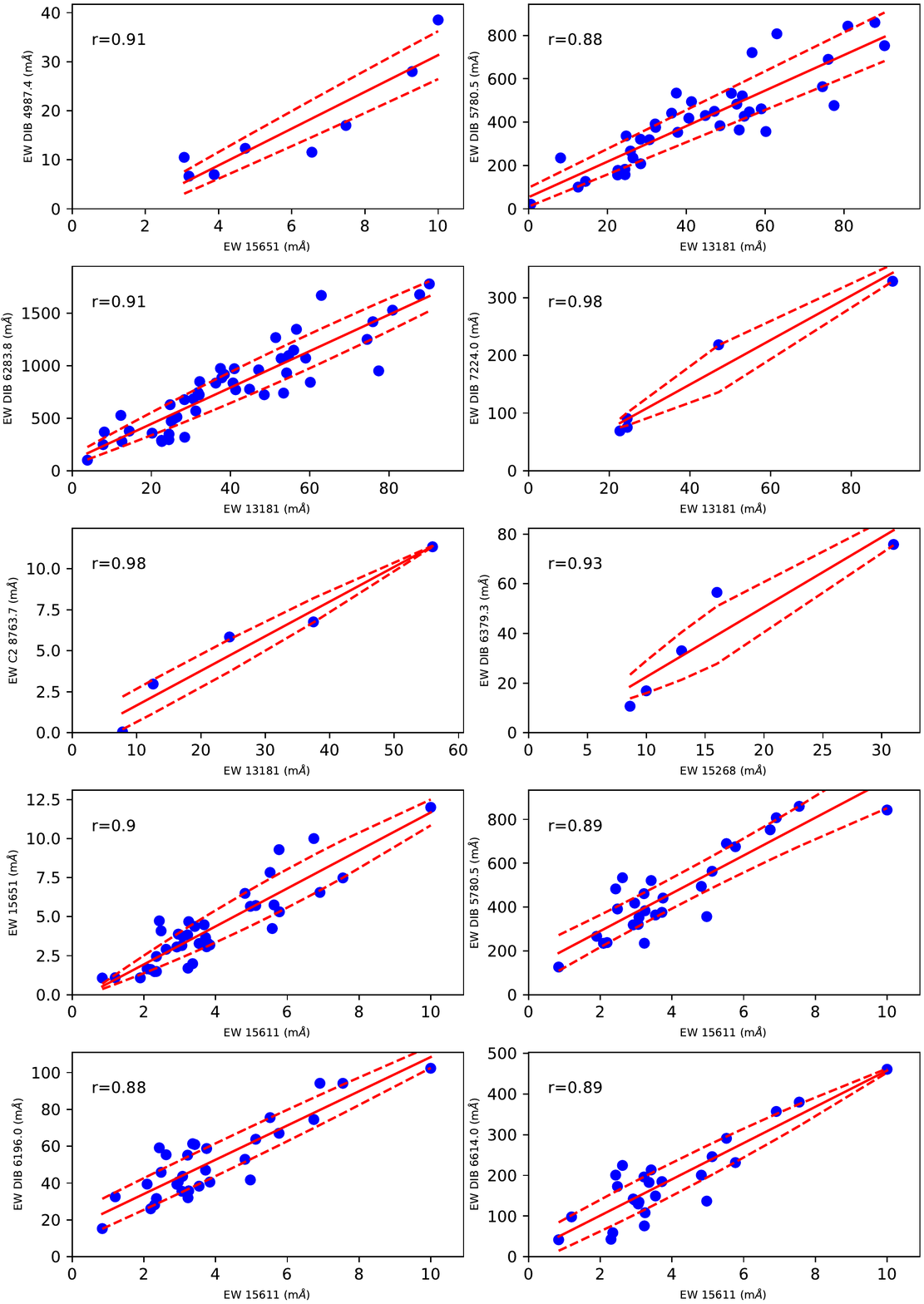}}
   \end{center}
\caption{Correlations ctd.}
\end{figure*}

\setcounter{figure}{2}
\begin{figure*}
\begin{center}
   \resizebox{17cm}{!}{\includegraphics{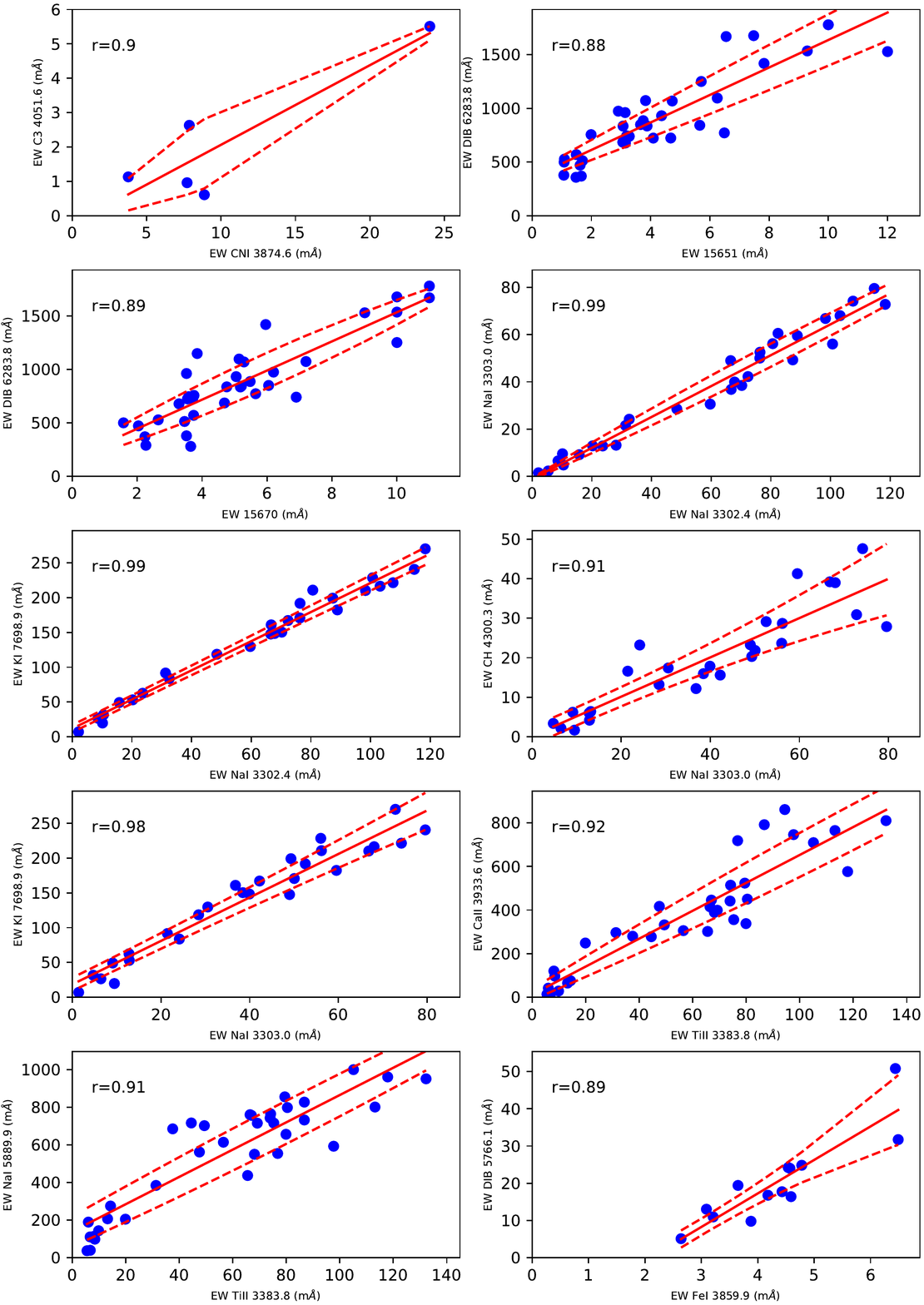}}
   \end{center}
\caption{Correlations ctd.}
\end{figure*}

\setcounter{figure}{2}
\begin{figure*}
\begin{center}
   \resizebox{17cm}{!}{\includegraphics{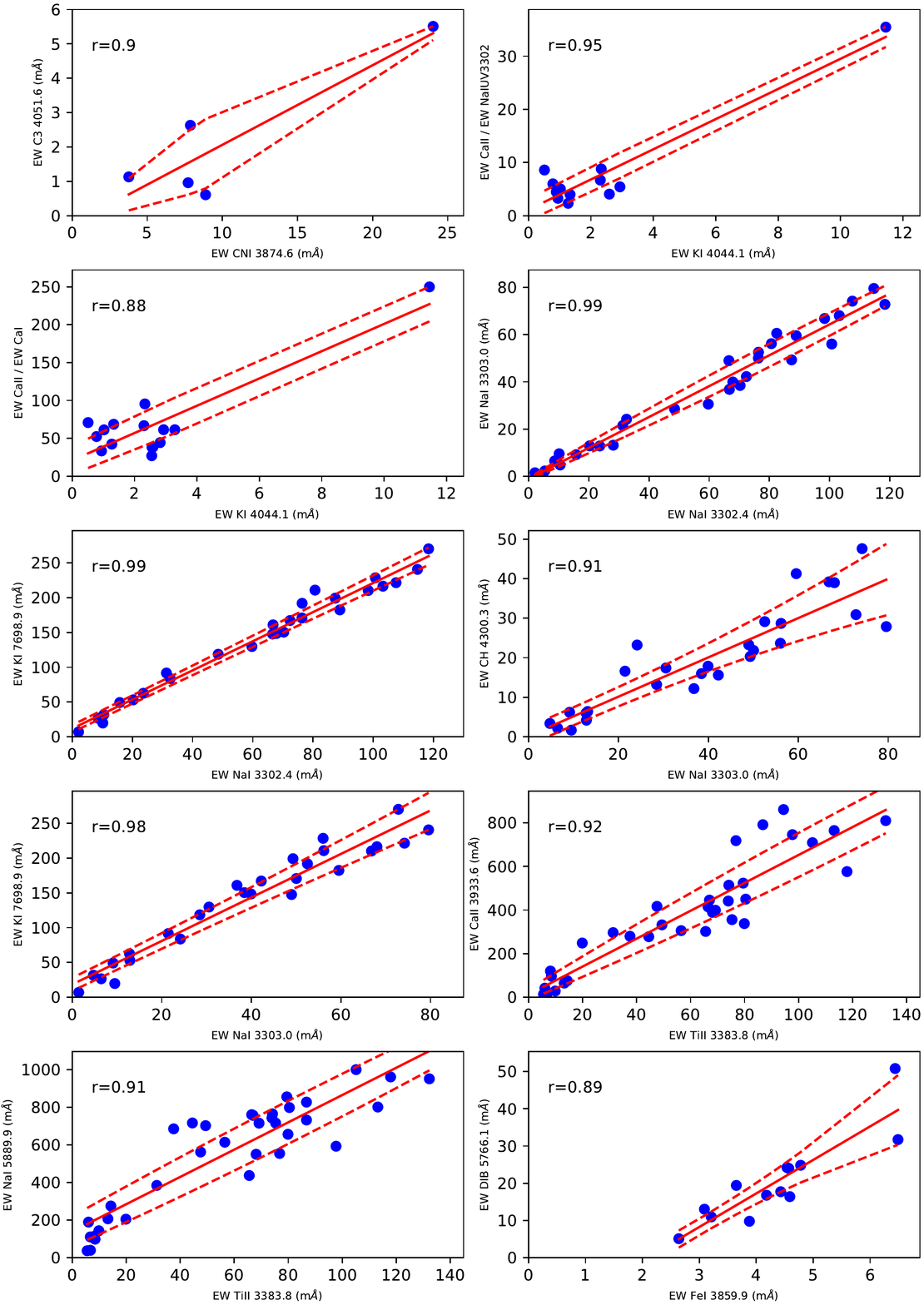}}
   \end{center}
\caption{Correlations ctd.}
\end{figure*}

\setcounter{figure}{2}
\begin{figure*}
\begin{center}
   \resizebox{17cm}{!}{\includegraphics{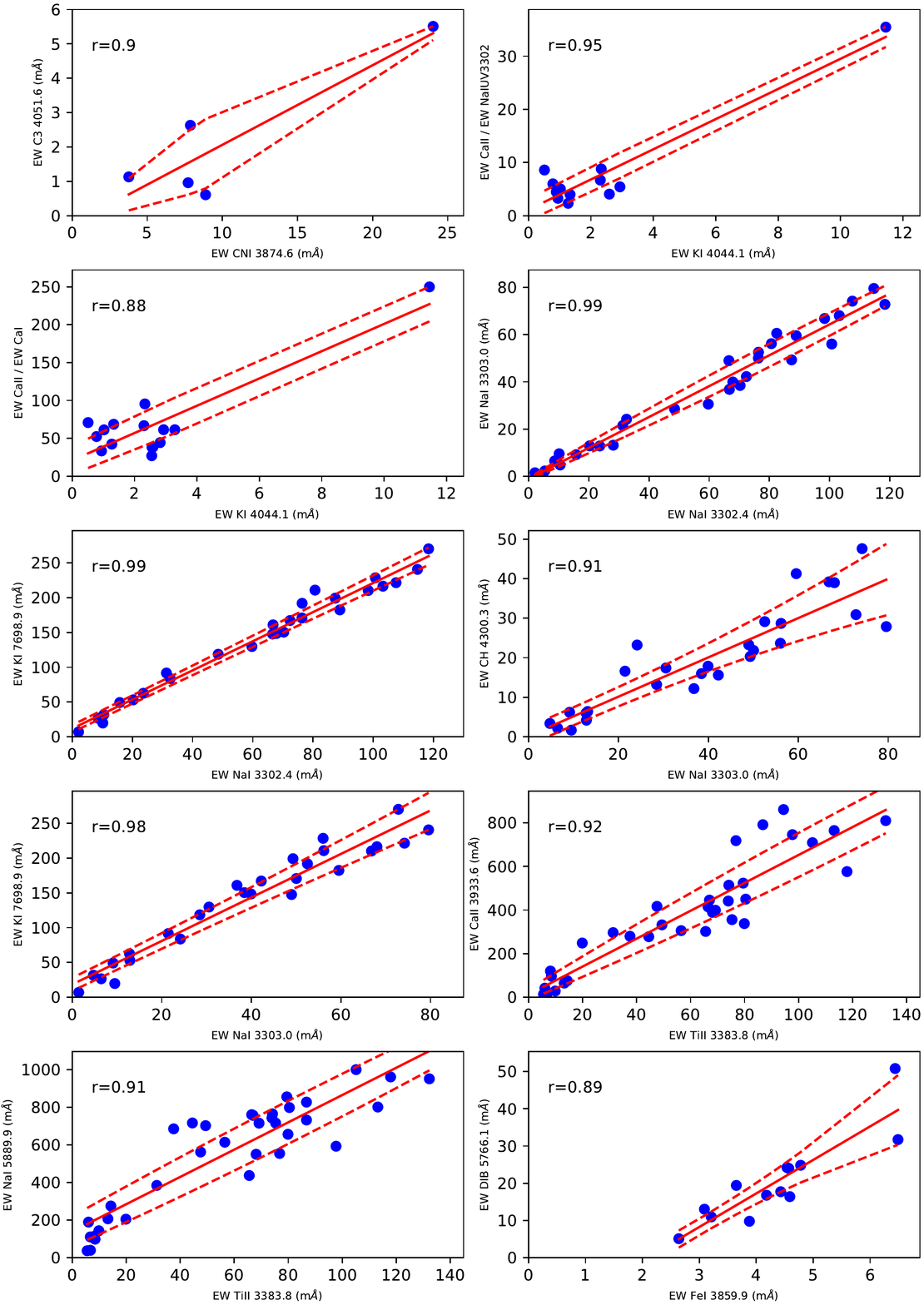}}
   \end{center}
\caption{Correlations ctd.}
\end{figure*}

\setcounter{figure}{2}
\begin{figure*}
\begin{center}
   \resizebox{17cm}{!}{\includegraphics{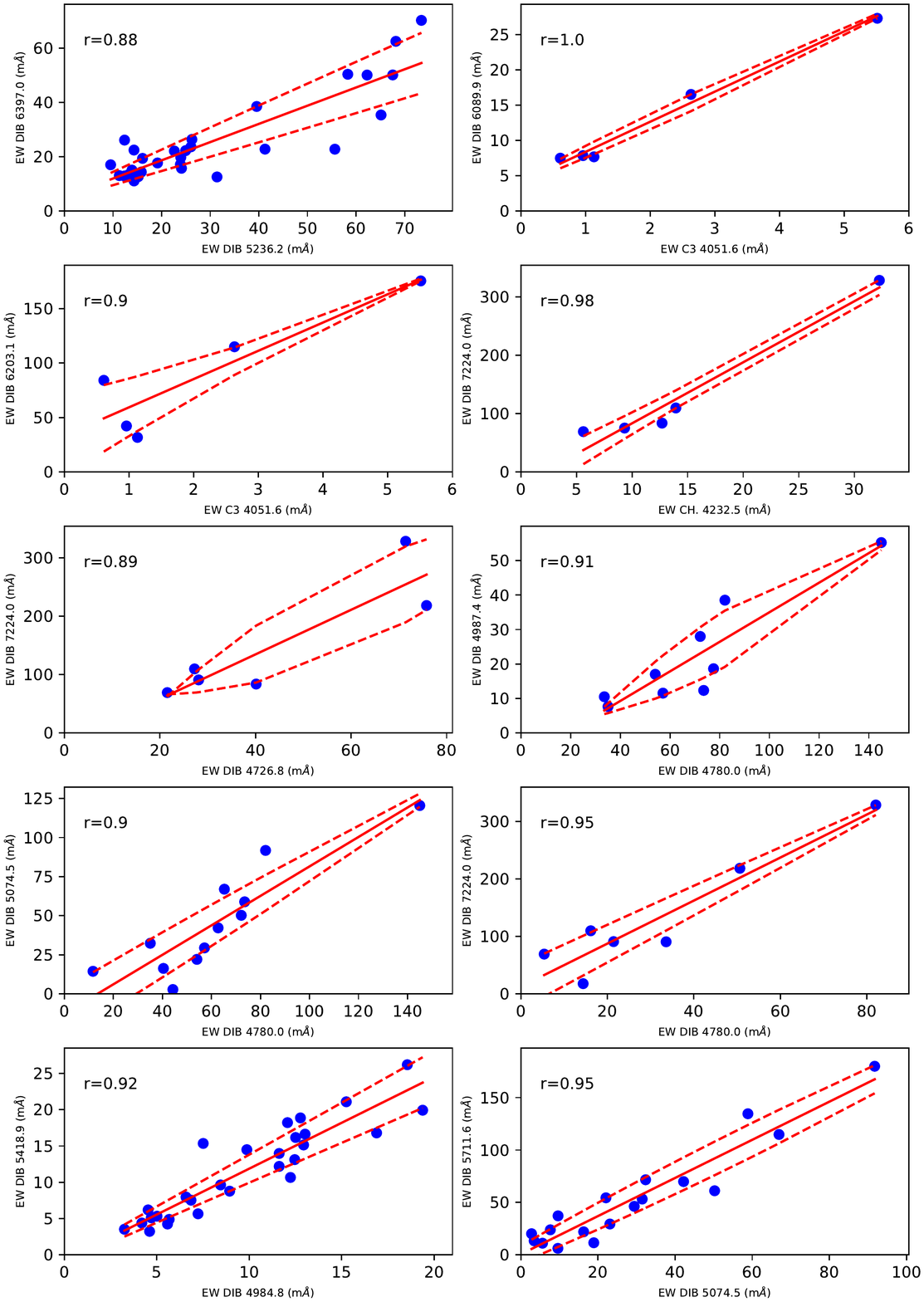}}
   \end{center}
\caption{Correlations ctd.}
\end{figure*}

\clearpage
\newpage

\setcounter{figure}{2}
\begin{figure*}
\begin{center}
   \resizebox{17cm}{!}{\includegraphics{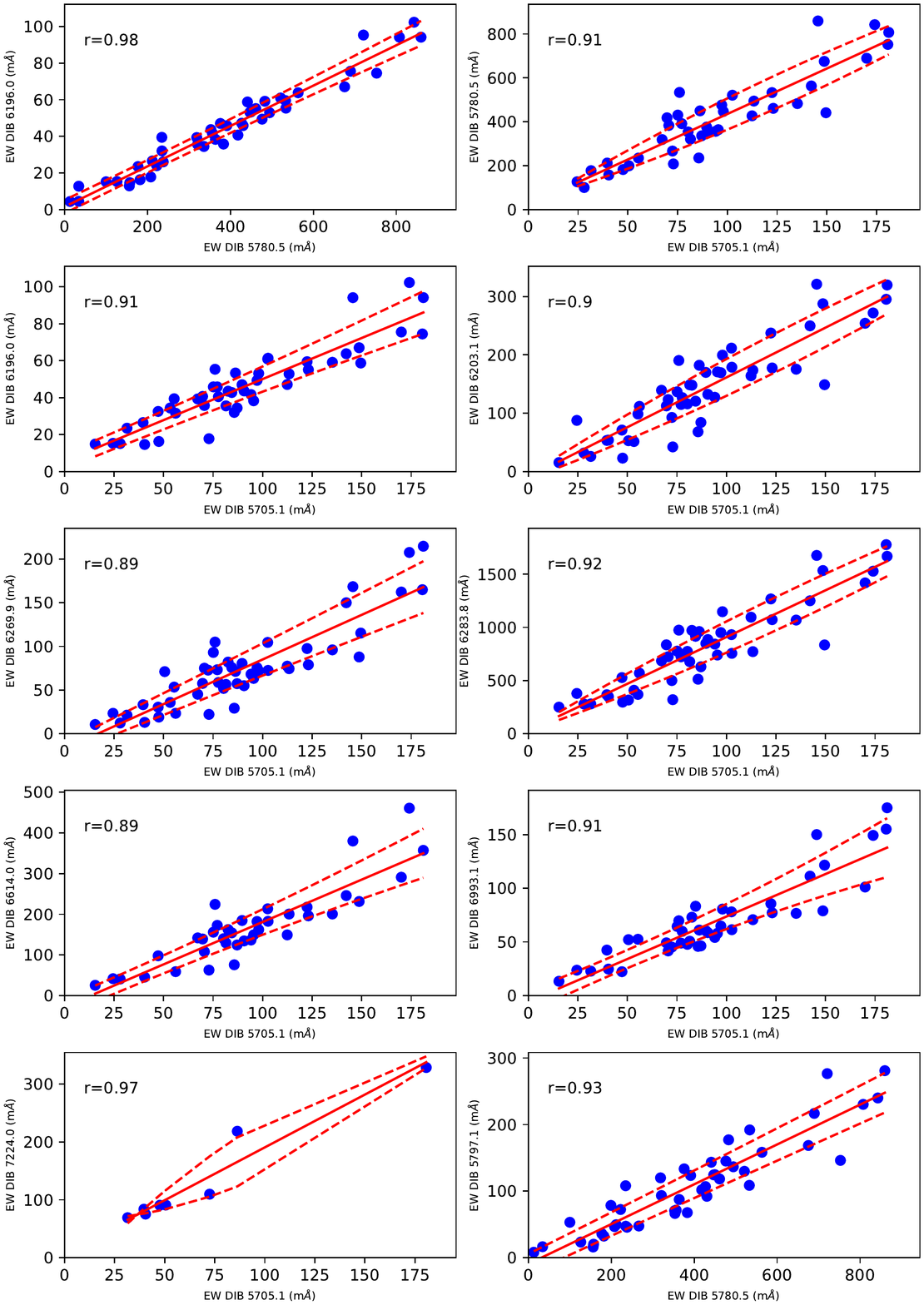}}
   \end{center}
\caption{Correlations ctd.}
\end{figure*}

\setcounter{figure}{2}
\begin{figure*}
\begin{center}
   \resizebox{17cm}{!}{\includegraphics{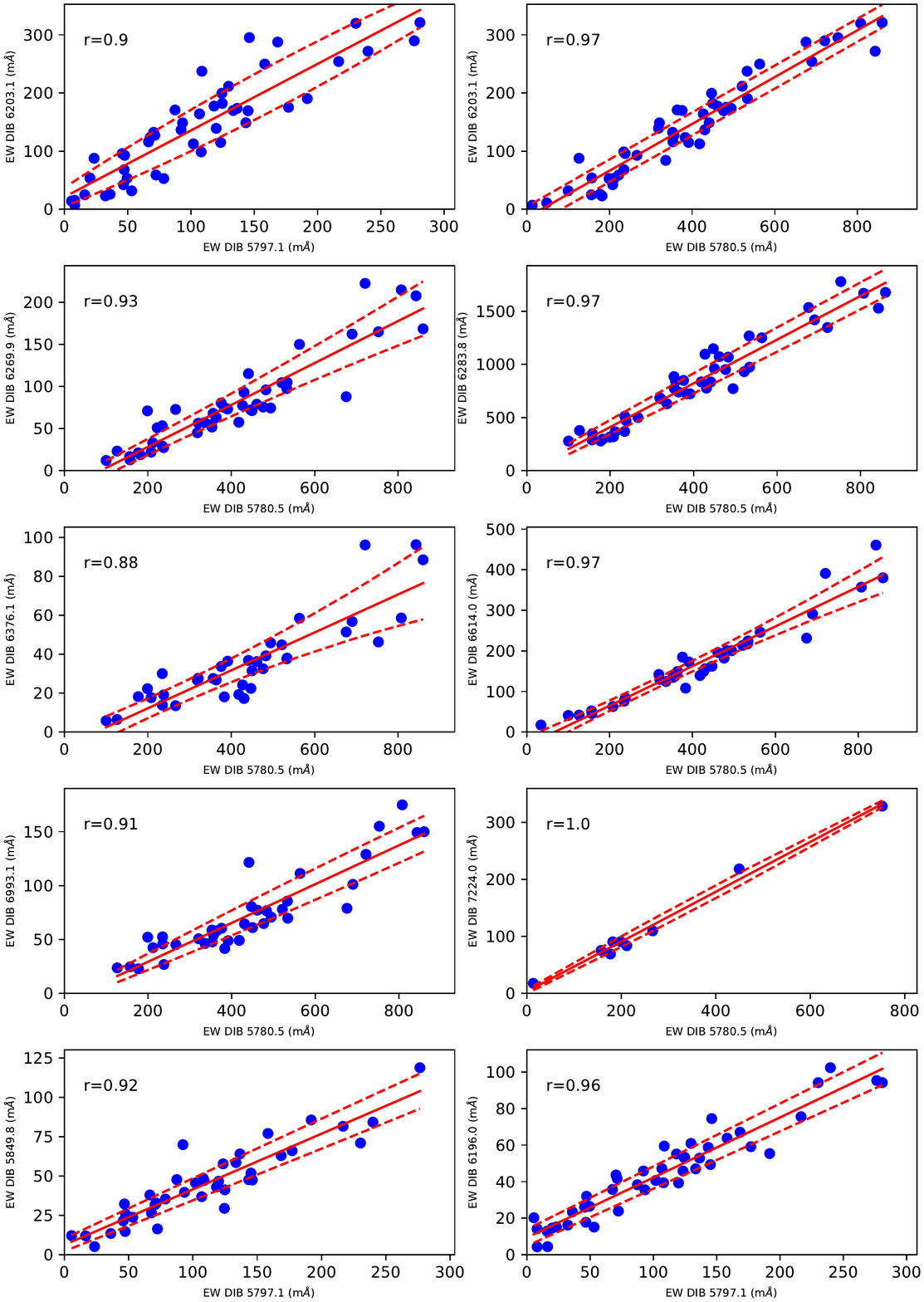}}
   \end{center}
\caption{Correlations ctd.}
\end{figure*}

\setcounter{figure}{2}
\begin{figure*}
\begin{center}
   \resizebox{17cm}{!}{\includegraphics{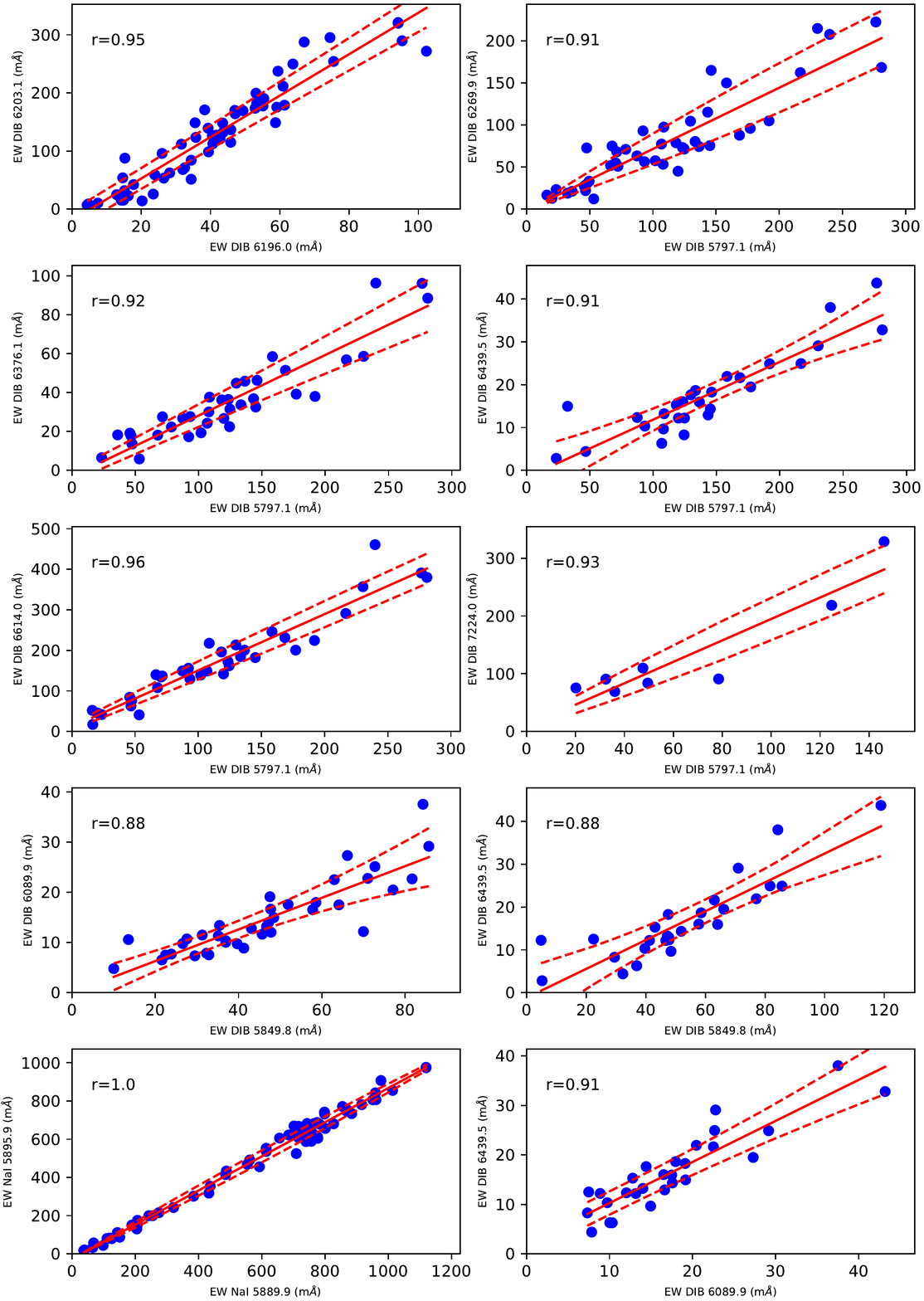}}
   \end{center}
\caption{Correlations ctd.}
\end{figure*}

\setcounter{figure}{2}
\begin{figure*}
\begin{center}
   \resizebox{17cm}{!}{\includegraphics{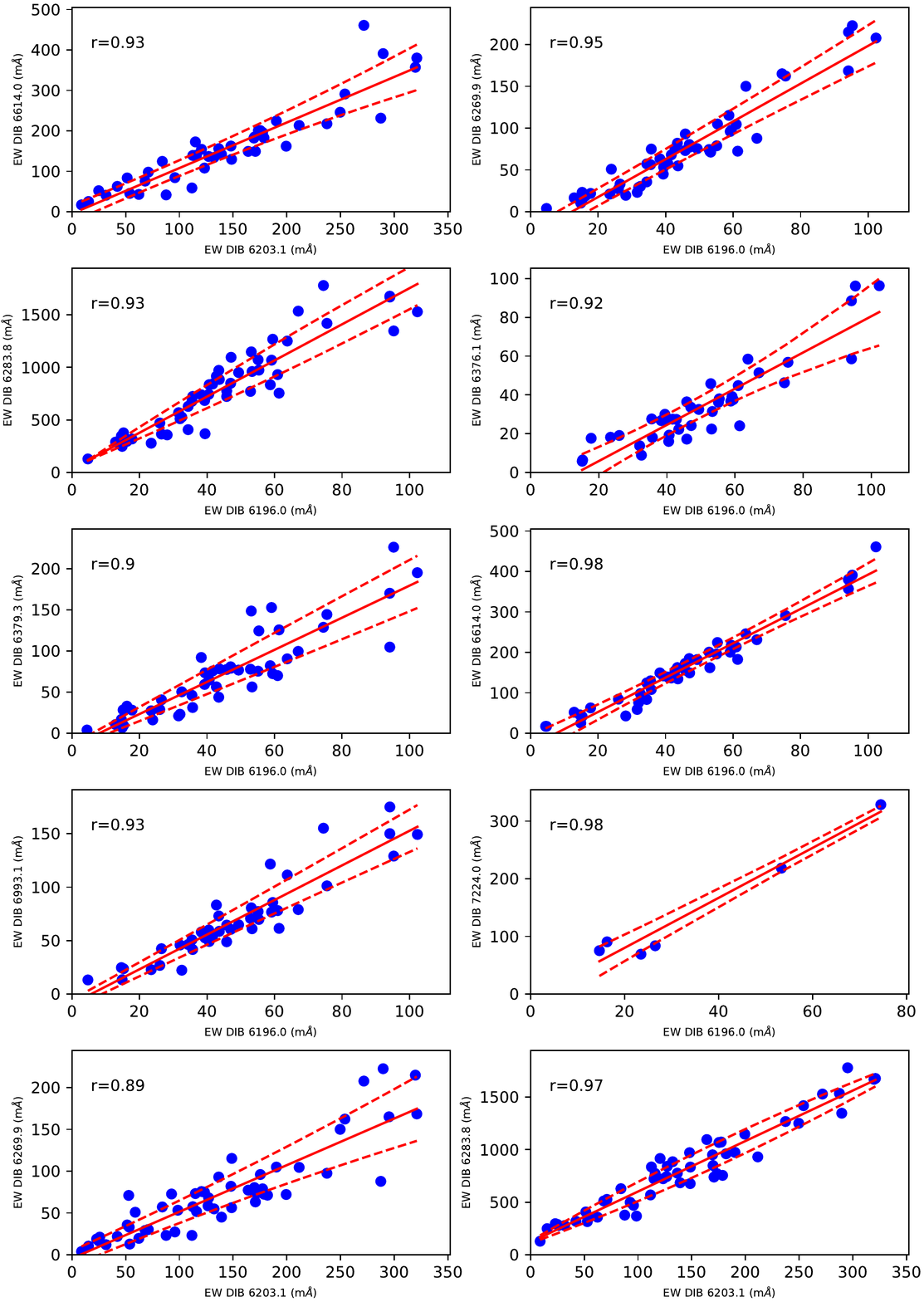}}
   \end{center}
\caption{Correlations ctd.}
\end{figure*}




%
%
\setcounter{table}{2}
\begin{sidewaystable*}
\caption{Correlation coefficients. The number of sightlines used per correlation is shown in brackets. Both the standard pearson correlation coefficient (top row per parameter) and 95 percent confidence range (second row) for the partial correlation coefficients are presented. }
\begin{center}
\begin{tabular}{rrrrrrrrrrr}
\hline
           &    E(B-V)           & FeI         & DIB         & DIB         & DIB         & DIB         & DIB     )       & DIB             & DIB         & DIB         \\
           &                  &    (3859.9) &    (5074.5) &    (5418.9) &    (5705.1) &    (5711.6) &    (5780.5)       &    (5797.1)     &    (5849.8) &    (6089.9) \\
\hline
\label{DIB_corr_coeff}
E(B-V)  &  0.49 (9)                 & 0.44 (17)                  & 0.39 (20)                  & 0.55 (32)                  & 0.80 (45)                  & 0.50 (27)                   & 0.91 (47)                  & 0.89 (44)                  & 0.80 (48)                  & 0.66 (40)                   \\
 &  -0.47$\rightarrow$0.9 & 0.06$\rightarrow$0.83 & -0.06$\rightarrow$0.74 & 0.16$\rightarrow$0.72 & 0.62$\rightarrow$0.87 & 0.29$\rightarrow$0.81 & 0.82$\rightarrow$0.94 & 0.81$\rightarrow$0.94 & 0.65$\rightarrow$0.88 & 0.33$\rightarrow$0.77 \\
Na(3302.4)  &  0.84 (30)                 & -0.31 (23)                  & 0.73 (13)                  & 0.05 (26)                  & 0.74 (26)                  & -0.01 (39)                   & 0.79 (20)                  & -0.01 (37)                  & 0.19 (24)                  & -0.01 (42)                   \\
 &  0.79$\rightarrow$0.95 & -0.64$\rightarrow$0.16 & -0.17$\rightarrow$0.84 & -0.33$\rightarrow$0.48 & -0.27$\rightarrow$0.52 & -0.38$\rightarrow$0.28 & -0.15$\rightarrow$0.7 & -0.31$\rightarrow$0.37 & -0.38$\rightarrow$0.47 & -0.27$\rightarrow$0.36 \\
FeI(3859.9)  &  0.44 (17)                 & -0.34 (42)                  & 0.73 (13)                  & 0.05 (26)                  & 0.30 (15)                  & -0.01 (39)                   & 0.39 (16)                  & -0.01 (37)                  & 0.19 (24)                  & -0.01 (42)                   \\
 &  0.06$\rightarrow$0.83 & -0.57$\rightarrow$0.01 & -0.17$\rightarrow$0.84 & -0.33$\rightarrow$0.48 & -0.13$\rightarrow$0.8 & -0.38$\rightarrow$0.28 & -0.11$\rightarrow$0.79 & -0.31$\rightarrow$0.37 & -0.38$\rightarrow$0.47 & -0.27$\rightarrow$0.36 \\
CaII(3933.6)  &  0.66 (59)                 & 0.23 (16)                  & 0.73 (13)                  & 0.05 (26)                  & 0.45 (45)                  & -0.01 (39)                   & 0.61 (47)                  & -0.01 (37)                  & 0.19 (24)                  & -0.01 (42)                   \\
 &  0.55$\rightarrow$0.82 & -0.58$\rightarrow$0.48 & -0.17$\rightarrow$0.84 & -0.33$\rightarrow$0.48 & -0.24$\rightarrow$0.38 & -0.38$\rightarrow$0.28 & -0.04$\rightarrow$0.53 & -0.31$\rightarrow$0.37 & -0.38$\rightarrow$0.47 & -0.27$\rightarrow$0.36 \\
CaI(4226.7)  &  0.57 (44)                 & 0.49 (15)                  & 0.73 (13)                  & 0.05 (26)                  & 0.47 (39)                  & -0.01 (39)                   & 0.55 (37)                  & -0.01 (37)                  & 0.19 (24)                  & -0.01 (42)                   \\
 &  0.41$\rightarrow$0.79 & -0.42$\rightarrow$0.66 & -0.17$\rightarrow$0.84 & -0.33$\rightarrow$0.48 & -0.05$\rightarrow$0.56 & -0.38$\rightarrow$0.28 & -0.07$\rightarrow$0.56 & -0.31$\rightarrow$0.37 & -0.38$\rightarrow$0.47 & -0.27$\rightarrow$0.36 \\
CH+(4232.5)  &  0.84 (49)                 & -0.43 (45)                  & 0.73 (13)                  & 0.05 (26)                  & 0.75 (43)                  & -0.01 (39)                   & 0.81 (39)                  & -0.01 (37)                  & 0.19 (24)                  & -0.01 (42)                   \\
 &  0.73$\rightarrow$0.91 & -0.72$\rightarrow$-0.27 & -0.17$\rightarrow$0.84 & -0.33$\rightarrow$0.48 & -0.11$\rightarrow$0.5 & -0.38$\rightarrow$0.28 & 0.08$\rightarrow$0.65 & -0.31$\rightarrow$0.37 & -0.38$\rightarrow$0.47 & -0.27$\rightarrow$0.36 \\
CH(4300.3)  &  0.90 (51)                 & 0.04 (41)                  & 0.73 (13)                  & 0.05 (26)                  & 0.70 (44)                  & -0.01 (39)                   & 0.78 (44)                  & -0.01 (37)                  & 0.19 (24)                  & -0.01 (42)                   \\
 &  0.78$\rightarrow$0.92 & -0.23$\rightarrow$0.4 & -0.17$\rightarrow$0.84 & -0.33$\rightarrow$0.48 & -0.26$\rightarrow$0.36 & -0.38$\rightarrow$0.28 & -0.36$\rightarrow$0.26 & -0.31$\rightarrow$0.37 & -0.38$\rightarrow$0.47 & -0.27$\rightarrow$0.36 \\
DIB(4726.8)  &  0.81 (48)                 & 0.01 (42)                  & 0.73 (13)                  & 0.05 (26)                  & 0.62 (44)                  & -0.01 (39)                   & 0.71 (43)                  & -0.01 (37)                  & 0.19 (24)                  & -0.01 (42)                   \\
 &  0.67$\rightarrow$0.89 & -0.17$\rightarrow$0.45 & -0.17$\rightarrow$0.84 & -0.33$\rightarrow$0.48 & -0.21$\rightarrow$0.41 & -0.38$\rightarrow$0.28 & -0.28$\rightarrow$0.35 & -0.31$\rightarrow$0.37 & -0.38$\rightarrow$0.47 & -0.27$\rightarrow$0.36 \\
DIB(4984.8)  &  0.49 (36)                 & 0.34 (29)                  & 0.73 (13)                  & 0.05 (26)                  & 0.34 (33)                  & -0.01 (39)                   & 0.25 (31)                  & -0.01 (37)                  & 0.19 (24)                  & -0.01 (42)                   \\
 &  0.1$\rightarrow$0.66 & -0.26$\rightarrow$0.48 & -0.17$\rightarrow$0.84 & -0.33$\rightarrow$0.48 & -0.44$\rightarrow$0.27 & -0.38$\rightarrow$0.28 & -0.63$\rightarrow$0.04 & -0.31$\rightarrow$0.37 & -0.38$\rightarrow$0.47 & -0.27$\rightarrow$0.36 \\
DIB(5074.5)  &  0.39 (20)                 & 0.73 (13)                  & -0.34 (42)                  & 0.05 (26)                  & 0.53 (16)                  & -0.01 (39)                   & 0.34 (17)                  & -0.01 (37)                  & 0.19 (24)                  & -0.01 (42)                   \\
 &  -0.06$\rightarrow$0.74 & -0.17$\rightarrow$0.84 & -0.57$\rightarrow$0.01 & -0.33$\rightarrow$0.48 & -0.13$\rightarrow$0.83 & -0.38$\rightarrow$0.28 & -0.22$\rightarrow$0.77 & -0.31$\rightarrow$0.37 & -0.38$\rightarrow$0.47 & -0.27$\rightarrow$0.36 \\
DIB(5418.9)  &  0.55 (32)                 & 0.05 (26)                  & 0.05 (26)                  & -0.34 (42)                  & 0.22 (30)                  & -0.01 (39)                   & 0.31 (27)                  & -0.01 (37)                  & 0.19 (24)                  & -0.01 (42)                   \\
 &  0.16$\rightarrow$0.72 & -0.33$\rightarrow$0.48 & -0.33$\rightarrow$0.48 & -0.57$\rightarrow$0.01 & -0.59$\rightarrow$0.11 & -0.38$\rightarrow$0.28 & -0.62$\rightarrow$0.11 & -0.31$\rightarrow$0.37 & -0.38$\rightarrow$0.47 & -0.27$\rightarrow$0.36 \\
DIB(5705.1)  &  0.80 (45)                 & 0.30 (15)                  & 0.53 (16)                  & 0.22 (30)                  & -0.34 (42)                  & -0.01 (39)                   & 0.91 (41)                  & -0.01 (37)                  & 0.19 (24)                  & -0.01 (42)                   \\
 &  0.62$\rightarrow$0.87 & -0.13$\rightarrow$0.8 & -0.13$\rightarrow$0.83 & -0.59$\rightarrow$0.11 & -0.57$\rightarrow$0.01 & -0.38$\rightarrow$0.28 & 0.34$\rightarrow$0.78 & -0.31$\rightarrow$0.37 & -0.38$\rightarrow$0.47 & -0.27$\rightarrow$0.36 \\
DIB(5711.6)  &  0.50 (27)                 & -0.01 (39)                  & -0.01 (39)                  & -0.01 (39)                  & -0.01 (39)                  & -0.34 (42)                   & 0.50 (21)                  & -0.01 (37)                  & 0.19 (24)                  & -0.01 (42)                   \\
 &  0.29$\rightarrow$0.81 & -0.38$\rightarrow$0.28 & -0.38$\rightarrow$0.28 & -0.38$\rightarrow$0.28 & -0.38$\rightarrow$0.28 & -0.57$\rightarrow$0.01 & -0.21$\rightarrow$0.68 & -0.31$\rightarrow$0.37 & -0.38$\rightarrow$0.47 & -0.27$\rightarrow$0.36 \\
DIB(5780.5)  &  0.91 (47)                 & 0.39 (16)                  & 0.34 (17)                  & 0.31 (27)                  & 0.91 (41)                  & 0.50 (21)                   & -0.34 (42)                  & -0.01 (37)                  & 0.19 (24)                  & -0.01 (42)                   \\
 &  0.82$\rightarrow$0.94 & -0.11$\rightarrow$0.79 & -0.22$\rightarrow$0.77 & -0.62$\rightarrow$0.11 & 0.34$\rightarrow$0.78 & -0.21$\rightarrow$0.68 & -0.57$\rightarrow$0.01 & -0.31$\rightarrow$0.37 & -0.38$\rightarrow$0.47 & -0.27$\rightarrow$0.36 \\
DIB(5797.1)  &  0.89 (44)                 & -0.01 (37)                  & -0.01 (37)                  & -0.01 (37)                  & -0.01 (37)                  & -0.01 (37)                   & -0.01 (37)                  & -0.34 (42)                  & 0.19 (24)                  & -0.01 (42)                   \\
 &  0.81$\rightarrow$0.94 & -0.31$\rightarrow$0.37 & -0.31$\rightarrow$0.37 & -0.31$\rightarrow$0.37 & -0.31$\rightarrow$0.37 & -0.31$\rightarrow$0.37 & -0.31$\rightarrow$0.37 & -0.57$\rightarrow$0.01 & -0.38$\rightarrow$0.47 & -0.27$\rightarrow$0.36 \\
DIB(5849.8)  &  0.80 (48)                 & 0.19 (24)                  & 0.19 (24)                  & 0.19 (24)                  & 0.19 (24)                  & 0.19 (24)                   & 0.19 (24)                  & 0.19 (24)                  & -0.34 (42)                  & -0.01 (42)                   \\
 &  0.65$\rightarrow$0.88 & -0.38$\rightarrow$0.47 & -0.38$\rightarrow$0.47 & -0.38$\rightarrow$0.47 & -0.38$\rightarrow$0.47 & -0.38$\rightarrow$0.47 & -0.38$\rightarrow$0.47 & -0.38$\rightarrow$0.47 & -0.57$\rightarrow$0.01 & -0.27$\rightarrow$0.36 \\
DIB(6089.9)  &  0.66 (40)                 & -0.01 (42)                  & -0.01 (42)                  & -0.01 (42)                  & -0.01 (42)                  & -0.01 (42)                   & -0.01 (42)                  & -0.01 (42)                  & -0.01 (42)                  & -0.34 (42)                   \\
 &  0.33$\rightarrow$0.77 & -0.27$\rightarrow$0.36 & -0.27$\rightarrow$0.36 & -0.27$\rightarrow$0.36 & -0.27$\rightarrow$0.36 & -0.27$\rightarrow$0.36 & -0.27$\rightarrow$0.36 & -0.27$\rightarrow$0.36 & -0.27$\rightarrow$0.36 & -0.57$\rightarrow$0.01 \\
DIB(6196.0)  &  0.88 (54)                 & 0.39 (17)                  & 0.37 (21)                  & 0.35 (32)                  & 0.91 (46)                  & 0.53 (29)                   & 0.98 (45)                  & 0.95 (45)                  & 0.83 (45)                  & 0.75 (41)                   \\
 &  0.79$\rightarrow$0.93 & -0.25$\rightarrow$0.71 & -0.26$\rightarrow$0.65 & -0.46$\rightarrow$0.25 & 0.49$\rightarrow$0.82 & 0.02$\rightarrow$0.7 & 0.81$\rightarrow$0.94 & 0.61$\rightarrow$0.87 & 0.36$\rightarrow$0.76 & 0.22$\rightarrow$0.71 \\
DIB(6203.1)  &  0.91 (57)                 & -0.05 (44)                  & -0.05 (44)                  & -0.05 (44)                  & -0.05 (44)                  & -0.05 (44)                   & -0.05 (44)                  & -0.05 (44)                  & -0.05 (44)                  & -0.05 (44)                   \\
 &  0.87$\rightarrow$0.95 & -0.36$\rightarrow$0.25 & -0.36$\rightarrow$0.25 & -0.36$\rightarrow$0.25 & -0.36$\rightarrow$0.25 & -0.36$\rightarrow$0.25 & -0.36$\rightarrow$0.25 & -0.36$\rightarrow$0.25 & -0.36$\rightarrow$0.25 & -0.36$\rightarrow$0.25 \\
DIB(6269.9)  &  0.81 (51)                 & -0.15 (27)                  & -0.15 (27)                  & -0.15 (27)                  & -0.15 (27)                  & -0.15 (27)                   & -0.15 (27)                  & -0.15 (27)                  & -0.15 (27)                  & -0.15 (27)                   \\
 &  0.68$\rightarrow$0.89 & -0.56$\rightarrow$0.2 & -0.56$\rightarrow$0.2 & -0.56$\rightarrow$0.2 & -0.56$\rightarrow$0.2 & -0.56$\rightarrow$0.2 & -0.56$\rightarrow$0.2 & -0.56$\rightarrow$0.2 & -0.56$\rightarrow$0.2 & -0.56$\rightarrow$0.2 \\
\hline
\end{tabular}
\end{center}
\end{sidewaystable*}
\setcounter{table}{2}

\begin{sidewaystable*}
\caption{Correlation coefficients ctd.}
\begin{center}
\begin{tabular}{rrrrrrrrrrr}
\hline
           &    E(B-V)           & FeI         & DIB         & DIB         & DIB         & DIB         & DIB     )       & DIB             & DIB         & DIB         \\
           &                  &    (3859.9) &    (5074.5) &    (5418.9) &    (5705.1) &    (5711.6) &    (5780.5)       &    (5797.1)     &    (5849.8) &    (6089.9) \\
\hline
DIB(6283.8)  &  0.83 (51)                 & -0.10 (43)                  & -0.10 (43)                  & -0.10 (43)                  & -0.10 (43)                  & -0.10 (43)                   & -0.10 (43)                  & -0.10 (43)                  & -0.10 (43)                  & -0.10 (43)                   \\
 &  0.7$\rightarrow$0.9 & -0.38$\rightarrow$0.24 & -0.38$\rightarrow$0.24 & -0.38$\rightarrow$0.24 & -0.38$\rightarrow$0.24 & -0.38$\rightarrow$0.24 & -0.38$\rightarrow$0.24 & -0.38$\rightarrow$0.24 & -0.38$\rightarrow$0.24 & -0.38$\rightarrow$0.24 \\
DIB(6376.1)  &  0.84 (40)                 & -0.01 (42)                  & -0.01 (42)                  & -0.01 (42)                  & -0.01 (42)                  & -0.01 (42)                   & -0.01 (42)                  & -0.01 (42)                  & -0.01 (42)                  & -0.01 (42)                   \\
 &  0.66$\rightarrow$0.9 & -0.3$\rightarrow$0.33 & -0.3$\rightarrow$0.33 & -0.3$\rightarrow$0.33 & -0.3$\rightarrow$0.33 & -0.3$\rightarrow$0.33 & -0.3$\rightarrow$0.33 & -0.3$\rightarrow$0.33 & -0.3$\rightarrow$0.33 & -0.3$\rightarrow$0.33 \\
DIB(6439.5)  &  0.71 (30)                 & 0.18 (37)                  & 0.18 (37)                  & 0.18 (37)                  & 0.18 (37)                  & 0.18 (37)                   & 0.18 (37)                  & 0.18 (37)                  & 0.18 (37)                  & 0.18 (37)                   \\
 &  0.46$\rightarrow$0.86 & -0.01$\rightarrow$0.6 & -0.01$\rightarrow$0.6 & -0.01$\rightarrow$0.6 & -0.01$\rightarrow$0.6 & -0.01$\rightarrow$0.6 & -0.01$\rightarrow$0.6 & -0.01$\rightarrow$0.6 & -0.01$\rightarrow$0.6 & -0.01$\rightarrow$0.6 \\
DIB(6614.0)  &  0.87 (44)                 & 0.15 (27)                  & 0.15 (27)                  & 0.15 (27)                  & 0.15 (27)                  & 0.15 (27)                   & 0.15 (27)                  & 0.15 (27)                  & 0.15 (27)                  & 0.15 (27)                   \\
 &  0.74$\rightarrow$0.92 & -0.47$\rightarrow$0.32 & -0.47$\rightarrow$0.32 & -0.47$\rightarrow$0.32 & -0.47$\rightarrow$0.32 & -0.47$\rightarrow$0.32 & -0.47$\rightarrow$0.32 & -0.47$\rightarrow$0.32 & -0.47$\rightarrow$0.32 & -0.47$\rightarrow$0.32 \\
DIB(6993.1)  &  0.75 (41)                 & -0.22 (37)                  & -0.22 (37)                  & -0.22 (37)                  & -0.22 (37)                  & -0.22 (37)                   & -0.22 (37)                  & -0.22 (37)                  & -0.22 (37)                  & -0.22 (37)                   \\
 &  0.6$\rightarrow$0.87 & -0.53$\rightarrow$0.1 & -0.53$\rightarrow$0.1 & -0.53$\rightarrow$0.1 & -0.53$\rightarrow$0.1 & -0.53$\rightarrow$0.1 & -0.53$\rightarrow$0.1 & -0.53$\rightarrow$0.1 & -0.53$\rightarrow$0.1 & -0.53$\rightarrow$0.1 \\
KI(7698.9)  &  0.84 (51)                 & 0.49 (16)                  & 0.21 (18)                  & 0.48 (29)                  & 0.74 (47)                  & 0.38 (24)                   & 0.78 (43)                  & 0.85 (41)                  & 0.85 (44)                  & 0.57 (39)                   \\
 &  0.72$\rightarrow$0.9 & -0.3$\rightarrow$0.7 & -0.53$\rightarrow$0.53 & -0.02$\rightarrow$0.65 & -0.07$\rightarrow$0.51 & -0.32$\rightarrow$0.53 & 0.03$\rightarrow$0.6 & 0.35$\rightarrow$0.78 & 0.4$\rightarrow$0.78 & 0.12$\rightarrow$0.66 \\
DIB(13180)  &  0.88 (82)                 & -0.31 (23)                  & 0.73 (13)                  & 0.05 (26)                  & 0.84 (44)                  & -0.01 (39)                   & 0.88 (40)                  & -0.01 (37)                  & 0.19 (24)                  & -0.01 (42)                   \\
 &  0.85$\rightarrow$0.94 & -0.64$\rightarrow$0.16 & -0.17$\rightarrow$0.84 & -0.33$\rightarrow$0.48 & 0.33$\rightarrow$0.76 & -0.38$\rightarrow$0.28 & 0.38$\rightarrow$0.79 & -0.31$\rightarrow$0.37 & -0.38$\rightarrow$0.47 & -0.27$\rightarrow$0.36 \\
DIB(15650)  &  0.76 (43)                 & -0.31 (23)                  & 0.73 (13)                  & 0.05 (26)                  & 0.82 (33)                  & -0.01 (39)                   & 0.86 (29)                  & -0.01 (37)                  & 0.19 (24)                  & -0.01 (42)                   \\
 &  0.59$\rightarrow$0.87 & -0.64$\rightarrow$0.16 & -0.17$\rightarrow$0.84 & -0.33$\rightarrow$0.48 & 0.14$\rightarrow$0.71 & -0.38$\rightarrow$0.28 & 0.22$\rightarrow$0.78 & -0.31$\rightarrow$0.37 & -0.38$\rightarrow$0.47 & -0.27$\rightarrow$0.36 \\
DIB(15670)  &  0.71 (46)                 & -0.31 (23)                  & 0.73 (13)                  & 0.05 (26)                  & 0.80 (35)                  & -0.01 (39)                   & 0.85 (32)                  & -0.01 (37)                  & 0.19 (24)                  & -0.01 (42)                   \\
 &  0.59$\rightarrow$0.86 & -0.64$\rightarrow$0.16 & -0.17$\rightarrow$0.84 & -0.33$\rightarrow$0.48 & 0.14$\rightarrow$0.7 & -0.38$\rightarrow$0.28 & 0.19$\rightarrow$0.75 & -0.31$\rightarrow$0.37 & -0.38$\rightarrow$0.47 & -0.27$\rightarrow$0.36 \\
\hline
\end{tabular}
\end{center}
\end{sidewaystable*}
%
%
\setcounter{table}{2}
\begin{sidewaystable*}
\caption{continued.}
\begin{center}
\begin{tabular}{rrrrrrrrrrrr}
\hline
           & DIB         & DIB         & DIB         & DIB         & DIB         & DIB         & DIB         & DIB         & DIB              & DIB              & DIB       \\
           &    (6196.0) &    (6203.1) &    (6269.9) &    (6283.8) &    (6376.1) &    (6439.5) &    (6614.0) &    (6993.1) &    (13180)       &    (15650)       &    (1567) \\
\hline
E(B-V)  &  0.88 (54)                  & 0.91 (57)                  & 0.81 (51)                  & 0.83 (51)                  & 0.84 (40)                  & 0.71 (30)                   & 0.87 (44)                  & 0.75 (41)                  & 0.88 (82)                  & 0.76 (43)                  & 0.71 (46)                   \\
 &  0.79$\rightarrow$0.93 & 0.87$\rightarrow$0.95 & 0.68$\rightarrow$0.89 & 0.7$\rightarrow$0.9 & 0.66$\rightarrow$0.9 & 0.46$\rightarrow$0.86 & 0.74$\rightarrow$0.92 & 0.6$\rightarrow$0.87 & 0.85$\rightarrow$0.94 & 0.59$\rightarrow$0.87 & 0.59$\rightarrow$0.86 \\
Na(3302.4)  &  0.81 (30)                  & -0.05 (44)                  & -0.15 (27)                  & -0.10 (43)                  & -0.01 (42)                  & 0.18 (37)                   & 0.15 (27)                  & -0.22 (37)                  & -0.11 (31)                  & -0.11 (31)                  & -0.11 (31)                   \\
 &  0.21$\rightarrow$0.76 & -0.36$\rightarrow$0.25 & -0.56$\rightarrow$0.2 & -0.38$\rightarrow$0.24 & -0.3$\rightarrow$0.33 & -0.01$\rightarrow$0.6 & -0.47$\rightarrow$0.32 & -0.53$\rightarrow$0.1 & -0.49$\rightarrow$0.24 & -0.49$\rightarrow$0.24 & -0.49$\rightarrow$0.24 \\
FeI(3859.9)  &  0.39 (17)                  & -0.05 (44)                  & -0.15 (27)                  & -0.10 (43)                  & -0.01 (42)                  & 0.18 (37)                   & 0.15 (27)                  & -0.22 (37)                  & -0.31 (23)                  & -0.31 (23)                  & -0.31 (23)                   \\
 &  -0.25$\rightarrow$0.71 & -0.36$\rightarrow$0.25 & -0.56$\rightarrow$0.2 & -0.38$\rightarrow$0.24 & -0.3$\rightarrow$0.33 & -0.01$\rightarrow$0.6 & -0.47$\rightarrow$0.32 & -0.53$\rightarrow$0.1 & -0.64$\rightarrow$0.16 & -0.64$\rightarrow$0.16 & -0.64$\rightarrow$0.16 \\
CaII(3933.6)  &  0.67 (54)                  & -0.05 (44)                  & -0.15 (27)                  & -0.10 (43)                  & -0.01 (42)                  & 0.18 (37)                   & 0.15 (27)                  & -0.22 (37)                  & 0.60 (48)                  & 0.34 (32)                  & 0.20 (34)                   \\
 &  -0.01$\rightarrow$0.51 & -0.36$\rightarrow$0.25 & -0.56$\rightarrow$0.2 & -0.38$\rightarrow$0.24 & -0.3$\rightarrow$0.33 & -0.01$\rightarrow$0.6 & -0.47$\rightarrow$0.32 & -0.53$\rightarrow$0.1 & -0.05$\rightarrow$0.51 & -0.39$\rightarrow$0.35 & -0.55$\rightarrow$0.12 \\
CaI(4226.7)  &  0.63 (44)                  & -0.05 (44)                  & -0.15 (27)                  & -0.10 (43)                  & -0.01 (42)                  & 0.18 (37)                   & 0.15 (27)                  & -0.22 (37)                  & 0.42 (39)                  & 0.27 (28)                  & 0.42 (31)                   \\
 &  0.15$\rightarrow$0.65 & -0.36$\rightarrow$0.25 & -0.56$\rightarrow$0.2 & -0.38$\rightarrow$0.24 & -0.3$\rightarrow$0.33 & -0.01$\rightarrow$0.6 & -0.47$\rightarrow$0.32 & -0.53$\rightarrow$0.1 & -0.35$\rightarrow$0.3 & -0.52$\rightarrow$0.26 & -0.25$\rightarrow$0.49 \\
CH+(4232.5)  &  0.80 (47)                  & -0.05 (44)                  & -0.15 (27)                  & -0.10 (43)                  & -0.01 (42)                  & 0.18 (37)                   & 0.15 (27)                  & -0.22 (37)                  & -0.43 (45)                  & -0.43 (45)                  & -0.43 (45)                   \\
 &  0.04$\rightarrow$0.58 & -0.36$\rightarrow$0.25 & -0.56$\rightarrow$0.2 & -0.38$\rightarrow$0.24 & -0.3$\rightarrow$0.33 & -0.01$\rightarrow$0.6 & -0.47$\rightarrow$0.32 & -0.53$\rightarrow$0.1 & -0.72$\rightarrow$-0.27 & -0.72$\rightarrow$-0.27 & -0.72$\rightarrow$-0.27 \\
CH(4300.3)  &  0.76 (47)                  & -0.05 (44)                  & -0.15 (27)                  & -0.10 (43)                  & -0.01 (42)                  & 0.18 (37)                   & 0.15 (27)                  & -0.22 (37)                  & 0.04 (41)                  & 0.04 (41)                  & 0.04 (41)                   \\
 &  -0.27$\rightarrow$0.32 & -0.36$\rightarrow$0.25 & -0.56$\rightarrow$0.2 & -0.38$\rightarrow$0.24 & -0.3$\rightarrow$0.33 & -0.01$\rightarrow$0.6 & -0.47$\rightarrow$0.32 & -0.53$\rightarrow$0.1 & -0.23$\rightarrow$0.4 & -0.23$\rightarrow$0.4 & -0.23$\rightarrow$0.4 \\
DIB(4726.8)  &  0.74 (46)                  & -0.05 (44)                  & -0.15 (27)                  & -0.10 (43)                  & -0.01 (42)                  & 0.18 (37)                   & 0.15 (27)                  & -0.22 (37)                  & 0.01 (42)                  & 0.01 (42)                  & 0.01 (42)                   \\
 &  -0.11$\rightarrow$0.48 & -0.36$\rightarrow$0.25 & -0.56$\rightarrow$0.2 & -0.38$\rightarrow$0.24 & -0.3$\rightarrow$0.33 & -0.01$\rightarrow$0.6 & -0.47$\rightarrow$0.32 & -0.53$\rightarrow$0.1 & -0.17$\rightarrow$0.45 & -0.17$\rightarrow$0.45 & -0.17$\rightarrow$0.45 \\
DIB(4984.8)  &  0.34 (35)                  & -0.05 (44)                  & -0.15 (27)                  & -0.10 (43)                  & -0.01 (42)                  & 0.18 (37)                   & 0.15 (27)                  & -0.22 (37)                  & 0.34 (29)                  & 0.34 (29)                  & 0.34 (29)                   \\
 &  -0.51$\rightarrow$0.16 & -0.36$\rightarrow$0.25 & -0.56$\rightarrow$0.2 & -0.38$\rightarrow$0.24 & -0.3$\rightarrow$0.33 & -0.01$\rightarrow$0.6 & -0.47$\rightarrow$0.32 & -0.53$\rightarrow$0.1 & -0.26$\rightarrow$0.48 & -0.26$\rightarrow$0.48 & -0.26$\rightarrow$0.48 \\
DIB(5074.5)  &  0.37 (21)                  & -0.05 (44)                  & -0.15 (27)                  & -0.10 (43)                  & -0.01 (42)                  & 0.18 (37)                   & 0.15 (27)                  & -0.22 (37)                  & 0.73 (13)                  & 0.73 (13)                  & 0.73 (13)                   \\
 &  -0.26$\rightarrow$0.65 & -0.36$\rightarrow$0.25 & -0.56$\rightarrow$0.2 & -0.38$\rightarrow$0.24 & -0.3$\rightarrow$0.33 & -0.01$\rightarrow$0.6 & -0.47$\rightarrow$0.32 & -0.53$\rightarrow$0.1 & -0.17$\rightarrow$0.84 & -0.17$\rightarrow$0.84 & -0.17$\rightarrow$0.84 \\
DIB(5418.9)  &  0.35 (32)                  & -0.05 (44)                  & -0.15 (27)                  & -0.10 (43)                  & -0.01 (42)                  & 0.18 (37)                   & 0.15 (27)                  & -0.22 (37)                  & 0.05 (26)                  & 0.05 (26)                  & 0.05 (26)                   \\
 &  -0.46$\rightarrow$0.25 & -0.36$\rightarrow$0.25 & -0.56$\rightarrow$0.2 & -0.38$\rightarrow$0.24 & -0.3$\rightarrow$0.33 & -0.01$\rightarrow$0.6 & -0.47$\rightarrow$0.32 & -0.53$\rightarrow$0.1 & -0.33$\rightarrow$0.48 & -0.33$\rightarrow$0.48 & -0.33$\rightarrow$0.48 \\
DIB(5705.1)  &  0.91 (46)                  & -0.05 (44)                  & -0.15 (27)                  & -0.10 (43)                  & -0.01 (42)                  & 0.18 (37)                   & 0.15 (27)                  & -0.22 (37)                  & 0.84 (44)                  & 0.82 (33)                  & 0.80 (35)                   \\
 &  0.49$\rightarrow$0.82 & -0.36$\rightarrow$0.25 & -0.56$\rightarrow$0.2 & -0.38$\rightarrow$0.24 & -0.3$\rightarrow$0.33 & -0.01$\rightarrow$0.6 & -0.47$\rightarrow$0.32 & -0.53$\rightarrow$0.1 & 0.33$\rightarrow$0.76 & 0.14$\rightarrow$0.71 & 0.14$\rightarrow$0.7 \\
DIB(5711.6)  &  0.53 (29)                  & -0.05 (44)                  & -0.15 (27)                  & -0.10 (43)                  & -0.01 (42)                  & 0.18 (37)                   & 0.15 (27)                  & -0.22 (37)                  & -0.01 (39)                  & -0.01 (39)                  & -0.01 (39)                   \\
 &  0.02$\rightarrow$0.7 & -0.36$\rightarrow$0.25 & -0.56$\rightarrow$0.2 & -0.38$\rightarrow$0.24 & -0.3$\rightarrow$0.33 & -0.01$\rightarrow$0.6 & -0.47$\rightarrow$0.32 & -0.53$\rightarrow$0.1 & -0.38$\rightarrow$0.28 & -0.38$\rightarrow$0.28 & -0.38$\rightarrow$0.28 \\
DIB(5780.5)  &  0.98 (45)                  & -0.05 (44)                  & -0.15 (27)                  & -0.10 (43)                  & -0.01 (42)                  & 0.18 (37)                   & 0.15 (27)                  & -0.22 (37)                  & 0.88 (40)                  & 0.86 (29)                  & 0.85 (32)                   \\
 &  0.81$\rightarrow$0.94 & -0.36$\rightarrow$0.25 & -0.56$\rightarrow$0.2 & -0.38$\rightarrow$0.24 & -0.3$\rightarrow$0.33 & -0.01$\rightarrow$0.6 & -0.47$\rightarrow$0.32 & -0.53$\rightarrow$0.1 & 0.38$\rightarrow$0.79 & 0.22$\rightarrow$0.78 & 0.19$\rightarrow$0.75 \\
DIB(5797.1)  &  0.95 (45)                  & -0.05 (44)                  & -0.15 (27)                  & -0.10 (43)                  & -0.01 (42)                  & 0.18 (37)                   & 0.15 (27)                  & -0.22 (37)                  & -0.01 (37)                  & -0.01 (37)                  & -0.01 (37)                   \\
 &  0.61$\rightarrow$0.87 & -0.36$\rightarrow$0.25 & -0.56$\rightarrow$0.2 & -0.38$\rightarrow$0.24 & -0.3$\rightarrow$0.33 & -0.01$\rightarrow$0.6 & -0.47$\rightarrow$0.32 & -0.53$\rightarrow$0.1 & -0.31$\rightarrow$0.37 & -0.31$\rightarrow$0.37 & -0.31$\rightarrow$0.37 \\
DIB(5849.8)  &  0.83 (45)                  & -0.05 (44)                  & -0.15 (27)                  & -0.10 (43)                  & -0.01 (42)                  & 0.18 (37)                   & 0.15 (27)                  & -0.22 (37)                  & 0.19 (24)                  & 0.19 (24)                  & 0.19 (24)                   \\
 &  0.36$\rightarrow$0.76 & -0.36$\rightarrow$0.25 & -0.56$\rightarrow$0.2 & -0.38$\rightarrow$0.24 & -0.3$\rightarrow$0.33 & -0.01$\rightarrow$0.6 & -0.47$\rightarrow$0.32 & -0.53$\rightarrow$0.1 & -0.38$\rightarrow$0.47 & -0.38$\rightarrow$0.47 & -0.38$\rightarrow$0.47 \\
DIB(6089.9)  &  0.75 (41)                  & -0.05 (44)                  & -0.15 (27)                  & -0.10 (43)                  & -0.01 (42)                  & 0.18 (37)                   & 0.15 (27)                  & -0.22 (37)                  & -0.01 (42)                  & -0.01 (42)                  & -0.01 (42)                   \\
 &  0.22$\rightarrow$0.71 & -0.36$\rightarrow$0.25 & -0.56$\rightarrow$0.2 & -0.38$\rightarrow$0.24 & -0.3$\rightarrow$0.33 & -0.01$\rightarrow$0.6 & -0.47$\rightarrow$0.32 & -0.53$\rightarrow$0.1 & -0.27$\rightarrow$0.36 & -0.27$\rightarrow$0.36 & -0.27$\rightarrow$0.36 \\
DIB(6196.0)  &  -0.34 (42)                  & -0.05 (44)                  & -0.15 (27)                  & -0.10 (43)                  & -0.01 (42)                  & 0.18 (37)                   & 0.15 (27)                  & -0.22 (37)                  & 0.81 (47)                  & 0.80 (34)                  & 0.80 (36)                   \\
 &  -0.57$\rightarrow$0.01 & -0.36$\rightarrow$0.25 & -0.56$\rightarrow$0.2 & -0.38$\rightarrow$0.24 & -0.3$\rightarrow$0.33 & -0.01$\rightarrow$0.6 & -0.47$\rightarrow$0.32 & -0.53$\rightarrow$0.1 & 0.26$\rightarrow$0.71 & 0.07$\rightarrow$0.67 & 0.14$\rightarrow$0.7 \\
DIB(6203.1)  &  -0.05 (44)                  & -0.34 (42)                  & -0.15 (27)                  & -0.10 (43)                  & -0.01 (42)                  & 0.18 (37)                   & 0.15 (27)                  & -0.22 (37)                  & -0.05 (44)                  & -0.05 (44)                  & -0.05 (44)                   \\
 &  -0.36$\rightarrow$0.25 & -0.57$\rightarrow$0.01 & -0.56$\rightarrow$0.2 & -0.38$\rightarrow$0.24 & -0.3$\rightarrow$0.33 & -0.01$\rightarrow$0.6 & -0.47$\rightarrow$0.32 & -0.53$\rightarrow$0.1 & -0.36$\rightarrow$0.25 & -0.36$\rightarrow$0.25 & -0.36$\rightarrow$0.25 \\
DIB(6269.9)  &  -0.15 (27)                  & -0.15 (27)                  & -0.34 (42)                  & -0.10 (43)                  & -0.01 (42)                  & 0.18 (37)                   & 0.15 (27)                  & -0.22 (37)                  & -0.15 (27)                  & -0.15 (27)                  & -0.15 (27)                   \\
 &  -0.56$\rightarrow$0.2 & -0.56$\rightarrow$0.2 & -0.57$\rightarrow$0.01 & -0.38$\rightarrow$0.24 & -0.3$\rightarrow$0.33 & -0.01$\rightarrow$0.6 & -0.47$\rightarrow$0.32 & -0.53$\rightarrow$0.1 & -0.56$\rightarrow$0.2 & -0.56$\rightarrow$0.2 & -0.56$\rightarrow$0.2 \\
\hline
\end{tabular}
\end{center}
\end{sidewaystable*}

\setcounter{table}{1}
\begin{sidewaystable*}
\caption{continued.}
\begin{center}
\begin{tabular}{rrrrrrrrrrrr}
\hline

           & DIB         & DIB         & DIB         & DIB         & DIB         & DIB         & DIB         & DIB         & DIB              & DIB              & DIB       \\
           &    (6196.0) &    (6203.1) &    (6269.9) &    (6283.8) &    (6376.1) &    (6439.5) &    (6614.0) &    (6993.1) &    (13180)       &    (15650)       &    (1567) \\
\hline
DIB(6283.8)  &  -0.10 (43)                  & -0.10 (43)                  & -0.10 (43)                  & -0.34 (42)                  & -0.01 (42)                  & 0.18 (37)                   & 0.15 (27)                  & -0.22 (37)                  & -0.10 (43)                  & -0.10 (43)                  & -0.10 (43)                   \\
 &  -0.38$\rightarrow$0.24 & -0.38$\rightarrow$0.24 & -0.38$\rightarrow$0.24 & -0.57$\rightarrow$0.01 & -0.3$\rightarrow$0.33 & -0.01$\rightarrow$0.6 & -0.47$\rightarrow$0.32 & -0.53$\rightarrow$0.1 & -0.38$\rightarrow$0.24 & -0.38$\rightarrow$0.24 & -0.38$\rightarrow$0.24 \\
DIB(6376.1)  &  -0.01 (42)                  & -0.01 (42)                  & -0.01 (42)                  & -0.01 (42)                  & -0.34 (42)                  & 0.18 (37)                   & 0.15 (27)                  & -0.22 (37)                  & -0.01 (42)                  & -0.01 (42)                  & -0.01 (42)                   \\
 &  -0.3$\rightarrow$0.33 & -0.3$\rightarrow$0.33 & -0.3$\rightarrow$0.33 & -0.3$\rightarrow$0.33 & -0.57$\rightarrow$0.01 & -0.01$\rightarrow$0.6 & -0.47$\rightarrow$0.32 & -0.53$\rightarrow$0.1 & -0.3$\rightarrow$0.33 & -0.3$\rightarrow$0.33 & -0.3$\rightarrow$0.33 \\
DIB(6439.5)  &  0.18 (37)                  & 0.18 (37)                  & 0.18 (37)                  & 0.18 (37)                  & 0.18 (37)                  & -0.34 (42)                   & 0.15 (27)                  & -0.22 (37)                  & 0.18 (37)                  & 0.18 (37)                  & 0.18 (37)                   \\
 &  -0.01$\rightarrow$0.6 & -0.01$\rightarrow$0.6 & -0.01$\rightarrow$0.6 & -0.01$\rightarrow$0.6 & -0.01$\rightarrow$0.6 & -0.57$\rightarrow$0.01 & -0.47$\rightarrow$0.32 & -0.53$\rightarrow$0.1 & -0.01$\rightarrow$0.6 & -0.01$\rightarrow$0.6 & -0.01$\rightarrow$0.6 \\
DIB(6614.0)  &  0.15 (27)                  & 0.15 (27)                  & 0.15 (27)                  & 0.15 (27)                  & 0.15 (27)                  & 0.15 (27)                   & -0.34 (42)                  & -0.22 (37)                  & 0.15 (27)                  & 0.15 (27)                  & 0.15 (27)                   \\
 &  -0.47$\rightarrow$0.32 & -0.47$\rightarrow$0.32 & -0.47$\rightarrow$0.32 & -0.47$\rightarrow$0.32 & -0.47$\rightarrow$0.32 & -0.47$\rightarrow$0.32 & -0.57$\rightarrow$0.01 & -0.53$\rightarrow$0.1 & -0.47$\rightarrow$0.32 & -0.47$\rightarrow$0.32 & -0.47$\rightarrow$0.32 \\
DIB(6993.1)  &  -0.22 (37)                  & -0.22 (37)                  & -0.22 (37)                  & -0.22 (37)                  & -0.22 (37)                  & -0.22 (37)                   & -0.22 (37)                  & -0.34 (42)                  & -0.22 (37)                  & -0.22 (37)                  & -0.22 (37)                   \\
 &  -0.53$\rightarrow$0.1 & -0.53$\rightarrow$0.1 & -0.53$\rightarrow$0.1 & -0.53$\rightarrow$0.1 & -0.53$\rightarrow$0.1 & -0.53$\rightarrow$0.1 & -0.53$\rightarrow$0.1 & -0.57$\rightarrow$0.01 & -0.53$\rightarrow$0.1 & -0.53$\rightarrow$0.1 & -0.53$\rightarrow$0.1 \\
KI(7698.9)  &  0.81 (49)                  & 0.78 (52)                  & 0.75 (52)                  & 0.75 (53)                  & 0.71 (40)                  & 0.64 (29)                   & 0.81 (41)                  & 0.65 (44)                  & 0.67 (47)                  & 0.50 (33)                  & 0.52 (35)                   \\
 &  0.25$\rightarrow$0.7 & -0.06$\rightarrow$0.49 & 0.05$\rightarrow$0.57 & -0.05$\rightarrow$0.49 & 0.24$\rightarrow$0.72 & -0.$\rightarrow$0.66 & 0.4$\rightarrow$0.8 & 0.03$\rightarrow$0.59 & -0.32$\rightarrow$0.28 & -0.42$\rightarrow$0.3 & -0.26$\rightarrow$0.44 \\
DIB(13180)  &  0.81 (47)                  & -0.05 (44)                  & -0.15 (27)                  & -0.10 (43)                  & -0.01 (42)                  & 0.18 (37)                   & 0.15 (27)                  & -0.22 (37)                  & -0.34 (42)                  & -0.08 (25)                  & -0.15 (28)                   \\
 &  0.26$\rightarrow$0.71 & -0.36$\rightarrow$0.25 & -0.56$\rightarrow$0.2 & -0.38$\rightarrow$0.24 & -0.3$\rightarrow$0.33 & -0.01$\rightarrow$0.6 & -0.47$\rightarrow$0.32 & -0.53$\rightarrow$0.1 & -0.57$\rightarrow$0.01 & -0.39$\rightarrow$0.44 & -0.26$\rightarrow$0.51 \\
DIB(15650)  &  0.80 (34)                  & -0.05 (44)                  & -0.15 (27)                  & -0.10 (43)                  & -0.01 (42)                  & 0.18 (37)                   & 0.15 (27)                  & -0.22 (37)                  & -0.08 (25)                  & -0.34 (42)                  & -0.15 (28)                   \\
 &  0.07$\rightarrow$0.67 & -0.36$\rightarrow$0.25 & -0.56$\rightarrow$0.2 & -0.38$\rightarrow$0.24 & -0.3$\rightarrow$0.33 & -0.01$\rightarrow$0.6 & -0.47$\rightarrow$0.32 & -0.53$\rightarrow$0.1 & -0.39$\rightarrow$0.44 & -0.57$\rightarrow$0.01 & -0.26$\rightarrow$0.51 \\
DIB(15670)  &  0.80 (36)                  & -0.05 (44)                  & -0.15 (27)                  & -0.10 (43)                  & -0.01 (42)                  & 0.18 (37)                   & 0.15 (27)                  & -0.22 (37)                  & -0.15 (28)                  & -0.15 (28)                  & -0.34 (42)                   \\
 &  0.14$\rightarrow$0.7 & -0.36$\rightarrow$0.25 & -0.56$\rightarrow$0.2 & -0.38$\rightarrow$0.24 & -0.3$\rightarrow$0.33 & -0.01$\rightarrow$0.6 & -0.47$\rightarrow$0.32 & -0.53$\rightarrow$0.1 & -0.26$\rightarrow$0.51 & -0.26$\rightarrow$0.51 & -0.57$\rightarrow$0.01 \\
\hline
\end{tabular}
\end{center}
\end{sidewaystable*}

%
%
\begin{sidewaystable*}
\setcounter{table}{3}
\caption{Correlation coefficients for single cloud dominated sightlines. The number of sightlines used per correlation is shown in brackets. Both the standard pearson correlation coefficient (top row per parameter) and 95 percent confidence range (second row) for the partial correlation coefficients are presented. Blank cells correspond to datasets with less than 5 datapoints.}
\begin{center}
\begin{tabular}{rrrrrrrrrrr}
\hline
           &    E(B-V)           & FeI         & DIB         & DIB         & DIB         & DIB         & DIB     )       & DIB             & DIB         & DIB         \\
           &                  &    (3859.9) &    (5074.5) &    (5418.9) &    (5705.1) &    (5711.6) &    (5780.5)       &    (5797.1)     &    (5849.8) &    (6089.9) \\
\hline
\label{DIB_corr_coeff_SDC}
E(B-V)  &  0.90 (12)                 &                     & 0.84 (6)                  & 0.78 (7)                  & 0.92 (10)                  & 0.59 (7)                   & 0.95 (12)                  & 0.92 (12)                  & 0.84 (10)                  & 0.75 (9)                   \\
 &  0.75$\rightarrow$0.99 &               & -0.8$\rightarrow$0.99 & 0.36$\rightarrow$1. & -0.01$\rightarrow$0.94 & 0.02$\rightarrow$0.99 & 0.44$\rightarrow$0.96 & 0.89$\rightarrow$0.99 & 0.22$\rightarrow$0.96 & -0.38$\rightarrow$0.92 \\
Na(3302.4)  &  0.99 (7)                 & -0.52 (5)                  &                     &                     &                     & 0.35 (9)                   &                     & 0.39 (11)                  & 0.44 (5)                  & 0.41 (12)                   \\
 &  0.36$\rightarrow$1. & -1.$\rightarrow$1. &               &                       &               & -0.77$\rightarrow$0.74 &               & -0.78$\rightarrow$0.51 & -1.$\rightarrow$1. & -0.08$\rightarrow$0.89 \\
FeI(3859.9)  &                     & 0.21 (10)                  &                     &                     &                     & 0.35 (9)                   &                     & 0.39 (11)                  & 0.44 (5)                  & 0.41 (12)                   \\
 &               & -0.92$\rightarrow$0.15 &               &                       &               & -0.77$\rightarrow$0.74 &               & -0.78$\rightarrow$0.51 & -1.$\rightarrow$1. & -0.08$\rightarrow$0.89 \\
CaII(3933.6)  &  0.83 (16)                 &                     & 0.74 (6)                  & 0.46 (7)                  & 0.79 (11)                  & 0.35 (9)                   & 0.73 (13)                  & 0.39 (11)                  & 0.44 (5)                  & 0.41 (12)                   \\
 &  0.18$\rightarrow$0.88 &               & -0.93$\rightarrow$0.98 & -0.9$\rightarrow$0.87 & -0.49$\rightarrow$0.79 & -0.77$\rightarrow$0.74 & -0.6$\rightarrow$0.66 & -0.78$\rightarrow$0.51 & -1.$\rightarrow$1. & -0.08$\rightarrow$0.89 \\
CaI(4226.7)  &  0.80 (13)                 &                     & 0.75 (6)                  & 0.34 (7)                  & 0.82 (9)                  & 0.35 (9)                   & 0.90 (11)                  & 0.39 (11)                  & 0.44 (5)                  & 0.41 (12)                   \\
 &  0.17$\rightarrow$0.92 &               & -0.95$\rightarrow$0.97 & -0.73$\rightarrow$0.95 & 0.05$\rightarrow$0.96 & -0.77$\rightarrow$0.74 & 0.09$\rightarrow$0.93 & -0.78$\rightarrow$0.51 & -1.$\rightarrow$1. & -0.08$\rightarrow$0.89 \\
CH+(4232.5)  &  0.88 (12)                 & 0.10 (13)                  & 0.50 (5)                  & 0.69 (7)                  & 0.73 (10)                  & 0.35 (9)                   & 0.85 (10)                  & 0.39 (11)                  & 0.44 (5)                  & 0.41 (12)                   \\
 &  0.85$\rightarrow$0.99 & -0.65$\rightarrow$0.54 & -1.$\rightarrow$1. & -0.96$\rightarrow$0.65 & -0.88$\rightarrow$0.36 & -0.77$\rightarrow$0.74 & -0.86$\rightarrow$0.59 & -0.78$\rightarrow$0.51 & -1.$\rightarrow$1. & -0.08$\rightarrow$0.89 \\
CH(4300.3)  &  0.94 (12)                 & 0.63 (10)                  & 0.77 (5)                  & 0.80 (7)                  & 0.79 (10)                  & 0.35 (9)                   & 0.83 (12)                  & 0.39 (11)                  & 0.44 (5)                  & 0.41 (12)                   \\
 &  0.81$\rightarrow$0.99 & 0.1$\rightarrow$0.95 & -1.$\rightarrow$1. & -0.96$\rightarrow$0.65 & -0.84$\rightarrow$0.48 & -0.77$\rightarrow$0.74 & -0.86$\rightarrow$0.3 & -0.78$\rightarrow$0.51 & -1.$\rightarrow$1. & -0.08$\rightarrow$0.89 \\
DIB(4726.8)  &  0.80 (12)                 & 0.27 (11)                  & 0.52 (6)                  & 0.77 (7)                  & 0.63 (10)                  & 0.35 (9)                   & 0.56 (10)                  & 0.39 (11)                  & 0.44 (5)                  & 0.41 (12)                   \\
 &  0.46$\rightarrow$0.96 & -0.46$\rightarrow$0.8 & -0.99$\rightarrow$0.87 & -0.96$\rightarrow$0.68 & -0.81$\rightarrow$0.55 & -0.77$\rightarrow$0.74 & -0.88$\rightarrow$0.51 & -0.78$\rightarrow$0.51 & -1.$\rightarrow$1. & -0.08$\rightarrow$0.89 \\
DIB(4984.8)  &  0.87 (7)                 & 0.23 (8)                  &                     & 0.85 (7)                  & 0.79 (6)                  & 0.35 (9)                   & 0.68 (6)                  & 0.39 (11)                  & 0.44 (5)                  & 0.41 (12)                   \\
 &  -0.6$\rightarrow$0.97 & -0.88$\rightarrow$0.7 &               & -0.64$\rightarrow$0.97 & -0.99$\rightarrow$0.9 & -0.77$\rightarrow$0.74 & -0.87$\rightarrow$0.99 & -0.78$\rightarrow$0.51 & -1.$\rightarrow$1. & -0.08$\rightarrow$0.89 \\
DIB(5074.5)  &  0.84 (6)                 &                     & 0.21 (10)                  &                     &                     & 0.35 (9)                   &                     & 0.39 (11)                  & 0.44 (5)                  & 0.41 (12)                   \\
 &  -0.8$\rightarrow$0.99 &               & -0.92$\rightarrow$0.15 &               &                       & -0.77$\rightarrow$0.74 &               & -0.78$\rightarrow$0.51 & -1.$\rightarrow$1. & -0.08$\rightarrow$0.89 \\
DIB(5418.9)  &  0.78 (7)                 &                     &                     & 0.21 (10)                  & 0.70 (6)                  & 0.35 (9)                   & 0.89 (6)                  & 0.39 (11)                  & 0.44 (5)                  & 0.41 (12)                   \\
 &  0.36$\rightarrow$1. &               &                       & -0.92$\rightarrow$0.15 & -0.7$\rightarrow$1. & -0.77$\rightarrow$0.74 & -1.$\rightarrow$-1. & -0.78$\rightarrow$0.51 & -1.$\rightarrow$1. & -0.08$\rightarrow$0.89 \\
DIB(5705.1)  &  0.92 (10)                 &                     &                     & 0.70 (6)                  & 0.21 (10)                  & 0.35 (9)                   & 0.94 (9)                  & 0.39 (11)                  & 0.44 (5)                  & 0.41 (12)                   \\
 &  -0.01$\rightarrow$0.94 &               &                       & -0.7$\rightarrow$1. & -0.92$\rightarrow$0.15 & -0.77$\rightarrow$0.74 & 0.5$\rightarrow$0.99 & -0.78$\rightarrow$0.51 & -1.$\rightarrow$1. & -0.08$\rightarrow$0.89 \\
DIB(5711.6)  &  0.59 (7)                 & 0.35 (9)                  & 0.35 (9)                  & 0.35 (9)                  & 0.35 (9)                  & 0.21 (10)                   & 0.48 (5)                  & 0.39 (11)                  & 0.44 (5)                  & 0.41 (12)                   \\
 &  0.02$\rightarrow$0.99 & -0.77$\rightarrow$0.74 & -0.77$\rightarrow$0.74 & -0.77$\rightarrow$0.74 & -0.77$\rightarrow$0.74 & -0.92$\rightarrow$0.15 & -1.$\rightarrow$1. & -0.78$\rightarrow$0.51 & -1.$\rightarrow$1. & -0.08$\rightarrow$0.89 \\
DIB(5780.5)  &  0.95 (12)                 &                     &                     & 0.89 (6)                  & 0.94 (9)                  & 0.48 (5)                   & 0.21 (10)                  & 0.39 (11)                  & 0.44 (5)                  & 0.41 (12)                   \\
 &  0.44$\rightarrow$0.96 &               &                       & -1.$\rightarrow$-1. & 0.5$\rightarrow$0.99 & -1.$\rightarrow$1. & -0.92$\rightarrow$0.15 & -0.78$\rightarrow$0.51 & -1.$\rightarrow$1. & -0.08$\rightarrow$0.89 \\
DIB(5797.1)  &  0.92 (12)                 & 0.39 (11)                  & 0.39 (11)                  & 0.39 (11)                  & 0.39 (11)                  & 0.39 (11)                   & 0.39 (11)                  & 0.21 (10)                  & 0.44 (5)                  & 0.41 (12)                   \\
 &  0.89$\rightarrow$0.99 & -0.78$\rightarrow$0.51 & -0.78$\rightarrow$0.51 & -0.78$\rightarrow$0.51 & -0.78$\rightarrow$0.51 & -0.78$\rightarrow$0.51 & -0.78$\rightarrow$0.51 & -0.92$\rightarrow$0.15 & -1.$\rightarrow$1. & -0.08$\rightarrow$0.89 \\
DIB(5849.8)  &  0.84 (10)                 & 0.44 (5)                  & 0.44 (5)                  & 0.44 (5)                  & 0.44 (5)                  & 0.44 (5)                   & 0.44 (5)                  & 0.44 (5)                  & 0.21 (10)                  & 0.41 (12)                   \\
 &  0.22$\rightarrow$0.96 & -1.$\rightarrow$1. & -1.$\rightarrow$1. & -1.$\rightarrow$1. & -1.$\rightarrow$1. & -1.$\rightarrow$1. & -1.$\rightarrow$1. & -1.$\rightarrow$1. & -0.92$\rightarrow$0.15 & -0.08$\rightarrow$0.89 \\
DIB(6089.9)  &  0.75 (9)                 & 0.41 (12)                  & 0.41 (12)                  & 0.41 (12)                  & 0.41 (12)                  & 0.41 (12)                   & 0.41 (12)                  & 0.41 (12)                  & 0.41 (12)                  & 0.21 (10)                   \\
 &  -0.38$\rightarrow$0.92 & -0.08$\rightarrow$0.89 & -0.08$\rightarrow$0.89 & -0.08$\rightarrow$0.89 & -0.08$\rightarrow$0.89 & -0.08$\rightarrow$0.89 & -0.08$\rightarrow$0.89 & -0.08$\rightarrow$0.89 & -0.08$\rightarrow$0.89 & -0.92$\rightarrow$0.15 \\
DIB(6196.0)  &  0.93 (16)                 &                     & 0.87 (6)                  & 0.50 (7)                  & 0.91 (10)                  & 0.72 (7)                   & 0.98 (12)                  & 0.97 (12)                  & 0.91 (10)                  & 0.92 (9)                   \\
 &  0.66$\rightarrow$0.96 &               & -0.97$\rightarrow$0.95 & -0.73$\rightarrow$0.95 & -0.15$\rightarrow$0.92 & -1.$\rightarrow$-1. & 0.45$\rightarrow$0.96 & 0.01$\rightarrow$0.9 & -0.23$\rightarrow$0.91 & 0.66$\rightarrow$0.99 \\
DIB(6203.1)  &  0.96 (15)                 & 0.28 (13)                  & 0.28 (13)                  & 0.28 (13)                  & 0.28 (13)                  & 0.28 (13)                   & 0.28 (13)                  & 0.28 (13)                  & 0.28 (13)                  & 0.28 (13)                   \\
 &  0.9$\rightarrow$0.99 & -0.61$\rightarrow$0.58 & -0.61$\rightarrow$0.58 & -0.61$\rightarrow$0.58 & -0.61$\rightarrow$0.58 & -0.61$\rightarrow$0.58 & -0.61$\rightarrow$0.58 & -0.61$\rightarrow$0.58 & -0.61$\rightarrow$0.58 & -0.61$\rightarrow$0.58 \\
DIB(6269.9)  &  0.90 (13)                 & 0.30 (7)                  & 0.30 (7)                  & 0.30 (7)                  & 0.30 (7)                  & 0.30 (7)                   & 0.30 (7)                  & 0.30 (7)                  & 0.30 (7)                  & 0.30 (7)                   \\
 &  0.56$\rightarrow$0.97 & -0.83$\rightarrow$0.92 & -0.83$\rightarrow$0.92 & -0.83$\rightarrow$0.92 & -0.83$\rightarrow$0.92 & -0.83$\rightarrow$0.92 & -0.83$\rightarrow$0.92 & -0.83$\rightarrow$0.92 & -0.83$\rightarrow$0.92 & -0.83$\rightarrow$0.92 \\
\hline
\end{tabular}
\end{center}
\end{sidewaystable*}
\setcounter{table}{3}

\begin{sidewaystable*}
\caption{continued. }
\begin{center}
\begin{tabular}{rrrrrrrrrrr}
\hline
           &    E(B-V)           & FeI         & DIB         & DIB         & DIB         & DIB         & DIB     )       & DIB             & DIB         & DIB         \\
           &                  &    (3859.9) &    (5074.5) &    (5418.9) &    (5705.1) &    (5711.6) &    (5780.5)       &    (5797.1)     &    (5849.8) &    (6089.9) \\
\hline
DIB(6283.8)  &  0.92 (12)                 & 0.65 (11)                  & 0.65 (11)                  & 0.65 (11)                  & 0.65 (11)                  & 0.65 (11)                   & 0.65 (11)                  & 0.65 (11)                  & 0.65 (11)                  & 0.65 (11)                   \\
 &  0.39$\rightarrow$0.96 & -0.31$\rightarrow$0.86 & -0.31$\rightarrow$0.86 & -0.31$\rightarrow$0.86 & -0.31$\rightarrow$0.86 & -0.31$\rightarrow$0.86 & -0.31$\rightarrow$0.86 & -0.31$\rightarrow$0.86 & -0.31$\rightarrow$0.86 & -0.31$\rightarrow$0.86 \\
DIB(6376.1)  &  0.88 (8)                 & 0.52 (10)                  & 0.52 (10)                  & 0.52 (10)                  & 0.52 (10)                  & 0.52 (10)                   & 0.52 (10)                  & 0.52 (10)                  & 0.52 (10)                  & 0.52 (10)                   \\
 &  0.15$\rightarrow$0.98 & -0.36$\rightarrow$0.88 & -0.36$\rightarrow$0.88 & -0.36$\rightarrow$0.88 & -0.36$\rightarrow$0.88 & -0.36$\rightarrow$0.88 & -0.36$\rightarrow$0.88 & -0.36$\rightarrow$0.88 & -0.36$\rightarrow$0.88 & -0.36$\rightarrow$0.88 \\
DIB(6439.5)  &  0.81 (5)                 & 0.37 (10)                  & 0.37 (10)                  & 0.37 (10)                  & 0.37 (10)                  & 0.37 (10)                   & 0.37 (10)                  & 0.37 (10)                  & 0.37 (10)                  & 0.37 (10)                   \\
 &  -1.$\rightarrow$1. & -0.56$\rightarrow$0.81 & -0.56$\rightarrow$0.81 & -0.56$\rightarrow$0.81 & -0.56$\rightarrow$0.81 & -0.56$\rightarrow$0.81 & -0.56$\rightarrow$0.81 & -0.56$\rightarrow$0.81 & -0.56$\rightarrow$0.81 & -0.56$\rightarrow$0.81 \\
DIB(6614.0)  &  0.92 (12)                 &                     & 0.92 (5)                  & 0.55 (7)                  & 0.90 (8)                  & 0.66 (6)                   & 0.97 (9)                  & 0.96 (9)                  & 0.96 (9)                  & 0.88 (7)                   \\
 &  0.57$\rightarrow$0.97 &               & -1.$\rightarrow$1. & -0.73$\rightarrow$0.95 & -0.12$\rightarrow$0.97 & -1.$\rightarrow$-1. & 0.54$\rightarrow$0.99 & -0.39$\rightarrow$0.91 & 0.15$\rightarrow$0.97 & 0.37$\rightarrow$1. \\
DIB(6993.1)  &  0.96 (8)                 & 0.21 (10)                  & 0.21 (10)                  & 0.21 (10)                  & 0.21 (10)                  & 0.21 (10)                   & 0.21 (10)                  & 0.21 (10)                  & 0.21 (10)                  & 0.21 (10)                   \\
 &  0.3$\rightarrow$0.99 & -0.65$\rightarrow$0.75 & -0.65$\rightarrow$0.75 & -0.65$\rightarrow$0.75 & -0.65$\rightarrow$0.75 & -0.65$\rightarrow$0.75 & -0.65$\rightarrow$0.75 & -0.65$\rightarrow$0.75 & -0.65$\rightarrow$0.75 & -0.65$\rightarrow$0.75 \\
KI(7698.9)  &  0.90 (12)                 &                     &                     & 0.79 (6)                  & 0.89 (11)                  & 0.58 (5)                   & 0.85 (10)                  & 0.95 (10)                  & 0.98 (9)                  & 0.66 (10)                   \\
 &  0.75$\rightarrow$0.99 &               &                       & -1.$\rightarrow$-1. & -0.3$\rightarrow$0.86 & -1.$\rightarrow$1. & -0.53$\rightarrow$0.82 & -0.29$\rightarrow$0.9 & -0.03$\rightarrow$0.96 & -0.66$\rightarrow$0.74 \\
DIB(13180)  &  0.80 (15)                 & -0.52 (5)                  &                     & 0.69 (5)                  & 0.84 (9)                  & 0.35 (9)                   & 0.90 (8)                  & 0.39 (11)                  & 0.44 (5)                  & 0.41 (12)                   \\
 &  -0.14$\rightarrow$0.87 & -1.$\rightarrow$1. &               & -1.$\rightarrow$1. & -0.21$\rightarrow$0.94 & -0.77$\rightarrow$0.74 & -0.08$\rightarrow$0.97 & -0.78$\rightarrow$0.51 & -1.$\rightarrow$1. & -0.08$\rightarrow$0.89 \\
DIB(15650)  &  0.96 (5)                 & -0.52 (5)                  &                     &                     & 0.84 (5)                  & 0.35 (9)                   &                     & 0.39 (11)                  & 0.44 (5)                  & 0.41 (12)                   \\
 &  -1.$\rightarrow$1. & -1.$\rightarrow$1. &               &                       & -1.$\rightarrow$1. & -0.77$\rightarrow$0.74 &               & -0.78$\rightarrow$0.51 & -1.$\rightarrow$1. & -0.08$\rightarrow$0.89 \\
DIB(15670)  &  0.88 (8)                 & -0.52 (5)                  &                     &                     & 0.87 (6)                  & 0.35 (9)                   & 0.86 (6)                  & 0.39 (11)                  & 0.44 (5)                  & 0.41 (12)                   \\
 &  0.11$\rightarrow$0.98 & -1.$\rightarrow$1. &               &                       & -0.7$\rightarrow$1. & -0.77$\rightarrow$0.74 & -0.99$\rightarrow$0.86 & -0.78$\rightarrow$0.51 & -1.$\rightarrow$1. & -0.08$\rightarrow$0.89 \\
\hline
\end{tabular}
\end{center}
\end{sidewaystable*}
%
%
\setcounter{table}{3}
\begin{sidewaystable*}
\caption{continued.}
\begin{center}
\begin{tabular}{rrrrrrrrrrrr}
\hline
           & DIB         & DIB         & DIB         & DIB         & DIB         & DIB         & DIB         & DIB         & DIB              & DIB              & DIB       \\
           &    (6196.0) &    (6203.1) &    (6269.9) &    (6283.8) &    (6376.1) &    (6439.5) &    (6614.0) &    (6993.1) &    (13180)       &    (15650)       &    (1567) \\
\hline
E(B-V)  &  0.93 (16)                  & 0.96 (15)                  & 0.90 (13)                  & 0.92 (12)                  & 0.88 (8)                  & 0.81 (5)                   & 0.92 (12)                  & 0.96 (8)                  & 0.80 (15)                  & 0.96 (5)                  & 0.88 (8)                   \\
 &  0.66$\rightarrow$0.96 & 0.9$\rightarrow$0.99 & 0.56$\rightarrow$0.97 & 0.39$\rightarrow$0.96 & 0.15$\rightarrow$0.98 & -1.$\rightarrow$1. & 0.57$\rightarrow$0.97 & 0.3$\rightarrow$0.99 & -0.14$\rightarrow$0.87 & -1.$\rightarrow$1. & 0.11$\rightarrow$0.98 \\
Na(3302.4)  &  0.89 (7)                  & 0.28 (13)                  & 0.30 (7)                  & 0.65 (11)                  & 0.52 (10)                  & 0.37 (10)                   & 0.90 (6)                  & 0.21 (10)                  & 0.12 (7)                  & 0.12 (7)                  & 0.12 (7)                   \\
 &  -0.94$\rightarrow$0.79 & -0.61$\rightarrow$0.58 & -0.83$\rightarrow$0.92 & -0.31$\rightarrow$0.86 & -0.36$\rightarrow$0.88 & -0.56$\rightarrow$0.81 & -0.93$\rightarrow$0.98 & -0.65$\rightarrow$0.75 & -0.95$\rightarrow$0.73 & -0.95$\rightarrow$0.73 & -0.95$\rightarrow$0.73 \\
FeI(3859.9)  &                      & 0.28 (13)                  & 0.30 (7)                  & 0.65 (11)                  & 0.52 (10)                  & 0.37 (10)                   &                     & 0.21 (10)                  & -0.52 (5)                  & -0.52 (5)                  & -0.52 (5)                   \\
 &               & -0.61$\rightarrow$0.58 & -0.83$\rightarrow$0.92 & -0.31$\rightarrow$0.86 & -0.36$\rightarrow$0.88 & -0.56$\rightarrow$0.81 &               & -0.65$\rightarrow$0.75 & -1.$\rightarrow$1. & -1.$\rightarrow$1. & -1.$\rightarrow$1. \\
CaII(3933.6)  &  0.83 (16)                  & 0.28 (13)                  & 0.30 (7)                  & 0.65 (11)                  & 0.52 (10)                  & 0.37 (10)                   & 0.81 (12)                  & 0.21 (10)                  & 0.83 (10)                  & 0.74 (5)                  & 0.55 (7)                   \\
 &  -0.41$\rightarrow$0.63 & -0.61$\rightarrow$0.58 & -0.83$\rightarrow$0.92 & -0.31$\rightarrow$0.86 & -0.36$\rightarrow$0.88 & -0.56$\rightarrow$0.81 & -0.5$\rightarrow$0.73 & -0.65$\rightarrow$0.75 & -0.58$\rightarrow$0.8 & -1.$\rightarrow$1. & -0.9$\rightarrow$0.86 \\
CaI(4226.7)  &  0.91 (13)                  & 0.28 (13)                  & 0.30 (7)                  & 0.65 (11)                  & 0.52 (10)                  & 0.37 (10)                   & 0.92 (10)                  & 0.21 (10)                  & 0.52 (9)                  & 0.59 (5)                  & 0.69 (7)                   \\
 &  0.25$\rightarrow$0.93 & -0.61$\rightarrow$0.58 & -0.83$\rightarrow$0.92 & -0.31$\rightarrow$0.86 & -0.36$\rightarrow$0.88 & -0.56$\rightarrow$0.81 & -0.16$\rightarrow$0.92 & -0.65$\rightarrow$0.75 & -0.74$\rightarrow$0.76 & -1.$\rightarrow$1. & -0.93$\rightarrow$0.8 \\
CH+(4232.5)  &  0.77 (12)                  & 0.28 (13)                  & 0.30 (7)                  & 0.65 (11)                  & 0.52 (10)                  & 0.37 (10)                   & 0.74 (11)                  & 0.21 (10)                  & 0.10 (13)                  & 0.10 (13)                  & 0.10 (13)                   \\
 &  -0.6$\rightarrow$0.65 & -0.61$\rightarrow$0.58 & -0.83$\rightarrow$0.92 & -0.31$\rightarrow$0.86 & -0.36$\rightarrow$0.88 & -0.56$\rightarrow$0.81 & -0.68$\rightarrow$0.64 & -0.65$\rightarrow$0.75 & -0.65$\rightarrow$0.54 & -0.65$\rightarrow$0.54 & -0.65$\rightarrow$0.54 \\
CH(4300.3)  &  0.80 (12)                  & 0.28 (13)                  & 0.30 (7)                  & 0.65 (11)                  & 0.52 (10)                  & 0.37 (10)                   & 0.81 (10)                  & 0.21 (10)                  & 0.63 (10)                  & 0.63 (10)                  & 0.63 (10)                   \\
 &  -0.83$\rightarrow$0.28 & -0.61$\rightarrow$0.58 & -0.83$\rightarrow$0.92 & -0.31$\rightarrow$0.86 & -0.36$\rightarrow$0.88 & -0.56$\rightarrow$0.81 & -0.86$\rightarrow$0.44 & -0.65$\rightarrow$0.75 & 0.1$\rightarrow$0.95 & 0.1$\rightarrow$0.95 & 0.1$\rightarrow$0.95 \\
DIB(4726.8)  &  0.66 (12)                  & 0.28 (13)                  & 0.30 (7)                  & 0.65 (11)                  & 0.52 (10)                  & 0.37 (10)                   & 0.61 (10)                  & 0.21 (10)                  & 0.27 (11)                  & 0.27 (11)                  & 0.27 (11)                   \\
 &  -0.82$\rightarrow$0.32 & -0.61$\rightarrow$0.58 & -0.83$\rightarrow$0.92 & -0.31$\rightarrow$0.86 & -0.36$\rightarrow$0.88 & -0.56$\rightarrow$0.81 & -0.83$\rightarrow$0.51 & -0.65$\rightarrow$0.75 & -0.46$\rightarrow$0.8 & -0.46$\rightarrow$0.8 & -0.46$\rightarrow$0.8 \\
DIB(4984.8)  &  0.53 (7)                  & 0.28 (13)                  & 0.30 (7)                  & 0.65 (11)                  & 0.52 (10)                  & 0.37 (10)                   & 0.56 (7)                  & 0.21 (10)                  & 0.23 (8)                  & 0.23 (8)                  & 0.23 (8)                   \\
 &  -0.94$\rightarrow$0.79 & -0.61$\rightarrow$0.58 & -0.83$\rightarrow$0.92 & -0.31$\rightarrow$0.86 & -0.36$\rightarrow$0.88 & -0.56$\rightarrow$0.81 & -0.94$\rightarrow$0.79 & -0.65$\rightarrow$0.75 & -0.88$\rightarrow$0.7 & -0.88$\rightarrow$0.7 & -0.88$\rightarrow$0.7 \\
DIB(5074.5)  &  0.87 (6)                  & 0.28 (13)                  & 0.30 (7)                  & 0.65 (11)                  & 0.52 (10)                  & 0.37 (10)                   & 0.92 (5)                  & 0.21 (10)                  &                     &                     &                      \\
 &  -0.97$\rightarrow$0.95 & -0.61$\rightarrow$0.58 & -0.83$\rightarrow$0.92 & -0.31$\rightarrow$0.86 & -0.36$\rightarrow$0.88 & -0.56$\rightarrow$0.81 & -1.$\rightarrow$1. & -0.65$\rightarrow$0.75 &               &                       &       \\
DIB(5418.9)  &  0.50 (7)                  & 0.28 (13)                  & 0.30 (7)                  & 0.65 (11)                  & 0.52 (10)                  & 0.37 (10)                   & 0.55 (7)                  & 0.21 (10)                  & 0.69 (5)                  &                     &                      \\
 &  -0.73$\rightarrow$0.95 & -0.61$\rightarrow$0.58 & -0.83$\rightarrow$0.92 & -0.31$\rightarrow$0.86 & -0.36$\rightarrow$0.88 & -0.56$\rightarrow$0.81 & -0.73$\rightarrow$0.95 & -0.65$\rightarrow$0.75 & -1.$\rightarrow$1. &               &       \\
DIB(5705.1)  &  0.91 (10)                  & 0.28 (13)                  & 0.30 (7)                  & 0.65 (11)                  & 0.52 (10)                  & 0.37 (10)                   & 0.90 (8)                  & 0.21 (10)                  & 0.84 (9)                  & 0.84 (5)                  & 0.87 (6)                   \\
 &  -0.15$\rightarrow$0.92 & -0.61$\rightarrow$0.58 & -0.83$\rightarrow$0.92 & -0.31$\rightarrow$0.86 & -0.36$\rightarrow$0.88 & -0.56$\rightarrow$0.81 & -0.12$\rightarrow$0.97 & -0.65$\rightarrow$0.75 & -0.21$\rightarrow$0.94 & -1.$\rightarrow$1. & -0.7$\rightarrow$1. \\
DIB(5711.6)  &  0.72 (7)                  & 0.28 (13)                  & 0.30 (7)                  & 0.65 (11)                  & 0.52 (10)                  & 0.37 (10)                   & 0.66 (6)                  & 0.21 (10)                  & 0.35 (9)                  & 0.35 (9)                  & 0.35 (9)                   \\
 &  -1.$\rightarrow$-1. & -0.61$\rightarrow$0.58 & -0.83$\rightarrow$0.92 & -0.31$\rightarrow$0.86 & -0.36$\rightarrow$0.88 & -0.56$\rightarrow$0.81 & -1.$\rightarrow$-1. & -0.65$\rightarrow$0.75 & -0.77$\rightarrow$0.74 & -0.77$\rightarrow$0.74 & -0.77$\rightarrow$0.74 \\
DIB(5780.5)  &  0.98 (12)                  & 0.28 (13)                  & 0.30 (7)                  & 0.65 (11)                  & 0.52 (10)                  & 0.37 (10)                   & 0.97 (9)                  & 0.21 (10)                  & 0.90 (8)                  &                     & 0.86 (6)                   \\
 &  0.45$\rightarrow$0.96 & -0.61$\rightarrow$0.58 & -0.83$\rightarrow$0.92 & -0.31$\rightarrow$0.86 & -0.36$\rightarrow$0.88 & -0.56$\rightarrow$0.81 & 0.54$\rightarrow$0.99 & -0.65$\rightarrow$0.75 & -0.08$\rightarrow$0.97 &               & -0.99$\rightarrow$0.86 \\
DIB(5797.1)  &  0.97 (12)                  & 0.28 (13)                  & 0.30 (7)                  & 0.65 (11)                  & 0.52 (10)                  & 0.37 (10)                   & 0.96 (9)                  & 0.21 (10)                  & 0.39 (11)                  & 0.39 (11)                  & 0.39 (11)                   \\
 &  0.01$\rightarrow$0.9 & -0.61$\rightarrow$0.58 & -0.83$\rightarrow$0.92 & -0.31$\rightarrow$0.86 & -0.36$\rightarrow$0.88 & -0.56$\rightarrow$0.81 & -0.39$\rightarrow$0.91 & -0.65$\rightarrow$0.75 & -0.78$\rightarrow$0.51 & -0.78$\rightarrow$0.51 & -0.78$\rightarrow$0.51 \\
DIB(5849.8)  &  0.91 (10)                  & 0.28 (13)                  & 0.30 (7)                  & 0.65 (11)                  & 0.52 (10)                  & 0.37 (10)                   & 0.96 (9)                  & 0.21 (10)                  & 0.44 (5)                  & 0.44 (5)                  & 0.44 (5)                   \\
 &  -0.23$\rightarrow$0.91 & -0.61$\rightarrow$0.58 & -0.83$\rightarrow$0.92 & -0.31$\rightarrow$0.86 & -0.36$\rightarrow$0.88 & -0.56$\rightarrow$0.81 & 0.15$\rightarrow$0.97 & -0.65$\rightarrow$0.75 & -1.$\rightarrow$1. & -1.$\rightarrow$1. & -1.$\rightarrow$1. \\
DIB(6089.9)  &  0.92 (9)                  & 0.28 (13)                  & 0.30 (7)                  & 0.65 (11)                  & 0.52 (10)                  & 0.37 (10)                   & 0.88 (7)                  & 0.21 (10)                  & 0.41 (12)                  & 0.41 (12)                  & 0.41 (12)                   \\
 &  0.66$\rightarrow$0.99 & -0.61$\rightarrow$0.58 & -0.83$\rightarrow$0.92 & -0.31$\rightarrow$0.86 & -0.36$\rightarrow$0.88 & -0.56$\rightarrow$0.81 & 0.37$\rightarrow$1. & -0.65$\rightarrow$0.75 & -0.08$\rightarrow$0.89 & -0.08$\rightarrow$0.89 & -0.08$\rightarrow$0.89 \\
DIB(6196.0)  &  0.21 (10)                  & 0.28 (13)                  & 0.30 (7)                  & 0.65 (11)                  & 0.52 (10)                  & 0.37 (10)                   & 0.99 (12)                  & 0.21 (10)                  & 0.73 (10)                  & 0.83 (5)                  & 0.82 (7)                   \\
 &  -0.92$\rightarrow$0.15 & -0.61$\rightarrow$0.58 & -0.83$\rightarrow$0.92 & -0.31$\rightarrow$0.86 & -0.36$\rightarrow$0.88 & -0.56$\rightarrow$0.81 & 0.44$\rightarrow$0.96 & -0.65$\rightarrow$0.75 & -0.53$\rightarrow$0.82 & -1.$\rightarrow$1. & -0.93$\rightarrow$0.79 \\
DIB(6203.1)  &  0.28 (13)                  & 0.21 (10)                  & 0.30 (7)                  & 0.65 (11)                  & 0.52 (10)                  & 0.37 (10)                   & 0.94 (11)                  & 0.21 (10)                  & 0.28 (13)                  & 0.28 (13)                  & 0.28 (13)                   \\
 &  -0.61$\rightarrow$0.58 & -0.92$\rightarrow$0.15 & -0.83$\rightarrow$0.92 & -0.31$\rightarrow$0.86 & -0.36$\rightarrow$0.88 & -0.56$\rightarrow$0.81 & 0.56$\rightarrow$0.98 & -0.65$\rightarrow$0.75 & -0.61$\rightarrow$0.58 & -0.61$\rightarrow$0.58 & -0.61$\rightarrow$0.58 \\
DIB(6269.9)  &  0.30 (7)                  & 0.30 (7)                  & 0.21 (10)                  & 0.65 (11)                  & 0.52 (10)                  & 0.37 (10)                   & 0.94 (11)                  & 0.21 (10)                  & 0.30 (7)                  & 0.30 (7)                  & 0.30 (7)                   \\
 &  -0.83$\rightarrow$0.92 & -0.83$\rightarrow$0.92 & -0.92$\rightarrow$0.15 & -0.31$\rightarrow$0.86 & -0.36$\rightarrow$0.88 & -0.56$\rightarrow$0.81 & 0.38$\rightarrow$0.96 & -0.65$\rightarrow$0.75 & -0.83$\rightarrow$0.92 & -0.83$\rightarrow$0.92 & -0.83$\rightarrow$0.92 \\
\hline
\end{tabular}
\end{center}
\end{sidewaystable*}

\setcounter{table}{2}
\begin{sidewaystable*}
\caption{continued.}
\begin{center}
\begin{tabular}{rrrrrrrrrrrr}
\hline

           & DIB         & DIB         & DIB         & DIB         & DIB         & DIB         & DIB         & DIB         & DIB              & DIB              & DIB       \\
           &    (6196.0) &    (6203.1) &    (6269.9) &    (6283.8) &    (6376.1) &    (6439.5) &    (6614.0) &    (6993.1) &    (13180)       &    (15650)       &    (1567) \\
\hline
DIB(6283.8)  &  0.65 (11)                  & 0.65 (11)                  & 0.65 (11)                  & 0.21 (10)                  & 0.52 (10)                  & 0.37 (10)                   & 0.92 (10)                  & 0.21 (10)                  & 0.65 (11)                  & 0.65 (11)                  & 0.65 (11)                   \\
 &  -0.31$\rightarrow$0.86 & -0.31$\rightarrow$0.86 & -0.31$\rightarrow$0.86 & -0.92$\rightarrow$0.15 & -0.36$\rightarrow$0.88 & -0.56$\rightarrow$0.81 & 0.69$\rightarrow$0.99 & -0.65$\rightarrow$0.75 & -0.31$\rightarrow$0.86 & -0.31$\rightarrow$0.86 & -0.31$\rightarrow$0.86 \\
DIB(6376.1)  &  0.52 (10)                  & 0.52 (10)                  & 0.52 (10)                  & 0.52 (10)                  & 0.21 (10)                  & 0.37 (10)                   & 0.98 (7)                  & 0.21 (10)                  & 0.52 (10)                  & 0.52 (10)                  & 0.52 (10)                   \\
 &  -0.36$\rightarrow$0.88 & -0.36$\rightarrow$0.88 & -0.36$\rightarrow$0.88 & -0.36$\rightarrow$0.88 & -0.92$\rightarrow$0.15 & -0.56$\rightarrow$0.81 & -0.27$\rightarrow$0.99 & -0.65$\rightarrow$0.75 & -0.36$\rightarrow$0.88 & -0.36$\rightarrow$0.88 & -0.36$\rightarrow$0.88 \\
DIB(6439.5)  &  0.37 (10)                  & 0.37 (10)                  & 0.37 (10)                  & 0.37 (10)                  & 0.37 (10)                  & 0.21 (10)                   &                     & 0.21 (10)                  & 0.37 (10)                  & 0.37 (10)                  & 0.37 (10)                   \\
 &  -0.56$\rightarrow$0.81 & -0.56$\rightarrow$0.81 & -0.56$\rightarrow$0.81 & -0.56$\rightarrow$0.81 & -0.56$\rightarrow$0.81 & -0.92$\rightarrow$0.15 &               & -0.65$\rightarrow$0.75 & -0.56$\rightarrow$0.81 & -0.56$\rightarrow$0.81 & -0.56$\rightarrow$0.81 \\
DIB(6614.0)  &  0.99 (12)                  & 0.94 (11)                  & 0.94 (11)                  & 0.92 (10)                  & 0.98 (7)                  &                      & 0.21 (10)                  & 0.21 (10)                  & 0.71 (8)                  & 0.80 (5)                  & 0.82 (6)                   \\
 &  0.44$\rightarrow$0.96 & 0.56$\rightarrow$0.98 & 0.38$\rightarrow$0.96 & 0.69$\rightarrow$0.99 & -0.27$\rightarrow$0.99 &               & -0.92$\rightarrow$0.15 & -0.65$\rightarrow$0.75 & -0.69$\rightarrow$0.89 & -1.$\rightarrow$1. & -0.7$\rightarrow$1. \\
DIB(6993.1)  &  0.21 (10)                  & 0.21 (10)                  & 0.21 (10)                  & 0.21 (10)                  & 0.21 (10)                  & 0.21 (10)                   & 0.21 (10)                  & 0.21 (10)                  & 0.21 (10)                  & 0.21 (10)                  & 0.21 (10)                   \\
 &  -0.65$\rightarrow$0.75 & -0.65$\rightarrow$0.75 & -0.65$\rightarrow$0.75 & -0.65$\rightarrow$0.75 & -0.65$\rightarrow$0.75 & -0.65$\rightarrow$0.75 & -0.65$\rightarrow$0.75 & -0.92$\rightarrow$0.15 & -0.65$\rightarrow$0.75 & -0.65$\rightarrow$0.75 & -0.65$\rightarrow$0.75 \\
KI(7698.9)  &  0.91 (12)                  & 0.87 (13)                  & 0.88 (13)                  & 0.83 (13)                  & 0.91 (9)                  & 0.58 (6)                   & 0.94 (10)                  & 0.96 (9)                  & 0.65 (10)                  & 0.66 (5)                  & 0.61 (7)                   \\
 &  -0.6$\rightarrow$0.66 & -0.62$\rightarrow$0.58 & -0.7$\rightarrow$0.48 & -0.63$\rightarrow$0.57 & -0.62$\rightarrow$0.85 & -0.85$\rightarrow$0.99 & -0.59$\rightarrow$0.79 & -0.51$\rightarrow$0.88 & -0.78$\rightarrow$0.61 & -1.$\rightarrow$1. & -0.94$\rightarrow$0.77 \\
DIB(13180)  &  0.73 (10)                  & 0.28 (13)                  & 0.30 (7)                  & 0.65 (11)                  & 0.52 (10)                  & 0.37 (10)                   & 0.71 (8)                  & 0.21 (10)                  & 0.21 (10)                  & 0.29 (5)                  & 0.26 (5)                   \\
 &  -0.53$\rightarrow$0.82 & -0.61$\rightarrow$0.58 & -0.83$\rightarrow$0.92 & -0.31$\rightarrow$0.86 & -0.36$\rightarrow$0.88 & -0.56$\rightarrow$0.81 & -0.69$\rightarrow$0.89 & -0.65$\rightarrow$0.75 & -0.92$\rightarrow$0.15 & -1.$\rightarrow$1. & -1.$\rightarrow$1. \\
DIB(15650)  &  0.83 (5)                  & 0.28 (13)                  & 0.30 (7)                  & 0.65 (11)                  & 0.52 (10)                  & 0.37 (10)                   & 0.80 (5)                  & 0.21 (10)                  & 0.29 (5)                  & 0.21 (10)                  & 0.26 (5)                   \\
 &  -1.$\rightarrow$1. & -0.61$\rightarrow$0.58 & -0.83$\rightarrow$0.92 & -0.31$\rightarrow$0.86 & -0.36$\rightarrow$0.88 & -0.56$\rightarrow$0.81 & -1.$\rightarrow$1. & -0.65$\rightarrow$0.75 & -1.$\rightarrow$1. & -0.92$\rightarrow$0.15 & -1.$\rightarrow$1. \\
DIB(15670)  &  0.82 (7)                  & 0.28 (13)                  & 0.30 (7)                  & 0.65 (11)                  & 0.52 (10)                  & 0.37 (10)                   & 0.82 (6)                  & 0.21 (10)                  & 0.26 (5)                  & 0.26 (5)                  & 0.21 (10)                   \\
 &  -0.93$\rightarrow$0.79 & -0.61$\rightarrow$0.58 & -0.83$\rightarrow$0.92 & -0.31$\rightarrow$0.86 & -0.36$\rightarrow$0.88 & -0.56$\rightarrow$0.81 & -0.7$\rightarrow$1. & -0.65$\rightarrow$0.75 & -1.$\rightarrow$1. & -1.$\rightarrow$1. & -0.92$\rightarrow$0.15 \\
\hline
\end{tabular}
\end{center}
\end{sidewaystable*}





\end{document}